\documentclass[longauth]{aa}  

\usepackage{graphicx}
\usepackage{txfonts}
\usepackage{natbib}
\usepackage{longtable}
\usepackage{multirow}
\usepackage{color}
\usepackage{url}
\usepackage{mathtools}
\usepackage{amsmath}
\usepackage{soul}

\usepackage[pdfpagemode={UseOutlines},bookmarks=true,bookmarksopen=true,
   bookmarksopenlevel=0,bookmarksnumbered=true,hypertexnames=false,
   colorlinks,linkcolor={red},citecolor={blue},urlcolor={red},
   pdfstartview={FitV},unicode,breaklinks=true]{hyperref}
\usepackage[switch]{lineno}

\def\phcms{photons~cm$^{-2}$~s$^{-1}$}

\newcommand{\myemail}{angioni@mpifr-bonn.mpg.de}

\begin{document}

   \title{Gamma-ray emission in radio galaxies under the VLBI scope}
   \subtitle{I. Parsec-scale jet kinematics and high-energy properties of $\gamma$-ray-detected TANAMI radio galaxies}

   \author{R.~Angioni
          \inst{1,2}
	  \and
          E.~Ros
          \inst{1}
          \and
          M.~Kadler
          \inst{2}
          \and
          R.~Ojha
          \inst{3,4,5}
          \and
          C.~M\"uller
          \inst{6,1}
          \and
          P.~G.~Edwards
          \inst{7}
          \and
          P.~R.~Burd
          \inst{2}
          \and
          B.~Carpenter
          \inst{3,4,5}
          \and
          M.~S.~Dutka
          \inst{3,8}
          \and
          S.~Gulyaev
          \inst{9}
          \and
          H.~Hase
          \inst{10}
          \and
          S.~Horiuchi
          \inst{11}
          \and
          F.~Krau\ss
          \inst{12}
          \and
          J.~E.~J.~Lovell
          \inst{13}
          \and
          T.~Natusch
          \inst{9}
          \and
          C.~Phillips
          \inst{7}
          \and
          C.~Pl\"otz
          \inst{10}
          \and
          J.~F.~H.~Quick
          \inst{14}
          \and
          F.~R\"osch
          \inst{2}
          \and
          R.~Schulz
          \inst{15,2,16}
          \and
          J.~Stevens
          \inst{7}
          \and
          A.~K.~Tzioumis
          \inst{7}
          \and
          S.~Weston
          \inst{9}
          \and
          J.~Wilms
          \inst{16}
          \and
          J.~A.~Zensus
          \inst{1}
                      }

   \institute{Max-Planck-Institut f\"ur Radioastronomie, Auf dem H\"ugel 69, 53121 Bonn, Germany, email: \myemail
   \and
   Institut f\"ur Theoretische Physik und Astrophysik, Universit\"at
   W\"urzburg, Emil-Fischer-Str. 31, 97074 W\"urzburg, Germany
  \and
   NASA Goddard Space Flight Center, Greenbelt, MD 20771, USA
   \and
   Catholic University of America, Washington, DC 20064, USA
   \and
   University of Maryland, Baltimore County, 1000 Hilltop Cir, Baltimore, MD 21250, USA
   \and
   Department of Astrophysics/IMAPP, Radboud University Nijmegen, PO Box 9010, 6500 GL Nijmegen, the Netherlands
   \and
   CSIRO Astronomy and Space Science, PO Box 76, Epping, NSW 1710, Australia
   \and
   Wyle  Science,  Technology  and  Engineering  Group,  Greenbelt, MD 20771, USA
   \and
   Institute for Radio Astronomy \& Space Research, AUT University, 1010 Auckland, New Zealand
   \and
   Bundesamt  f\"ur  Kartographie  und  Geod\"asie,  93444  Bad  K\"otzting, Germany
   \and
   CSIRO  Astronomy  and  Space  Science,  Canberra  Deep  Space Communications Complex, PO Box 1035, Tuggeranong, ACT 2901, Australia
   \and
   GRAPPA \& Anton Pannekoek Institute for Astronomy, University of Amsterdam, Science Park 904, 1098 XH Amsterdam, The Netherlands
   \and
   School of Mathematics \& Physics, University of Tasmania, Private Bag 37, Hobart, 7001 Tasmania, Australia
   \and
   Hartebeesthoek Radio Astronomy Observatory, PO Box 443, 1740 Krugersdorp, South Africa
   \and
   ASTRON,  the  Netherlands  Institute  for  Radio  Astronomy,  Post-bus 2, 7990 AA Dwingeloo, The Netherlands
   \and
   Dr.~Remeis-Sternwarte  \&  ECAP,  Universit\"at  Erlangen-N\"urnberg, Sternwartstra\ss e 7, 96049 Bamberg, Germany
   }

   \date{Received 15 April 2019; Accepted 19 June 2019}

   \abstract
   {}
     {In the framework of the TANAMI multi-wavelength and VLBI monitoring, we study the evolution of the parsec-scale radio emission in radio galaxies in the southern hemisphere and their relationship to the $\gamma$-ray properties of the sources. Our study investigates systematically, for the first time, the relationship between the two energy regimes in radio galaxies. In this first paper, we focus on \textit{Fermi}-LAT-detected sources.}
   {The TANAMI program monitors a large sample of radio-loud AGN at 8.4 GHz and 22.3 GHz with the Australian
   Long Baseline Array (LBA) and associated telescopes in Antarctica,
   Chile, New Zealand and South Africa. We perform a kinematic
   analysis for five $\gamma$-ray detected radio galaxies using multi-epoch 8.4~GHz VLBI images, deriving limits on
   intrinsic jet parameters such as speed and viewing angle. We analyzed 103 months of \textit{Fermi}-LAT data in order to study possible connections between the $\gamma$-ray properties and the pc-scale jets of \textit{Fermi}-LAT-detected radio galaxies, both in terms of variability and average properties. We discuss the individual source results and draw preliminary conclusions on sample properties including published VLBI results from the MOJAVE survey, with a total of fifteen sources.}
   {We find that the first $\gamma$-ray detection of Pictor~A might be associated with the
   passage of a new VLBI component through the radio core, which appears to be a defining feature of high-energy emitting Fanaroff-Riley type II radio galaxies. For the peculiar AGN PKS~0521$-$36, we detect subluminal parsec-scale
   jet motions, and we confirm the presence of fast $\gamma$-ray
   variability in the source down to timescales of 6 hours, which is
   not accompanied by variations in the VLBI jet. We robustly confirm the presence of significant superluminal motion, up to $\beta_{\mathrm{app}}\sim3$, in the jet of the TeV radio galaxy
   PKS~0625$-$35. Our VLBI results constrain the jet viewing angle to
   be $\theta<53^{\circ}$, allowing for the possibility of a closely aligned jet. Finally, by analyzing the first pc-scale multi-epoch images of the prototypical Compact Symmetric Object (CSO) PKS~1718$-$649, we place an upper limit on the separation speed between the two mini-lobes, which in turn allows us to derive a lower limit on the age of the source.}
     {We can draw some preliminary conclusions on the relationship between pc-scale jets and $\gamma$-ray emission in radio galaxies, based on \textit{Fermi}-LAT-detected sources with available multi-epoch VLBI measurements. We find that the VLBI core flux density correlates with the $\gamma$-ray flux, as seen in blazars. On the other hand, the $\gamma$-ray luminosity does not show any dependence on the core brightness temperature and core dominance, two common indicators of jet Doppler boosting. This seems to indicate that $\gamma$-ray emission in radio galaxies is not driven by orientation-dependent effects, as in blazars, which is consistent with the unified model of jetted AGN.}

   \maketitle
%

\section{Introduction}
Radio-loud active galactic nuclei (AGN) host symmetric, highly relativistic,
well-collimated jets, ejected from the vicinity of the central
supermassive black hole (SMBH). These jets produce bright non-thermal
radiation across the whole electromagnetic spectrum. AGN dominate the $\gamma$-ray sky, as revealed by the \textit{Fermi} Large Area Telescope (LAT) \citep{2015ApJS..218...23A}. The largest class of identified
$\gamma$-ray sources is blazars, i.e, radio-loud AGN with jets
oriented at small angles ($\theta<10^{\circ}$) to the observer's line of
sight \citep{2015ApJ...810...14A}. The alignment and the relativistic bulk speed of the jet
imply that its radiation is beamed in the direction of motion and
boosted via relativistic Doppler effects, which make these sources
extremely bright and shortens their variability time scales.

After decades of studies, there are still important unanswered
questions on jet physics and their high-energy emission. For example,
the angular resolution of the currently best $\gamma$-ray instrument, the \textit{Fermi}-LAT, can only reach down to $\theta\sim0.1^{\circ}$. Therefore it is rarely possible to associate the $\gamma$-ray emission to a specific
morphological component of the AGN, such as e.g., the radio core, jet,
or diffuse extended lobes.

According to the most popular models, the unbeamed parent population
of blazars is made by classic radio galaxies \citep[e.g., ][]{1995PASP..107..803U,2016A&ARv..24...10T}. In these sources, the jet is oriented at larger angles to our line of sight,
therefore its radiation is much less affected by Doppler boosting. For
this reason, they are unfavored $\gamma$-ray emitters, and indeed they make up only about 1-2\% of identified sources in the
\textit{Fermi}-LAT catalogs, with a total of $\sim20$ objects \citep{2015ApJ...810...14A}. The
faint fluxes and small sample size create a challenge to
understanding $\gamma$-ray radio galaxies as a population.
Nonetheless, studying these sources provides us with a view of
jets which is less masked by orientation-based
relativistic effects, allowing us to better reconstruct the intrinsic
properties \citep{2010ApJ...720..912A}. 

The combination of $\gamma$-ray data with high-resolution radio
observations can be a powerful tool in investigating these questions.
Very Long Baseline Interferometry (VLBI) radio observations allow us to achieve an angular resolution which is
orders of magnitude better than that of any $\gamma$-ray instrument, down to
milliarcsecond scales. Since AGN are variable sources,
it is possible to identify the $\gamma$-ray emission region on VLBI
scales by looking for correlated variability in radio morphology or
flux, and high-energy emission \citep[e.g., ][]{2015ApJ...808..162C}. Additionally, multi-epoch VLBI observations
provide the only direct measure of relativistic jet motion, allowing
us to derive relevant jet parameters such as apparent speed and jet
orientation angle. This motivation led to the development of long-term
radio VLBI programs monitoring a large number of AGN \citep[e.g., MOJAVE, ][]{2009AJ....138.1874L}.

TANAMI (Tracking Active galactic Nuclei with Austral Milliarcsecond
Interferometry) conducts the only such large monitoring program for sources
south of $-30^{\circ}$ declination, with a sample of about 130 AGN. First-epoch 8.4 GHz images were presented for an initial sample of 43 sources
by \citet{2010A&A...519A..45O}. The sample was regularly expanded with new
$\gamma$-ray detected sources or otherwise interesting new objects, and
first-epoch images of the first-ever high-resolution observations, for
most sources, was recently published \citep{2017arXiv170903091M}. The radio VLBI monitoring is complemented by excellent multi-wavelength coverage, including NIR, optical, UV, X-ray and $\gamma$-ray data, providing a quasi-simultaneous broadband view of
the sources, as is required for detailed studies of variable sources
such as AGN \citep[see e.g., ][]{2016A&A...591A.130K}. An overview of the multi-wavelength program and selected
TANAMI results can be found in \citet{2015AN....336..499K}.

In this paper, we focus on the radio VLBI properties of the $\gamma$-ray detected TANAMI radio galaxies, i.e.
Pictor~A (a.k.a. 0518$-$458), PKS~0625$-$35 (a.k.a. 0625$-$354, 0625$-$35), Centaurus~B (a.k.a. Cen~B, 1343$-$601) and PKS~1718$-$649 (a.k.a. 1718$-$649, NGC~6328), and the peculiar AGN PKS~0521$-$36~\footnote{This source is included in the study since several studies suggest its jet has a viewing angle larger than 10$^\circ$.} (a.k.a. PKS~0521$-$365, 0521$-$365). We produce high-resolution images at 8.4 GHz for all available
epochs between 2007 and 2013. We perform a kinematic analysis by
fitting Gaussian components to the jet and tracking their motion across the
epochs. To complement this, we produce $\gamma$-ray light curves and spectra from
\textit{Fermi}-LAT data, investigating a possible connection
between the two bands.

In a subsequent paper, we will focus on the $\gamma$-ray faint
TANAMI radio galaxies, and present the radio kinematic results, along
with updated \textit{Fermi}-LAT flux upper limits, to investigate if and how the two subsamples
differ from each other.

The paper is organized as follows. In Section~\ref{sample} we review the TANAMI sample definition,
investigate the completeness of the TANAMI radio galaxy sample, and
give an overview of the source sample. In Section~\ref{data} we
illustrate the data reduction procedures for the radio VLBI (Section~\ref{vlbi}) and the
$\gamma$-ray data (Section~\ref{fermi}). In Section~\ref{results} we present the
multi-frequency VLBI imaging results (Section~\ref{img_results}), the
VLBI kinematics results (Section~\ref{kin_results}), and the results
of the \textit{Fermi}-LAT analysis (Section~\ref{lat_results}).
Finally, in Section~\ref{disc} we discuss the scientific implications
of the radio and $\gamma$-ray data, and draw our conclusions in Section~\ref{conc}. All the results are presented on a source-by-source basis. Throughout the paper we assume a cosmology with $H_0 =  73$ km s$^{-1}$ Mpc$^{-1}$, $\Omega_{\mathrm{m}}$ =   0.27, $\Omega_{\mathrm{\Lambda}}$ =   0.73~\citep{2011ApJS..192...18K}, and the radio spectral indices refer to the convention $S\propto\nu^{+\alpha}$.


\section{The TANAMI radio galaxy sample}
\label{sample}
The TANAMI sample was defined starting from two subsamples of sources south of $\delta=-30^{\circ}$, i.e. a radio-selected sample and a $\gamma$-ray selected sample. The
radio-selected subsample was based on the catalogue of
\citet{1994A&AS..105..211S}\footnote{In turn, this catalogue is based
  on \citet{1981A&AS...45..367K}.}, with a flux density cut at $S_{\mathrm{5GHz}}>2$ Jy and a
spectral index cut at $\alpha>-0.5$ ($S\propto\nu^{+\alpha}$) between 2.7 and
5 GHz. The $\gamma$-ray selected subsample included all identified EGRET blazars
in the given declination range. The sample also includes a few
additional sources of interest which did not satisfy this selection,
and were added manually to the monitoring program.

We investigated whether the sample of radio galaxies in TANAMI, including sixteen sources is
representative of a complete sample of radio galaxies in our
declination range. To do this, we cross-matched the
V\'eron-V\'eron 13$^\mathrm{th}$ edition AGN catalog \citep{2010A&A...518A..10V} with
the Parkes radio catalog \citep{1990PKS...C......0W}. While the V\'eron-V\'eron catalog is not complete in a statistical sense, it is a comprehensive compendium of known AGN. The Parkes catalog on the other hand provides extensive information on the radio properties of the sources.

Starting from this cross-matched catalog, we first applied the declination cut for TANAMI, i.e., $\delta<30^\circ$. Since we are only interested in radio galaxies for the purpose of this study, we
had to clean the sample by excluding known BL Lacs and QSOs. Physically, the only criterion distinguishing radio galaxies from blazars should be the jet viewing angle. However, it can be challenging to obtain this information, and estimates often suffer from a large uncertainty. We therefore made this selection based on the optical spectral classification provided in the V\'eron-V\'eron catalog. 

This left us with a total of 83 sources. Twelve out of sixteen TANAMI radio galaxies are included in this sample~\footnote{The missing sources
are Centaurus~A, which is misclassified as BL Lac in the V\'eron-V\'eron
catalog; Centaurus~B, and PMN~J1603$-$4904 which are not included in the
catalog probably due to their location in the galactic plane; as well as
PKS~1258$-$321 which is also missing from the V\'eron-V\'eron catalog.}.
We looked for existing VLBI measurements for each of these 83
sources individually on the \href{http://ned.ipac.caltech.edu/}{NASA/IPAC Extragalactic Database (NED)}, and found
existing data for six non-TANAMI radio galaxies, from the Very Long Baseline Array (VLBA) at 8.6 GHz
\citep{2005AJ....129.1163P} or the Long Baseline Array (LBA) at 8.4 GHz or 4.8 GHz
\citep{2004AJ....128.2593F,2009MNRAS.397.2030H}.  Therefore, in total we found 22 southern radio galaxies with compact radio
emission on VLBI scales. The TANAMI radio galaxy sample, with sixteen sources, can be considered representative of the sub-population of radio galaxies with a bright compact core. 

Our sample includes well-known sources for which TANAMI provides the highest-resolution data
available. The full list of TANAMI radio galaxies can be found in
Table~\ref{rgs}. Several radio galaxy subclasses are present, from
classic FR~I (e.g., Centaurus~A) and FR~II (e.g., Pictor~A) to young radio sources
(e.g., PKS~1718$-$649) and peculiar or misclassified AGN (e.g., PKS~0521$-$36). In this work we focus on $\gamma$-ray detected sources.
Only half of the sample has been associated to a \textit{Fermi}-LAT source so far. This is
not surprising, given that radio galaxies are faint $\gamma$-ray emitters.
Since TANAMI is, to date, the largest VLBI program monitoring
southern-hemisphere sources at milliarcsecond resolution, including a considerable number of radio galaxies, it provides
the first data set suitable for kinematic studies on these scales.
The most notable member of the TANAMI radio galaxy sample is the closest radio-loud AGN, Centaurus~A, which has been studied extensively using TANAMI data, revealing the
complex dynamics of the jet's inner parsec. A recent kinematic analysis indicates downstream jet
acceleration \citep{2014A&A...569A.115M}, and the spectral index map between 8.4 GHz and 22.3 GHz
suggests a transverse structure that can be explained within
the spine-sheath scenario \citep{2005A&A...432..401G}.

The peculiar source PMN~J1603$-$4904 has also been extensively studied within
the TANAMI collaboration before
\citep{2014A&A...562A...4M,2015A&A...574A.117M,2016A&A...593L..19M, 2016A&A...586L...2G, 2018A&A...610L...8K}. Therefore these two sources are not included in this study.


\begin{table*}[h!tbp]
\caption{TANAMI radio galaxies. Redshifts are from the \href{http://ned.ipac.caltech.edu/}{NASA/IPAC Extragalactic Database (NED)}, unless otherwise indicated.}
\begin{center}
\begin{tabular}{llllllc}
\hline
\hline
B1950 name & Catalog name & Class$^a$ & Redshift & RA(J2000) & Dec(J2000) & LAT$^b$\\
\hline
0518$-$458 & Pictor~A & FR II & 0.035 & 79.957 & $-$45.779 & yes\\
0521$-$365 & PKS~0521$-$36	& RG/SSRQ & 0.057 & 80.742 & $-$36.459 & yes\\
0625$-$354 & PKS~0625$-$35 & FR I/BLL & 0.055 & 96.778 & $-$35.487 & yes\\
0825$-$500 & PKS~0823$-$500 & RG & - & 126.362 & $-$50.178 & no\\
1258$-$321 & PKS~1258$-$321 &  FR I & 0.017 & 195.253 & $-$32.441 & no\\
1322$-$428 & Centaurus~A & FR I & 0.0018 & 201.365 & $-$43.019 & yes\\
1333$-$337 & IC~4296 &  FR I & 0.013 & 204.162 &	$-$33.966 & no\\
1343$-$601 & Centaurus~B &  FR I & 0.013 & 206.704 & $-$60.408 & yes\\
1549$-$790 & PKS~1549$-$79	&  RG/CFS & 0.15 & 239.245 &	$-$79.234 & no\\
1600$-$489 & PMN~J1603$-$4904 & MSO$^c$ & 0.23$^d$ & 240.961 & $-$49.068 & yes\\
1718$-$649 & PKS~1718$-$649 &  GPS/CSO & 0.014 & 260.921 & $-$65.010 & yes$^e$\\
1733$-$565 & PKS~1733$-$56 &  FR II & 0.098 & 264.399 & $-$56.567 & no\\
1814$-$637 & PKS~1814$-$63 &  CSS/CSO & 0.065 & 274.896 & $-$63.763 & no\\
2027$-$308 & PKS~2027$-$308 &  RG & 0.54 & 307.741 & $-$30.657 & no\\
2152$-$699 & PKS~2153$-$69 &  FR II & 0.028 & 329.275 & $-$69.690 & no\\
\hline
\hline
\end{tabular}
\end{center}
$^a$ FR I: Fanaroff-Riley type 1; FR II: Fanaroff-Riley type 2; BLL:
BL Lac; RG: Radio galaxy; SSRQ: Steep Spectrum Radio Quasar; CFS: Compact Flat Spectrum; MSO: Medium-size
Symmetric Object; GPS: Gigahertz Peaked Spectrum; CSO: Compact
Symmetric Object; CSS: Compact Steep Spectrum.\\
$^b$ Associated with a LAT $\gamma$-ray source from the 3FGL~\citep{2015ApJS..218...23A}, unless otherwise indicated.\\
$^c$ Originally misclassified as BL Lac, this source has been classified as a young radio galaxy based on multi-wavelength
studies \citep{2014A&A...562A...4M,2015A&A...574A.117M,2016A&A...593L..19M}.\\ 
$^d$ \citet{2016A&A...586L...2G}.\\
$^e$ First $\gamma$-ray detection reported by \citet{2016ApJ...821L..31M}.\\
\label{rgs}
\end{table*}

\section{Observations and data reduction}
\label{data}
\subsection{Radio data}
\label{vlbi}

TANAMI monitors the sources in its sample with a cadence of about two observations per year
since 2007, at 8.4 GHz and 22.3 GHz. The VLBI array consists of the Australian LBA,
supported by associated antennas in South Africa, New Zealand,
Antarctica and Chile (a full list of the participating antennas is reported in Table~\ref{array}).

Details on the array configuration for each observation are provided
in
Tables~\ref{pica_tab},~\ref{0521_tab},~\ref{0625_tab},~\ref{cenb_tab}
and~\ref{1718_tab}. The data were calibrated using the National Radio Astronomy Observatory's Astronomical Image Processing System~\citep[AIPS,][]{aips}  and imaged using the \texttt{CLEAN} algorithm~\citep{1974A&AS...15..417H} as implemented in the Difmap package~\citep{difmap} following the procedures described in \cite{2010A&A...519A..45O}.

\begin{table*}[!htbp]
\caption{List of radio telescopes forming the TANAMI array.}
\begin{center}
\begin{tabular}{llll}

\hline
\hline
Antenna & Code & Diameter (m) & Location\\
\hline
Parkes & PKS & 64 & Parkes, New South Wales, Australia\\
ATCA & AT & 5$\times$22 & Narrabri, New South Wales, Australia\\
Mopra & MP & 22 & Coonabarabran, New South Wales, Australia\\
Hobart & HO & 26 & Mt. Pleasant, Tasmania, Australia\\
Ceduna & CD & 30 & Ceduna, South Australia, Australia\\
Hartebeesthoek$^a$ & HH & 26 & Hartebeesthoek, South Africa\\
DSS 43$^b$ & - & 70 & Tidbinbilla, Australia\\
DSS 45$^b$ & - & 34 & Tidbinbilla, Australia\\
DSS 34$^b$ & - & 34 & Tidbinbilla, Australia\\
O'Higgins$^c$ & OH & 9 & O'Higgins, Antarctica\\
TIGO$^{c,d}$ & TC & 6 & Concepci\'on, Chile\\
Warkworth$^e$ & WW & 12 & Auckland, New Zealand\\
Katherine & KE & 12 & Northern Territory, Australia\\
Yarragadee & YG & 12 & Yarragadee, Western Australia\\
ASKAP$^f$ & AK & 12 & Murchinson, Western Australia\\
\hline
\end{tabular}
\end{center}
$^a$ Unavailable between Sept. 2008 and Sept. 2010.\\
$^b$ Operated by the Deep Space Network of the USA National Aeronautics
and Space Administration (NASA). DSS 45 was decommissioned in November 2016.\\
$^c$ Operated by the German Bundesamt f\"ur Kartographie und Geodesie
(BKG).\\
$^d$ Now in La Plata, Argentina.\\
$^e$ Operated by the Institute for Radio Astronomy and Space Research (IRASR).\\
$^f$ Contributing with a single antenna of the 36-element array.
\label{array}
\end{table*}

Many more antennas are available at 8.4 GHz (than at 22.3 GHz), yielding better angular resolution in this band, despite the lower frequency. Observations in
this band are also more frequent. Hence, 8.4~GHz data are used for the
kinematic analysis. We have used selected epochs at 22.3~GHz, simultaneous with the 8.4~GHz ones, to produce spectral index maps of our targets. The spectral information is crucial in order to identify the VLBI core component, which usually presents a flat spectral index, and as an input for estimating the jet viewing angle. Since the absolute position information is lost in the imaging process due to the phase self-calibration, it is necessary to properly align the maps before computing the spectral index. This is done through a 2D cross-correlation procedure \citep{2013A&A...557A.105F}, referenced on an optically thin region of the jet, whose position should not vary with frequency. After aligning the maps and convolving them with the same beam (typically the one of the map with lowest resolution), the spectral index is calculated for each point as 

\begin{equation}
    \alpha = \frac{\log(S_1/S_2)}{\log(\nu_1/\nu_2)}.
\end{equation}

A full estimate of the uncertainty of spectral index maps requires taking into account frequency-dependent flux density errors, pixel-by-pixel signal-to-noise ratios, and an analysis of the error in the shift between the two maps through Monte Carlo simulations~\citep[see Appendix D in][]{2013A&A...557A.105F}. Since we did not perform a detailed study and in-depth interpretation of our spectral index maps, such an extensive error analysis is beyond the scope of this paper.

In order to study the evolution of the jet, we fit the self-calibrated visibilities with circular Gaussian components using the \texttt{Modelfit} task in Difmap. We then cross
identify them in the different epochs by selecting a component in the first epoch, and searching for the closest component in the following epochs. This selection is then corroborated by visual inspection of the maps and the component properties, e.g., the evolution of their flux density. We then fit the motion of the
components which are robustly detected in at least 5 epochs,
separately in RA and Dec, to derive the two components of the velocity
vector, following \citet{2009AJ....138.1874L}:

\begin{equation}
\begin{gathered}
x(t)=\mu_x(t-t_{0x})\\
y(t)=\mu_y(t-t_{0y})\\
\mu=\sqrt{\mu_x^2+\mu_y^2}\\
\beta_\mathrm{app}=\mu D_\mathrm{L}/c(1+z)
\end{gathered}
\end{equation}

\noindent where $\mu_x$ and $\mu_y$ are the angular speeds in RA and Dec, $\mu$
is the resulting vector modulus, and $\beta_\mathrm{app}$ is the apparent
speed in units of speed of light, obtained using the luminosity
distance $D_\mathrm{L}$ and the redshift $z$. We do not fit accelerations since
second-order terms are difficult to constrain without a long enough
monitoring. \citet{2009AJ....138.1874L} require a component to be detected in at
least 10 epochs in order to fit an acceleration, and
the maximum number of TANAMI epochs for the sources in our sample is
9.
For the uncertainty on the component position, we adopt the approach described in \cite{2012A&A...545A.117L}. The error is calculated as the component size divided by its signal-to-noise ratio (SNR). The latter is calculated as the ratio between the peak brightness of the component and the map noise. For bright components, this estimate can yield unrealistically small uncertainties. As a lower limit for this case, we take the largest value between half the component size and 1/5 of the beam major axis. The uncertainty on the flux is given by the calibration errors, which we estimate to be 15\%. 

Since the TANAMI array composition can vary significantly across
epochs, the $(u,v)$ coverage, and consequently the beam will also be
inhomogeneous across the epochs for a single source. This complicates
the identification of components across the epochs, since one
component detected in a low-resolution epoch may be resolved into
multiple components in another epoch. To overcome this problem, we re-imaged
the sources applying a Gaussian taper to the highest-resolution maps
in order to downweight the visibilities from the longest baselines, and
approximately match the beam size of the epoch with lowest resolution along the jet direction.

Using the apparent speed measured from kinematics ($\beta_\mathrm{app}$)
together with the jet-to-counterjet flux density ratio $R=S_\mathrm{jet}/S_\mathrm{counterjet}$, it is possible to set some limits on the intrinsic jet speed $\beta=v/c$ and
viewing angle $\theta$.

$R$ and $\beta_\mathrm{app}$ can be expressed as a function of these two parameters:
\begin{equation}
\noindent
\beta_\mathrm{app}=\frac{\beta\sin\theta}{1-\beta\cos\theta}\\
R=\left(\frac{1+\beta\cos\theta}{1-\beta\cos\theta}\right)^{p-\alpha}
\end{equation}

\noindent where $\alpha$
is the jet spectral index, and the index $p$ takes a value of 2 or 3 depending on whether $R$ is calculated
integrating the flux density over the jet or using a single component,
respectively. We used the former method for cases where a counterjet is detected, while we used a single component in the sources with no detectable counterjet. In the cases where no counterjet emission
is detected, we place a lower limit on $R$. To do this, we take the ratio between the peak flux of the innermost jet component and the peak flux on the counterjet side. Out of the different measurements coming from the multiple available epochs, we use the most constraining one, given by the map with the best quality. These relations allow us to set limits in the space of intrinsic
parameters $\beta$ and $\theta$ (see Section~\ref{kin_results}).

\subsection{\textit{Fermi}-LAT data}
\label{fermi}
The LAT is a pair-conversion telescope, launched on 2008 June 11 as the
main scientific instrument on board the \textit{Fermi Gamma-ray
Space Telescope} \citep{2009ApJ...697.1071A}. Its main energy range is
0.1-100 GeV, but its sensitivity extends down to 60 MeV and up to 2
TeV \citep{2017ApJS..232...18A}. The telescope operates almost exclusively in survey mode, scanning the entire $\gamma$-ray sky approximately every three hours.

We use the Python package \texttt{Fermipy} \citep{2017arXiv170709551W}
throughout the analysis, and assume as starting model the latest
\textit{Fermi}-LAT source catalog, i.e, the third source catalog
~\citep[3FGL, ][]{2015ApJS..218...23A}. We take
a Region of Interest (ROI) of 10$^\circ$ around the target position, and
include in the model all sources from the 3FGL within 15$^{\circ}$ from the
ROI center, together with the latest model for the galactic and
isotropic diffuse emission (\texttt{gll\_iem\_v06.fits} and
\texttt{iso\_P8R2\_SOURCE\_V6\_v06.txt}, respectively). We perform a binned analysis with 10 bins per decade in
energy and 0.1$^{\circ}$ binning in space. We use the source event class,
and the latest Pass 8 response function \texttt{P8R2\_SOURCE\_V6}. We
enable an energy dispersion correction to take into account the degradation in
energy resolution at low energies. We select only times when
the zenith angle of the incoming events with respect to the telescope is smaller than 100$^{\circ}$, to avoid
contamination by the Earth's limb. We take advantage of the new Pass 8
characterization of events in different PSF quartiles, based on the
quality of the direction reconstruction, from the worse quartile
(PSF0) to the best (PSF3). We model each type separately and combine
the resulting likelihood using the summed-likelihood method. Since the
angular resolution of the LAT degrades at low energy, we progressively increase the
low-energy cut for the worse PSF quartiles. The details of the
selection of the different components are listed in Table~\ref{comps}.
\begin{table}
\caption{Details of the selection cuts for different components in the
  LAT analysis.}
\label{comps}  
\begin{center}    
\begin{tabular}{cccc}    
\hline\hline
E$_{min}$ & E$_{max}$ & PSF quartile & Event type\\
\hline
100 MeV & 100 GeV & PSF3 & 32\\
400 MeV & 100 GeV & PSF2 & 16\\
500 MeV & 100 GeV & PSF1 & 8\\
800 MeV & 100 GeV & PSF0 & 4\\
\hline
\end{tabular}
\end{center}
\end{table}

We fit the ROI with the initial 3FGL model, freeing all the parameters
of the target source and the normalization of all sources within 5$^{\circ}$
of the ROI center. Since our data set more than doubles the
integration time with respect to the 3FGL catalog, we look for new
sources with an iterative procedure. We produce a map of excess Test
Statistic (TS). The TS is defined as $2\log(L/L_0)$ where
$L$ is the likelihood of the model with a point source at the target
position, and $L_0$ is the likelihood without the source. A value of TS=25
corresponds to a significance of
4.2$\sigma$~\citep{1996ApJ...461..396M}. A TS map is produced by
inserting a test source at each map pixel and evaluating its
significance over the current model. We look for TS$>$9 peaks in
the TS map, with a minimum separation of 0.3$^{\circ}$, and add a new point source to the model for each peak,
assuming a power-law spectrum. We then fit again the ROI, and
produce a new TS map. This process is iterated until all significant
excesses are modeled out. We also perform a localization analysis on
the target source and all new sources with TS$>$25 found in the ROI.

As listed in Table~\ref{rgs}, roughly half of the TANAMI radio
galaxies are detected in the 3FGL catalog. For those that were not
already extensively studied in previous TANAMI papers (see
Section~\ref{sample}) we consider 103 months of LAT data ($\sim8.5$ years). We produced 0.1-100 GeV Spectral Energy Distributions (SEDs) with a binning optimized for each source in order to balance spectral coverage with sufficient statistics in each bin. We also computed light curves with monthly bins and, when possible, a finer binning to reveal more detailed flaring activity. We also report the average source properties in the 1-100 GeV range in order to compare with those presented in the 3FGL.

For the variability analysis, we look at the energy range
0.1-300 GeV. All the other parameters of the multi-component analysis are the same
as those described above. We first perform a standard analysis over the
whole time range, then we perform a full likelihood fit in each time bin. The number of free parameters in the latter fit depends on the statistics in each bin. We first attempt a fit with the normalization of all sources in the ROI left free to vary. If the statistics do not allow the fit to converge, we progressively reduce the radius including sources with free normalization, first to 3\degr, then to 1\degr. If all these methods fail to result in a successful fit, we free only the normalization of the target source.

We consider the source detected in each bin if TS$>$10
and the signal to noise ratio is higher than 2,
otherwise we place a 95\% confidence flux upper limit.

\section{Results}
\label{results}

\subsection{Radio imaging results}
\label{img_results}
\begin{figure*}
    \centering
    \includegraphics[width=0.32\linewidth]{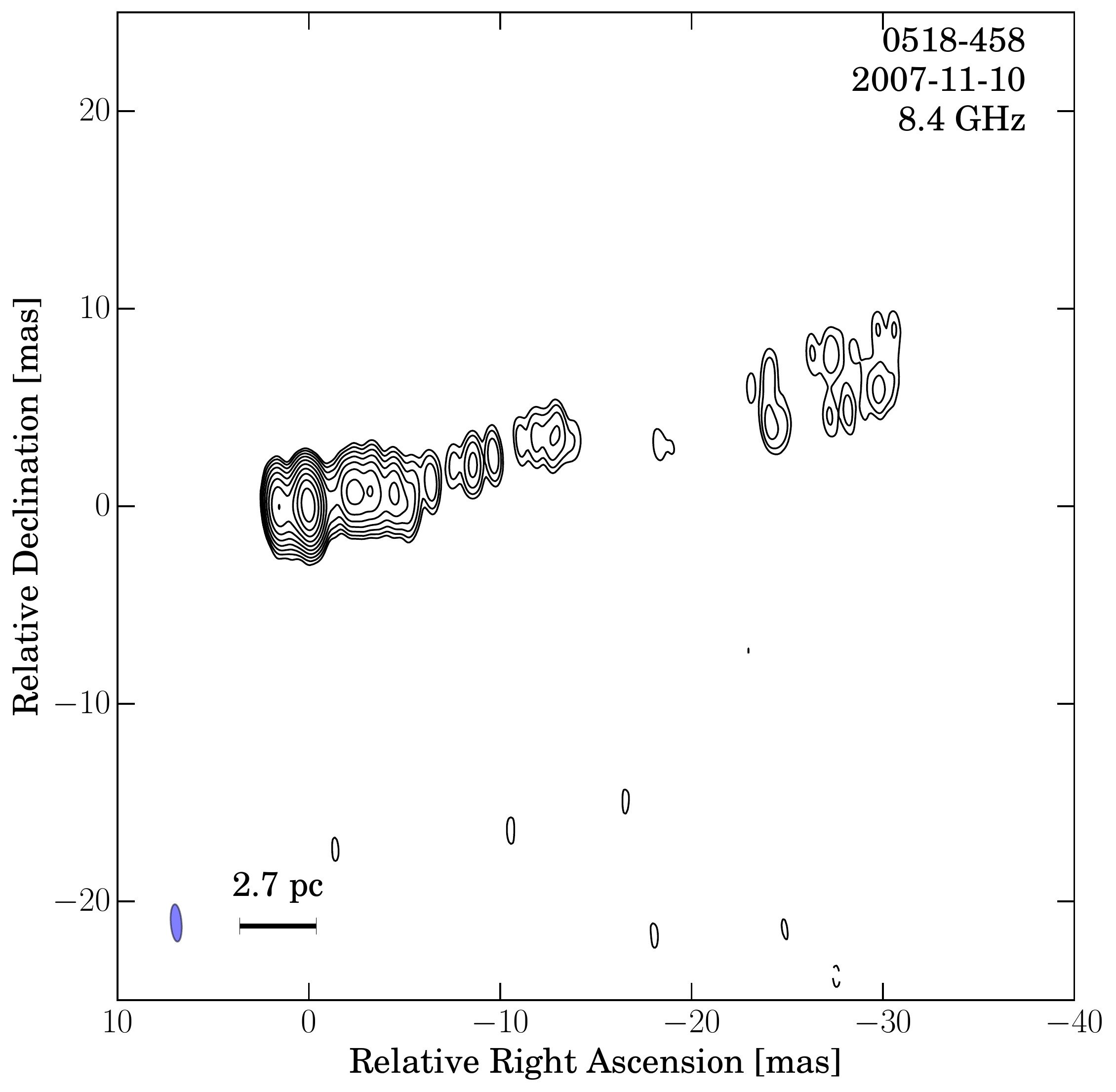}
    \includegraphics[width=0.32\linewidth]{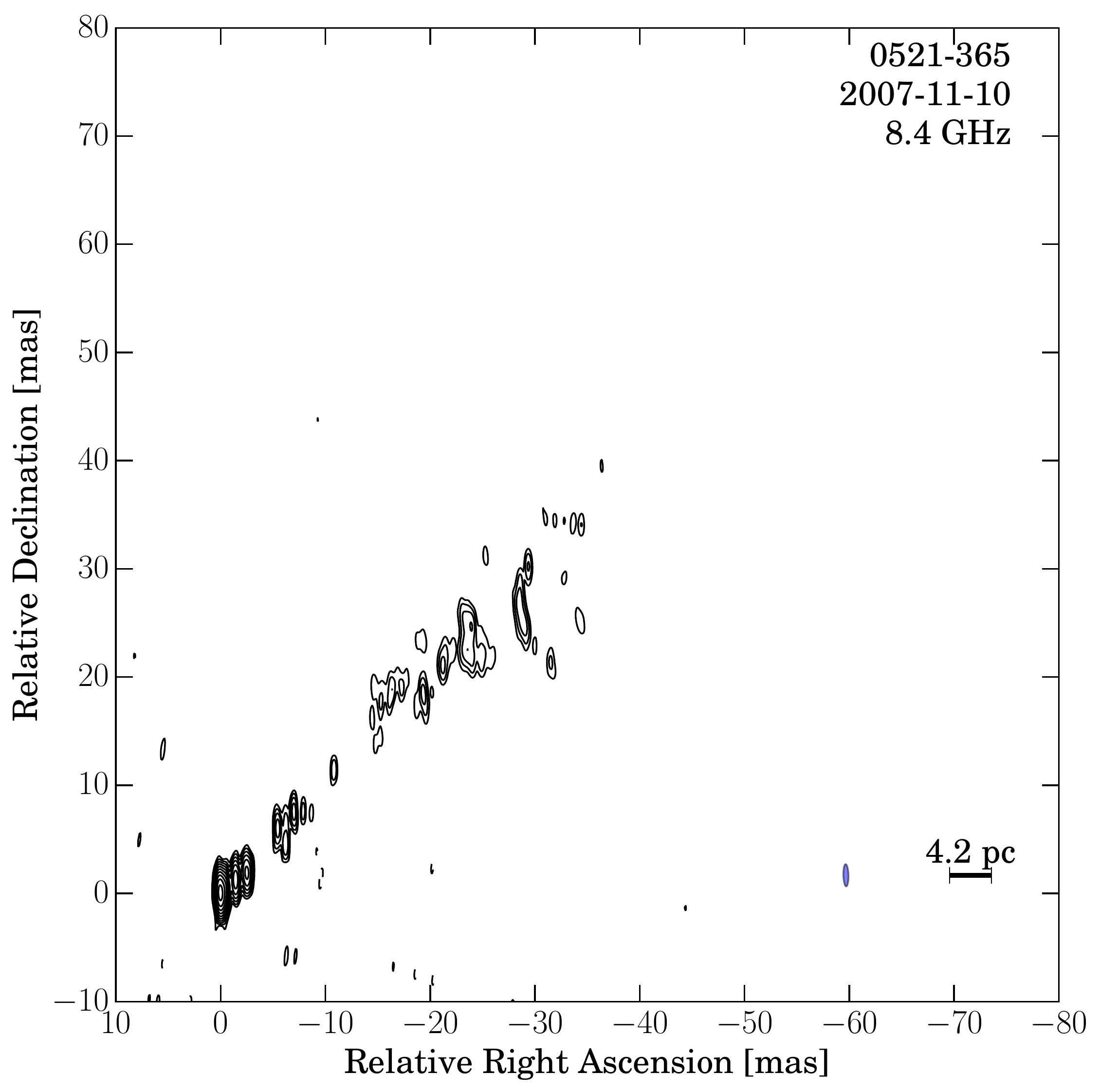}
    \includegraphics[width=0.32\linewidth]{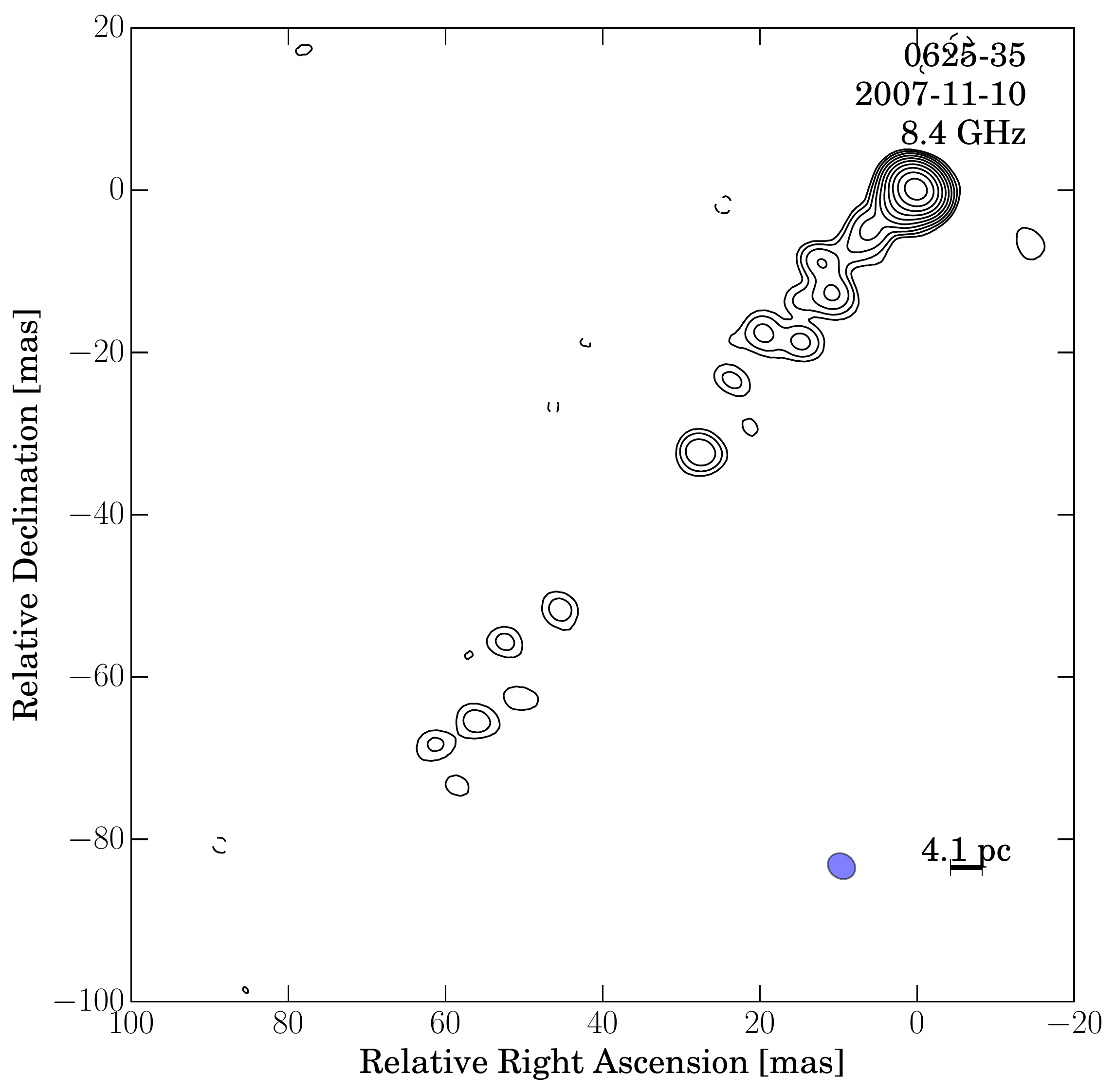}
    \includegraphics[width=0.32\linewidth]{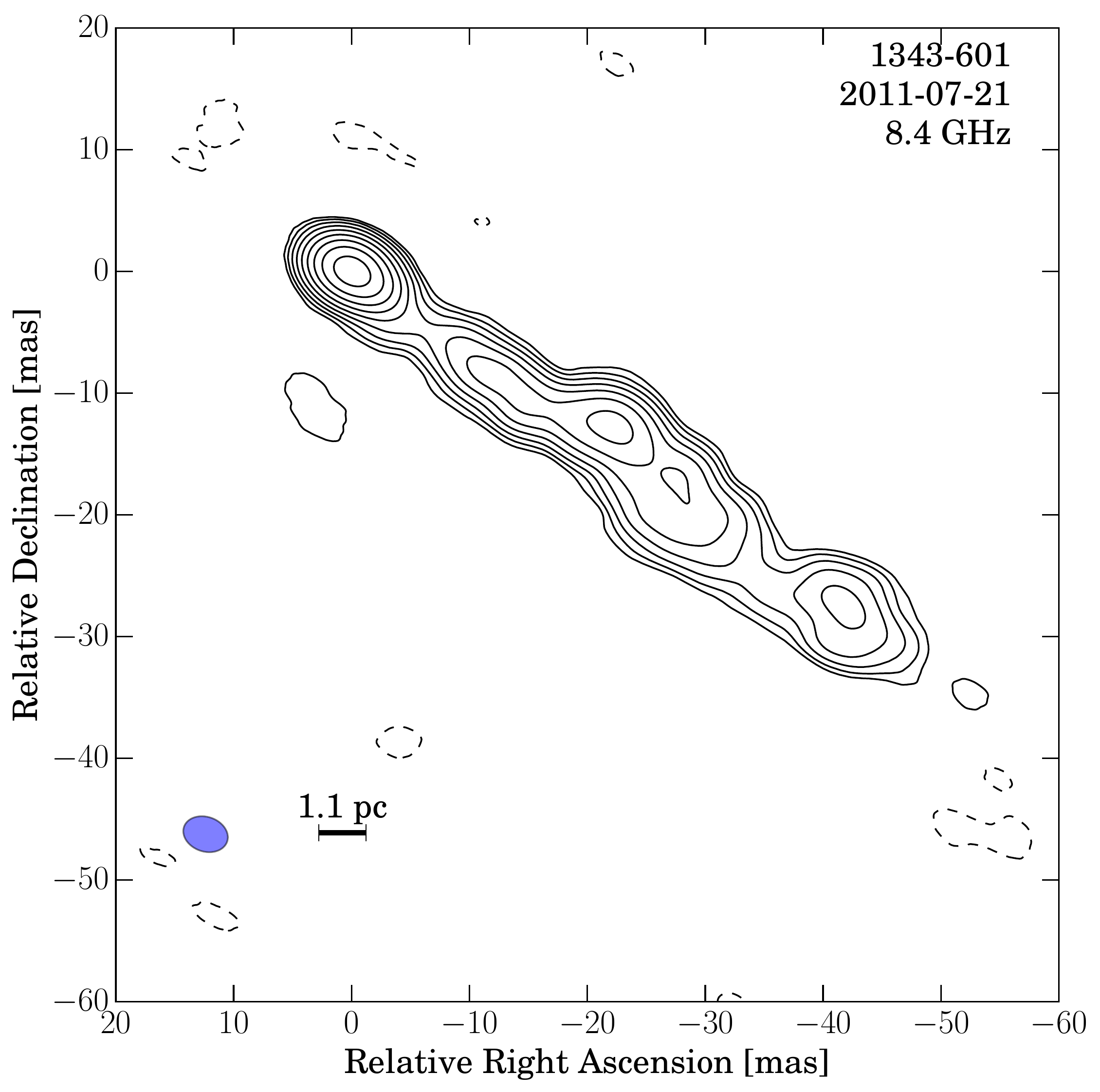}
    \includegraphics[width=0.32\linewidth]{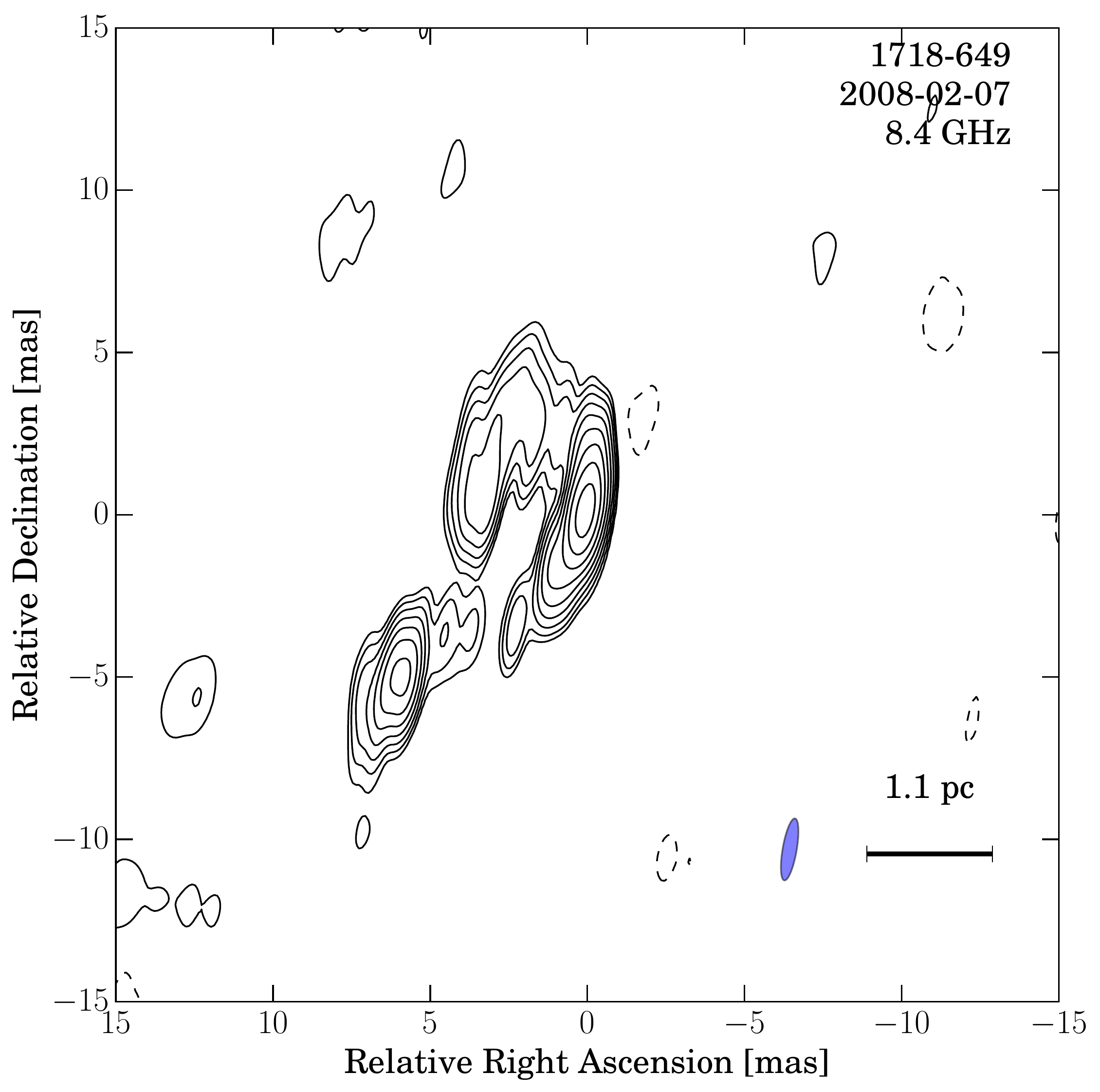}
    \caption{First-epoch 8.4\,GHz contour maps of $\gamma$-ray detected TANAMI radio galaxies. The blue ellipse indicates the convolving beam, while the black bar shows the linear scale. Top left to bottom right: Pictor~A, PKS~0521$-$36, PKS~0625$-$35, Centaurus~B, PKS~1718$-$649 (the B1950 name is indicated in the top right corner of each image). The full set of multi-epoch images and the associated map parameter tables are presented in Appendix~\ref{app:maps}.}
    \label{fig:wallpaper}
\end{figure*}

In Fig.~\ref{fig:wallpaper} we present first-epoch VLBI images of the $\gamma$-ray loud TANAMI radio galaxies, while the full set of multi-epoch images and related map parameters is presented in Appendix~\ref{app:maps}. The imaging results for each individual source are discussed in this subsection.

\paragraph{0518$-$458 (Pictor~A)}
Pictor~A is a classical powerful FR~II radio galaxy. In a previous
kinematic study \citet{2000AJ....119.1695T} characterized the pc-scale jet of
the source with three components, with a fastest apparent motion of
$\beta_\mathrm{app}=1.1\pm0.5$. They did not detect any counterjet at this
scale, and additionally they found an apparent bend in the jet at
$\sim10$ mas from the core.


TANAMI monitoring has provided 5 epochs for this source. As shown in Fig.~\ref{fig:wallpaper}, the jet extends for $\sim30$ mas westward from the brightest component, and
several features can be identified and tracked. We consistently detect
emission eastward of the brightest component, which we assumed to be the VLBI core,
in all epochs. This feature is not present in the first epoch map of
\citet{2010A&A...519A..45O}, but its
detection in multiple epochs in this work indicate that it is real. Additionally, this emission feature is also detected at 22.3~GHz.

To test if this emission feature can be considered as counterjet emission, we produce a
spectral index map of the source between the quasi-simultaneous 8.4~GHz (epoch
2008 Nov 27) and 22.3~GHz (epoch 2008 Nov 29) images. 

\begin{figure}[h!tbp]
\begin{center}
\includegraphics[width=\linewidth]{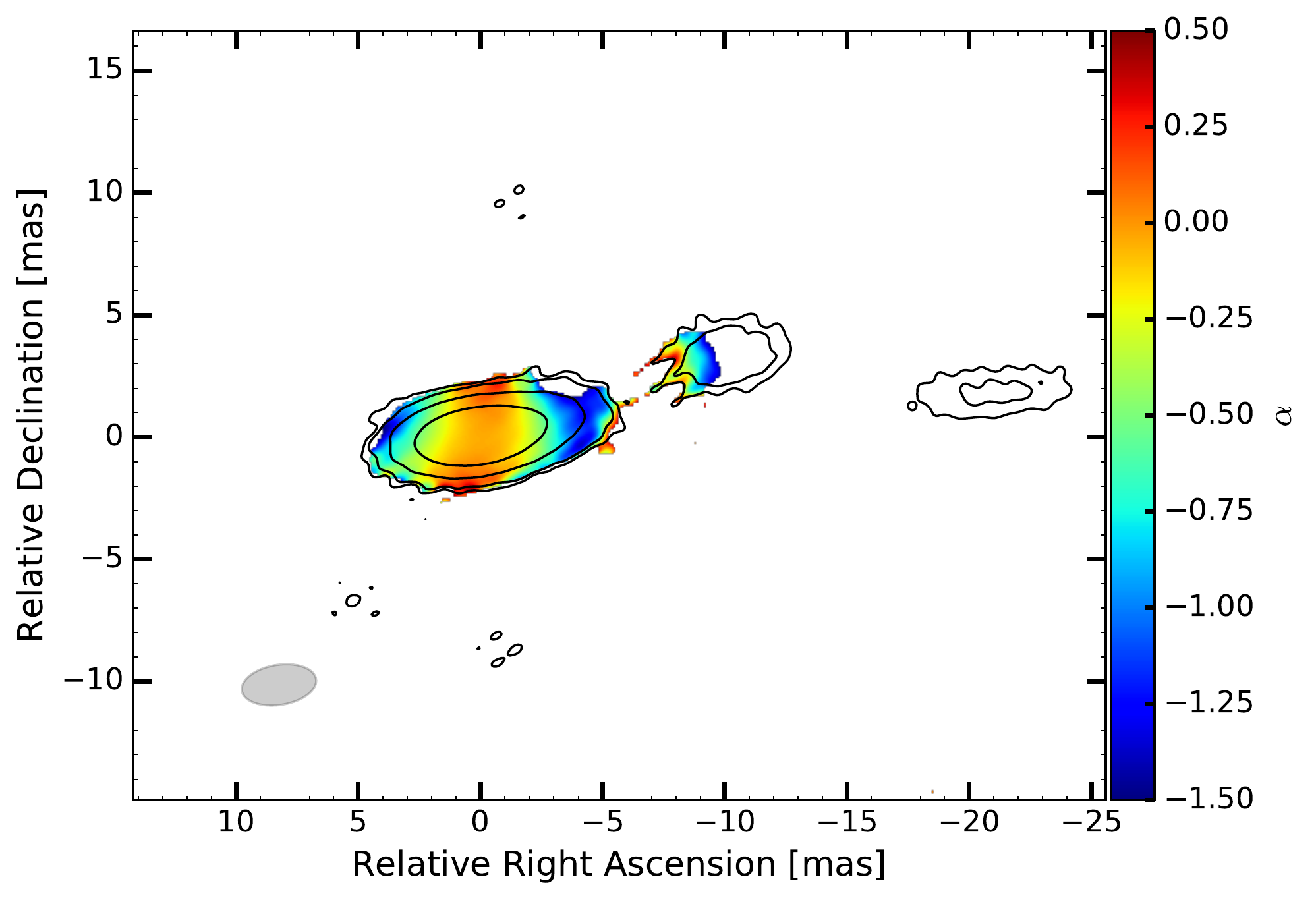}
\end{center}
\caption{Spectral index map of Pictor~A (0518$-$458) between 8.4 GHz and 22.3 GHz for epoch 2008 Nov 27. Black contours are from the 8.4 GHz image. The convolving beam is represented in grey in the lower-left corner.}
\label{0518_spix}
\end{figure}

The resulting map is presented in Fig.~\ref{0518_spix}. We see that the spectrum flattens around the brightest component, and becomes optically thin again in the eastward region. This indicates that this is counterjet emission, and should
therefore be taken into account when computing jet-to-counterjet flux density
ratios.

\begin{figure}[h!tbp]
\begin{center}
\includegraphics[width=\linewidth]{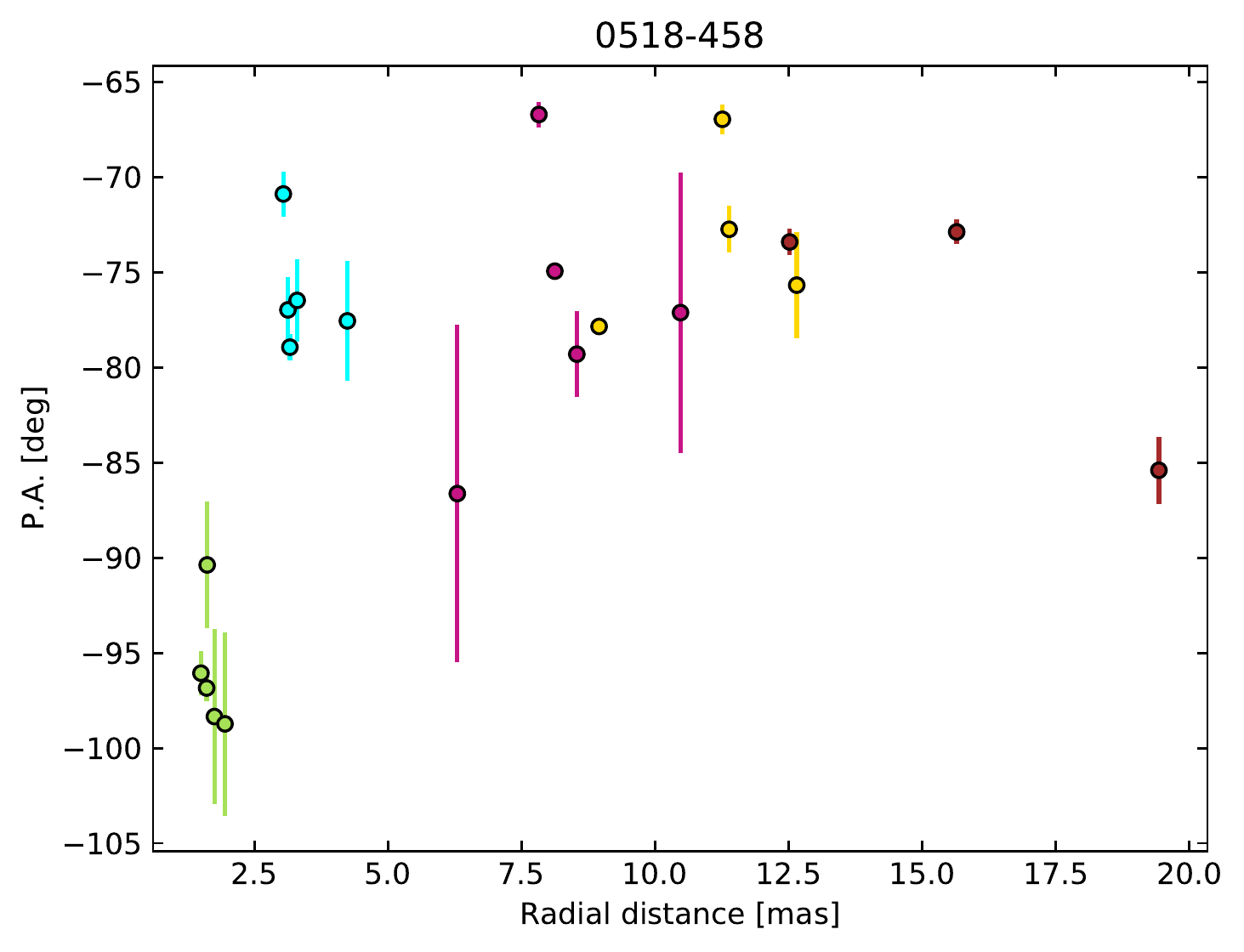}
\end{center}
\caption{Position angle of Gaussian jet components for Pictor~A (0518$-$458) versus
  radial distance from the core. The color coding indicates different
  components, and is the same as the upper left panel of Fig.~\ref{kin_dist}. The sign of the  counter-jet component CJ1 has been reversed to improve visualization.}
\label{pica_psi}
\end{figure}

We investigated the significance of a putative jet bending by plotting
the position angle of the Gaussian components versus the radial
distance. This is shown in Fig.~\ref{pica_psi}. The error on the position angle $\psi$ has been calculated adapting Eq.\,9 from \cite{2012A&A...537A..70S}, as:
\begin{equation}
    \Delta\psi = \arctan \left (
    \frac{a/2r}{\mathrm{SNR}} \times \frac{180}{\pi}
    \right )
\end{equation}

where $a$ is the component size, $r$ is the component radial distance and the SNR is defined as in Section~\ref{vlbi}.
There is no obvious
break in the distribution around $\sim10$ mas, in contrast with the
findings of \citet{2000AJ....119.1695T}.

\paragraph{0521$-$365}

This is a nearby AGN with an uncertain classification. \citet{2016A&A...586A..70L} derive limits on the jet viewing angle, speed, and Doppler
factor using the Atacama Large Millimeter Array (ALMA). Their results suggest a jet viewing angle in the range $16^{\circ}\le\theta\le38^{\circ}$, from the jet-to-counterjet ratio. Their detection of a large scale double structure already suggests that Doppler boosting effects in this source are not dominant on the kpc scale. \cite{2004ApJ...613..752G} studied PKS~0521$-$36 as part of a sample study on low-redshift BL Lacs. Using 5~GHz VLBA observations, the authors constrain the viewing angle of the source to lie in the range $21^{\circ}\le\theta\le27^{\circ}$.
\citet{2015MNRAS.450.3975D} constrain the same parameters
using SED modeling including $\gamma$-ray data, obtaining a more
aligned jet viewing angle of $6^{\circ}\le\theta\le15^{\circ}$. These results
point to an intermediate jet viewing angle between a blazar and a
steep spectrum radio quasar (SSRQ) or Broad Line Radio Galaxy (BLRG). However, due to its uncertain nature, PKS~0521$-$36 was not included in the large study of $\gamma$-ray properties of misaligned AGN by \cite{2010ApJ...720..912A}.

Previous VLBI observations performed with the VLBA and with the Southern Hemisphere
VLBI Experiment (SHEVE) at 4.9 GHz and 8.4 GHz provided an upper limit on the apparent speed of
jet components $\beta_\mathrm{app}<1.2$ \citep{2002AJ....124..652T}. This is also consistent
with the hypothesis that the jet of PKS~0521$-$36 is not strongly beamed.

Due to its known variability at $\gamma$-ray energies \citep{2015MNRAS.450.3975D}, PKS~0521$-$36 is one of the more densely monitored sources in the sample, with
nine TANAMI epochs, which provide an excellent data set for a kinematic
analysis. The full resolution maps and the corresponding image parameters are presented in Appendix~\ref{app:maps}.

Our TANAMI images show a faint jet
flow extending out to $\sim60$ mas from the core to the north-west (see Fig.~\ref{fig:wallpaper}).
There is a drop in the jet brightness around 10-15 mas from the core,
which is seen consistently in all epochs. The jet structure is
remarkably consistent across the epochs, hinting at a slow speed. The
extended jet is not detected in the last epoch due to the lack of the
shortest baseline ATCA-Mopra which provides the necessary sensitivity
to the larger scale emission.

\begin{figure}[h!tbp]
\begin{center}
\includegraphics[width=\linewidth]{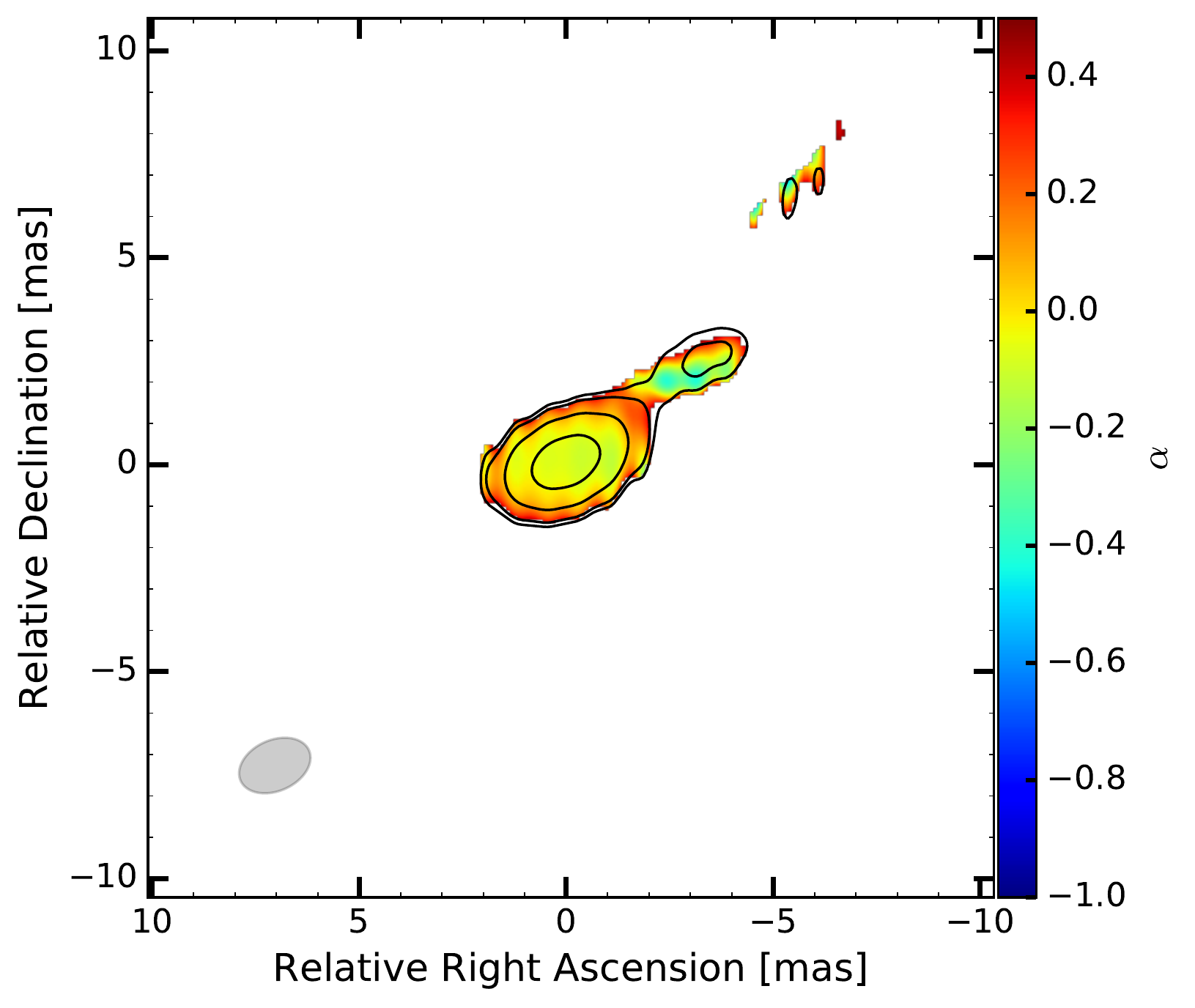}
\end{center}
\caption{Spectral index map of PKS~0521$-$36 (0521$-$365) between 8.4 GHz and 22.3 GHz for epoch 2008 Mar 28. Black contours are from the 8.4 GHz image. The
  convolving beam is represented in
  grey in the lower-left corner.}
\label{0521_spix}
\end{figure}

We produced a spectral index map between the quasi-simultaneous 8.4~GHz (epoch 2008 Mar 28) and 22.3~GHz (epoch
2008 Mar 26) images, which is presented in Fig.~\ref{0521_spix}.

\paragraph{0625$-$354}

This is an FR~I radio galaxy, but shows an optical spectrum similar to a
BL Lac object. Its $\gamma$-ray properties (see Section~\ref{lat_results}) also suggest a moderately
aligned jet, similar (but less extreme) to the case of IC~310 \citep{2014Sci...346.1080A}. PKS~0625$-$35 is also one of the very few radio galaxies to have been detected at TeV energies by Cherenkov telescopes \citep{2018MNRAS.476.4187A}.

TANAMI is the first multi-epoch VLBI data set for this source,
providing 9 epochs. A previous
single-epoch observation with the LBA at 2.29 GHz by \citet{2000A&A...363...84V} provided
constraints on the jet angle to the line of sight and intrinsic jet speed
using a lower limit on the jet-to-counterjet ratio and an estimate of
the core dominance. The latter method gives the most constraining
estimates of $\theta\le43^{\circ}$ and $\beta\ge0.74$.

\begin{figure}[h!tbp]
\begin{center}
\includegraphics[width=\linewidth]{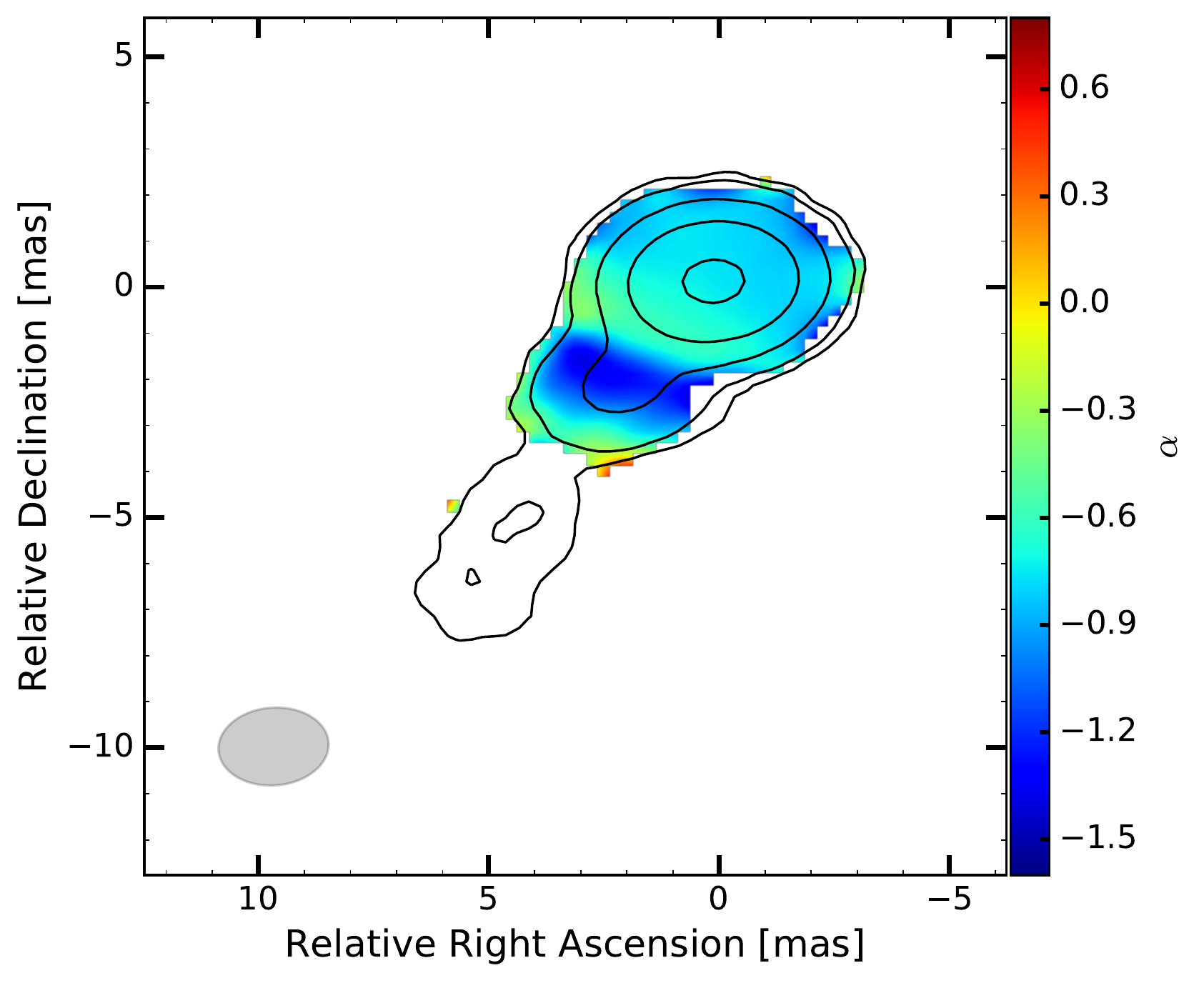}
\end{center}
\caption{Spectral index map of PKS~0625$-$35 (0625$-$354) between 8.4 GHz and 22.3 GHz for epoch 2008 Nov 27. The black contours are from the 8.4 GHz image. The convolving beam is represented in grey in the lower-left corner.}
\label{0625_spix}
\end{figure}

Our TANAMI images show a one sided core-jet structure with a faint extended jet in the south-east direction (see Fig.~\ref{fig:wallpaper}). We produce a spectral index map between 8.4~GHz (epoch 2008 Nov 27) and
22.3~GHz (epoch 2008 Nov 29), which is presented in Fig.~\ref{0625_spix}. The index is
relatively steep even in the brightest region.

  \paragraph{1343$-$601 (Centaurus B)}
This classic FR~I radio galaxy was added to the TANAMI sample after
being detected by \textit{Fermi}-LAT in the second source catalog \citep{2012ApJS..199...31N}. TANAMI imaging reveals a smooth, one-sided jet extending in the south-west direction (see Fig.~\ref{fig:wallpaper}). Only two epochs are currently
available, which is insufficient to perform a robust kinematic
analysis. Using one well-defined jet component, we can obtain a rough upper limit on the jet apparent speed, which is lower than $0.89c$. 

\paragraph{1718$-$649}
This is one of the most classic examples of the Compact Symmetric Object
(CSO) and Gigahertz-Peaked Spectrum (GPS) source classes \citep[see e.g.,][]{1998PASP..110..493O,2009AN....330..193G}, i.e, a young
radio galaxy. These sources are typically small (linear size $<$ 1 kpc), and the radio emission is
dominated by symmetric mini-lobes which can show hot-spots similar to
the large scale lobes of FR~II radio galaxies. The advance speed of
these hot-spots provides a kinematical age estimate for these sources, which is
typically $t_\mathrm{age}<10^5$ yr \citep{1997AJ....113.2025T}.

TANAMI provides the first multi-epoch data set for this well-studied
source, and therefore the first opportunity for a measurement of its
kinematical age. The TANAMI images shows a clear structure of two components separated by $\sim8$ mas (see Fig.~\ref{fig:wallpaper}), which is consistent across the epochs.

\begin{figure}[h!tbp]
\begin{center}
\includegraphics[width=\linewidth]{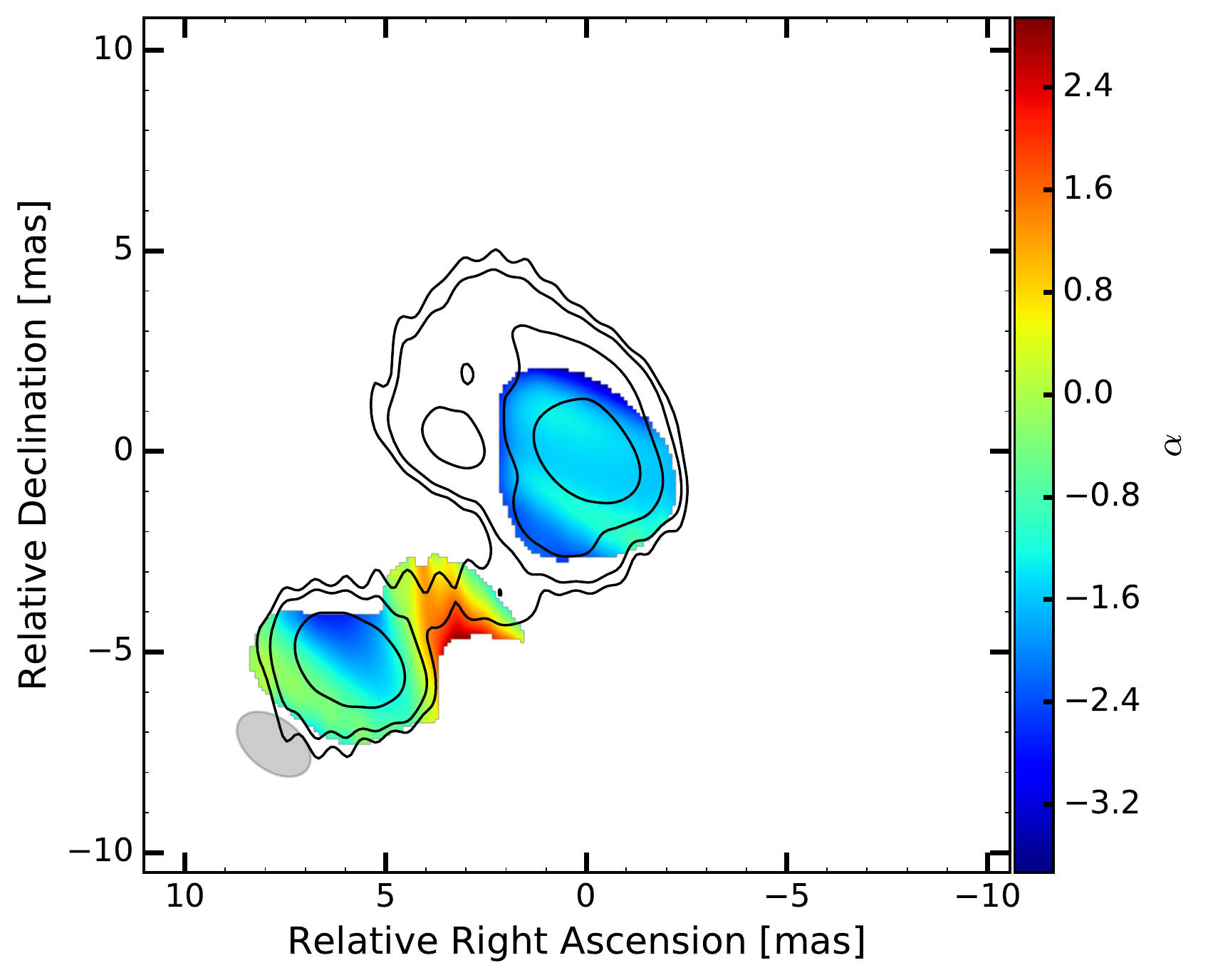}
\end{center}
\caption{Spectral index map of PKS~1718$-$649 between 8.4 GHz and 22.3 GHz for epoch 2008 Feb 07. The black contours are from the 8.4 GHz image. The convolving beam is represented in grey in the lower-left corner.}
\label{1718_spix}
\end{figure}

We present a spectral index map between 8.4~GHz and 22.3~GHz in Fig.~\ref{1718_spix}. The spectral morphology suggests that the core of the young radio source is strongly absorbed at these frequencies, and is located between the two emission components, corresponding to the region of highly inverted spectral index.

  \subsection{Radio kinematic analysis results}
  \label{kin_results}

\begin{figure*}[!htbp]
\begin{center}
\includegraphics[width=0.495\linewidth]{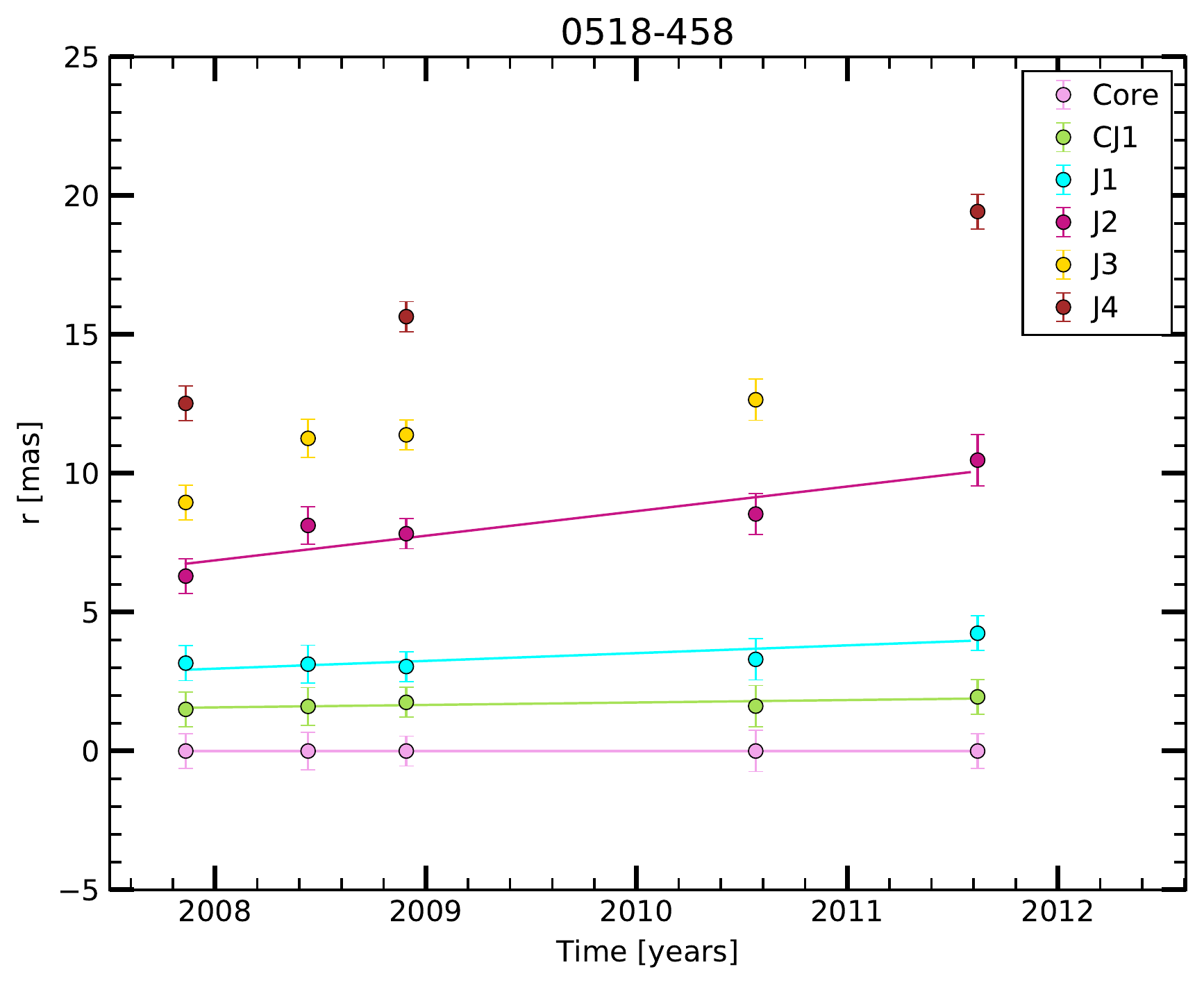}
\includegraphics[width=0.495\linewidth]{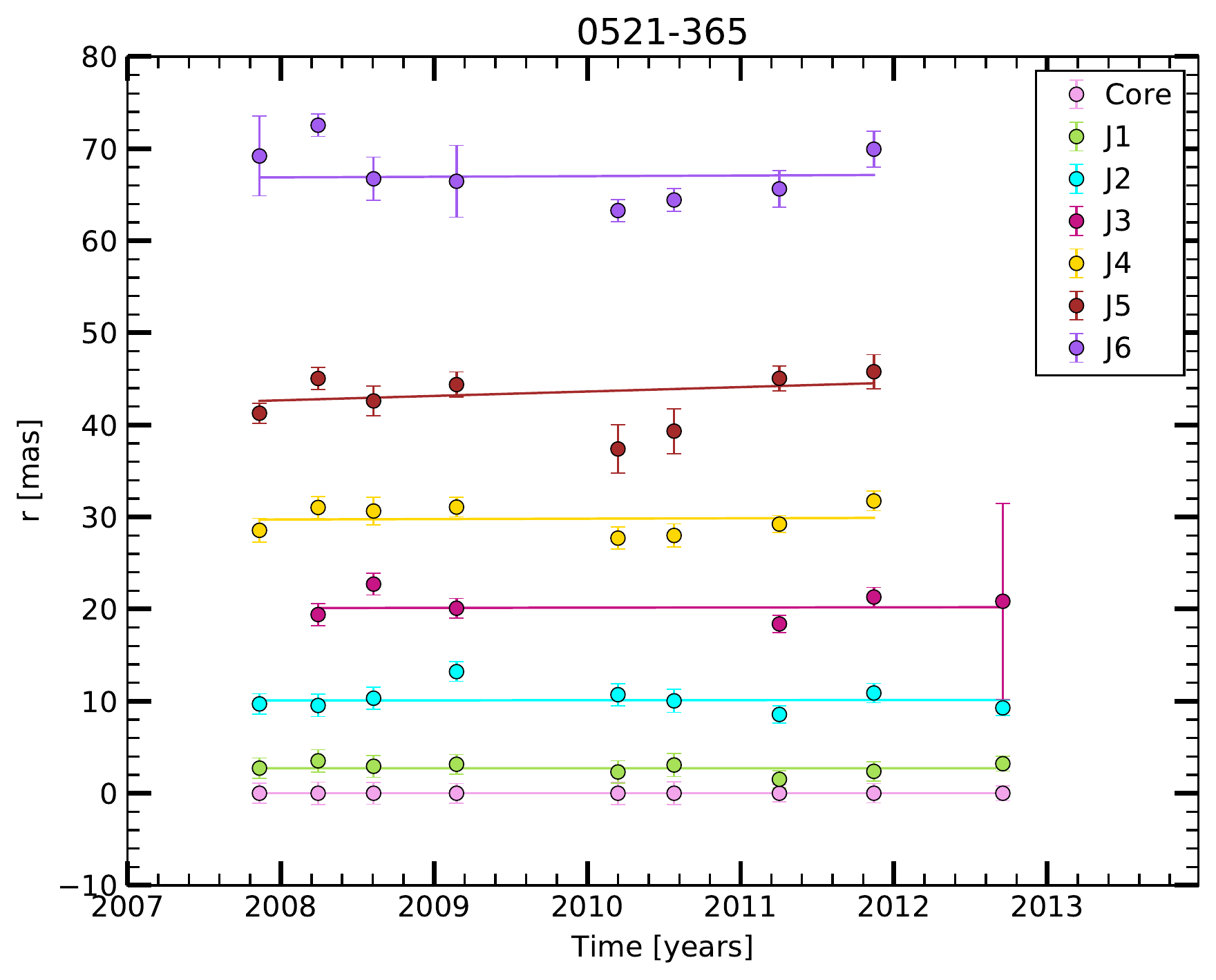}
\includegraphics[width=0.495\linewidth]{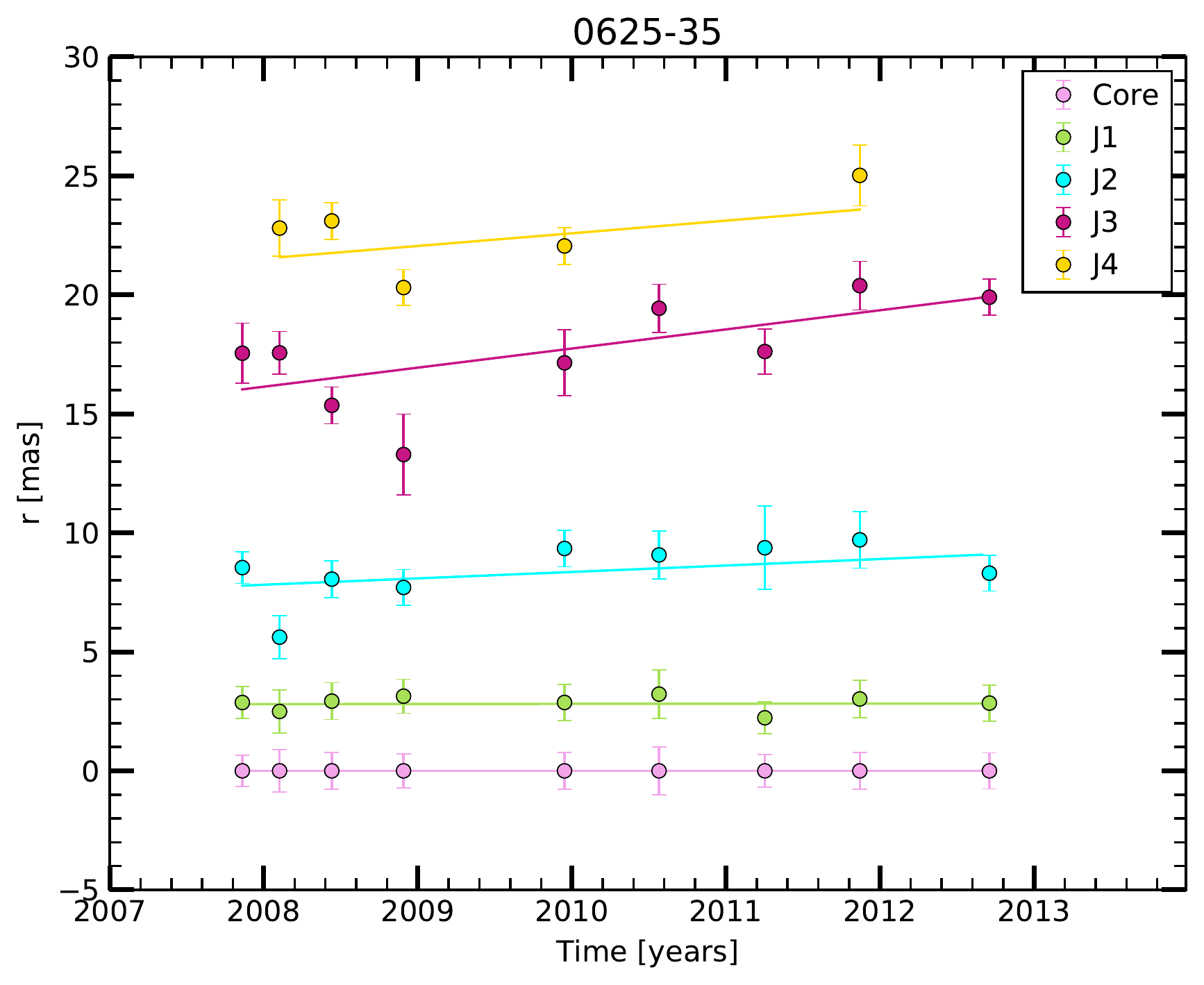}
\includegraphics[width=0.495\linewidth]{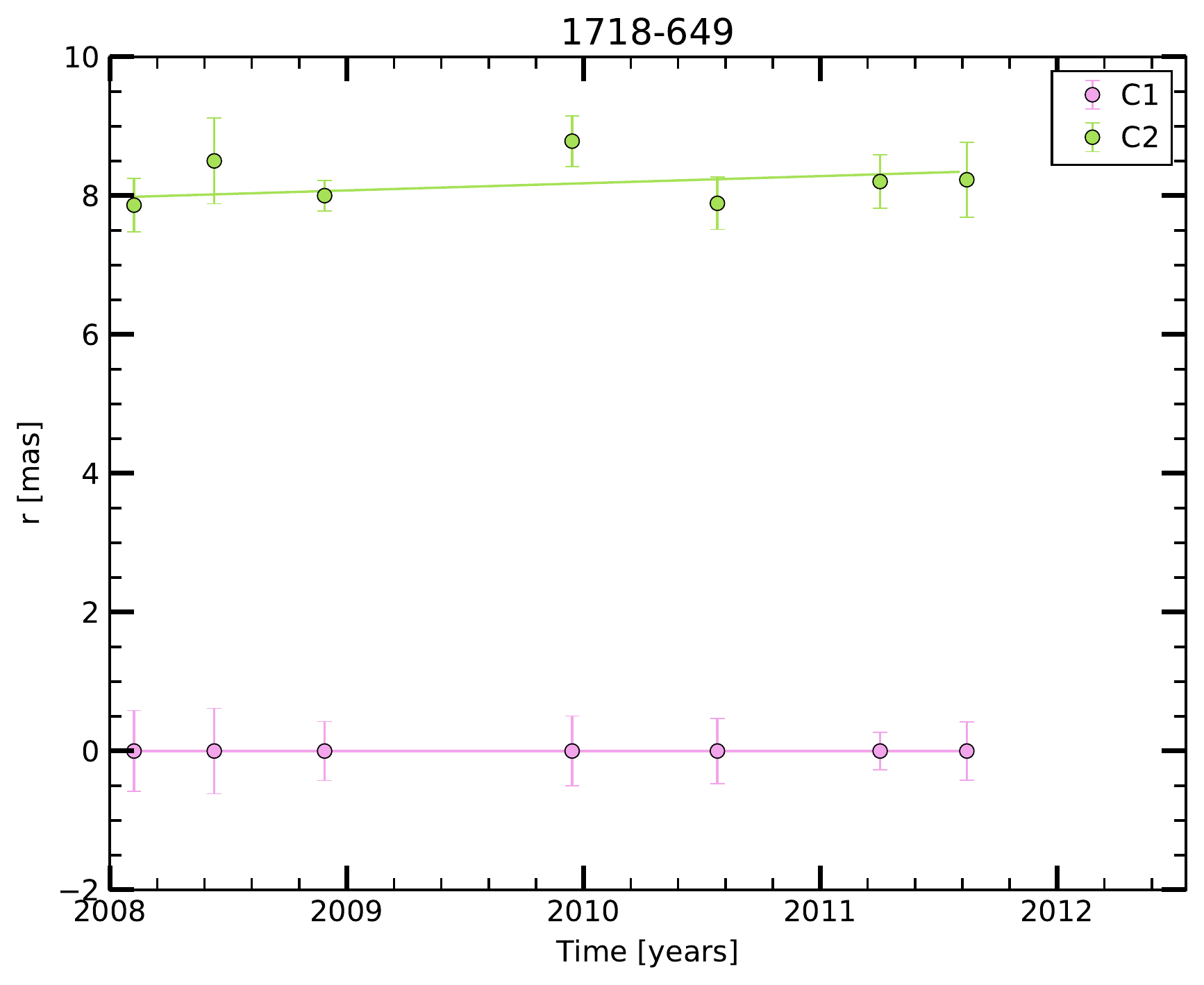}
\end{center}
\caption{Jet kinematics of our radio galaxies: core distance of jet features as a function of time. The solid lines represent a least squares fit to their positions (the slope is the apparent speed). Top left to bottom right: Pictor~A, PKS~0521$-$36, PKS~0625$-$35, PKS~1718$-$649. The B1950 name is given above each image.}
\label{kin_dist}
\end{figure*}

\begin{figure}[!htbp]
\begin{center}
\includegraphics[width=\linewidth]{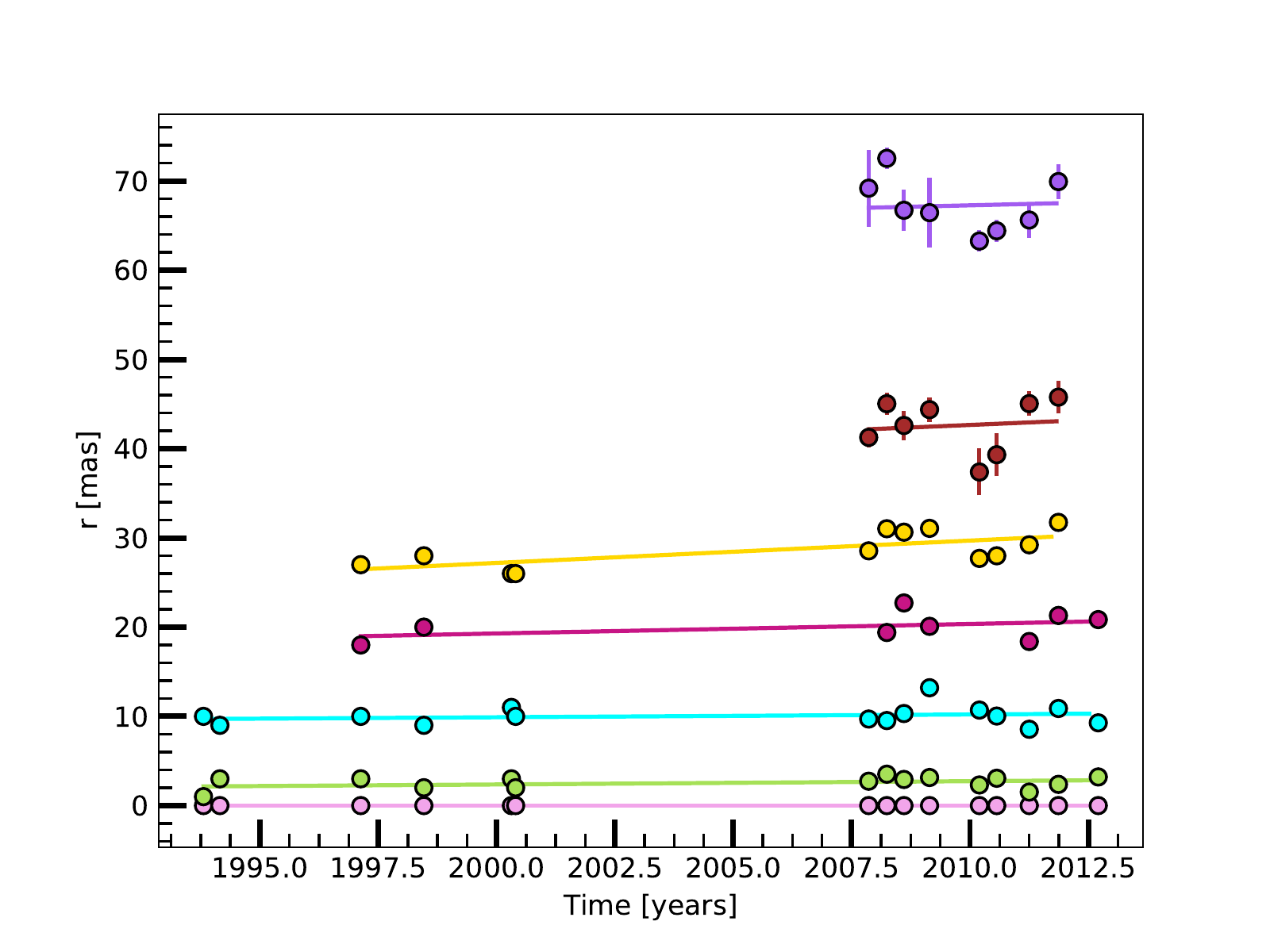}
\end{center}
\caption{Jet kinematics of PKS~0521$-$36 (0521$-$365): core distance of jet features as a function of time, including the previous VLBI dataset of \cite{2002AJ....124..652T}. The solid lines represent a least squares fit to their positions (the slope is the apparent speed). Compare with top-right panel of Fig.~\ref{kin_dist}.}
\label{0521_wold}
\end{figure}

\begin{figure*}[!htbp]
\begin{center}
\includegraphics[width=0.49\linewidth]{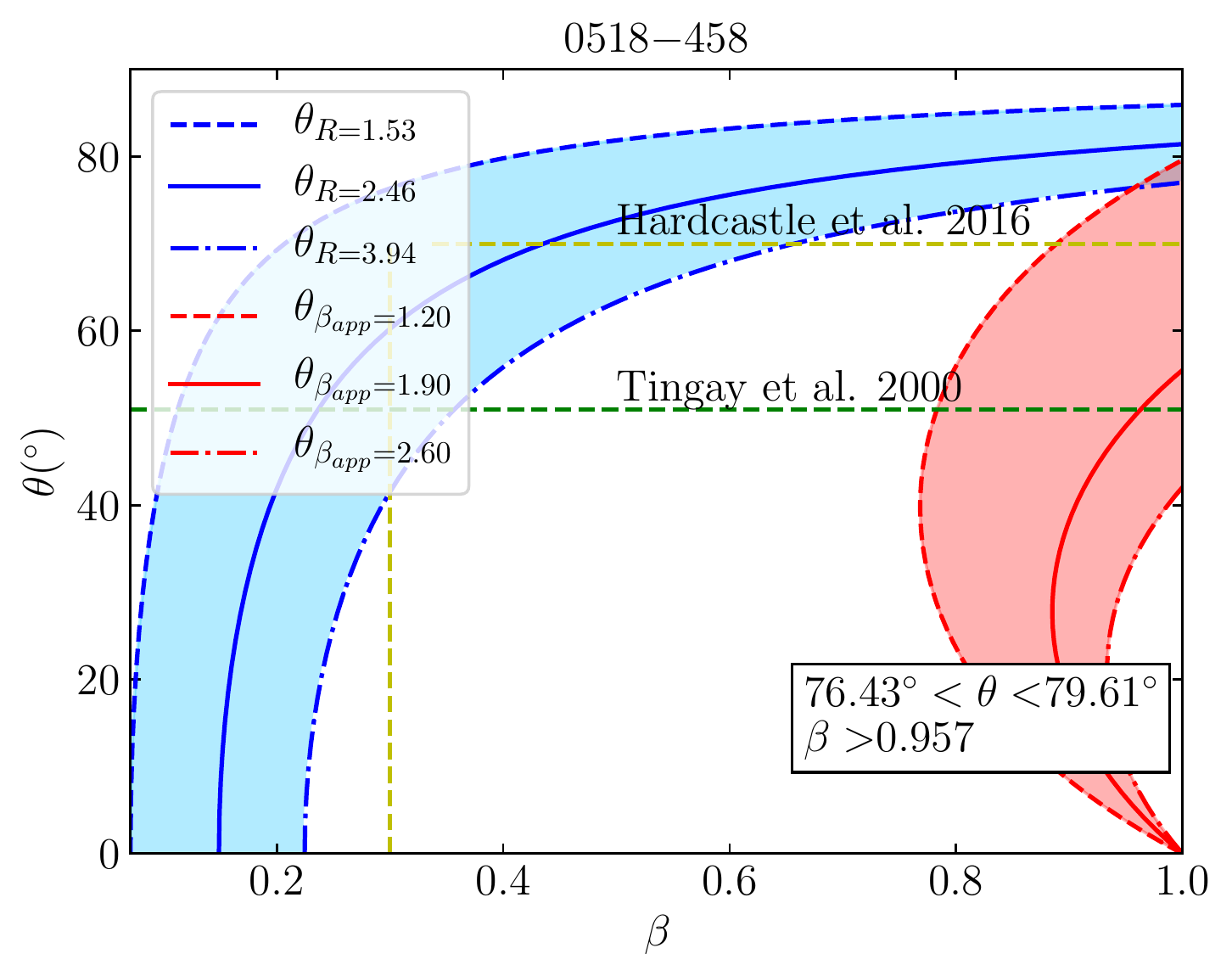}
\includegraphics[width=0.49\linewidth]{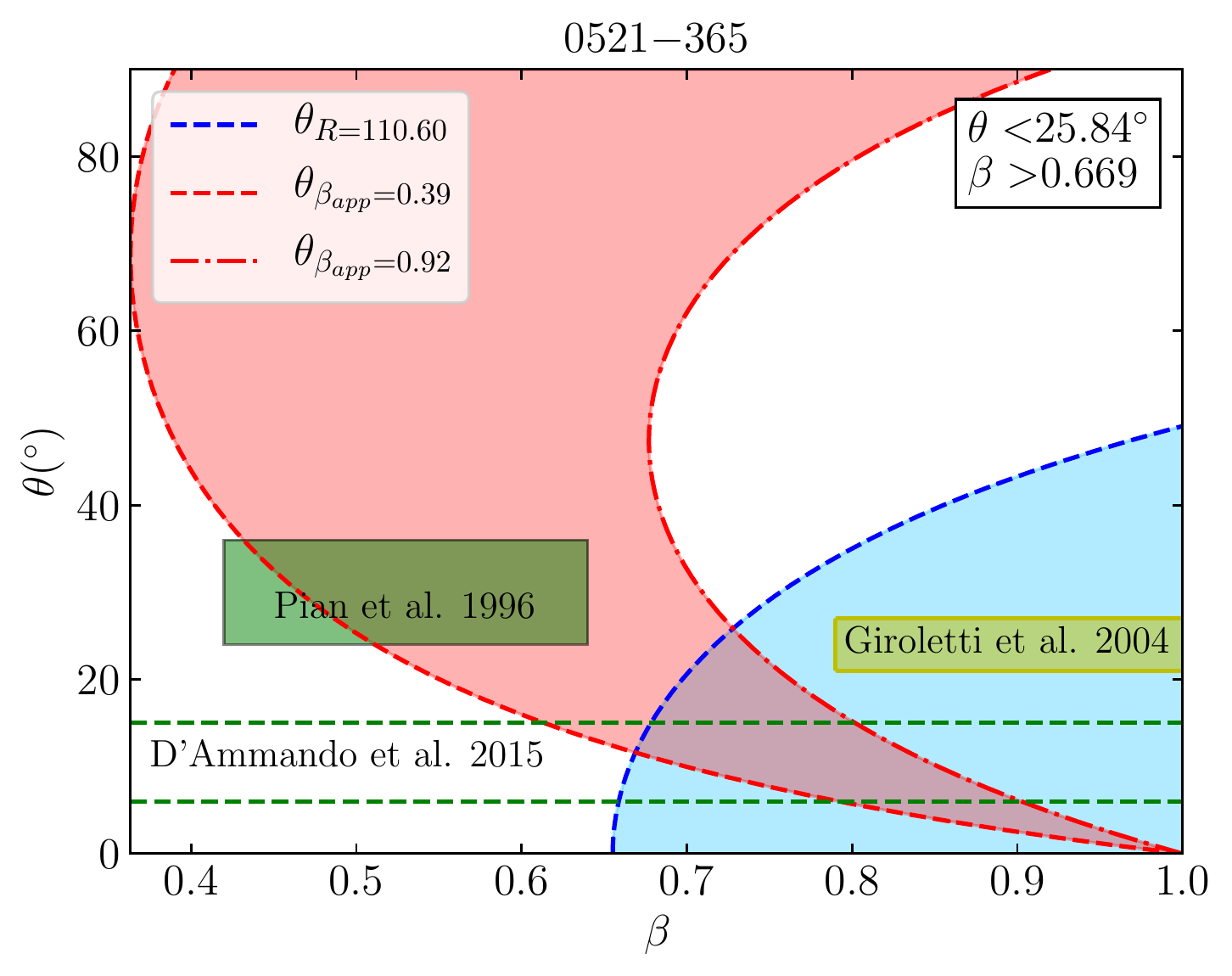}
\includegraphics[width=0.49\linewidth]{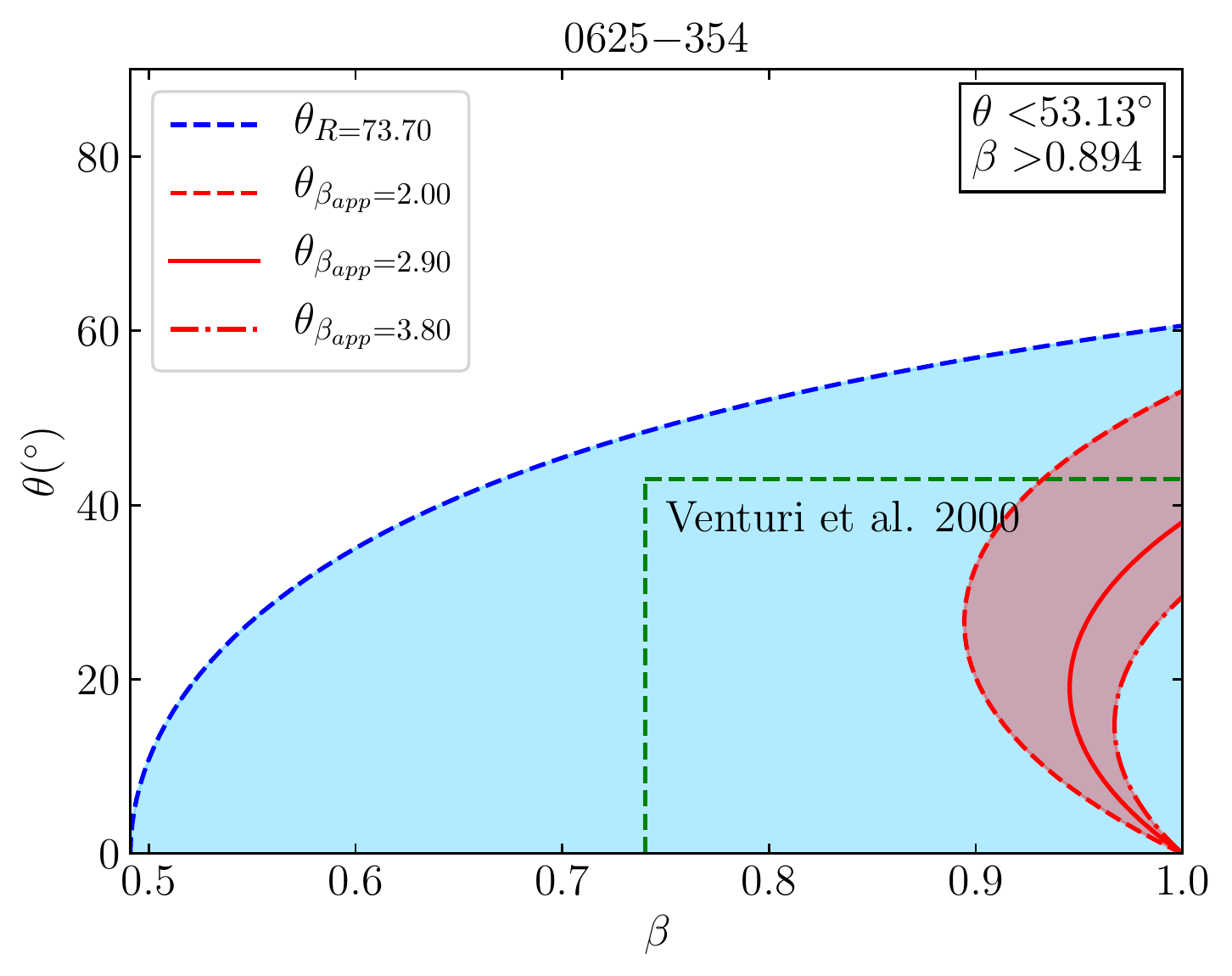}
\end{center}
\caption{Parameter space of intrinsic jet speed $\beta$ and viewing
  angle $\theta$ allowed by our observations. The blue shaded area is
  the one allowed by the measurement of $R$ ($\theta$ as function of $\beta$ given $R$, i.e., $\theta_R$), while the red shaded area
  is the one allowed by the observed $\beta_\mathrm{app}$  ($\theta$ as function of $\beta$ given $\beta_\mathrm{app}$, i.e., $\theta_{\beta_\mathrm{app}}$). For each source we provide a minimum, maximum and (except for PKS~0521$-$36) a central estimate of $R$ and $\beta_\mathrm{app}$. The top-right legend reports the resulting limits on $\theta$ and $\beta$. The dashed colored lines and boxes indicate constraints from previous works, namely \citet{2016MNRAS.455.3526H,2000AJ....119.1695T} for Pictor~A (0518$-$458), \citet{1996ApJ...459..169P,2004ApJ...613..752G,2015MNRAS.450.3975D} for PKS~0521$-$36 (0521$-$365), and \citet{2000A&A...363...84V} for PKS~0625$-$35 (0625$-$354).}
\label{beta_theta}
\end{figure*}

The main results of the kinematic analysis for Pictor~A, PKS~0521$-$36, PKS~0625$-$35 and PKS~1718$-$649, following the procedure described in Section~\ref{vlbi}, are summarized by the plots of the identified Gaussian component's core separation versus time (Fig.~\ref{kin_dist},~\ref{0521_wold}). The resulting values for the component speed and estimated ejection date are listed in Tables~\ref{kin_tab_pica} through~\ref{kin_tab_0625}. Only components with at least five epochs are listed, and ejection dates are given only for sources with speed not consistent with zero. Finally, the parameter space of intrinsic jet parameters $\beta_\mathrm{app}$ and $\theta$ allowed by our results is illustrated in Fig.~\ref{beta_theta}, for sources with significant component motion. Additional information illustrating the kinematic analysis results is provided in Appendix~\ref{app:kin}.

   \paragraph{0518$-$458 (Pictor A)}
We re-imaged the source applying a Gaussian taper in order to match
the resolution of epoch 2010 Jul 24 (see Fig.~\ref{pica_full}).


\begin{table}[h!tbp]
\caption{Results of the kinematic analysis of Pictor~A (0518$-$458).}             
\label{kin_tab_pica}  
\begin{center}    
\begin{tabular}{ccccc}    
\hline\hline 
ID & $\mu$ (mas/yr) & $\beta_\mathrm{app}$ & Ej. date & \# ep.\\
\hline
CJ1 & 0.1$\pm$0.2 & 0.2$\pm$0.4 & * & 5\\
J1 & 0.3$\pm$0.2 & 0.6$\pm$0.5 & 1997$\pm$9 & 5\\
J2 & 0.8$\pm$0.3 & 1.9$\pm$0.7 & 2000$\pm$3 &5\\
\hline  
\end{tabular}
\end{center}
\end{table}

We find at least mildly relativistic apparent speeds with a
minimum significant value of $0.1c$ (J1) and a maximum of $2.6c$
(J2), as allowed within the $1\sigma$ errors. The lower limits are
consistent with the estimates in \citet{2000AJ....119.1695T}, while the
increased number of epochs allows us to reveal one component which is not consistent with subluminal motion (currently at a significance level of $\sim1.3\,\sigma$) in the jet of Pictor~A.

The upper left panel of Fig.~\ref{beta_theta} shows the resulting limits on
the intrinsic jet speed and viewing angle for Pictor~A obtained via
the combination of the apparent speed information with the
jet-to-counterjet ratio. The estimates for $R$ represent the minimum (dashed blue line), mean (continuous
blue line) and maximum (dot-dashed blue line) values. The central estimate
for $\beta_\mathrm{app}$ is the one for the fastest component, while the
minimum and maximum values represent its error. The relatively small value of the jet ratio combined with the mildly superluminal apparent speed results in a tight constrain on the viewing angle, which lies in the range $76^{\circ}< \theta <80^{\circ}$. To account for the observed apparent speed with such a large jet angle, the intrinsic jet speed should be $\beta>0.96$.

   \paragraph{0521$-$365}
We re-imaged all epochs applying a Gaussian taper, in order to
approximately match the resolution of epoch 2010 Mar 12 (see Fig.~\ref{0521_full_a}).
   
\begin{table}[h!tbp]
\caption{Results of the kinematic analysis of PKS~0521$-$36 (0521$-$365).}             
\label{kin_tab_0521}  
\begin{center}    
\begin{tabular}{ccccc}    
\hline\hline 
ID & $\mu$ (mas/yr) & $\beta_\mathrm{app}$ & Ej. date & \# ep.\\
\hline
J1 & 0.003$\pm$0.006 & 0.01$\pm$0.02 & * & 9\\
J2 & 0.01$\pm$0.005 & 0.04$\pm$0.02 & * &  9\\
J3 & 0.020$\pm$0.009 & 0.07$\pm$0.03 & * &  6\\
J4 & 0.03$\pm$0.02 & 0.11$\pm$0.06 & * &  8\\
J5 & 0.5$\pm$0.2 & 1.9$\pm$0.9 & 1950$\pm$23 &  8\\
J6 & 0.0665$\pm$0.0009 & 0.242$\pm$0.003 & * &  8\\
\hline  
\end{tabular}
\end{center}
\end{table}

Since we obtain slow jet speeds, we attempted to fit our
TANAMI jet model together with the previous 8.4 GHz VLBI dataset from
\cite{2002AJ....124..652T}. It is possible to cross-identify and fit four
components between the two datasets. This is shown in
Fig.~\ref{0521_wold}. The updated values of $\mu$ and $\beta_\mathrm{app}$ for these
components are reported in Table~\ref{kin_tab_0521_wold}. We note that the measured apparent speed values are significantly larger when using a longer time range for the kinematic analysis. This indicates that the limited time coverage of our TANAMI observations may lead us to underestimate the apparent speed in the case of slow apparent motions, which typically require a larger time range to be adequately constrained~\citep[see e.g.,][]{2018ApJ...853...68P}.

\begin{table}[h!tbp]
\caption{Results of the kinematic analysis of PKS~0521$-$36 (0521$-$365) for the
  components cross-identified with the \cite{2002AJ....124..652T}
  dataset.}             
\label{kin_tab_0521_wold}  
\begin{center}    
\begin{tabular}{cccc}    
\hline\hline 
Component ID & $\mu$ (mas/yr) & $\beta_\mathrm{app}$ & Epochs\\
\hline
J1 & 0.04$\pm$0.05 & 0.16$\pm$0.16 & 15\\
J2 & 0.04$\pm$0.04 & 0.13$\pm$0.16 & 15\\
J3 & 0.11$\pm$0.07 & 0.4$\pm$0.3 & 8\\
J4 & 0.25$\pm$0.07 & 0.9$\pm$0.3 & 12\\
\hline  
\end{tabular}
\end{center}
\end{table}

The upper right panel of Fig.~\ref{beta_theta} shows the resulting limits on
the intrinsic jet speed and viewing angle for PKS~0521$-$36. Since
there is no counterjet, we place a lower limit on $R$ as described in Section~\ref{vlbi}. For $\beta_\mathrm{app}$ the values adopted are the minimum and
maximum observed values, considering the four components that are
cross-identified with the \cite{2002AJ....124..652T} dataset. We obtain a range of $\beta>0.67$ and $\theta<26^\circ$, respectively.

   \paragraph{0625$-$354}
The outer jet of this source is too faint to be modeled reliably with
Gaussian components, therefore we re-imaged only the inner $\sim20$
mas for the kinematic analysis.


\begin{table}[h!tbp]
\caption{Results of the kinematic analysis of PKS~0625$-$35 (0625$-$354).}             
\label{kin_tab_0625}  
\begin{center}    
\begin{tabular}{ccccc}    
\hline\hline 
ID & $\mu$ (mas/yr) & $\beta_\mathrm{app}$ & Ej. date & \# ep.\\
\hline
J1 & 0.01$\pm$0.18 & 0.0$\pm$0.6 & * & 9\\
J2 & 0.4$\pm$0.2 & 1.4$\pm$0.7 & 1991$\pm$8 & 9\\
J3 & 0.8$\pm$0.3 & 2.9$\pm$0.9 & 1989$\pm$5 & 9\\
J4 & 0.6$\pm$0.3 & 2.0$\pm$1.1 & 1971$\pm$14 & 5\\
\hline  
\end{tabular}
\end{center}
\end{table}

\cite{2012evn..confE..20M} presented a preliminary kinematic analysis
of the TANAMI data on PKS~0625$-$35, using the first six epochs. They
found a highest apparent speed of $\beta_\mathrm{app} = 3.0\pm0.5$, which is
consistent with our result (component J3), including three more
epochs. Our results provide a robust confirmation of superluminal component motion in
the pc-scale jet of PKS~0625$-$35. 

The lower panel of Fig.~\ref{beta_theta} shows the limits on
the intrinsic jet speed and viewing angle for PKS~0625$-$35 resulting
from our observations. In this case, again, we do not detect a counterjet, therefore we place a lower limit on $R$. The estimates for
$\beta_\mathrm{app}$ are given by the fastest observed speed and its uncertainty. Our observations limit the intrinsic jet parameters to $\beta>0.89$ and $\theta<53^\circ$.

\paragraph{1343$-$601 (Cen~B)}

\begin{figure}
\begin{center}
\includegraphics[width=0.9\linewidth]{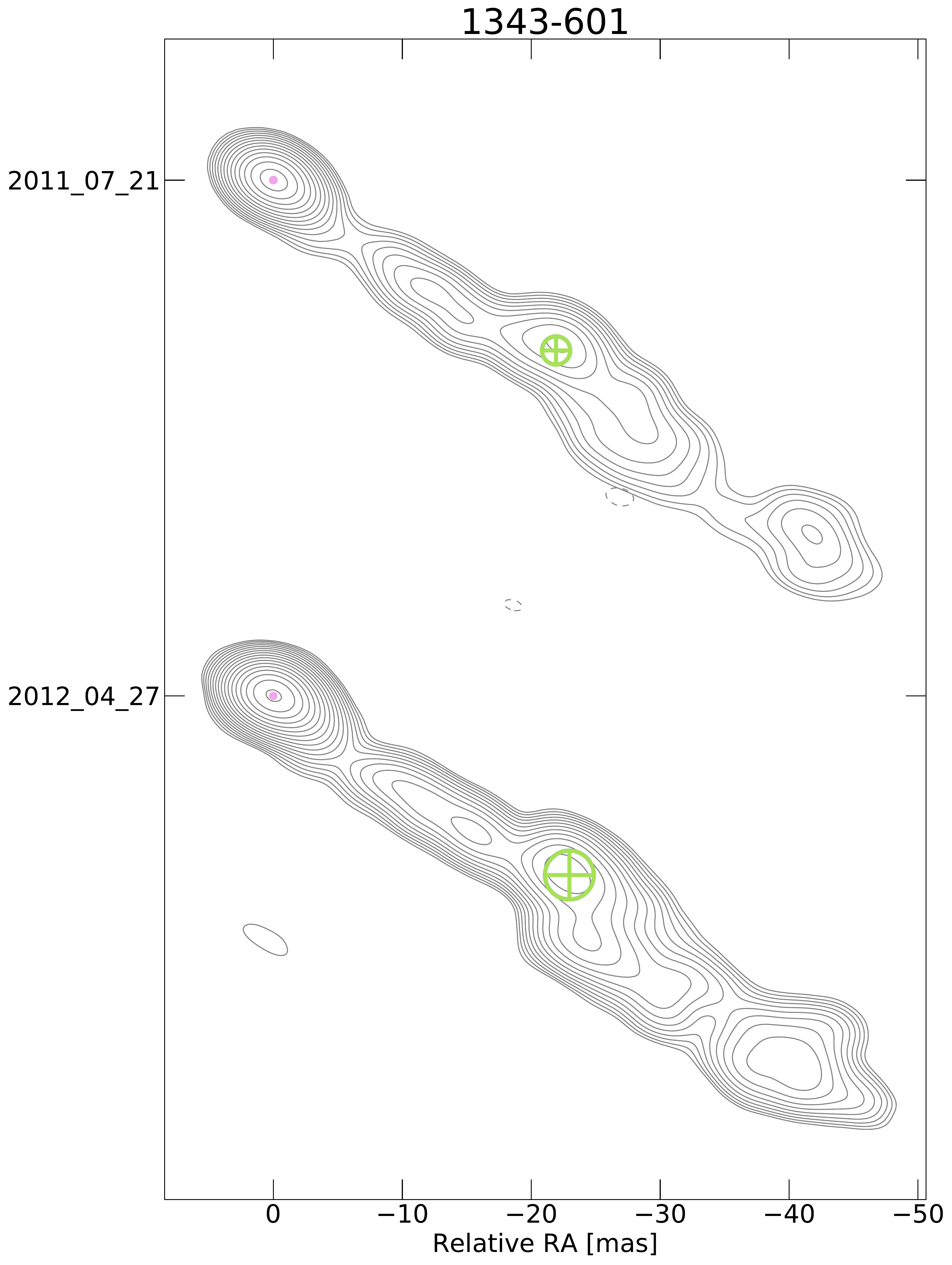}
\end{center}
\caption{Multi-epoch images of Centaurus~B (1343$-$601). The colored crossed circles represent the circular Gaussian components that have been fitted to the clean maps.}
\label{kina_cenb}
\end{figure}

As mentioned above, there are only two available TANAMI epochs for Cen~B, separated by about nine months. Although it is not possible to perform a detailed kinematic analysis, as for the other sources, the data can give some indications regarding the presence or absence of motions in the source. We tentatively identify a local maximum in the brightness distribution of the Cen~B VLBI jet at a distance of $\sim25$ mas downstream of the core in both our images, which were taken approximately nine months apart (see Fig.~\ref{kina_cenb}). If this association is correct, which will be tested by forthcoming TANAMI epochs, the apparent jet speed is likely subluminal or at most mildly superluminal.

   \paragraph{1718$-$649}

In this case the source is more complex than the classic core-jet morphology seen
in the other radio galaxies presented here. We reference the kinematic
analysis at the position of the brightest component (C1), even though it is
not the core of the radio source in this case. The resulting kinematic
values represent the evolution of the distance between the two
components consistently detected in all epochs. We have preferred such a simple two-component modeling of the brightness distribution, although finer structures are clearly seen in certain epochs. However, due to differences in image noise and dynamic range across the epochs, an attempt to identify and track such faint sub-structures in a kinematic analysis would yield unreliable results. We therefore chose a more simplistic modeling approach.

In the lower-right
panel of Fig.~\ref{kin_dist} we plot the distance between the two
components (referenced to the brightest one) as a function of time,
and the corresponding linear fit. The angular separation speed is
$\mu=(0.13\pm0.06)$ mas/yr, and the corresponding apparent linear
speed is $\beta_{app}=0.13\pm0.06$. We are therefore able to estimate when the young radio source first ejected its two symmetric component, and find a zero-separation epoch of 1963$\pm$22.


\clearpage

\clearpage
\subsection{\textit{Fermi}-LAT results}
\label{lat_results}

   \paragraph{0518$-$458}
Pictor~A was reported as a \textit{Fermi}-LAT detection for the first time by
\cite{2012MNRAS.421.2303B} based on three years of data, and confirmed
in the  3FGL. It is the faintest source among the 3FGL TANAMI radio
galaxies. While it has been established that diffuse structures such
as lobes and hot spots may give a significant or even dominant
contribution to $\gamma$-ray emission in radio galaxies
\citep{2010Sci...328..725A,2016ApJ...826....1A},
\cite{2012MNRAS.421.2303B} found that this is not the case for
Pictor~A. Through SED modeling of the western hot-spot, the authors
found that Synchrotron Self-Compton (SSC) and Inverse Compton with CMB
photons (IC/CMB) models failed to reproduce the X-ray and
$\gamma$-ray data at the same time, and therefore concluded that the
$\gamma$-ray emission must be dominated by the AGN jet. This is also
supported by the indications of variability observed in the LAT data.

Our analysis over $\sim8.5$ years results in a higher significance
w.r.t. the 3FGL, and the source properties are consistent with
the catalog results within the errors (see Table~\ref{rgs-lat}). 

\begin{figure*}[!htbp]
\begin{center}
\includegraphics[width=0.8\linewidth]{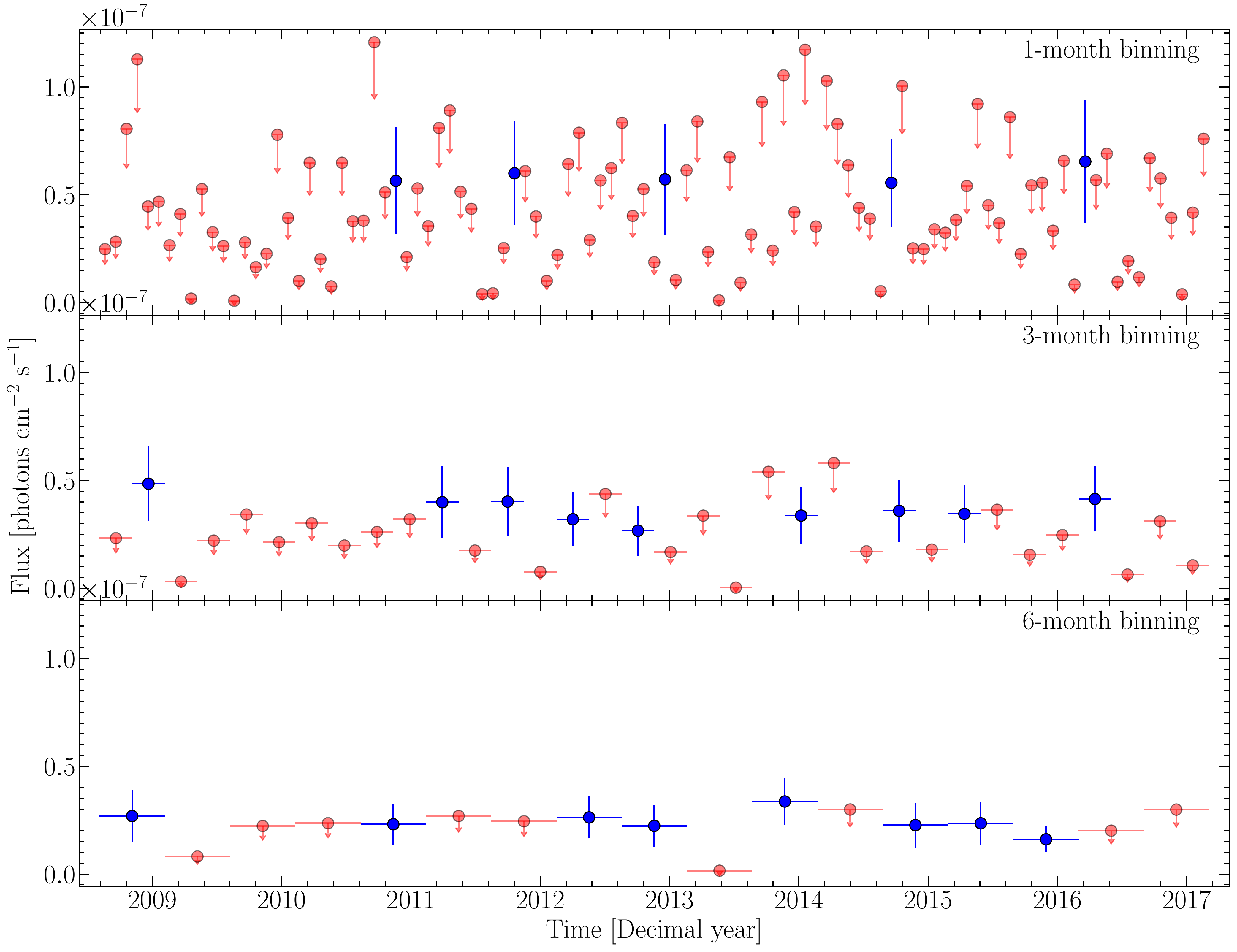}
\end{center}
\caption{Light curve of Pictor~A (0518$-$458) between 0.1-300 GeV over 103
  months of \textit{Fermi}-LAT data. Blue points are
  detections, red arrows are upper limits. Top to bottom: 1-month, 3-months and 6-months binning, respectively.}
\label{pica_lc}
\end{figure*}

We produced a monthly light curve over the whole $\sim8.5$ years time
period, as described in Section~\ref{fermi} (see top panel of Fig.~\ref{pica_lc}). Pictor~A is a relatively
faint $\gamma$-ray source, therefore it is not detected on monthly
timescales for most of this time range. The high state in
September 2010 provided the statistics to allow a detection
integrating over the first three years of \textit{Fermi}-LAT data, with a monthly significance of TS\,$\sim40$. In order to better characterize the variability of Pictor~A, we also produced 3-month and 6-month-binned light curves, shown in the center and bottom panel of Fig.~\ref{pica_lc}, respectively. Integrating over longer timescales allows us to detect the source for a larger fraction of the time range (27\% and 47\%, respectively).

\begin{figure*}[!htbp]
\begin{center}
\includegraphics[width=\linewidth]{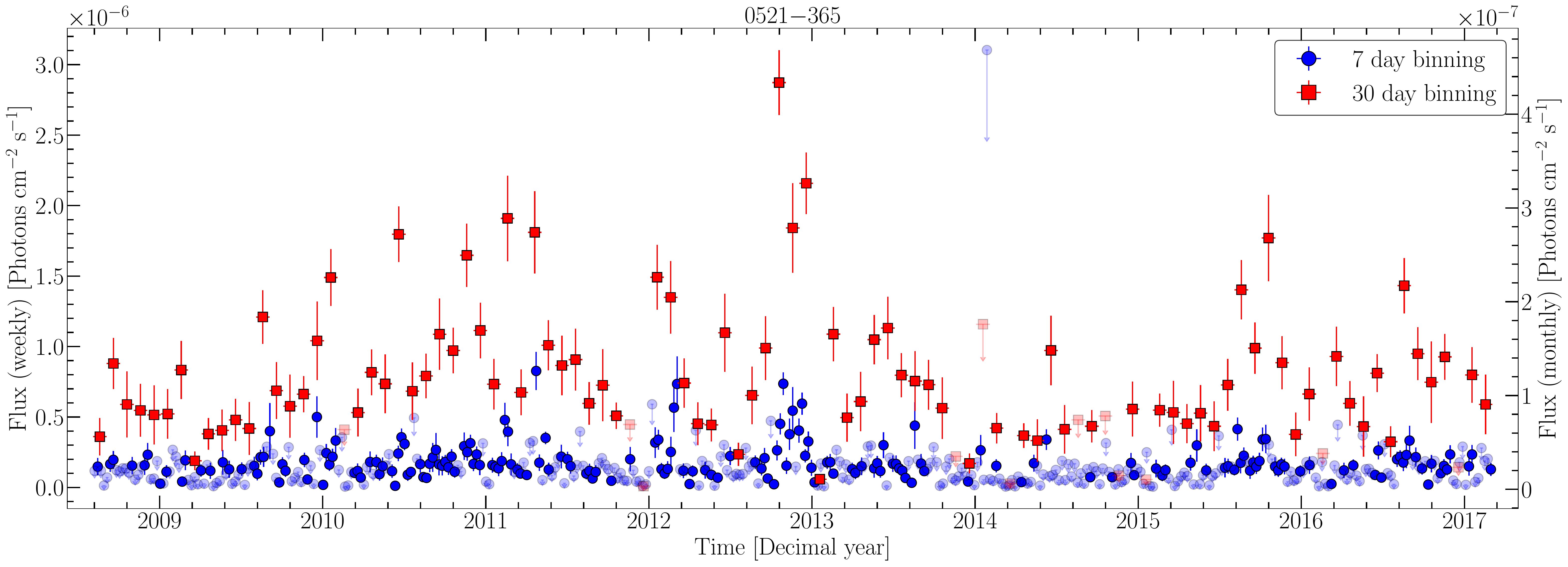}
\end{center}
\caption{Light curve of PKS~0521$-$36 (0521$-$365) between 0.1-300 GeV over 103
  months of \textit{Fermi}-LAT data, with weekly (blue points) and monthly (red
  points) binning. Upper limits are indicated by arrows of the
  respective colors. The left y-axis reports the weekly flux values,
  the right one reports the monthly flux values.}
\label{0521_lc}
\end{figure*}

\begin{figure*}[!htbp]
\begin{center}
\includegraphics[width=\linewidth]{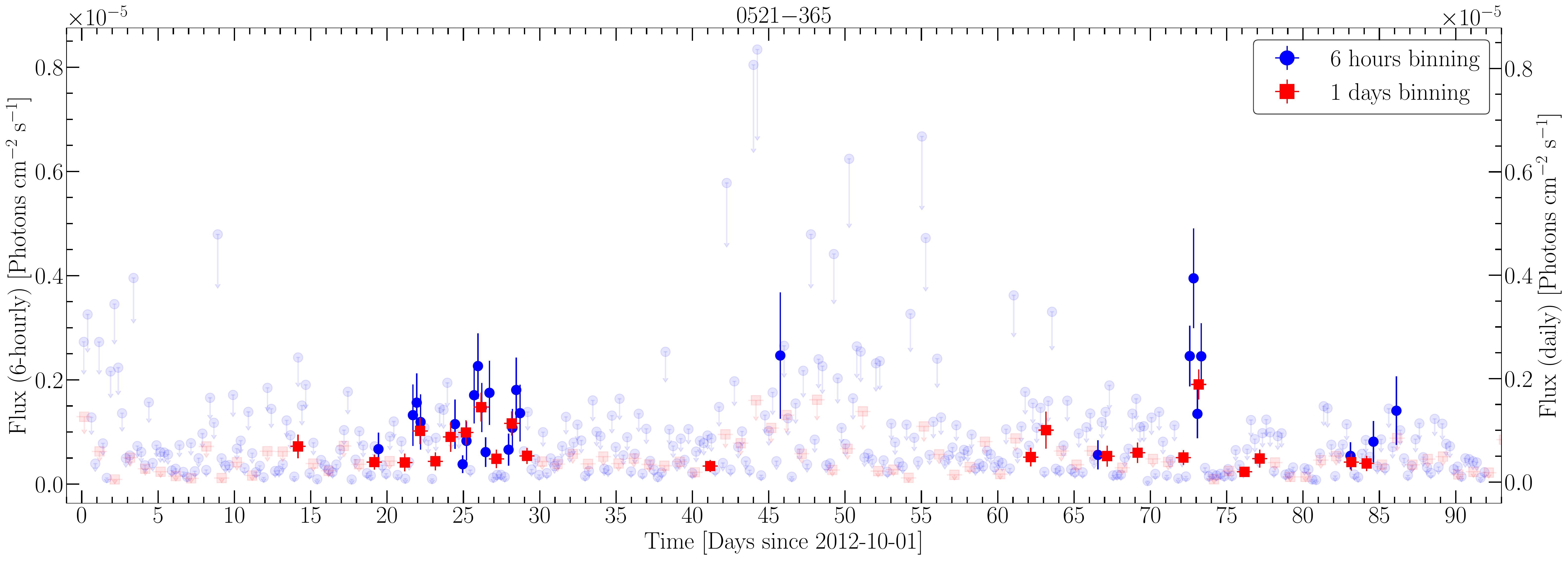}
\end{center}
\caption{Light curve of PKS~0521$-$36 (0521$-$365) between 0.1-300 GeV over 103
  months of \textit{Fermi}-LAT data, with 6-hours (blue points) and daily (red
  points) binning. Upper limits are indicated by arrows of the
  respective colors. The left y-axis reports the 6-hours flux values,
  the right one reports the daily flux values.}
\label{0521_lc2}
\end{figure*}

   \paragraph{0521$-$365}
This source is a very bright $\gamma$-ray emitter, showing significant
variability. Monthly and weekly light curves over $\sim8.5$ years of data are shown in
Fig.~\ref{0521_lc}. The flaring activity in 2010-2011
has been studied by \citet{2015MNRAS.450.3975D} down to 12-hour time
scales. We find a second
period of activity of comparable magnitude at the end of 2012. We
investigated the short variability in this period by producing daily
and 6-hours time scale light curves, which are presented in
Fig.~\ref{0521_lc2}. The source is especially variable on these time
scales in 2012 Dec. The maximum peak on 6-hours time scales is observed on 2012 Dec 12, with a flux of $(4.0\pm1.0)\times10^{-6}$~\phcms. The sharpest flux change is the decrease from this bin to the following 6-hours bin, with a variation of a factor $\sim3$.


   \paragraph{0625$-$354}
This source has been detected by \textit{Fermi}-LAT since the first source catalog
\citep{2010ApJS..188..405A}. It is a recent addition to the small sample of
six radio galaxies which have been detected by Cherenkov Telescopes in the
TeV range \citep{2018MNRAS.476.4187A}. This is mostly due to its
notably hard spectrum (see Table~\ref{rgs-lat-new}). The source properties in the 3FGL energy range
are consistent with the catalog values, and the significance is
increased.

\begin{figure*}[!htbp]
\begin{center}
\includegraphics[width=\linewidth]{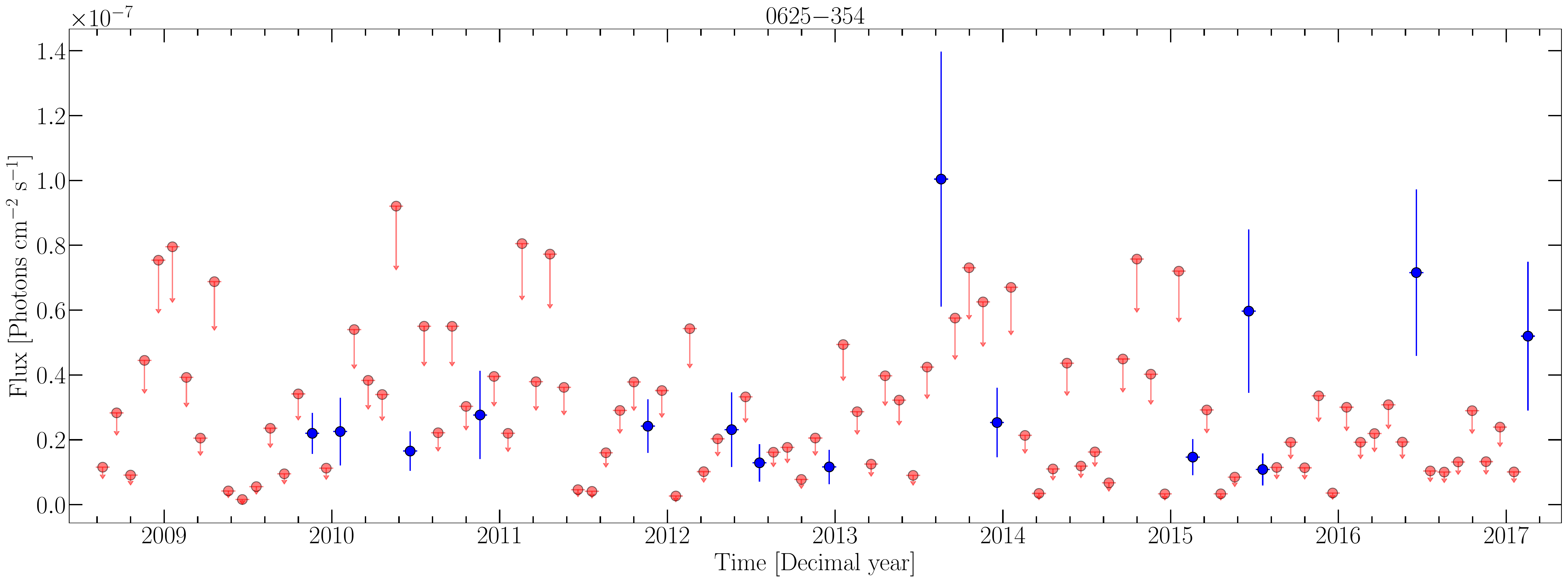}
\end{center}
\caption{Light curve of PKS~0625$-$35 (0625$-$354) between 0.1-300 GeV over 103
  months of \textit{Fermi}-LAT data, with monthly binning. Blue points are
  detections, red arrows are upper limits.}
\label{0625_lc}
\end{figure*}

A monthly-binned light curve of the source over 8 years is presented
in Fig.~\ref{0625_lc}. No significant activity is detected.

  \paragraph{1343$-$601}
Centaurus~B was detected by \textit{Fermi}-LAT in the 2FGL
\citep{2012ApJS..199...31N}. Our analysis over $\sim8.5$ years
in the 3FGL range gives results consistent with the catalog values
within the error. It is interesting to note that in spite of its
relatively steep spectral index, Cen~B has been indicated as a good
candidate for a TeV detection \citep{2017APh....92...42A}. This is
probably favored by its very low redshift (see Table~\ref{rgs}).

The analysis of this source is particularly challenging due to its
location behind the Galactic plane ($b=1.73^\circ$), which contains rich
diffuse emission structures, complicating point-source analysis. This ROI required the addition of many
sources in excess of the 3FGL, since we are using more than double the data
with respect to the catalog, some of which may actually be due to
contribution from improperly modeled Galactic diffuse emission.

\begin{figure*}[!htbp]
\begin{center}
\includegraphics[width=\linewidth]{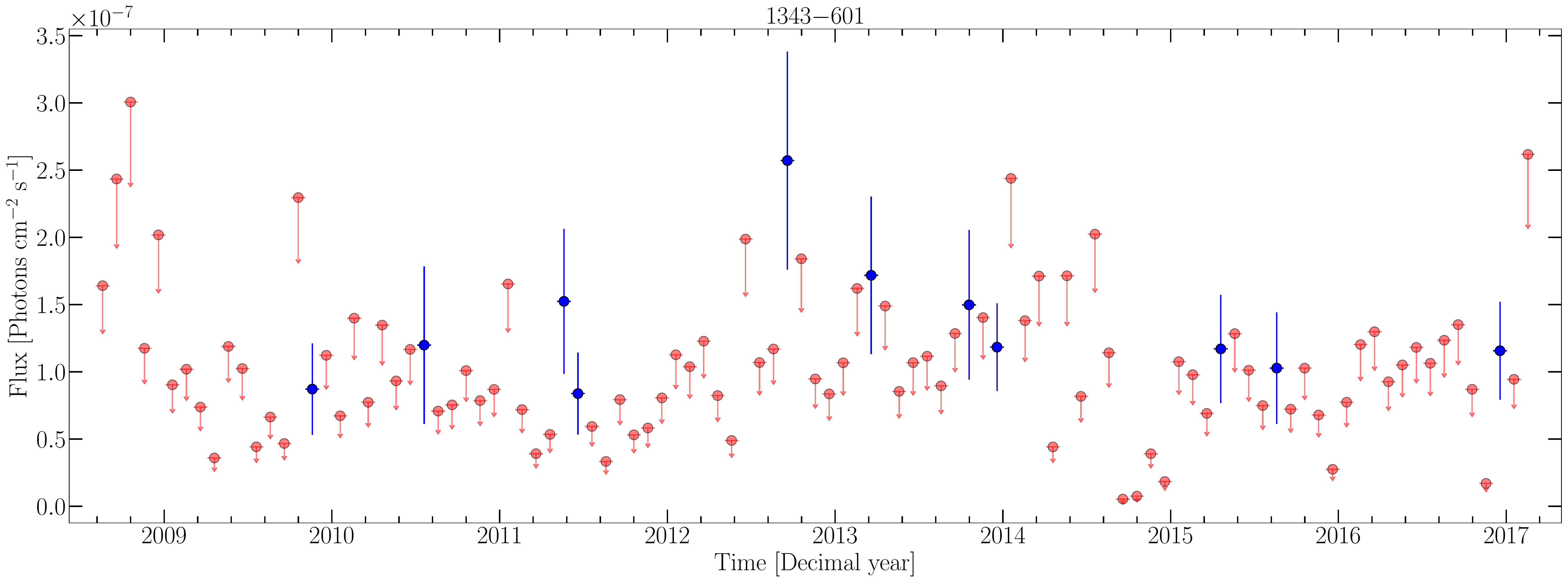}
\end{center}
\caption{Light curve of Centaurus~B between 0.1-300 GeV over 103
  months of \textit{Fermi}-LAT data, with monthly binning. Blue points are
  detections, red arrows are upper limits.}
\label{cenb_lc}
\end{figure*}

A monthly light curve of Cen~B is presented in Fig.~\ref{cenb_lc}. The
source is undetected for most of the time range, and does not show any
significant variability.

\begin{figure*}[!htbp]
\begin{center}
\includegraphics[width=0.75\linewidth]{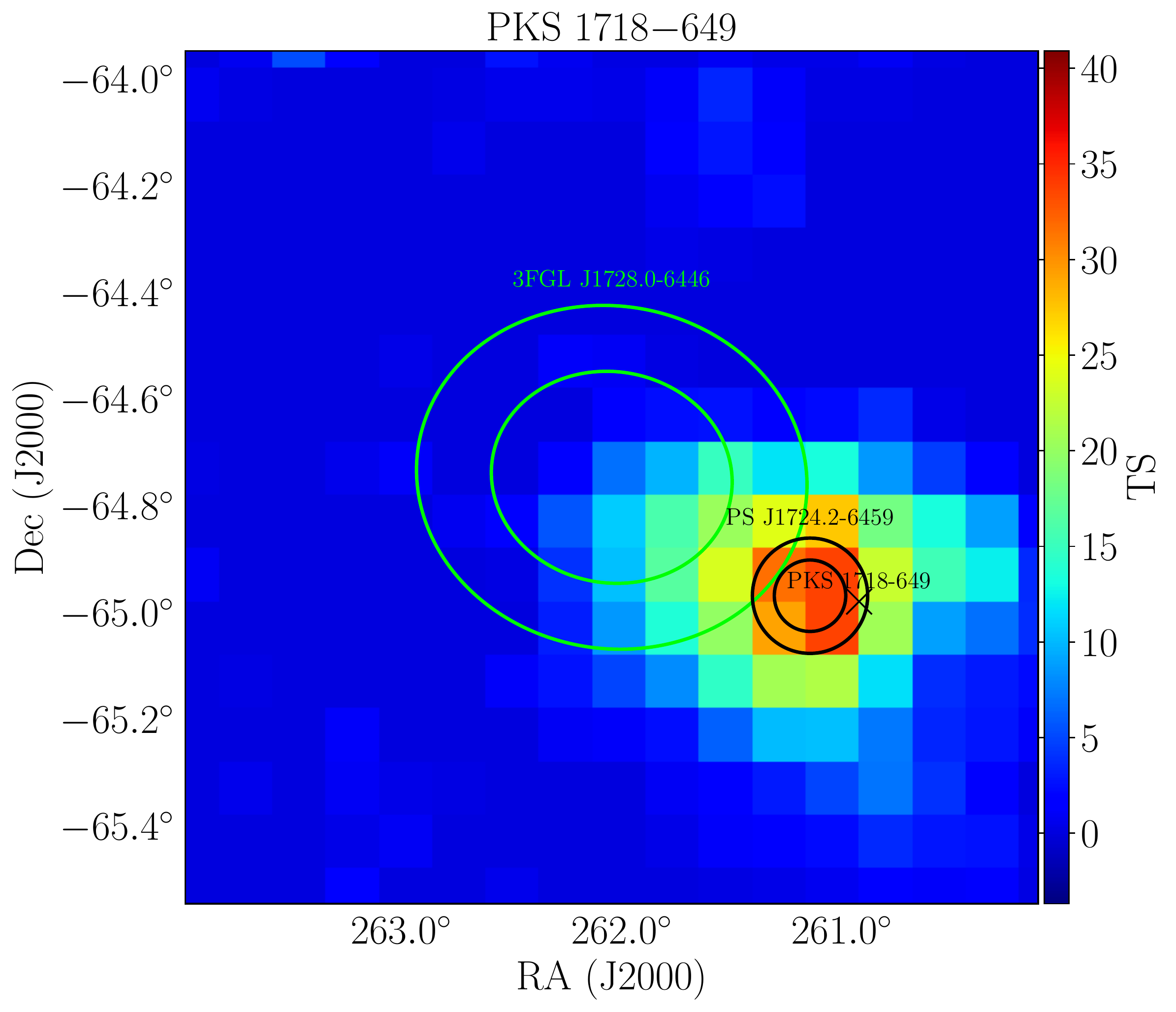}
\end{center}
\caption{\textit{Fermi}-LAT map of excess TS in the inner region of the ROI centered on
  PKS~1718$-$649, after removing the unidentified catalog source
  3FGL~J1728.0$-$6446. The ellipses represent the 68\% and 95\% confidence
  positional errors on the catalog source, while the black circles
  represent the same errors for the new source PS~J1724.2$-$6459. The map size is $1.6^{\circ}\times1.6^{\circ}$. Each pixel corresponds to 0.1$^{\circ}$.}
\label{1718_ts}
\end{figure*}

   \paragraph{1718$-$649}
This source has recently become the first $\gamma$-ray detected young
radio galaxy. \citet{2016ApJ...821L..31M} detected the source using 7 years of LAT
data, with TS\,$\sim36$, confirming it as the counterpart of the
unidentified catalog source 3FGL~J1728.0$-$6446. We double-checked this
result using $\sim8.5$ years of data. In order to avoid any bias towards either of the two hypotheses, i.e, an unidentified $\gamma$-ray source at the position of 3FGL~J1728.0$-$6446, or a $\gamma$-ray source associated to PKS~1718$-$649, we removed the catalog source from the model and evaluated the excess statistics in the ROI. Fig.~\ref{1718_ts} shows a TS excess map after subtracting
3FGL~J1728.0$-$6446. The excess is nicely coincident with the position
of PKS~1718$-$649. A new source is found by the source-finding algorithm
(dubbed PS~J1724.2$-$6459, for ``Point Source''), and after localization we found that the source coincides with the position of the target within the errors.


\begin{table*}[h!tbp]
\caption{$\gamma$-ray properties of \textit{Fermi}-LAT detected TANAMI radio
  galaxies in the same energy range of the 3FGL, over 103 months of data.}
\begin{center}
\begin{tabular}{lllcccc}
\hline
\hline
Source & Class & \textit{Fermi}-LAT name & Flux$^a$ & Spectral index$^b$ &
Curvature$^c$ & TS\\
\hline
0518$-$458 & FR~II & 3FGL J0519.2$-$4542 & $(4.69\pm0.68)\times10^{-10}$ &
2.71$\pm$0.21 & * & 104.5\\
0521$-$365 & RG/BLL & 3FGL J0522.9$-$3628 & $(4.32\pm0.14)\times10^{-9}$ &
2.84$\pm$0.20 & $-$0.04$\pm$0.04 & 3523.5\\
0625$-$354 & FR~I/BLL & 3FGL J0627.0$-$3529 &
$(1.41\pm0.09)\times10^{-9}$ & 1.90$\pm$0.06 & * & 938.6\\
1322$-$428 & FR~I & 3FGL J1325.4$-$4301 & $(3.46\pm0.14)\times10^{-9}$ & 2.46$\pm$0.05 & * & 1775.7\\
1343$-$601 & FR~I & 3FGL J1346.6$-$6027 & $(2.06\pm0.17)\times10^{-9}$ &
2.48$\pm$0.09 & * & 264.3\\

\hline
\hline
\end{tabular}
\end{center}
$^a$ \textit{Fermi}-LAT flux between 1-100 GeV in \phcms.\\
$^b$ \textit{Fermi}-LAT spectral index $\Gamma$, in case of power-law spectrum
$dN/dE = N_0\times(E/E_0)^{-\Gamma}$, or
$\alpha$ in case of logParabola spectrum $dN/dE =
N_0\times(E/E_b)^{-[\alpha+\beta log(E/E_b)]}$.\\
$^c$ Curvature parameter of logParabola spectrum $\beta$.\\

\label{rgs-lat}
\end{table*}


\begin{table*}[h!tbp]
\caption{0.1-100 GeV \textit{Fermi}-LAT results on TANAMI $\gamma$-ray radio galaxies.}
\begin{center}
\begin{tabular}{lllccc}
\hline
\hline
Source & Class & Flux$^a$ & Spectral index$^b$ & Curvature$^b$ & TS\\
\hline
0518$-$458 & FR~II & $(1.63\pm0.19)\times10^{-8}$ & 2.63$\pm$0.08 & * & 225.5\\
0521$-$365 & RG/BLL & $(1.17\pm0.03)\times10^{-7}$ & 2.33$\pm$0.02 & $-$0.074$\pm$0.011 & 11313.2\\
0625$-$354 & FR~I/BLL & $(1.07\pm0.10)\times10^{-8}$ & 1.88$\pm$0.04 & * & 1150.4\\
1322$-$428 & FR~I & $(1.73\pm0.04)\times10^{-7}$ & 2.68$\pm$0.02 & * & 7500.0\\
1343$-$601 & FR~I & $(6.85\pm0.66)\times10^{-8}$ & 2.58$\pm$0.05 & * & 416.0\\
1600$-$489 & MSO & $(6.8\pm0.4)\times10^{-8}$ & 2.06$\pm$0.02 & * & 2771.4\\
1718$-$649 & GPS/CSO & $(5.8\pm2.2)\times10^{-9}$ & 2.43$\pm$0.18 & * &44.3\\
\hline
\hline
\end{tabular}
\end{center}
$^a$ \textit{Fermi}-LAT flux between 0.1-100 GeV in \phcms.\\
$^b$ \textit{Fermi}-LAT spectral index $\Gamma$, in case of power-law spectrum
$dN/dE = N_0\times(E/E_0)^{-\Gamma}$, or
$\alpha$ in case of logParabola spectrum $dN/dE =
N_0\times(E/E_b)^{-[\alpha+\beta log(E/E_b)]}$.\\
$^c$ Curvature parameter of logParabola spectrum $\beta$.\\

\label{rgs-lat-new}
\end{table*}




\section{Discussion}
\label{disc}
We now discuss the results presented in the previous sections, focusing first on the individual sources, and then on the properties of $\gamma$-ray radio galaxies as a sample.

\subsection{Individual sources}

\paragraph{0518$-$458}

Based on their VLBI data, \cite{2000AJ....119.1695T} place an upper
limit on the jet viewing angle of $\theta<51^{\circ}$. This is consistent
with the allowed range of parameters from our TANAMI observations, i.e.~ $40^{\circ}<\theta<78^{\circ}$ and $\beta>0.78$.

\cite{2016MNRAS.455.3526H} observed the jet of Pictor~A in X-rays
using \textit{Chandra} observations, combined with ATCA observations
of the large scale radio structure. They detect an X-ray
counterjet, and place an upper limit on the
viewing angle of $\theta<70^{\circ}$ and a lower limit on the intrinsic
jet speed of $\beta>0.3$ from the jet sidedness. The authors noted that at the time there was
no direct evidence of bulk relativistic motion in the jet of Pictor~A,
and pointed out the value of future VLBI studies on the source. Our
results provide the first robust
measurement of component motion in the pc-scale jet of Pictor~A. Our
lower limit on the intrinsic jet speed does not constrain the jet to
be highly relativistic, but significantly reduces the parameter space
with respect to the estimate by \cite{2016MNRAS.455.3526H}. Our lower
limit on the viewing angle confirms that the X-ray emission mechanism
cannot be Inverse Compton on CMB photons (IC/CMB), since this would require $\theta$ smaller
than a few degrees and $\beta\sim1$.

The $\gamma$-ray light curve of Pictor~A shows that the source is
often undetected on monthly time scales. In a variability study of 12 radio galaxies using four years of \textit{Fermi}-LAT
data \cite{2013EPJWC..6104007G} found that FR~II sources were detected
only during short periods of activity \citep[e.g., 3C~111,][]{2012ApJ...751L...3G}, while FR~I sources were detected for a
larger fraction of the investigated time range. This seems to be
consistent with our results on Pictor~A, a classic FR~II radio galaxy,
which is detected for a fraction of $\sim5$\% of the time over 103 months.

Our TANAMI data, combined with the $\gamma$-ray variability allows us
to shed some light on the nature of the high-energy emission in
Pictor~A. We note that a new VLBI jet feature emerged from the radio
core of Pictor A between 2010 Jul 24 and 2011 Aug 13. Since the elevated
$\gamma$-ray state that allowed the first \textit{Fermi}-LAT detection was
within this time range, it is plausible that it was caused by the
passage of a new shock through the radio core, corresponding to the
appearance of this new radio feature. This hypothesis is in agreement
with the results of \cite{2012MNRAS.421.2303B}, who modeled the SED of
the western hot-spot of Pictor~A, and found that it cannot reproduce
the observed X-ray and $\gamma$-ray emission at the same time. Therefore the LAT detection is probably associated with emission from the
innermost part of the jet.

One can speculate whether the association with pc-scale jet activity is a defining feature of $\gamma$-ray emitting FR~II radio galaxies. 
If our inference is correct, Pictor~A would be yet another example of this kind of behavior, after 3C~111 \citep{2012ApJ...751L...3G} and 3C~120~\footnote{While 3C~120 shows an FR~I-like
  extended structure~\citep{1987ApJ...316..546W}, its innermost jet and accretion flow are much more similar to an FR~II source~\citep{2012arXiv1205.1691T}.} \citep{2015ApJ...808..162C}.

This would nicely fit with the findings of \citet{2013EPJWC..6104007G}, if we assume that FR~II sources can only be detected when there is significant activity in their innermost jet,
and/or  when the inner jet is temporarily well-aligned with our
line-of-sight~\citep[see][for the case of 3C~120]{2015ApJ...808..162C,
  2016MNRAS.458.2360J}. The presence of jet spine ``wiggling'' and precession is suggested by numerical simulations such as e.g., \citet{2017arXiv170706619L}. This would naturally lead to a low duty cycle for $\gamma$-ray activity, and consequently to fewer detected FR~IIs, exactly as observed~\citep[e.g.][and references therein]{2016MNRAS.457....2G}. 

\paragraph{0521$-$365}

Our VLBI results complete the multi-wavelength picture provided by
\cite{2002AJ....124..652T},~\cite{2015MNRAS.450.3975D}, and \cite{2016A&A...586A..70L},
confirming that the jet of PKS~0521$-$36 is not highly beamed, with
viewing angles larger than $10^{\circ}$ still compatible with our observations.

\citet{1996ApJ...459..169P} used multi-wavelength data to model the
broadband SED of PKS~0521$-$36, and found that the source is likely
not Doppler-boosting dominated, with an estimated viewing angle of
$\theta=30^{\circ}\pm6^{\circ}$ and a Doppler factor $\delta\sim1.5$.
This range of parameters is completely ruled out by our kinematic
analysis, as can be seen in Fig.~\ref{beta_theta}~\footnote{The Doppler factor is related to the viewing angle and intrinsic speed by $\delta = [\Gamma(1-\beta\cos{\theta})]^{-1}$, where $\Gamma = (1-\beta^2)^{-1/2}$ is the bulk jet Lorentz factor.}.

Fast $\gamma$-ray variability is typical of blazar jets, where the time
scales are strongly reduced by the large Doppler factors. However, as
shown by the VLBI results presented here, and by previous
multi-wavelength studies, it is unlikely that the jet of PKS~0521$-$36
is strongly affected by Doppler boosting. Therefore the jet morphology does
not appear to change in response to the strong $\gamma$-ray flaring activity. On the other
hand, we observe a doubling of the VLBI core flux density during the first
$\gamma$-ray flaring periods of 2010-2011 (see Fig.~\ref{kin_flux}), which suggests that the $\gamma$-ray emission region is
located inside the radio core, and not in the jet.

This combination of slow pc-scale jet and fast $\gamma$-ray
variability bears some resemblance to the case of IC~310, a
transitional FR~I-BL Lac object which shows minute-timescale
variability at VHE \citep{2014Sci...346.1080A}, but no fast jet motion
in VLBI images \citep{2016PhDT.......104S}. For this source,
\citet{2014Sci...346.1080A} favor a model where the fast $\gamma$-ray
variability is produced by charge depletion in the supermassive black
hole magnetosphere due to low accretion rate phases.

\paragraph{0625$-$354}

The estimated range of viewing angles ($\theta<38^{\circ}$) from our VLBI data does not allow us to settle the uncertain classification of PKS~0625$-$35, as it is consistent with jet orientations typical of radio galaxies \citep[see e.g.,][for the case of M\,87]{2018A&A...616A.188K}, but also allows for smaller viewing angles, more typical of blazar sources.
The transitional nature of this source is also supported by the
indications of a hard X-ray nuclear component and by its
location in the parameter space of radio core luminosity at 5
GHz and X-ray non-thermal luminosity, which places it exactly in the
region between FR Is and BL Lacs \citep{1999A&A...348..437T}. Finally, a small viewing angle would partially explain its TeV detection \citep{2018MNRAS.476.4187A}. In this respect, our finding of superluminal jet motions in PKS~0625$-$35 appears in contrast with what is observed for the bulk of the TeV blazars population, which typically shows slow jet speeds. However, a recent study by \cite{2018ApJ...853...68P} has shown for the first time a superluminal tail in the jet speed distribution of TeV sources. PKS~0625$-$35 thus appears to belong to this elusive class of TeV-bright AGN jets with fast parsec-scale motion.

\paragraph{1718$-$649}

Our kinematic analysis allows us to obtain a rough estimate on the age of this young radio source, albeit with a large relative error, as $t_\mathrm{age} = d_\mathrm{max}/\mu = (70\pm30)$ years. This limit is consistent with a quite young age for the
source, compared to usual estimates for young sources
\citep[$t_{\mathrm{age}}$$\sim10^5$ yr, ][and references therein]{1997AJ....113.2025T}. This is mostly due to the notably small
linear size of $\sim2.5$~pc, compared with the bulk of the young radio
sources population \citep[e.g.][]{2014MNRAS.438..463O}. Our estimates are also in broad agreement with the one reported in \cite{2009AN....330..193G}. According to young radio source evolutionary models \citep[e.g.,][]{2000MNRAS.319..445S,2012ApJ...760...77A,2014MNRAS.438..463O}, at the scales relevant to PKS~1718$-$649 the flux density should decrease going back in time in the life of the source. If our estimate is correct, this could potentially be compared to the earliest flux density measurements for this object.


\subsection{Sample properties}
In this subsection, we investigate the correlated radio and $\gamma$-ray properties of the LAT-detected radio galaxies in the TANAMI program. To increase the sample size, we added all radio galaxies with published VLBI results from the MOJAVE survey, and performed the same LAT analysis described in Section~\ref{fermi}. The resulting $\gamma$-ray properties of the sources with a significant $\gamma$-ray detection are listed in Table~\ref{mojave_lat}. We detect (TS$>$25) three sources not previously published in any \textit{Fermi}-LAT catalog, i.e, NGC~315, NRAO~128, and PKS~1514$+$00, in addition to the well-known $\gamma$-ray source 3C~120. PKS~1128$-$047 is not classified as radio galaxy in the \textit{Fermi}-LAT catalogs, but it is according to the \href{http://ned.ipac.caltech.edu/}{NASA/IPAC Extragalactic Database (NED)} and MOJAVE. We exclude NGC~1052 due to its lack of a clear VLBI core in the MOJAVE data, although we are aware that at higher frequencies the absorption by the surrounding torus is not present and a clear core is detected at 86~GHz \citep{2016A&A...593A..47B}.

The resulting subsample includes sixteen radio galaxies with measured VLBI kinematics and LAT properties. This is the largest sample of $\gamma$-ray detected radio galaxies studied with VLBI techniques so far. For the MOJAVE sources, the kinematics results were taken from \cite{2013AJ....146..120L} and \cite{2016AJ....152...12L}. The exact reference for each source is given in Table~\ref{mojave_lat}. 

In order to combine the VLBI data from the TANAMI and MOJAVE samples, we had to account for the different frequency in the two surveys, i.e. 8.4\,GHz and 15\,GHz, respectively. In order to do so, we have back-extrapolated the MOJAVE core fluxes down to 8.4\,GHz using the spectral index values provided in \cite{Hovatta2014}. 

In order to investigate possible correlations between the radio and $\gamma$-ray properties of the sources in this sample, we have used the Kendall's correlation coefficient ($\tau$). The correlation coefficient is equal to zero in the case of uncorrelated data, one in case of maximum correlation, and minus one in case of maximum anti-correlation. The resulting correlation coefficients with errors and the relative $p$-values are listed in Table~\ref{tab:corrs}.


\begin{table*}[h!tbp]
\caption{0.1-100 GeV \textit{Fermi}-LAT results on MOJAVE $\gamma$-ray radio galaxies. Sources reported as LAT detections here for the first time are highlighted in italic.}
\begin{center}
\begin{tabular}{llcccccc}
\hline
\hline
Source & Name & Redshift & Flux$^a$ & Spectral index$^b$ & Curvature$^c$ & TS & Ref.$^d$\\
\hline
  \textit{0055$+$300} & \textit{NGC~315} & 0.0165 & $(5.5\pm1.3)\times10^{-9}$ & 2.29$\pm$0.11 & * & 77.3 & [2]\\
  0305$+$039 & 3C~78 & 0.0287 & $(7.0\pm1.0)\times10^{-9}$ & 1.96$\pm$0.07 &  * &385 & [1]\\
  \textit{0309$+$411} & \textit{NRAO~128} & 0.136 & $(5.7\pm1.7)\times10^{-9}$ & 2.29$\pm$0.13 &  * & 53.6 & [2]\\
  0316$+$413 & 3C~84 & 0.018 & $(3.36\pm0.04)\times10^{-7}$ & 2.006$\pm$0.008 & 0.060$\pm$0.004 & 9.63$\times10^4$ & [1]\\
  0415$+$379 & 3C~111 & 0.0491 & $(3.4\pm0.3)\times10^{-8}$ & 2.75$\pm$0.07 & * & 186 & [1]\\
  0430$+$052 & 3C~120 & 0.033 & $(2.8\pm0.3)\times10^{-8}$ & 2.70$\pm$0.06 & * & 226 & [1]\\
  1128$-$047 & PKS 1128$-$047  & 0.27 & $(7.6\pm1.3)\times10^{-9}$ & 2.46$\pm$0.10 & * & 58.9 & [2]\\
  1228$+$126 & M87 & 0.00436 & $(1.9\pm0.2)\times10^{-8}$ & 2.08$\pm$0.04 & * & 1410 & [2]\\
  \textit{1514$+$004} & \textit{PKS~1514$+$00} & 0.052 & $(8.8\pm1.6)\times10^{-9}$ & 2.46$\pm$0.10 & * & 82.3 & [2]\\
  1637$+$826 & NGC~6251 & 0.0247 & $(2.2\pm0.2)\times10^{-8}$ & 2.28$\pm$0.04 & 0.09$\pm$0.02 & 1610 & [2]\\
\hline
\hline
\end{tabular}
\end{center}
$^a$ \textit{Fermi}-LAT flux between 0.1-100 GeV in \phcms.\\
$^b$ \textit{Fermi}-LAT spectral index $\Gamma$, in case of power-law spectrum
$dN/dE = N_0\times(E/E_0)^{-\Gamma}$, or
$\alpha$ in case of logParabola spectrum $dN/dE =
N_0\times(E/E_b)^{-[\alpha+\beta\log{(E/E_b)}]}$.\\
$^c$ Curvature parameter of logParabola spectrum $\beta$.\\
$^d$ MOJAVE reference paper for the VLBI results: [1] \cite{2013AJ....146..120L}; [2] \cite{2016AJ....152...12L}\\

\label{mojave_lat}
\end{table*}

\begin{table}[h!tbp]
\caption{Correlation coefficients between radio and $\gamma$-ray properties of our radio galaxy sample, and corresponding significance.}
\begin{center}
\begin{tabular}{lccc}
\hline
\hline
Variables & Kendall's $\tau$ & $p$-value & Sign. ($\sigma$)\\
\hline
  $<S_\mathrm{core}^\mathrm{VLBI}>$ vs. $F_\gamma$ & 0.55 & 3$\times10^{-3}$ & 3.0\\
  $<S_\mathrm{jet}^\mathrm{VLBI}>$ vs. $F_\gamma$ & 0.27 & 0.02 & 2.3\\
  $L_\gamma$ vs. $<T_\mathrm{b}^\mathrm{core}>$  & -0.23 & 0.05 & 2.0\\
  $L_\gamma$ vs. $<CD_\mathrm{VLBI}>$ & 0.10 & 0.41 & 0.8\\
\hline
\hline
\end{tabular}
\end{center}
\label{tab:corrs}
\end{table}

\begin{figure}
\begin{center}
\includegraphics[width=\linewidth]{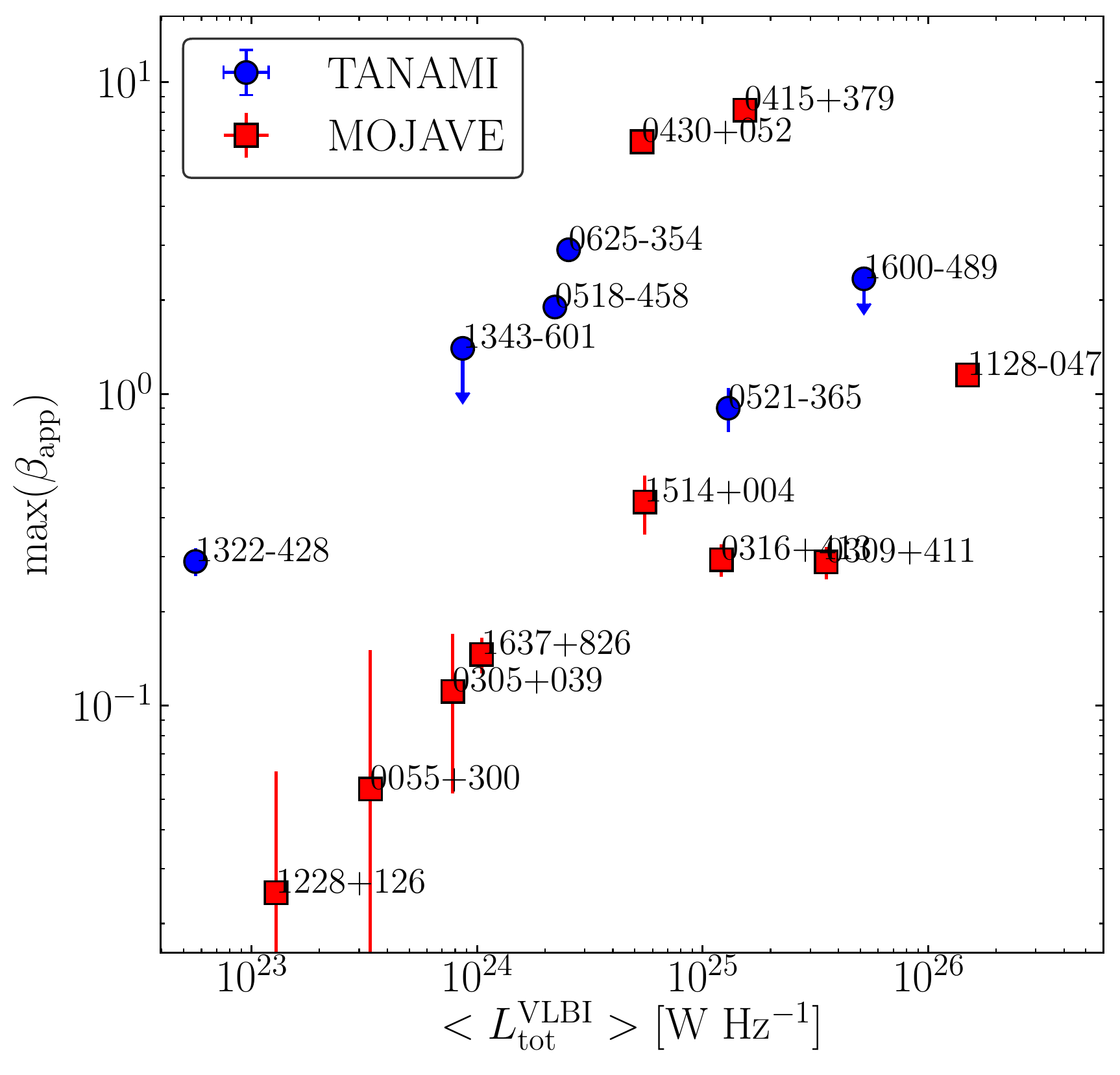}
\end{center}
\caption{Maximum measured apparent speed $\beta_{\mathrm{app}}$ as a function of median total VLBI luminosity for our TANAMI $\gamma$-ray radio galaxies (blue circles) and MOJAVE $\gamma$-ray radio galaxies (red squares). The downward arrows indicate upper limits on the apparent speed, for sources with speed consistent with zero within the uncertainties.}
\label{mojave_plot}
\end{figure}

In Fig.~\ref{mojave_plot} we show the maximum measured apparent speed versus the total VLBI luminosity of this subsample. It can be seen that there is a lack of low-luminosity sources ($\lesssim 10^{24}$ W/Hz) with high apparent speed ($\beta_\mathrm{app}\gtrsim 0.5$). This was already found to be true for blazars \citep[e.g.][]{2007ApJ...658..232C,2013AJ....146..120L}, but this is the first time that it is observed in radio galaxies. On the other hand, while for blazars a population of high-luminosity, low-speed sources is observed \citep{2007ApJ...658..232C}, this is not the case in our radio galaxy sample. \citet{2007ApJ...658..232C} interpreted this sub-population, mostly composed of BL Lac blazars, as sources where the measured VLBI apparent speed at 15~GHz is not representative of the jet bulk Lorentz factor. This is now a well-established fact for high-energy peaked BL Lacs \citep[HBLs, see e.g.,][and references therein]{2018ApJ...853...68P}. We do not find clear evidence for a misaligned parent population of these intrinsically luminous, low-speed blazars.

\begin{figure}
\begin{center}
\includegraphics[width=\linewidth]{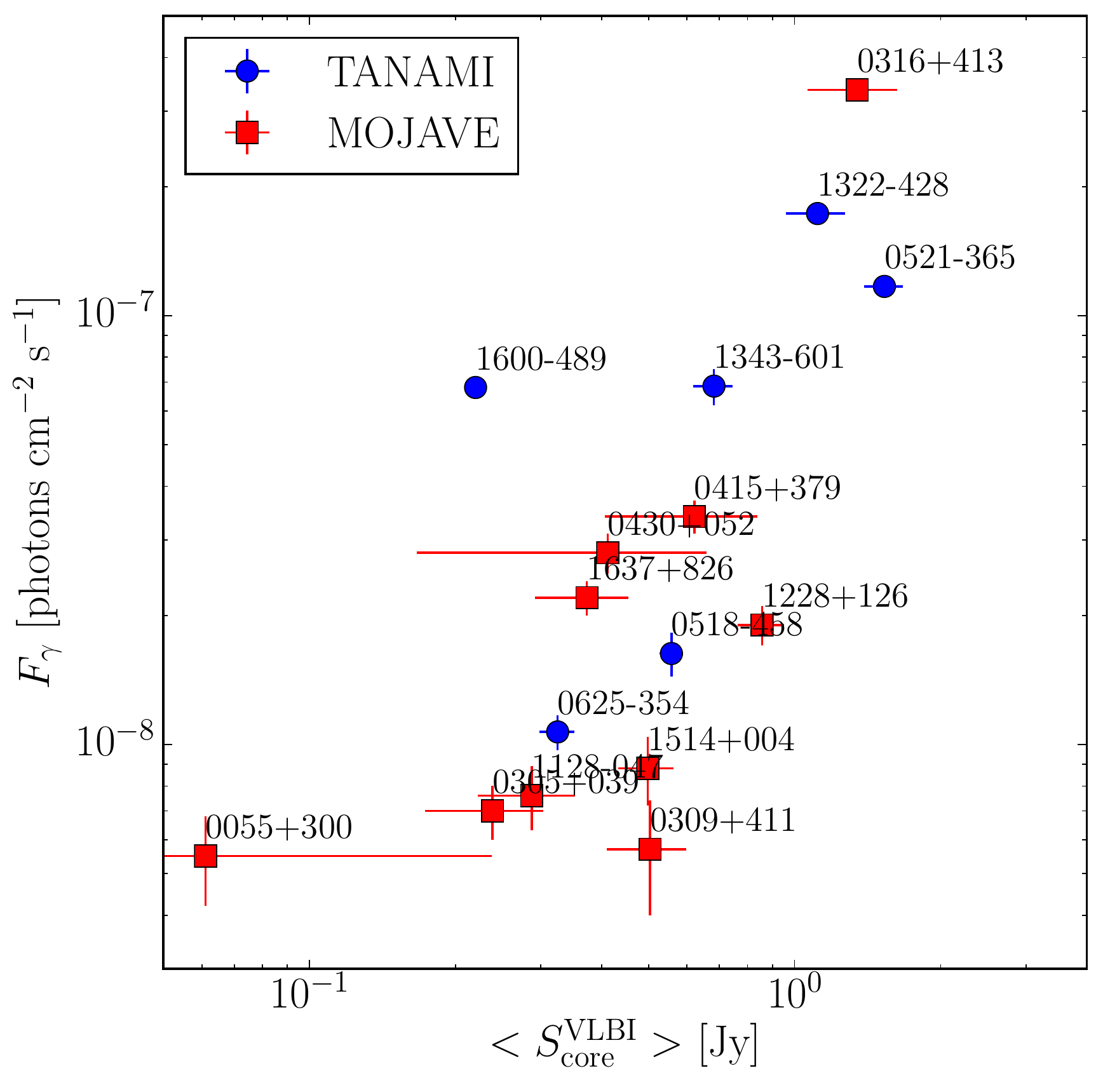}
\end{center}
\caption{\textit{Fermi}-LAT flux above 100 MeV as a function of median radio VLBI core flux density for our TANAMI $\gamma$-ray radio galaxies (blue circles) and MOJAVE $\gamma$-ray radio galaxies (red squares).} 
\label{fluxes}
\end{figure}

In Fig.~\ref{fluxes} we show the  LAT flux above 100 MeV versus the median VLBI core flux density. We test for a linear correlation between the log core flux density and the log of the $\gamma$-ray flux, and find that a Kendall's test yields a correlation coefficient $\tau = 0.55$ and a $p$-value of 0.3\%. This result suggests that in radio galaxies, the observed $\gamma$-ray flux is related to the radio flux density of the innermost pc-scale jet. Such a correlation has already been highlighted for large, blazar-dominated AGN samples \citep[see e.g., ][]{2009ApJ...696L..17K,2011ApJ...741...30A,2017A&A...606A.138L}. In a study of the radio and $\gamma$-ray properties of the full TANAMI sample using the first year of \textit{Fermi}-LAT data, \citet{2016A&A...590A..40B} found a similar correlation between radio and \textit{Fermi}-LAT fluxes. This sample was dominated by blazars, while our (smaller) sample is focused on radio galaxies alone. We can therefore confirm that a similar dependence of $\gamma$-ray flux on the radio core flux density holds for misaligned AGN jets as well. The idea that $\gamma$-ray emission in radio galaxies with compact VLBI emission is related to the innermost jet is supported by the recent detection of diffuse $\gamma$-ray emission from the large-scale lobes of Fornax~A \citep{2016ApJ...826....1A}. In this radio galaxy, the $\gamma$-ray emission has been determined to be spatially extended at $>5\sigma$ confidence level. The contribution of the extended lobes to the high-energy flux is dominant, while the radio core contribution is no more than $14\%$. Interestingly, Fornax~A has not been detected with VLBI observations, suggesting that it does not show a bright, compact radio jet on parsec scales. Therefore, the sub-dominant contribution of the radio core to the $\gamma$-ray emission in Fornax~A is to be expected, if the radio brightness of the innermost jet is indeed related to its $\gamma$-ray flux, as our results indicate.

\begin{figure}
\begin{center}
\includegraphics[width=\linewidth]{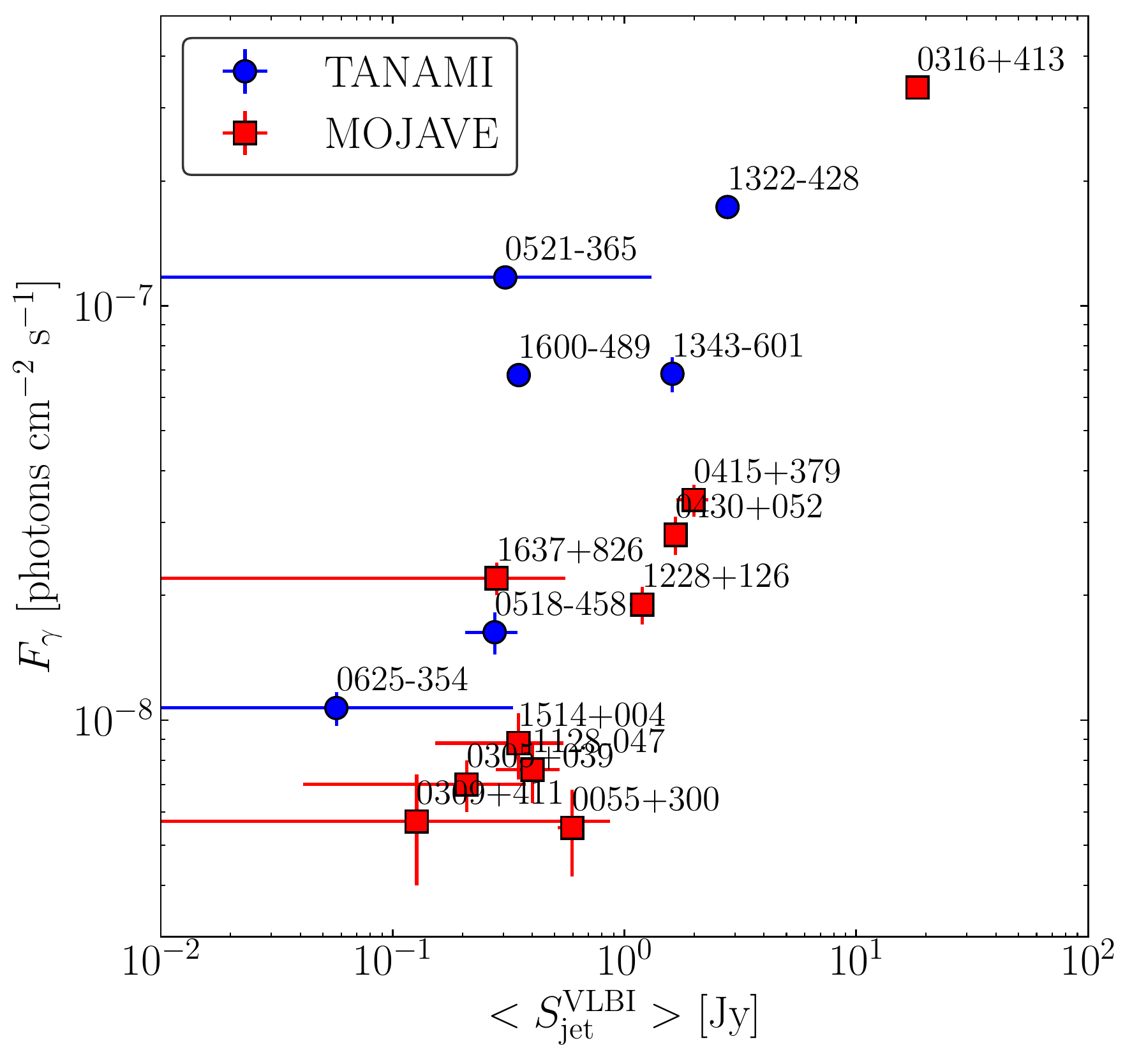}
\end{center}
\caption{\textit{Fermi}-LAT flux above 100 MeV as a function of median radio VLBI jet flux density for our TANAMI $\gamma$-ray radio galaxies (blue circles) and MOJAVE $\gamma$-ray radio galaxies (red squares).}
\label{jetflux}
\end{figure}

In Fig.~\ref{jetflux} we show the LAT flux versus the VLBI jet flux, i.e., the flux obtained subtracting the core modeled flux density from the total imaged flux density. A Kendall's correlation test for these two quantities yields a coefficient $\tau=0.27$ with a $p$-value of 2\%. Since this exceeds the 1\% threshold, we conclude that the $\gamma$-ray flux is not significantly related to the pc-scale jet flux density.

\begin{figure}
\begin{center}
\includegraphics[width=\linewidth]{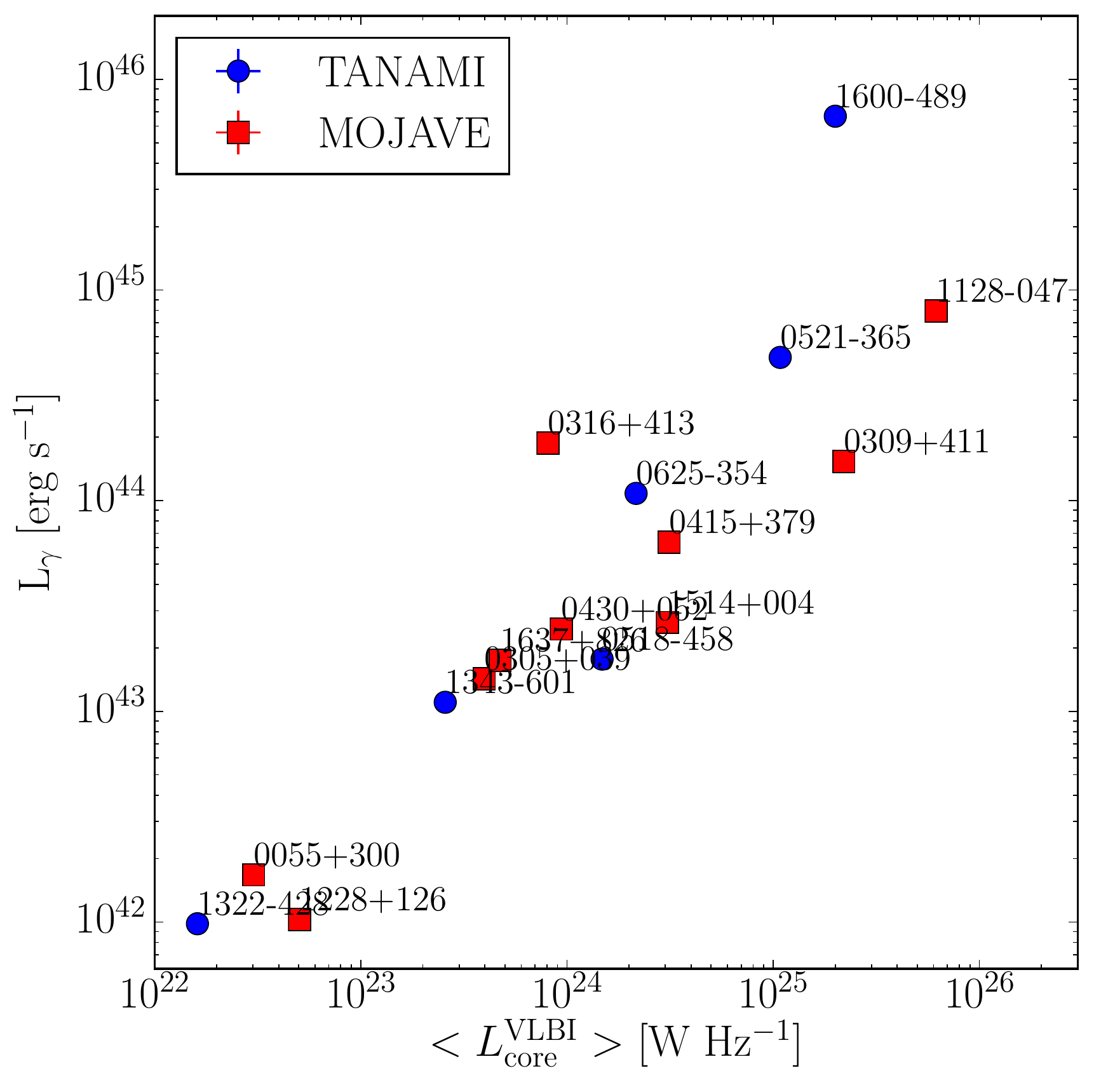}
\end{center}
\caption{\textit{Fermi}-LAT luminosity as a function of median radio VLBI core luminosity for our TANAMI $\gamma$-ray radio galaxies (blue circles) and MOJAVE $\gamma$-ray radio galaxies (red squares).}
\label{lums}
\end{figure}

In Fig.~\ref{lums} we show the LAT luminosity versus the median VLBI core luminosity. The clear linear correlation observed between the log quantities is consistent with a 1:1 correlation, which suggests that it is induced by the common dependence on redshift. The linear correlation results in $\tau=0.78$ and a $p$-value=2$\times10^{-5}$. When accounting for the redshift as a third parameter, the partial correlation analysis (see Section 2.1 of \citealt{Akritas1996}) yields $\tau=0.39$ and a $p$-value=0.05. One might assume that controlling for the effect of redshift when correlating luminosity with luminosity should yield the same result as the flux-flux density correlation. However, there are two effects that can explain why this is not the case. First, the flux correlation and the luminosity correlation would only yield exactly the same result if all the sources were the same, without any difference with redshift. Second, while the VLBI core luminosity has a simple dependence in the form  $L_\mathrm{core}^\mathrm{VLBI}\propto D_L^2\,S_\mathrm{core}^\mathrm{VLBI}$, the $\gamma$-ray luminosity is integrated over several orders of magnitude in energy, and therefore depends not only on flux and luminosity distance, but also on photon index.

\begin{figure}
\begin{center}
\includegraphics[width=\linewidth]{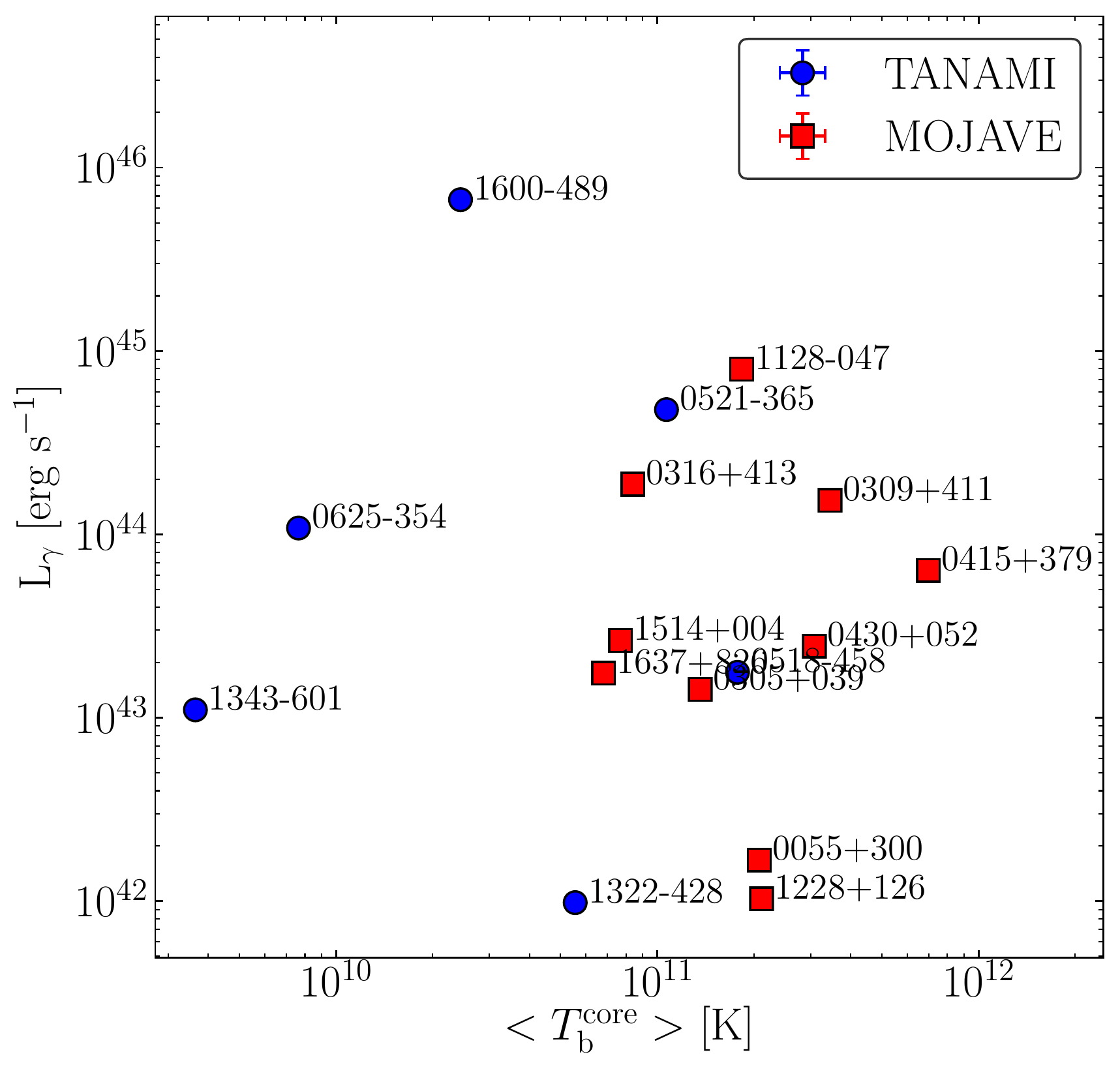}
\end{center}
\caption{Median radio VLBI core brightness temperature as a function of \textit{Fermi}-LAT luminosity above 100 MeV for our TANAMI $\gamma$-ray radio galaxies (blue circles) and MOJAVE $\gamma$-ray radio galaxies (red squares). A correction factor has been applied to the MOJAVE data to account for the difference in frequency, but this might still imply a bias.}
\label{temp}
\end{figure}

In Fig.~\ref{temp} we show the median VLBI core brightness temperature versus $\gamma$-ray luminosity. The brightness temperature is calculated as 

\begin{equation}
    T_b = \frac{2\ln{2}}{\pi k_b}\frac{S\lambda^2(1+z)}{\theta^2},
\end{equation}

\noindent where $k_b$ is Boltzmann's constant, $S$ is the radio flux density, $\lambda$ is the observing wavelength, $z$ is the source redshift, and $\theta$ is the FWHM of the circular Gaussian model fitted to the component (in our case, the core). Since the component sizes resulting from modeling can be much smaller than the actual resolution, we have calculated an upper limit on the component size based on its signal-to-noise ratio (SNR), following \cite{2005AJ....130.2473K}:

\begin{equation}
    \theta_{\mathrm{lim}} = b\sqrt{\frac{4\,\ln{2}}{\pi}\ln{\left(\frac{\mathrm{SNR}}{\mathrm{SNR}-1}\right)}}
\end{equation}

\noindent where $b$ is the beam size, and the SNR was calculated as the ratio between the peak flux density of the residual map after subtracting the given component, in an area defined by the component size, and the post-fit rms in the same residual map area. When the fitted component size was $\theta<\theta_{\mathrm{lim}}$, we have used $\theta_{\mathrm{lim}}$ as an upper limit, and therefore the obtained $T_b$ is a lower limit estimate. We did not compute the resolution limit for the MOJAVE sources, since we only made use of publicly available data.

A correction factor given by the ratio of the two wavelengths has been applied to the MOJAVE data to account for the different observing wavelength. \citet{2016A&A...590A..40B} found, for the full TANAMI sample, an indication of increasing core brightness temperature with increasing $\gamma$-ray luminosity. Interestingly, our results focused on $\gamma$-ray radio galaxies show that the $\gamma$-ray luminosity appears to be completely uncorrelated with the core brightness temperature for misaligned jets. This is confirmed by the correlation analysis (see Table~\ref{tab:corrs}). Since the core brightness temperature is often considered as an indicator of Doppler boosting \citep[e.g., ][]{2005AJ....130.2473K}, this indicates that while the core flux density does reflect higher $\gamma$-ray fluxes, high-energy emission in radio galaxies is not Doppler boosting dominated.

\begin{figure}
\begin{center}
\includegraphics[width=\linewidth]{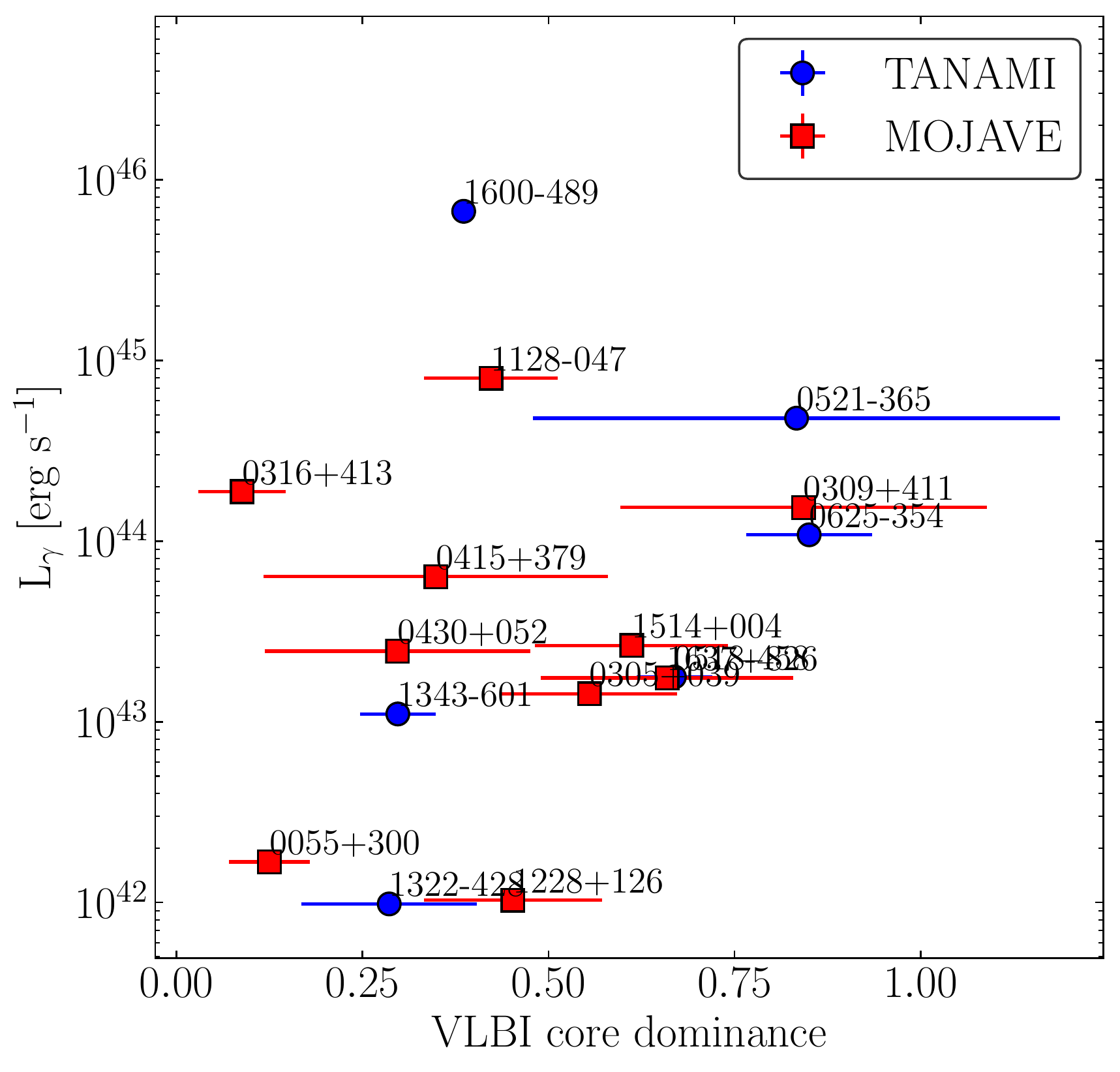}
\end{center}
\caption{Median radio VLBI core dominance as a function of \textit{Fermi}-LAT luminosity above 100 MeV for our TANAMI $\gamma$-ray radio galaxies (blue circles) and MOJAVE $\gamma$-ray radio galaxies (red squares).}
\label{CD}
\end{figure}

Another indication supporting this inference comes from the VLBI core dominance (CD), another indicator of Doppler boosting. We define this parameter as a ratio between VLBI core flux density and total flux density, i.e., $CD = S_\mathrm{core}^\mathrm{VLBI}/S_\mathrm{tot}^\mathrm{VLBI}$. As can be seen in Fig.~\ref{CD}, the observed $\gamma$-ray luminosity is not correlated with the CD (see also Table~\ref{tab:corrs}), supporting the finding that it is not driven by Doppler boosting. In the first systematic study of the $\gamma$-ray properties of misaligned AGN, \cite{2010ApJ...720..912A} noted how $\gamma$-ray detected radio galaxies from the revised Third Cambridge catalog of radio sources \citep[3CRR,][]{1983MNRAS.204..151L} showed preferentially higher CD than their undetected counterparts. This might indicate that radio galaxies with more aligned jets are preferentially detected by the \textit{Fermi}-LAT. However, the definition of CD in \cite{2010ApJ...720..912A} and the one we adopted are quite distinct. \cite{2010ApJ...720..912A} define it as CD=log[$S_\mathrm{core}/(S_\mathrm{tot}-S_\mathrm{core})$], where $S_\mathrm{core}$ is the core flux density at 5~GHz measured with the Very Large Array (VLA), i.e, at arcsecond scale, and $S_\mathrm{tot}$ is the total flux at 178~MHz on arcminute scale. It is clear how our definition of CD on pc-scales differs significantly from the one in \cite{2010ApJ...720..912A}, and that therefore our results do not necessarily contradict the ones presented in this previous work.

Overall, the comparison of several VLBI and \textit{Fermi}-LAT properties of our sample of $\gamma$-ray radio galaxies consistently suggests that high-energy emission in misaligned jet sources is not driven by orientation-dependent Doppler boosting effects, in agreement with the unified model of jetted AGN.


\section{Conclusions}
\label{conc}
In this paper, we presented the first part of a study investigating the radio and $\gamma$-ray properties of radio galaxies in the TANAMI monitoring program. We presented the pc-scale jet
kinematic properties of four well-known $\gamma$-ray emitting radio
galaxies. High-resolution VLBI results were
indicated by studies in different wavelengths as an important piece of
the puzzle in order to complete our understanding of this subpopulation of AGN \citep[e.g.][]{2002AJ....124..652T,2016MNRAS.455.3526H}. Our results fit
ideally in this context. The TANAMI program provides the best multi-epoch milliarcsecond
resolution images of these radio galaxies, allowing us to robustly study
variations in the pc-scale jet structure for the first time. We complemented the VLBI analysis with \textit{Fermi}-LAT $\gamma$-ray light curves, and investigated whether there are physical connections
between the emission observed at the two ends of the spectrum. Our main results on individual sources can be summarized as follows:
\begin{itemize}
    \item \textbf{Pictor~A}: we find that the first $\gamma$-ray detection of this source was coincident with the passage of a new VLBI component through the compact core, an association that appears to be a defining feature of $\gamma$-ray FR~II radio galaxies. Additionally, we detect a pc-scale counterjet for the first time in this source, which allows us to obtain a much better constraint on the viewing angle, which should lie approximately between 76$^{\circ}$~and 79$^{\circ}$.
    \item \textbf{PKS~0521$-$36}: our VLBI results show subluminal motions on a $\sim20$ years time range, while the \textit{Fermi}-LAT light curve shows fast variability down to 6-hours time scales. Such a combination of fast high-energy flaring activity and slow jet motions bears some resemblance to the case of the hybrid FR~I/BL~Lac object IC~310~\citep{2014Sci...346.1080A}.
    \item \textbf{PKS~0625$-$35}: our TANAMI monitoring confirms the presence of superluminal motion in this jet, up to $\beta_{\mathrm{app}}\sim3$. 
    Our results constrain the jet viewing angle to values $\theta<53^{\circ}$, leaving open the possibility of a transitional jet orientation between the radio galaxy and blazar classes. A small viewing angle would be consistent with the $\gamma$-ray properties of the source, i.e., a hard \textit{Fermi}-LAT spectrum and the observed TeV emission.
    \item \textbf{Centaurus~B}: The small number of epochs does not allow a full kinematic analysis. However, using the two available images, we tentatively constrain the jet speed to be likely subluminal or at most mildly superluminal.
    \item \textbf{PKS~1718$-$649}: TANAMI provides the first multi-epoch pc-scale maps of the source. Our kinematic analysis yields a rough estimate of the age of this young radio object, which is of the order of 70 years. This is the first kinematic age measurement of this archetypal CSO. Moreover, a spectral index map between 8.4~GHz and 22.3~GHz suggests that the core of this young radio source is strongly absorbed at these frequencies.
\end{itemize}

In order to study the connection between pc-scale properties and high-energy emission in radio galaxies in a more general fashion, we included public results from the MOJAVE sources, which resulted in a total sample of fifteen sources with VLBI monitoring and detected by \textit{Fermi}-LAT, the largest sample of $\gamma$-ray detected radio galaxies studied with VLBI techniques so far. We find that the VLBI core flux density correlates with the observed $\gamma$-ray flux, as observed in blazars, while the $\gamma$-ray luminosity does not correlate with typical Doppler boosting indicators such as core brightness temperature and core dominance. This indicates that while the compact pc-scale emission does drive the observed high-energy flux, the observed $\gamma$-ray luminosity is not driven by Doppler boosting effects, as is observed in blazars. This difference reinforces the orientation-based unified model of jetted AGN, since radio galaxies are not expected to be Doppler boosting-dominated, having misaligned jets w.r.t. our line of sight.

The TANAMI program is still ongoing, as new observing epochs are
added and analyzed. This will allow us to obtain more
robust kinematic results, including a first
measurement of (or limit to) the separation speed between the two
lobes of PKS~1718$-$649, and further investigation of
the new VLBI component 
possibly associated with the LAT detection of Pictor~A.
\textit{Fermi}-LAT monitoring will also continue to provide
$\gamma$-ray data in the coming years.

In the second paper of this series, we will report on the VLBI
properties and updated $\gamma$-ray flux upper limits of LAT-undetected
TANAMI radio galaxies. We will compare the two sub-populations, and also investigate in a more general fashion the possible connections between VLBI and high-energy properties in radio galaxies.

\begin{acknowledgements}
We thank Laura Vega Garc\'ia for the development of the Python
GUI-based code that was used for the kinematic analysis and the
spectral index maps. We thank Bindu Rani as the \textit{Fermi}-LAT collaboration internal referee, and Rocco Lico as the MPIfR internal referee. We also thank Marcello Giroletti, David Thompson and Nicola Omodei for providing additional feedback on the manuscript.

This research has made use of the NASA/IPAC Extragalactic Database (NED),
which is operated by the Jet Propulsion Laboratory, California Institute of Technology, under contract with the National Aeronautics and Space Administration; data from the MOJAVE database that is maintained by the MOJAVE team \citep{2018ApJS..234...12L}; APLpy, an open-source plotting package for Python \citep{aplpy}; Astropy,\footnote{http://www.astropy.org} a community-developed core Python package for Astronomy \citep{astropy:2013, astropy:2018}

C.M. acknowledges support from the ERC
Synergy  Grant  “BlackHoleCam:  Imaging  the  Event  Horizon  of  Black  Holes”
(Grant 610058).

R.S.
gratefully acknowledge support from the European Research Council under
the European Union's Seventh Framework Programme (FP/2007-2013)/ERC
Advanced Grant RADIOLIFE-320745.

The \textit{Fermi}-LAT Collaboration acknowledges generous ongoing support from a
number of agencies and institutes that have supported both the
development and the operation of the LAT, as well as scientific data
analysis. These include the National Aeronautics and Space
Administration and the Department of Energy in the United States; the
Commissariat \'a l’Energie Atomique and the Centre National de la
Recherche Scientifique/Institut National de Physique Nucl\'eaire et de
Physique des Particules in France; the Agenzia Spaziale Italiana and
the Istituto Nazionale di Fisica Nucleare in Italy; the Ministry of
Education, Culture, Sports, Science and Technology (MEXT), High Energy
Accelerator Research Organization (KEK), and Japan Aerospace
Exploration Agency (JAXA) in Japan; and the K. A. Wallenberg
Foundation, the Swedish Research  Council,
and  the  Swedish  National  Space  Board in Sweden.

Additional support for science analysis during the operations
phase is gratefully acknowledged from the Istituto Nazionale di
Astrofisica in Italy and the Centre National d'Etudes Spatiales in
France.

The Long Baseline Array is part of the Australia Telescope National
Facility which is funded by the Commonwealth of Australia for
operation as a National Facility managed by CSIRO. This study made use
of data collected through the AuScope initiative. AuScope Ltd is
funded under the National Collaborative Research Infrastructure
Strategy (NCRIS), an Australian Commonwealth Government Programme.
This work made use of the Swinburne University of Technology software
correlator, developed as part of the Australian Major National
Research Facilities Programme. This work was supported by resources
provided by the Pawsey Supercomputing Centre with funding from the
Australian Government and the Government of Western Australia.
Hartebeesthoek Radio Astronomy Observatory (HartRAO) is a facility of
the National Research Foundation (NRF) of South Africa. This research
was funded in part by NASA through Fermi Guest Investigator grants
NNH10ZDA001N, NNH12ZDA001N, and NNH13ZDA001N-FERMI (proposal numbers
41213, 61089, and 71326, respectively). This research was supported by
an appointment to the NASA Postdoctoral Program at the Goddard Space
Flight Center, administered by Universities Space Research Association
through a contract with NASA. This work was performed in part under DOE Contract DE-AC02-76SF00515. We would like to thank the Institute of Radio Astronomy and Space Research, AUT University, New Zealand for the use and operational support of their radio telescopes in the collection of data for this work.

\end{acknowledgements}

\bibliographystyle{aa}
\bibliography{kin_v2}

\begin{thebibliography}{87}
\expandafter\ifx\csname natexlab\endcsname\relax\def\natexlab#1{#1}\fi

\bibitem[{{Abdalla} {et~al.}(2018){Abdalla}, {Abramowski}, {Aharonian}, {Ait
  Benkhali}, {Akhperjanian}, {Andersson}, {Ang{\"u}ner}, {Arrieta}, {Aubert},
  {Backes}, \& et~al.}]{2018MNRAS.476.4187A}
{Abdalla}, H., {Abramowski}, A., {Aharonian}, F., {et~al.} 2018, \mnras, 476,
  4187

\bibitem[{{Abdo} {et~al.}(2010{\natexlab{a}}){Abdo}, {Ackermann}, {Ajello},
  {Allafort}, {Antolini}, {Atwood}, {Axelsson}, {Baldini}, {Ballet},
  {Barbiellini}, \& et~al.}]{2010ApJS..188..405A}
{Abdo}, A.~A., {Ackermann}, M., {Ajello}, M., {et~al.} 2010{\natexlab{a}},
  \apjs, 188, 405

\bibitem[{{Abdo} {et~al.}(2010{\natexlab{b}}){Abdo}, {Ackermann}, {Ajello},
  {Atwood}, {Baldini}, {Ballet}, {Barbiellini}, {Bastieri}, {Baughman},
  {Bechtol}, {Bellazzini}, {Berenji}, {Blandford}, {Bloom}, {Bonamente},
  {Borgland}, {Bregeon}, {Brez}, {Brigida}, {Bruel}, {Burnett}, {Buson},
  {Caliandro}, {Cameron}, {Caraveo}, {Casandjian}, {Cavazzuti}, {Cecchi}, {{\c
  C}elik}, {Chekhtman}, {Cheung}, {Chiang}, {Ciprini}, {Claus}, {Cohen-Tanugi},
  {Colafrancesco}, {Cominsky}, {Conrad}, {Costamante}, {Cutini}, {Davis},
  {Dermer}, {de Angelis}, {de Palma}, {Digel}, {do Couto e Silva}, {Drell},
  {Dubois}, {Dumora}, {Farnier}, {Favuzzi}, {Fegan}, {Finke}, {Focke},
  {Fortin}, {Fukazawa}, {Funk}, {Fusco}, {Gargano}, {Gasparrini}, {Gehrels},
  {Georganopoulos}, {Germani}, {Giebels}, {Giglietto}, {Giordano}, {Giroletti},
  {Glanzman}, {Godfrey}, {Grenier}, {Grove}, {Guillemot}, {Guiriec},
  {Hanabata}, {Harding}, {Hayashida}, {Hays}, {Hughes}, {Jackson},
  {J{\'o}hannesson G.}, {Johnson}, {Johnson}, {Johnson}, {Kamae}, {Katagiri},
  {Kataoka}, {Kawai}, {Kerr}, {Kn{\"o}dlseder}, {Kocian}, {Kuss}, {Lande},
  {Latronico}, {Lemoine-Goumard}, {Longo}, {Loparco}, {Lott}, {Lovellette},
  {Lubrano}, {Madejski}, {Makeev}, {Mazziotta}, {McConville}, {McEnery},
  {Meurer}, {Michelson}, {Mitthumsiri}, {Mizuno}, {Moiseev}, {Monte},
  {Monzani}, {Morselli}, {Moskalenko}, {Murgia}, {Nolan}, {Norris}, {Nuss},
  {Ohsugi}, {Omodei}, {Orlando}, {Ormes}, {Paneque}, {Parent}, {Pelassa},
  {Pepe}, {Pesce-Rollins}, {Piron}, {Porter}, {Rain{\`o}}, {Rando}, {Razzano},
  {Razzaque}, {Reimer}, {Reimer}, {Reposeur}, {Ritz}, {Rochester}, {Rodriguez},
  {Romani}, {Roth}, {Ryde}, {Sadrozinski}, {Sambruna}, {Sanchez}, {Sander},
  {Saz Parkinson}, {Scargle}, {Sgr{\`o}}, {Siskind}, {Smith}, {Smith},
  {Spandre}, {Spinelli}, {Starck}, {Stawarz}, {Strickman}, {Suson}, {Tajima},
  {Takahashi}, {Takahashi}, {Tanaka}, {Thayer}, {Thayer}, {Thompson},
  {Tibaldo}, {Torres}, {Tosti}, {Tramacere}, {Uchiyama}, {Vasileiou},
  {Vilchez}, {Vitale}, {Waite}, {Wallace}, {Wang}, {Winer}, {Wood}, {Ylinen},
  {Ziegler}, {Hardcastle}, {Kazanas}, \& {Fermi LAT
  Collaboration}}]{2010Sci...328..725A}
{Abdo}, A.~A., {Ackermann}, M., {Ajello}, M., {et~al.} 2010{\natexlab{b}},
  Science, 328, 725

\bibitem[{{Abdo} {et~al.}(2010{\natexlab{c}}){Abdo}, {Ackermann}, {Ajello},
  {Baldini}, {Ballet}, {Barbiellini}, {Bastieri}, {Bechtol}, {Bellazzini},
  {Berenji}, {Blandford}, {Bloom}, {Bonamente}, {Borgland}, {Bouvier},
  {Brandt}, {Bregeon}, {Brez}, {Brigida}, {Bruel}, {Buehler}, {Burnett},
  {Buson}, {Caliandro}, {Cameron}, {Cannon}, {Caraveo}, {Carrigan},
  {Casandjian}, {Cavazzuti}, {Cecchi}, {{\c C}elik}, {Celotti}, {Charles},
  {Chekhtman}, {Chen}, {Cheung}, {Chiang}, {Ciprini}, {Claus}, {Cohen-Tanugi},
  {Colafrancesco}, {Conrad}, {Davis}, {Dermer}, {de Angelis}, {de Palma},
  {Silva}, {Drell}, {Dubois}, {Favuzzi}, {Fegan}, {Ferrara}, {Fortin},
  {Frailis}, {Fukazawa}, {Fusco}, {Gargano}, {Gasparrini}, {Gehrels},
  {Germani}, {Giglietto}, {Giommi}, {Giordano}, {Giroletti}, {Glanzman},
  {Godfrey}, {Grandi}, {Grenier}, {Grove}, {Guillemot}, {Guiriec}, {Hadasch},
  {Hayashida}, {Hays}, {Horan}, {Hughes}, {Jackson}, {J{\'o}hannesson},
  {Johnson}, {Johnson}, {Kamae}, {Katagiri}, {Kataoka}, {Kn{\"o}dlseder},
  {Kuss}, {Lande}, {Latronico}, {Lee}, {Lemoine-Goumard}, {Llena Garde},
  {Longo}, {Loparco}, {Lott}, {Lovellette}, {Lubrano}, {Madejski}, {Makeev},
  {Malaguti}, {Mazziotta}, {McConville}, {McEnery}, {Michelson}, {Migliori},
  {Mitthumsiri}, {Mizuno}, {Monte}, {Monzani}, {Morselli}, {Moskalenko},
  {Murgia}, {Naumann-Godo}, {Nestoras}, {Nolan}, {Norris}, {Nuss}, {Ohsugi},
  {Okumura}, {Omodei}, {Orlando}, {Ormes}, {Paneque}, {Panetta}, {Parent},
  {Pelassa}, {Pepe}, {Persic}, {Pesce-Rollins}, {Piron}, {Porter}, {Rain{\`o}},
  {Rando}, {Razzano}, {Razzaque}, {Reimer}, {Reimer}, {Reyes}, {Roth},
  {Sadrozinski}, {Sanchez}, {Sander}, {Scargle}, {Sgr{\`o}}, {Siskind},
  {Smith}, {Spandre}, {Spinelli}, {Stawarz}, {Stecker}, {Strickman}, {Suson},
  {Takahashi}, {Tanaka}, {Thayer}, {Thayer}, {Thompson}, {Tibaldo}, {Torres},
  {Torresi}, {Tosti}, {Tramacere}, {Uchiyama}, {Usher}, {Vandenbroucke},
  {Vasileiou}, {Vilchez}, {Villata}, {Vitale}, {Waite}, {Wang}, {Winer},
  {Wood}, {Yang}, {Ylinen}, \& {Ziegler}}]{2010ApJ...720..912A}
{Abdo}, A.~A., {Ackermann}, M., {Ajello}, M., {et~al.} 2010{\natexlab{c}},
  \apj, 720, 912

\bibitem[{{Acero} {et~al.}(2015){Acero}, {Ackermann}, {Ajello}, {Albert},
  {Atwood}, {Axelsson}, {Baldini}, {Ballet}, {Barbiellini}, {Bastieri},
  {Belfiore}, {Bellazzini}, {Bissaldi}, {Blandford}, {Bloom}, {Bogart},
  {Bonino}, {Bottacini}, {Bregeon}, {Britto}, {Bruel}, {Buehler}, {Burnett},
  {Buson}, {Caliandro}, {Cameron}, {Caputo}, {Caragiulo}, {Caraveo},
  {Casandjian}, {Cavazzuti}, {Charles}, {Chaves}, {Chekhtman}, {Cheung},
  {Chiang}, {Chiaro}, {Ciprini}, {Claus}, {Cohen-Tanugi}, {Cominsky}, {Conrad},
  {Cutini}, {D'Ammando}, {de Angelis}, {DeKlotz}, {de Palma}, {Desiante},
  {Digel}, {Di Venere}, {Drell}, {Dubois}, {Dumora}, {Favuzzi}, {Fegan},
  {Ferrara}, {Finke}, {Franckowiak}, {Fukazawa}, {Funk}, {Fusco}, {Gargano},
  {Gasparrini}, {Giebels}, {Giglietto}, {Giommi}, {Giordano}, {Giroletti},
  {Glanzman}, {Godfrey}, {Grenier}, {Grondin}, {Grove}, {Guillemot}, {Guiriec},
  {Hadasch}, {Harding}, {Hays}, {Hewitt}, {Hill}, {Horan}, {Iafrate}, {Jogler},
  {J{\'o}hannesson}, {Johnson}, {Johnson}, {Johnson}, {Johnson}, {Kamae},
  {Kataoka}, {Katsuta}, {Kuss}, {La Mura}, {Landriu}, {Larsson}, {Latronico},
  {Lemoine-Goumard}, {Li}, {Li}, {Longo}, {Loparco}, {Lott}, {Lovellette},
  {Lubrano}, {Madejski}, {Massaro}, {Mayer}, {Mazziotta}, {McEnery},
  {Michelson}, {Mirabal}, {Mizuno}, {Moiseev}, {Mongelli}, {Monzani},
  {Morselli}, {Moskalenko}, {Murgia}, {Nuss}, {Ohno}, {Ohsugi}, {Omodei},
  {Orienti}, {Orlando}, {Ormes}, {Paneque}, {Panetta}, {Perkins},
  {Pesce-Rollins}, {Piron}, {Pivato}, {Porter}, {Racusin}, {Rando}, {Razzano},
  {Razzaque}, {Reimer}, {Reimer}, {Reposeur}, {Rochester}, {Romani},
  {Salvetti}, {S{\'a}nchez-Conde}, {Saz Parkinson}, {Schulz}, {Siskind},
  {Smith}, {Spada}, {Spandre}, {Spinelli}, {Stephens}, {Strong}, {Suson},
  {Takahashi}, {Takahashi}, {Tanaka}, {Thayer}, {Thayer}, {Thompson},
  {Tibaldo}, {Tibolla}, {Torres}, {Torresi}, {Tosti}, {Troja}, {Van Klaveren},
  {Vianello}, {Winer}, {Wood}, {Wood}, {Zimmer}, \& {Fermi-LAT
  Collaboration}}]{2015ApJS..218...23A}
{Acero}, F., {Ackermann}, M., {Ajello}, M., {et~al.} 2015, \apjs, 218, 23

\bibitem[{{Ackermann} {et~al.}(2011){Ackermann}, {Ajello}, {Allafort},
  {Angelakis}, {Axelsson}, {Baldini}, {Ballet}, {Barbiellini}, {Bastieri},
  {Bellazzini}, {Berenji}, {Blandford}, {Bloom}, {Bonamente}, {Borgland},
  {Bouvier}, {Bregeon}, {Brez}, {Brigida}, {Bruel}, {Buehler}, {Buson},
  {Caliandro}, {Cameron}, {Cannon}, {Caraveo}, {Casandjian}, {Cavazzuti},
  {Cecchi}, {Charles}, {Chekhtman}, {Cheung}, {Ciprini}, {Claus},
  {Cohen-Tanugi}, {Cutini}, {de Palma}, {Dermer}, {Silva}, {Drell}, {Dubois},
  {Dumora}, {Escande}, {Favuzzi}, {Fegan}, {Focke}, {Fortin}, {Frailis},
  {Fuhrmann}, {Fukazawa}, {Fusco}, {Gargano}, {Gasparrini}, {Gehrels},
  {Giglietto}, {Giommi}, {Giordano}, {Giroletti}, {Glanzman}, {Godfrey},
  {Grandi}, {Grenier}, {Guiriec}, {Hadasch}, {Hayashida}, {Hays}, {Healey},
  {J{\'o}hannesson}, {Johnson}, {Kamae}, {Katagiri}, {Kataoka},
  {Kn{\"o}dlseder}, {Kuss}, {Lande}, {Lee}, {Longo}, {Loparco}, {Lott},
  {Lovellette}, {Lubrano}, {Makeev}, {Max-Moerbeck}, {Mazziotta}, {McEnery},
  {Mehault}, {Michelson}, {Mizuno}, {Monte}, {Monzani}, {Morselli},
  {Moskalenko}, {Murgia}, {Naumann-Godo}, {Nishino}, {Nolan}, {Norris}, {Nuss},
  {Ohsugi}, {Okumura}, {Omodei}, {Orlando}, {Ormes}, {Ozaki}, {Paneque},
  {Pavlidou}, {Pelassa}, {Pepe}, {Pesce-Rollins}, {Pierbattista}, {Piron},
  {Porter}, {Rain{\`o}}, {Razzano}, {Readhead}, {Reimer}, {Reimer}, {Richards},
  {Romani}, {Sadrozinski}, {Scargle}, {Sgr{\`o}}, {Siskind}, {Smith},
  {Spandre}, {Spinelli}, {Strickman}, {Suson}, {Takahashi}, {Tanaka}, {Taylor},
  {Thayer}, {Thayer}, {Thompson}, {Torres}, {Tosti}, {Tramacere}, {Troja},
  {Vandenbroucke}, {Vianello}, {Vitale}, {Waite}, {Wang}, {Winer}, {Wood},
  {Yang}, \& {Ziegler}}]{2011ApJ...741...30A}
{Ackermann}, M., {Ajello}, M., {Allafort}, A., {et~al.} 2011, \apj, 741, 30

\bibitem[{{Ackermann} {et~al.}(2015){Ackermann}, {Ajello}, {Atwood}, {Baldini},
  {Ballet}, {Barbiellini}, {Bastieri}, {Becerra Gonzalez}, {Bellazzini},
  {Bissaldi}, {Blandford}, {Bloom}, {Bonino}, {Bottacini}, {Brandt}, {Bregeon},
  {Britto}, {Bruel}, {Buehler}, {Buson}, {Caliandro}, {Cameron}, {Caragiulo},
  {Caraveo}, {Carpenter}, {Casandjian}, {Cavazzuti}, {Cecchi}, {Charles},
  {Chekhtman}, {Cheung}, {Chiang}, {Chiaro}, {Ciprini}, {Claus},
  {Cohen-Tanugi}, {Cominsky}, {Conrad}, {Cutini}, {D'Abrusco}, {D'Ammando}, {de
  Angelis}, {Desiante}, {Digel}, {Di Venere}, {Drell}, {Favuzzi}, {Fegan},
  {Ferrara}, {Finke}, {Focke}, {Franckowiak}, {Fuhrmann}, {Fukazawa},
  {Furniss}, {Fusco}, {Gargano}, {Gasparrini}, {Giglietto}, {Giommi},
  {Giordano}, {Giroletti}, {Glanzman}, {Godfrey}, {Grenier}, {Grove},
  {Guiriec}, {Hewitt}, {Hill}, {Horan}, {Itoh}, {J{\'o}hannesson}, {Johnson},
  {Johnson}, {Kataoka}, {Kawano}, {Krauss}, {Kuss}, {La Mura}, {Larsson},
  {Latronico}, {Leto}, {Li}, {Li}, {Longo}, {Loparco}, {Lott}, {Lovellette},
  {Lubrano}, {Madejski}, {Mayer}, {Mazziotta}, {McEnery}, {Michelson},
  {Mizuno}, {Moiseev}, {Monzani}, {Morselli}, {Moskalenko}, {Murgia}, {Nuss},
  {Ohno}, {Ohsugi}, {Ojha}, {Omodei}, {Orienti}, {Orlando}, {Paggi}, {Paneque},
  {Perkins}, {Pesce-Rollins}, {Piron}, {Pivato}, {Porter}, {Rain{\`o}},
  {Rando}, {Razzano}, {Razzaque}, {Reimer}, {Reimer}, {Romani}, {Salvetti},
  {Schaal}, {Schinzel}, {Schulz}, {Sgr{\`o}}, {Siskind}, {Sokolovsky}, {Spada},
  {Spandre}, {Spinelli}, {Stawarz}, {Suson}, {Takahashi}, {Takahashi},
  {Tanaka}, {Thayer}, {Thayer}, {Tibaldo}, {Torres}, {Torresi}, {Tosti},
  {Troja}, {Uchiyama}, {Vianello}, {Winer}, {Wood}, \&
  {Zimmer}}]{2015ApJ...810...14A}
{Ackermann}, M., {Ajello}, M., {Atwood}, W.~B., {et~al.} 2015, \apj, 810, 14

\bibitem[{{Ackermann} {et~al.}(2016){Ackermann}, {Ajello}, {Baldini}, {Ballet},
  {Barbiellini}, {Bastieri}, {Bellazzini}, {Bissaldi}, {Blandford}, {Bloom},
  {Bonino}, {Brandt}, {Bregeon}, {Bruel}, {Buehler}, {Buson}, {Caliandro},
  {Cameron}, {Caragiulo}, {Caraveo}, {Cavazzuti}, {Cecchi}, {Charles},
  {Chekhtman}, {Cheung}, {Chiaro}, {Ciprini}, {Cohen}, {Cohen-Tanugi},
  {Costanza}, {Cutini}, {D'Ammando}, {Davis}, {de Angelis}, {de Palma},
  {Desiante}, {Digel}, {Di Lalla}, {Di Mauro}, {Di Venere}, {Favuzzi}, {Fegan},
  {Ferrara}, {Focke}, {Fukazawa}, {Funk}, {Fusco}, {Gargano}, {Gasparrini},
  {Georganopoulos}, {Giglietto}, {Giordano}, {Giroletti}, {Godfrey}, {Green},
  {Grenier}, {Guiriec}, {Hays}, {Hewitt}, {Hill}, {Jogler}, {J{\'o}hannesson},
  {Kensei}, {Kuss}, {Larsson}, {Latronico}, {Li}, {Li}, {Longo}, {Loparco},
  {Lubrano}, {Magill}, {Maldera}, {Manfreda}, {Mayer}, {Mazziotta},
  {McConville}, {McEnery}, {Michelson}, {Mitthumsiri}, {Mizuno}, {Monzani},
  {Morselli}, {Moskalenko}, {Murgia}, {Negro}, {Nuss}, {Ohno}, {Ohsugi},
  {Orienti}, {Orlando}, {Ormes}, {Paneque}, {Perkins}, {Pesce-Rollins},
  {Piron}, {Pivato}, {Porter}, {Rain{\`o}}, {Rando}, {Razzano}, {Reimer},
  {Reimer}, {Schmid}, {Sgr{\`o}}, {Simone}, {Siskind}, {Spada}, {Spandre},
  {Spinelli}, {Stawarz}, {Takahashi}, {Thayer}, {Thompson}, {Torres}, {Tosti},
  {Troja}, {Vianello}, {Wood}, {Wood}, {Zimmer}, \& {Fermi LAT
  Collaboration}}]{2016ApJ...826....1A}
{Ackermann}, M., {Ajello}, M., {Baldini}, L., {et~al.} 2016, \apj, 826, 1

\bibitem[{{Ajello} {et~al.}(2017){Ajello}, {Atwood}, {Baldini}, {Ballet},
  {Barbiellini}, {Bastieri}, {Bellazzini}, {Bissaldi}, {Blandford}, {Bloom},
  {Bonino}, {Bregeon}, {Britto}, {Bruel}, {Buehler}, {Buson}, {Cameron},
  {Caputo}, {Caragiulo}, {Caraveo}, {Cavazzuti}, {Cecchi}, {Charles},
  {Chekhtman}, {Cheung}, {Chiaro}, {Ciprini}, {Cohen}, {Costantin}, {Costanza},
  {Cuoco}, {Cutini}, {D'Ammando}, {de Palma}, {Desiante}, {Digel}, {Di Lalla},
  {Di Mauro}, {Di Venere}, {Dom{\'{\i}}nguez}, {Drell}, {Dumora}, {Favuzzi},
  {Fegan}, {Ferrara}, {Fortin}, {Franckowiak}, {Fukazawa}, {Funk}, {Fusco},
  {Gargano}, {Gasparrini}, {Giglietto}, {Giommi}, {Giordano}, {Giroletti},
  {Glanzman}, {Green}, {Grenier}, {Grondin}, {Grove}, {Guillemot}, {Guiriec},
  {Harding}, {Hays}, {Hewitt}, {Horan}, {J{\'o}hannesson}, {Kensei}, {Kuss},
  {La Mura}, {Larsson}, {Latronico}, {Lemoine-Goumard}, {Li}, {Longo},
  {Loparco}, {Lott}, {Lubrano}, {Magill}, {Maldera}, {Manfreda}, {Mazziotta},
  {McEnery}, {Meyer}, {Michelson}, {Mirabal}, {Mitthumsiri}, {Mizuno},
  {Moiseev}, {Monzani}, {Morselli}, {Moskalenko}, {Negro}, {Nuss}, {Ohsugi},
  {Omodei}, {Orienti}, {Orlando}, {Palatiello}, {Paliya}, {Paneque}, {Perkins},
  {Persic}, {Pesce-Rollins}, {Piron}, {Porter}, {Principe}, {Rain{\`o}},
  {Rando}, {Razzano}, {Razzaque}, {Reimer}, {Reimer}, {Reposeur}, {Saz
  Parkinson}, {Sgr{\`o}}, {Simone}, {Siskind}, {Spada}, {Spandre}, {Spinelli},
  {Stawarz}, {Suson}, {Takahashi}, {Tak}, {Thayer}, {Thayer}, {Thompson},
  {Torres}, {Torresi}, {Troja}, {Vianello}, {Wood}, \&
  {Wood}}]{2017ApJS..232...18A}
{Ajello}, M., {Atwood}, W.~B., {Baldini}, L., {et~al.} 2017, \apjs, 232, 18

\bibitem[{{Akritas} \& {Siebert}(1996)}]{Akritas1996}
{Akritas}, M.~G. \& {Siebert}, J. 1996, \mnras, 278, 919

\bibitem[{{Aleksi{\'c}} {et~al.}(2014){Aleksi{\'c}}, {Ansoldi}, {Antonelli},
  {Antoranz}, {Babic}, {Bangale}, {Barrio}, {Gonz{\'a}lez}, {Bednarek},
  {Bernardini}, {Biasuzzi}, {Biland}, {Blanch}, {Bonnefoy}, {Bonnoli},
  {Borracci}, {Bretz}, {Carmona}, {Carosi}, {Colin}, {Colombo}, {Contreras},
  {Cortina}, {Covino}, {Da Vela}, {Dazzi}, {De Angelis}, {De Caneva}, {De
  Lotto}, {Wilhelmi}, {Mendez}, {Prester}, {Dorner}, {Doro}, {Einecke},
  {Eisenacher}, {Elsaesser}, {Fonseca}, {Font}, {Frantzen}, {Fruck}, {Galindo},
  {L{\'o}pez}, {Garczarczyk}, {Terrats}, {Gaug}, {Godinovi{\'c}}, {Mu{\~n}oz},
  {Gozzini}, {Hadasch}, {Hanabata}, {Hayashida}, {Herrera}, {Hildebrand},
  {Hose}, {Hrupec}, {Idec}, {Kadenius}, {Kellermann}, {Kodani}, {Konno},
  {Krause}, {Kubo}, {Kushida}, {La Barbera}, {Lelas}, {Lewandowska},
  {Lindfors}, {Lombardi}, {Longo}, {L{\'o}pez}, {L{\'o}pez-Coto},
  {L{\'o}pez-Oramas}, {Lorenz}, {Lozano}, {Makariev}, {Mallot}, {Maneva},
  {Mankuzhiyil}, {Mannheim}, {Maraschi}, {Marcote}, {Mariotti},
  {Mart{\'{\i}}nez}, {Mazin}, {Menzel}, {Miranda}, {Mirzoyan}, {Moralejo},
  {Munar-Adrover}, {Nakajima}, {Niedzwiecki}, {Nilsson}, {Nishijima}, {Noda},
  {Orito}, {Overkemping}, {Paiano}, {Palatiello}, {Paneque}, {Paoletti},
  {Paredes}, {Paredes-Fortuny}, {Persic}, {Poutanen}, {Moroni}, {Prandini},
  {Puljak}, {Reinthal}, {Rhode}, {Rib{\'o}}, {Rico}, {Garcia}, {R{\"u}gamer},
  {Saito}, {Saito}, {Satalecka}, {Scalzotto}, {Scapin}, {Schultz}, {Schweizer},
  {Shore}, {Sillanp{\"a}{\"a}}, {Sitarek}, {Snidaric}, {Sobczynska}, {Spanier},
  {Stamatescu}, {Stamerra}, {Steinbring}, {Storz}, {Strzys}, {Takalo},
  {Takami}, {Tavecchio}, {Temnikov}, {Terzi{\'c}}, {Tescaro}, {Teshima},
  {Thaele}, {Tibolla}, {Torres}, {Toyama}, {Treves}, {Uellenbeck}, {Vogler},
  {Zanin}, {Kadler}, {Schulz}, {Ros}, {Bach}, {Krau{\ss}}, \&
  {Wilms}}]{2014Sci...346.1080A}
{Aleksi{\'c}}, J., {Ansoldi}, S., {Antonelli}, L.~A., {et~al.} 2014, Science,
  346, 1080

\bibitem[{{An} \& {Baan}(2012)}]{2012ApJ...760...77A}
{An}, T. \& {Baan}, W.~A. 2012, \apj, 760, 77

\bibitem[{{Angioni} {et~al.}(2017){Angioni}, {Grandi}, {Torresi}, {Vignali}, \&
  {Kn{\"o}dlseder}}]{2017APh....92...42A}
{Angioni}, R., {Grandi}, P., {Torresi}, E., {Vignali}, C., \& {Kn{\"o}dlseder},
  J. 2017, Astroparticle Physics, 92, 42

\bibitem[{{Astropy Collaboration} {et~al.}(2018){Astropy Collaboration},
  {Price-Whelan}, {Sip{\H o}cz}, {G{\"u}nther}, {Lim}, {Crawford}, {Conseil},
  {Shupe}, {Craig}, {Dencheva}, {Ginsburg}, {VanderPlas}, {Bradley},
  {P{\'e}rez-Su{\'a}rez}, {de Val-Borro}, {Aldcroft}, {Cruz}, {Robitaille},
  {Tollerud}, {Ardelean}, {Babej}, {Bach}, {Bachetti}, {Bakanov}, {Bamford},
  {Barentsen}, {Barmby}, {Baumbach}, {Berry}, {Biscani}, {Boquien}, {Bostroem},
  {Bouma}, {Brammer}, {Bray}, {Breytenbach}, {Buddelmeijer}, {Burke},
  {Calderone}, {Cano Rodr{\'{\i}}guez}, {Cara}, {Cardoso}, {Cheedella},
  {Copin}, {Corrales}, {Crichton}, {D'Avella}, {Deil}, {Depagne}, {Dietrich},
  {Donath}, {Droettboom}, {Earl}, {Erben}, {Fabbro}, {Ferreira}, {Finethy},
  {Fox}, {Garrison}, {Gibbons}, {Goldstein}, {Gommers}, {Greco}, {Greenfield},
  {Groener}, {Grollier}, {Hagen}, {Hirst}, {Homeier}, {Horton}, {Hosseinzadeh},
  {Hu}, {Hunkeler}, {Ivezi{\'c}}, {Jain}, {Jenness}, {Kanarek}, {Kendrew},
  {Kern}, {Kerzendorf}, {Khvalko}, {King}, {Kirkby}, {Kulkarni}, {Kumar},
  {Lee}, {Lenz}, {Littlefair}, {Ma}, {Macleod}, {Mastropietro}, {McCully},
  {Montagnac}, {Morris}, {Mueller}, {Mumford}, {Muna}, {Murphy}, {Nelson},
  {Nguyen}, {Ninan}, {N{\"o}the}, {Ogaz}, {Oh}, {Parejko}, {Parley}, {Pascual},
  {Patil}, {Patil}, {Plunkett}, {Prochaska}, {Rastogi}, {Reddy Janga},
  {Sabater}, {Sakurikar}, {Seifert}, {Sherbert}, {Sherwood-Taylor}, {Shih},
  {Sick}, {Silbiger}, {Singanamalla}, {Singer}, {Sladen}, {Sooley},
  {Sornarajah}, {Streicher}, {Teuben}, {Thomas}, {Tremblay}, {Turner},
  {Terr{\'o}n}, {van Kerkwijk}, {de la Vega}, {Watkins}, {Weaver}, {Whitmore},
  {Woillez}, {Zabalza}, \& {Astropy Contributors}}]{astropy:2018}
{Astropy Collaboration}, {Price-Whelan}, A.~M., {Sip{\H o}cz}, B.~M., {et~al.}
  2018, \aj, 156, 123

\bibitem[{{Astropy Collaboration} {et~al.}(2013){Astropy Collaboration},
  {Robitaille}, {Tollerud}, {Greenfield}, {Droettboom}, {Bray}, {Aldcroft},
  {Davis}, {Ginsburg}, {Price-Whelan}, {Kerzendorf}, {Conley}, {Crighton},
  {Barbary}, {Muna}, {Ferguson}, {Grollier}, {Parikh}, {Nair}, {Unther},
  {Deil}, {Woillez}, {Conseil}, {Kramer}, {Turner}, {Singer}, {Fox}, {Weaver},
  {Zabalza}, {Edwards}, {Azalee Bostroem}, {Burke}, {Casey}, {Crawford},
  {Dencheva}, {Ely}, {Jenness}, {Labrie}, {Lim}, {Pierfederici}, {Pontzen},
  {Ptak}, {Refsdal}, {Servillat}, \& {Streicher}}]{astropy:2013}
{Astropy Collaboration}, {Robitaille}, T.~P., {Tollerud}, E.~J., {et~al.} 2013,
  \aap, 558, A33

\bibitem[{{Atwood} {et~al.}(2009){Atwood}, {Abdo}, {Ackermann}, {Althouse},
  {Anderson}, {Axelsson}, {Baldini}, {Ballet}, {Band}, {Barbiellini}, \&
  et~al.}]{2009ApJ...697.1071A}
{Atwood}, W.~B., {Abdo}, A.~A., {Ackermann}, M., {et~al.} 2009, \apj, 697, 1071

\bibitem[{{Baczko} {et~al.}(2016){Baczko}, {Schulz}, {Kadler}, {Ros},
  {Perucho}, {Krichbaum}, {B{\"o}ck}, {Bremer}, {Grossberger}, {Lindqvist},
  {Lobanov}, {Mannheim}, {Mart{\'{\i}}-Vidal}, {M{\"u}ller}, {Wilms}, \&
  {Zensus}}]{2016A&A...593A..47B}
{Baczko}, A.-K., {Schulz}, R., {Kadler}, M., {et~al.} 2016, \aap, 593, A47

\bibitem[{{B{\"o}ck} {et~al.}(2016){B{\"o}ck}, {Kadler}, {M{\"u}ller}, {Tosti},
  {Ojha}, {Wilms}, {Bastieri}, {Burnett}, {Carpenter}, {Cavazzuti}, {Dutka},
  {Blanchard}, {Edwards}, {Hase}, {Horiuchi}, {Jauncey}, {Krau{\ss}}, {Lister},
  {Lovell}, {Lott}, {Murphy}, {Phillips}, {Pl{\"o}tz}, {Pursimo}, {Quick},
  {Ros}, {Taylor}, {Thompson}, {Tingay}, {Tzioumis}, \&
  {Zensus}}]{2016A&A...590A..40B}
{B{\"o}ck}, M., {Kadler}, M., {M{\"u}ller}, C., {et~al.} 2016, \aap, 590, A40

\bibitem[{{Brown} \& {Adams}(2012)}]{2012MNRAS.421.2303B}
{Brown}, A.~M. \& {Adams}, J. 2012, \mnras, 421, 2303

\bibitem[{{Casadio} {et~al.}(2015){Casadio}, {G{\'o}mez}, {Grandi}, {Jorstad},
  {Marscher}, {Lister}, {Kovalev}, {Savolainen}, \&
  {Pushkarev}}]{2015ApJ...808..162C}
{Casadio}, C., {G{\'o}mez}, J.~L., {Grandi}, P., {et~al.} 2015, \apj, 808, 162

\bibitem[{{Cohen} {et~al.}(2007){Cohen}, {Lister}, {Homan}, {Kadler},
  {Kellermann}, {Kovalev}, \& {Vermeulen}}]{2007ApJ...658..232C}
{Cohen}, M.~H., {Lister}, M.~L., {Homan}, D.~C., {et~al.} 2007, \apj, 658, 232

\bibitem[{{D'Ammando} {et~al.}(2015){D'Ammando}, {Orienti}, {Tavecchio},
  {Ghisellini}, {Torresi}, {Giroletti}, {Raiteri}, {Grandi}, {Aller}, {Aller},
  {Gurwell}, {Malaguti}, {Pian}, \& {Tosti}}]{2015MNRAS.450.3975D}
{D'Ammando}, F., {Orienti}, M., {Tavecchio}, F., {et~al.} 2015, \mnras, 450,
  3975

\bibitem[{{Fey} {et~al.}(2004){Fey}, {Ojha}, {Reynolds}, {Ellingsen},
  {McCulloch}, {Jauncey}, \& {Johnston}}]{2004AJ....128.2593F}
{Fey}, A.~L., {Ojha}, R., {Reynolds}, J.~E., {et~al.} 2004, \aj, 128, 2593

\bibitem[{{Fromm} {et~al.}(2013){Fromm}, {Ros}, {Perucho}, {Savolainen},
  {Mimica}, {Kadler}, {Lobanov}, \& {Zensus}}]{2013A&A...557A.105F}
{Fromm}, C.~M., {Ros}, E., {Perucho}, M., {et~al.} 2013, \aap, 557, A105

\bibitem[{{Ghisellini} {et~al.}(2005){Ghisellini}, {Tavecchio}, \&
  {Chiaberge}}]{2005A&A...432..401G}
{Ghisellini}, G., {Tavecchio}, F., \& {Chiaberge}, M. 2005, \aap, 432, 401

\bibitem[{{Giroletti} {et~al.}(2004){Giroletti}, {Giovannini}, {Taylor}, \&
  {Falomo}}]{2004ApJ...613..752G}
{Giroletti}, M., {Giovannini}, G., {Taylor}, G.~B., \& {Falomo}, R. 2004, \apj,
  613, 752

\bibitem[{{Giroletti} \& {Polatidis}(2009)}]{2009AN....330..193G}
{Giroletti}, M. \& {Polatidis}, A. 2009, Astronomische Nachrichten, 330, 193

\bibitem[{{Goldoni} {et~al.}(2016){Goldoni}, {Pita}, {Boisson}, {M{\"u}ller},
  {Dauser}, {Jung}, {Krau{\ss}}, {Lenain}, \& {Sol}}]{2016A&A...586L...2G}
{Goldoni}, P., {Pita}, S., {Boisson}, C., {et~al.} 2016, \aap, 586, L2

\bibitem[{{Grandi} {et~al.}(2016){Grandi}, {Capetti}, \&
  {Baldi}}]{2016MNRAS.457....2G}
{Grandi}, P., {Capetti}, A., \& {Baldi}, R.~D. 2016, \mnras, 457, 2

\bibitem[{{Grandi} {et~al.}(2013){Grandi}, {Torresi}, {De Rosa}, {Rain{\'o}},
  \& {Malaguti}}]{2013EPJWC..6104007G}
{Grandi}, P., {Torresi}, E., {De Rosa}, A., {Rain{\'o}}, S., \& {Malaguti}, G.
  2013, in European Physical Journal Web of Conferences, Vol.~61, European
  Physical Journal Web of Conferences, 04007

\bibitem[{{Grandi} {et~al.}(2012){Grandi}, {Torresi}, \&
  {Stanghellini}}]{2012ApJ...751L...3G}
{Grandi}, P., {Torresi}, E., \& {Stanghellini}, C. 2012, \apjl, 751, L3

\bibitem[{{Greisen}(1998)}]{aips}
{Greisen}, E.~W. 1998, in Astronomical Society of the Pacific Conference
  Series, Vol. 145, Astronomical Data Analysis Software and Systems VII, ed.
  R.~{Albrecht}, R.~N. {Hook}, \& H.~A. {Bushouse}, 204

\bibitem[{{Hancock} {et~al.}(2009){Hancock}, {Tingay}, {Sadler}, {Phillips}, \&
  {Deller}}]{2009MNRAS.397.2030H}
{Hancock}, P.~J., {Tingay}, S.~J., {Sadler}, E.~M., {Phillips}, C., \&
  {Deller}, A.~T. 2009, \mnras, 397, 2030

\bibitem[{{Hardcastle} {et~al.}(2016){Hardcastle}, {Lenc}, {Birkinshaw},
  {Croston}, {Goodger}, {Marshall}, {Perlman}, {Siemiginowska}, {Stawarz}, \&
  {Worrall}}]{2016MNRAS.455.3526H}
{Hardcastle}, M.~J., {Lenc}, E., {Birkinshaw}, M., {et~al.} 2016, \mnras, 455,
  3526

\bibitem[{{H{\"o}gbom}(1974)}]{1974A&AS...15..417H}
{H{\"o}gbom}, J.~A. 1974, \aaps, 15, 417

\bibitem[{{Hovatta} {et~al.}(2014){Hovatta}, {Aller}, {Aller}, {Clausen-Brown},
  {Homan}, {Kovalev}, {Lister}, {Pushkarev}, \& {Savolainen}}]{Hovatta2014}
{Hovatta}, T., {Aller}, M.~F., {Aller}, H.~D., {et~al.} 2014, \aj, 147, 143

\bibitem[{{Janiak} {et~al.}(2016){Janiak}, {Sikora}, \&
  {Moderski}}]{2016MNRAS.458.2360J}
{Janiak}, M., {Sikora}, M., \& {Moderski}, R. 2016, \mnras, 458, 2360

\bibitem[{{Kadler} {et~al.}(2015){Kadler}, {Ojha}, \& {TANAMI
  Collaboration}}]{2015AN....336..499K}
{Kadler}, M., {Ojha}, R., \& {TANAMI Collaboration}. 2015, Astronomische
  Nachrichten, 336, 499

\bibitem[{{Kim} {et~al.}(2018){Kim}, {Krichbaum}, {Lu}, {Ros}, {Bach},
  {Bremer}, {de Vicente}, {Lindqvist}, \& {Zensus}}]{2018A&A...616A.188K}
{Kim}, J.-Y., {Krichbaum}, T.~P., {Lu}, R.-S., {et~al.} 2018, \aap, 616, A188

\bibitem[{{Komatsu} {et~al.}(2011){Komatsu}, {Smith}, {Dunkley}, {Bennett},
  {Gold}, {Hinshaw}, {Jarosik}, {Larson}, {Nolta}, {Page}, {Spergel},
  {Halpern}, {Hill}, {Kogut}, {Limon}, {Meyer}, {Odegard}, {Tucker}, {Weiland},
  {Wollack}, \& {Wright}}]{2011ApJS..192...18K}
{Komatsu}, E., {Smith}, K.~M., {Dunkley}, J., {et~al.} 2011, \apjs, 192, 18

\bibitem[{{Kovalev} {et~al.}(2009){Kovalev}, {Aller}, {Aller}, {Homan},
  {Kadler}, {Kellermann}, {Kovalev}, {Lister}, {McCormick}, {Pushkarev}, {Ros},
  \& {Zensus}}]{2009ApJ...696L..17K}
{Kovalev}, Y.~Y., {Aller}, H.~D., {Aller}, M.~F., {et~al.} 2009, \apjl, 696,
  L17

\bibitem[{{Kovalev} {et~al.}(2005){Kovalev}, {Kellermann}, {Lister}, {Homan},
  {Vermeulen}, {Cohen}, {Ros}, {Kadler}, {Lobanov}, {Zensus}, {Kardashev},
  {Gurvits}, {Aller}, \& {Aller}}]{2005AJ....130.2473K}
{Kovalev}, Y.~Y., {Kellermann}, K.~I., {Lister}, M.~L., {et~al.} 2005, \aj,
  130, 2473

\bibitem[{{Krau{\ss}} {et~al.}(2018){Krau{\ss}}, {Kreter}, {M{\"u}ller},
  {Markowitz}, {B{\"o}ck}, {Burnett}, {Dauser}, {Kadler}, {Kreikenbohm},
  {Ojha}, \& {Wilms}}]{2018A&A...610L...8K}
{Krau{\ss}}, F., {Kreter}, M., {M{\"u}ller}, C., {et~al.} 2018, \aap, 610, L8

\bibitem[{{Krau{\ss}} {et~al.}(2016){Krau{\ss}}, {Wilms}, {Kadler}, {Ojha},
  {Schulz}, {Tr{\"u}stedt}, {Edwards}, {Stevens}, {Ros}, {Baumgartner},
  {Beuchert}, {Blanchard}, {Buson}, {Carpenter}, {Dauser}, {Falkner},
  {Gehrels}, {Gr{\"a}fe}, {Gulyaev}, {Hase}, {Horiuchi}, {Kreikenbohm},
  {Kreykenbohm}, {Langejahn}, {Leiter}, {Lovell}, {M{\"u}ller}, {Natusch},
  {Nesci}, {Pursimo}, {Phillips}, {Pl{\"o}tz}, {Quick}, {Tzioumis}, \&
  {Weston}}]{2016A&A...591A.130K}
{Krau{\ss}}, F., {Wilms}, J., {Kadler}, M., {et~al.} 2016, \aap, 591, A130

\bibitem[{{Kuehr} {et~al.}(1981){Kuehr}, {Witzel}, {Pauliny-Toth}, \&
  {Nauber}}]{1981A&AS...45..367K}
{Kuehr}, H., {Witzel}, A., {Pauliny-Toth}, I.~I.~K., \& {Nauber}, U. 1981,
  \aaps, 45, 367

\bibitem[{{Laing} {et~al.}(1983){Laing}, {Riley}, \&
  {Longair}}]{1983MNRAS.204..151L}
{Laing}, R.~A., {Riley}, J.~M., \& {Longair}, M.~S. 1983, \mnras, 204, 151

\bibitem[{{Leon} {et~al.}(2016){Leon}, {Cortes}, {Guerard}, {Villard},
  {Hidayat}, {Oca{\~n}a Flaquer}, \& {Vila-Vilaro}}]{2016A&A...586A..70L}
{Leon}, S., {Cortes}, P.~C., {Guerard}, M., {et~al.} 2016, \aap, 586, A70

\bibitem[{{Lico} {et~al.}(2017){Lico}, {Giroletti}, {Orienti}, {Costamante},
  {Pavlidou}, {D'Ammando}, \& {Tavecchio}}]{2017A&A...606A.138L}
{Lico}, R., {Giroletti}, M., {Orienti}, M., {et~al.} 2017, \aap, 606, A138

\bibitem[{{Lico} {et~al.}(2012){Lico}, {Giroletti}, {Orienti}, {Giovannini},
  {Cotton}, {Edwards}, {Fuhrmann}, {Krichbaum}, {Sokolovsky}, {Kovalev},
  {Jorstad}, {Marscher}, {Kino}, {Paneque}, {Perez-Torres}, \&
  {Piner}}]{2012A&A...545A.117L}
{Lico}, R., {Giroletti}, M., {Orienti}, M., {et~al.} 2012, \aap, 545, A117

\bibitem[{{Liska} {et~al.}(2018){Liska}, {Hesp}, {Tchekhovskoy}, {Ingram}, {van
  der Klis}, \& {Markoff}}]{2017arXiv170706619L}
{Liska}, M., {Hesp}, C., {Tchekhovskoy}, A., {et~al.} 2018, \mnras, 474, L81

\bibitem[{{Lister} {et~al.}(2018){Lister}, {Aller}, {Aller}, {Hodge}, {Homan},
  {Kovalev}, {Pushkarev}, \& {Savolainen}}]{2018ApJS..234...12L}
{Lister}, M.~L., {Aller}, M.~F., {Aller}, H.~D., {et~al.} 2018, \apjs, 234, 12

\bibitem[{{Lister} {et~al.}(2013){Lister}, {Aller}, {Aller}, {Homan},
  {Kellermann}, {Kovalev}, {Pushkarev}, {Richards}, {Ros}, \&
  {Savolainen}}]{2013AJ....146..120L}
{Lister}, M.~L., {Aller}, M.~F., {Aller}, H.~D., {et~al.} 2013, \aj, 146, 120

\bibitem[{{Lister} {et~al.}(2016){Lister}, {Aller}, {Aller}, {Homan},
  {Kellermann}, {Kovalev}, {Pushkarev}, {Richards}, {Ros}, \&
  {Savolainen}}]{2016AJ....152...12L}
{Lister}, M.~L., {Aller}, M.~F., {Aller}, H.~D., {et~al.} 2016, \aj, 152, 12

\bibitem[{{Lister} {et~al.}(2009){Lister}, {Cohen}, {Homan}, {Kadler},
  {Kellermann}, {Kovalev}, {Ros}, {Savolainen}, \&
  {Zensus}}]{2009AJ....138.1874L}
{Lister}, M.~L., {Cohen}, M.~H., {Homan}, D.~C., {et~al.} 2009, \aj, 138, 1874

\bibitem[{{Mattox} {et~al.}(1996){Mattox}, {Bertsch}, {Chiang}, {Dingus},
  {Digel}, {Esposito}, {Fierro}, {Hartman}, {Hunter}, {Kanbach}, {Kniffen},
  {Lin}, {Macomb}, {Mayer-Hasselwander}, {Michelson}, {von Montigny},
  {Mukherjee}, {Nolan}, {Ramanamurthy}, {Schneid}, {Sreekumar}, {Thompson}, \&
  {Willis}}]{1996ApJ...461..396M}
{Mattox}, J.~R., {Bertsch}, D.~L., {Chiang}, J., {et~al.} 1996, \apj, 461, 396

\bibitem[{{Migliori} {et~al.}(2016){Migliori}, {Siemiginowska}, {Sobolewska},
  {Loh}, {Corbel}, {Ostorero}, \& {Stawarz}}]{2016ApJ...821L..31M}
{Migliori}, G., {Siemiginowska}, A., {Sobolewska}, M., {et~al.} 2016, \apjl,
  821, L31

\bibitem[{{M{\"u}ller} {et~al.}(2016){M{\"u}ller}, {Burd}, {Schulz},
  {Coppejans}, {Falcke}, {Intema}, {Kadler}, {Krau{\ss}}, \&
  {Ojha}}]{2016A&A...593L..19M}
{M{\"u}ller}, C., {Burd}, P.~R., {Schulz}, R., {et~al.} 2016, \aap, 593, L19

\bibitem[{{M{\"u}ller} {et~al.}(2014{\natexlab{a}}){M{\"u}ller}, {Kadler},
  {Ojha}, {B{\"o}ck}, {Krau{\ss}}, {Taylor}, {Wilms}, {Blanchard}, {Carpenter},
  {Dauser}, {Dutka}, {Edwards}, {Gehrels}, {Gro{\ss}berger}, {Hase},
  {Horiuchi}, {Kreikenbohm}, {Lovell}, {McConville}, {Phillips}, {Pl{\"o}tz},
  {Pursimo}, {Quick}, {Ros}, {Schulz}, {Stevens}, {Tingay}, {Tr{\"u}stedt},
  {Tzioumis}, \& {Zensus}}]{2014A&A...562A...4M}
{M{\"u}ller}, C., {Kadler}, M., {Ojha}, R., {et~al.} 2014{\natexlab{a}}, \aap,
  562, A4

\bibitem[{{M{\"u}ller} {et~al.}(2014{\natexlab{b}}){M{\"u}ller}, {Kadler},
  {Ojha}, {Perucho}, {Gro{\ss}berger}, {Ros}, {Wilms}, {Blanchard}, {B{\"o}ck},
  {Carpenter}, {Dutka}, {Edwards}, {Hase}, {Horiuchi}, {Kreikenbohm}, {Lovell},
  {Markowitz}, {Phillips}, {Pl{\"o}tz}, {Pursimo}, {Quick}, {Rothschild},
  {Schulz}, {Steinbring}, {Stevens}, {Tr{\"u}stedt}, \&
  {Tzioumis}}]{2014A&A...569A.115M}
{M{\"u}ller}, C., {Kadler}, M., {Ojha}, R., {et~al.} 2014{\natexlab{b}}, \aap,
  569, A115

\bibitem[{{M{\"u}ller} {et~al.}(2018){M{\"u}ller}, {Kadler}, {Ojha}, {Schulz},
  {Tr{\"u}stedt}, {Edwards}, {Ros}, {Carpenter}, {Angioni}, {Blanchard},
  {B{\"o}ck}, {Burd}, {D{\"o}rr}, {Dutka}, {Eberl}, {Gulyaev}, {Hase},
  {Horiuchi}, {Katz}, {Krau{\ss}}, {Lovell}, {Natusch}, {Nesci}, {Phillips},
  {Pl{\"o}tz}, {Pursimo}, {Quick}, {Stevens}, {Thompson}, {Tingay}, {Tzioumis},
  {Weston}, {Wilms}, \& {Zensus}}]{2017arXiv170903091M}
{M{\"u}ller}, C., {Kadler}, M., {Ojha}, R., {et~al.} 2018, \aap, 610, A1

\bibitem[{{M{\"u}ller} {et~al.}(2015){M{\"u}ller}, {Krau{\ss}}, {Dauser},
  {Kreikenbohm}, {Beuchert}, {Kadler}, {Ojha}, {Wilms}, {B{\"o}ck},
  {Carpenter}, {Dutka}, {Markowitz}, {McConville}, {Pottschmidt}, {Stawarz}, \&
  {Taylor}}]{2015A&A...574A.117M}
{M{\"u}ller}, C., {Krau{\ss}}, F., {Dauser}, T., {et~al.} 2015, \aap, 574, A117

\bibitem[{{M\"uller} {et~al.}(2012){M\"uller}, {Krauss}, {Kadler}, {Truestedt},
  {Ojha}, {Ros}, {Wilms}, {Boeck}, {Dutka}, \&
  {Carpenter}}]{2012evn..confE..20M}
{M\"uller}, C., {Krauss}, F., {Kadler}, M., {et~al.} 2012, in Proceedings of
  the 11th European VLBI Network Symposium \& Users Meeting. 9-12 October,
  2012. Bordeaux (France), 20

\bibitem[{{Nolan} {et~al.}(2012){Nolan}, {Abdo}, {Ackermann}, {Ajello},
  {Allafort}, {Antolini}, {Atwood}, {Axelsson}, {Baldini}, {Ballet}, \&
  et~al.}]{2012ApJS..199...31N}
{Nolan}, P.~L., {Abdo}, A.~A., {Ackermann}, M., {et~al.} 2012, \apjs, 199, 31

\bibitem[{{O'Dea}(1998)}]{1998PASP..110..493O}
{O'Dea}, C.~P. 1998, \pasp, 110, 493

\bibitem[{{Ojha} {et~al.}(2010){Ojha}, {Kadler}, {B{\"o}ck}, {Booth}, {Dutka},
  {Edwards}, {Fey}, {Fuhrmann}, {Gaume}, {Hase}, {Horiuchi}, {Jauncey},
  {Johnston}, {Katz}, {Lister}, {Lovell}, {M{\"u}ller}, {Pl{\"o}tz}, {Quick},
  {Ros}, {Taylor}, {Thompson}, {Tingay}, {Tosti}, {Tzioumis}, {Wilms}, \&
  {Zensus}}]{2010A&A...519A..45O}
{Ojha}, R., {Kadler}, M., {B{\"o}ck}, M., {et~al.} 2010, \aap, 519, A45

\bibitem[{{Orienti} \& {Dallacasa}(2014)}]{2014MNRAS.438..463O}
{Orienti}, M. \& {Dallacasa}, D. 2014, \mnras, 438, 463

\bibitem[{{Petrov} {et~al.}(2005){Petrov}, {Kovalev}, {Fomalont}, \&
  {Gordon}}]{2005AJ....129.1163P}
{Petrov}, L., {Kovalev}, Y.~Y., {Fomalont}, E., \& {Gordon}, D. 2005, \aj, 129,
  1163

\bibitem[{{Pian} {et~al.}(1996){Pian}, {Falomo}, {Ghisellini}, {Maraschi},
  {Sambruna}, {Scarpa}, \& {Treves}}]{1996ApJ...459..169P}
{Pian}, E., {Falomo}, R., {Ghisellini}, G., {et~al.} 1996, \apj, 459, 169

\bibitem[{{Piner} \& {Edwards}(2018)}]{2018ApJ...853...68P}
{Piner}, B.~G. \& {Edwards}, P.~G. 2018, \apj, 853, 68

\bibitem[{{Robitaille} \& {Bressert}(2012)}]{aplpy}
{Robitaille}, T. \& {Bressert}, E. 2012, {APLpy: Astronomical Plotting Library
  in Python}, Astrophysics Source Code Library

\bibitem[{{Schinzel} {et~al.}(2012){Schinzel}, {Lobanov}, {Taylor}, {Jorstad},
  {Marscher}, \& {Zensus}}]{2012A&A...537A..70S}
{Schinzel}, F.~K., {Lobanov}, A.~P., {Taylor}, G.~B., {et~al.} 2012, \aap, 537,
  A70

\bibitem[{{Schulz}(2016)}]{2016PhDT.......104S}
{Schulz}, R.~F. 2016, PhD thesis, Julius-Maximilians-Universit{\"a}t
  W{\"a}rzburg

\bibitem[{{Shepherd} {et~al.}(1994){Shepherd}, {Pearson}, \& {Taylor}}]{difmap}
{Shepherd}, M.~C., {Pearson}, T.~J., \& {Taylor}, G.~B. 1994, in \baas,
  Vol.~26, Bulletin of the American Astronomical Society, 987--989

\bibitem[{{Snellen} {et~al.}(2000){Snellen}, {Schilizzi}, {Miley}, {de Bruyn},
  {Bremer}, \& {R{\"o}ttgering}}]{2000MNRAS.319..445S}
{Snellen}, I.~A.~G., {Schilizzi}, R.~T., {Miley}, G.~K., {et~al.} 2000, \mnras,
  319, 445

\bibitem[{{Stickel} {et~al.}(1994){Stickel}, {Meisenheimer}, \&
  {Kuehr}}]{1994A&AS..105..211S}
{Stickel}, M., {Meisenheimer}, K., \& {Kuehr}, H. 1994, \aaps, 105

\bibitem[{{Tadhunter}(2016)}]{2016A&ARv..24...10T}
{Tadhunter}, C. 2016, \aapr, 24, 10

\bibitem[{{Tingay} \& {Edwards}(2002)}]{2002AJ....124..652T}
{Tingay}, S.~J. \& {Edwards}, P.~G. 2002, \aj, 124, 652

\bibitem[{{Tingay} {et~al.}(1997){Tingay}, {Jauncey}, {Reynolds}, {Tzioumis},
  {King}, {Preston}, {Lovell}, {McCulloch}, {Costa}, {Nicolson}, {Koekemoer},
  {Tornikoski}, {Kedziora-Chudczer}, \&
  {Campbell-Wilson}}]{1997AJ....113.2025T}
{Tingay}, S.~J., {Jauncey}, D.~L., {Reynolds}, J.~E., {et~al.} 1997, \aj, 113,
  2025

\bibitem[{{Tingay} {et~al.}(2000){Tingay}, {Jauncey}, {Reynolds}, {Tzioumis},
  {McCulloch}, {Ellingsen}, {Costa}, {Lovell}, {Preston}, \&
  {Simkin}}]{2000AJ....119.1695T}
{Tingay}, S.~J., {Jauncey}, D.~L., {Reynolds}, J.~E., {et~al.} 2000, \aj, 119,
  1695

\bibitem[{{Torresi}(2012)}]{2012arXiv1205.1691T}
{Torresi}, E. 2012, in 2012 Fermi and Jansky Proceedings - eConf C1111101

\bibitem[{{Trussoni} {et~al.}(1999){Trussoni}, {Vagnetti}, {Massaglia},
  {Feretti}, {Parma}, {Morganti}, {Fanti}, \& {Padovani}}]{1999A&A...348..437T}
{Trussoni}, E., {Vagnetti}, F., {Massaglia}, S., {et~al.} 1999, \aap, 348, 437

\bibitem[{{Urry} \& {Padovani}(1995)}]{1995PASP..107..803U}
{Urry}, C.~M. \& {Padovani}, P. 1995, \pasp, 107, 803

\bibitem[{{Venturi} {et~al.}(2000){Venturi}, {Morganti}, {Tzioumis}, \&
  {Reynolds}}]{2000A&A...363...84V}
{Venturi}, T., {Morganti}, R., {Tzioumis}, T., \& {Reynolds}, J. 2000, \aap,
  363, 84

\bibitem[{{V{\'e}ron-Cetty} \& {V{\'e}ron}(2010)}]{2010A&A...518A..10V}
{V{\'e}ron-Cetty}, M.-P. \& {V{\'e}ron}, P. 2010, \aap, 518, A10

\bibitem[{{Walker} {et~al.}(1987){Walker}, {Benson}, \&
  {Unwin}}]{1987ApJ...316..546W}
{Walker}, R.~C., {Benson}, J.~M., \& {Unwin}, S.~C. 1987, \apj, 316, 546

\bibitem[{{Wood} {et~al.}(2017){Wood}, {Caputo}, {Charles}, {Di Mauro},
  {Magill}, \& {Jeremy Perkins for the Fermi-LAT
  Collaboration}}]{2017arXiv170709551W}
{Wood}, M., {Caputo}, R., {Charles}, E., {et~al.} 2017, ArXiv e-prints
  [\eprint[arXiv]{1707.09551}]

\bibitem[{{Wright} \& {Otrupcek}(1990)}]{1990PKS...C......0W}
{Wright}, A. \& {Otrupcek}, R. 1990, in PKS Catalog (1990)

\end{thebibliography}

\begin{appendix}
\section{Full resolution VLBI maps and image parameters}
\label{app:maps}
Here we present the full-resolution VLBI images for the sources studied in this paper, along with the tables including the details of each observation, for each source.

\begin{figure*}[!htbp]
\begin{center}
\includegraphics[width=0.43\linewidth]{0518-458_2007-11-10.pdf}
\includegraphics[width=0.43\linewidth]{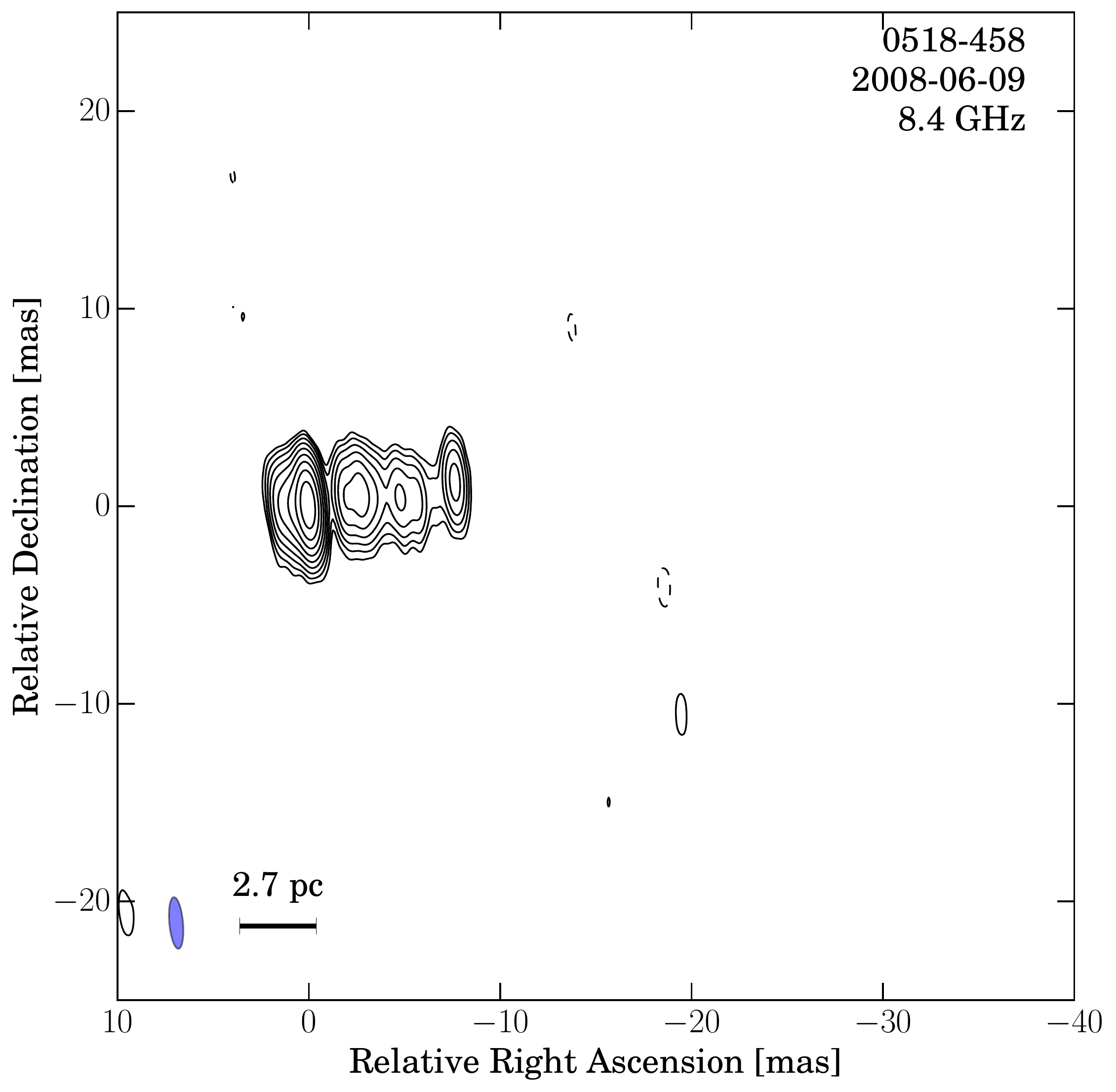}
\includegraphics[width=0.43\linewidth]{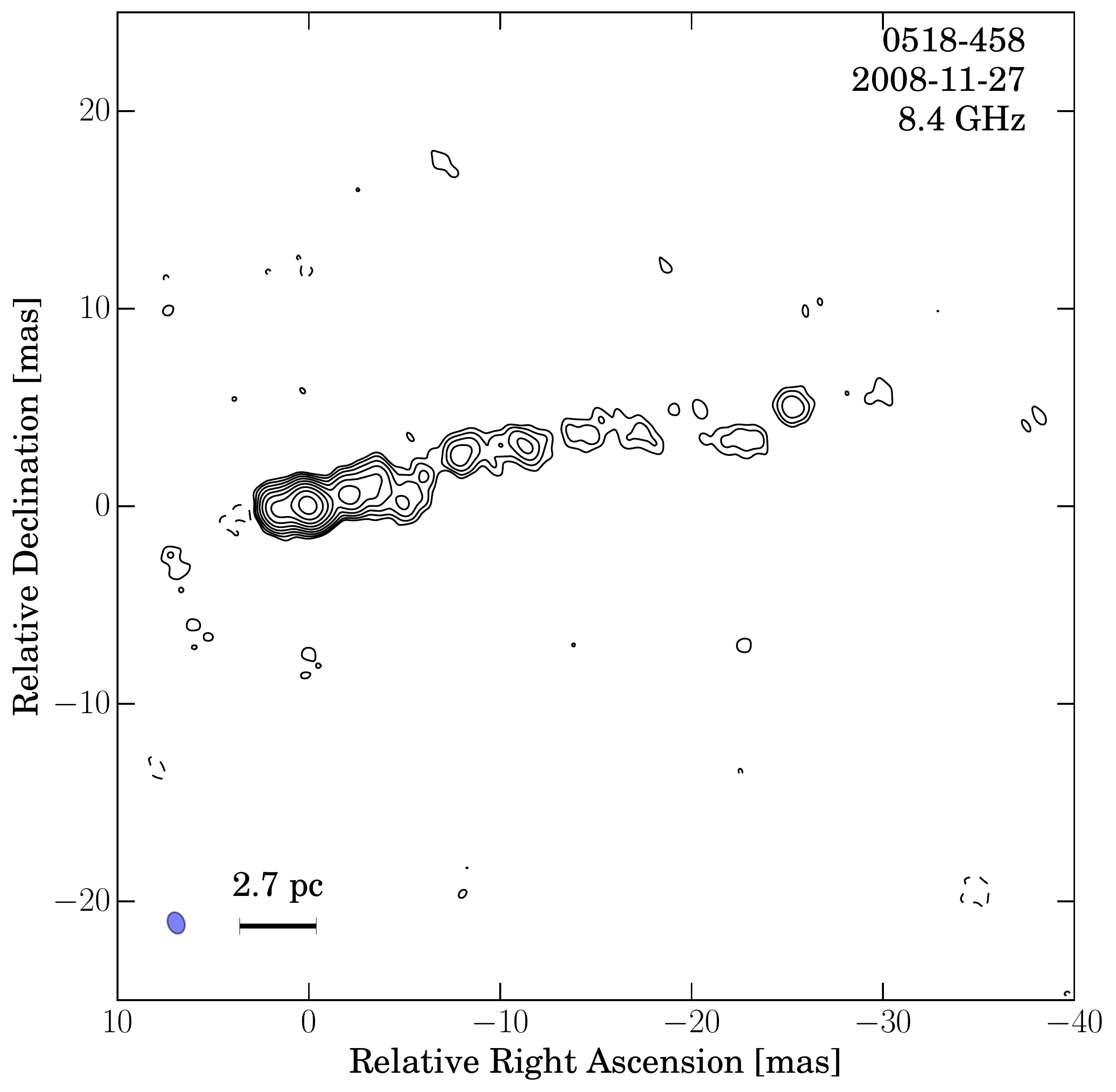}
\includegraphics[width=0.43\linewidth]{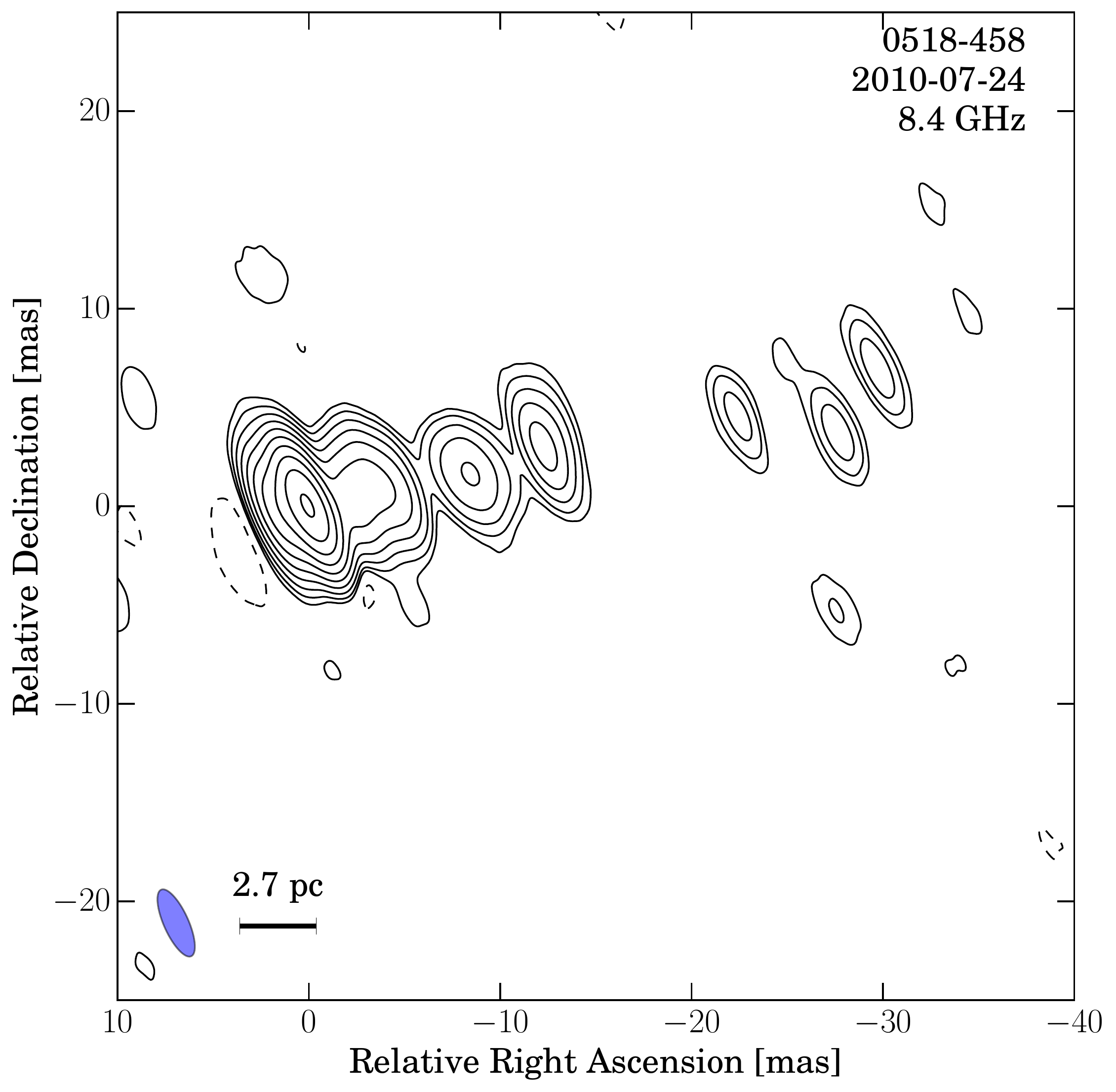}
\includegraphics[width=0.43\linewidth]{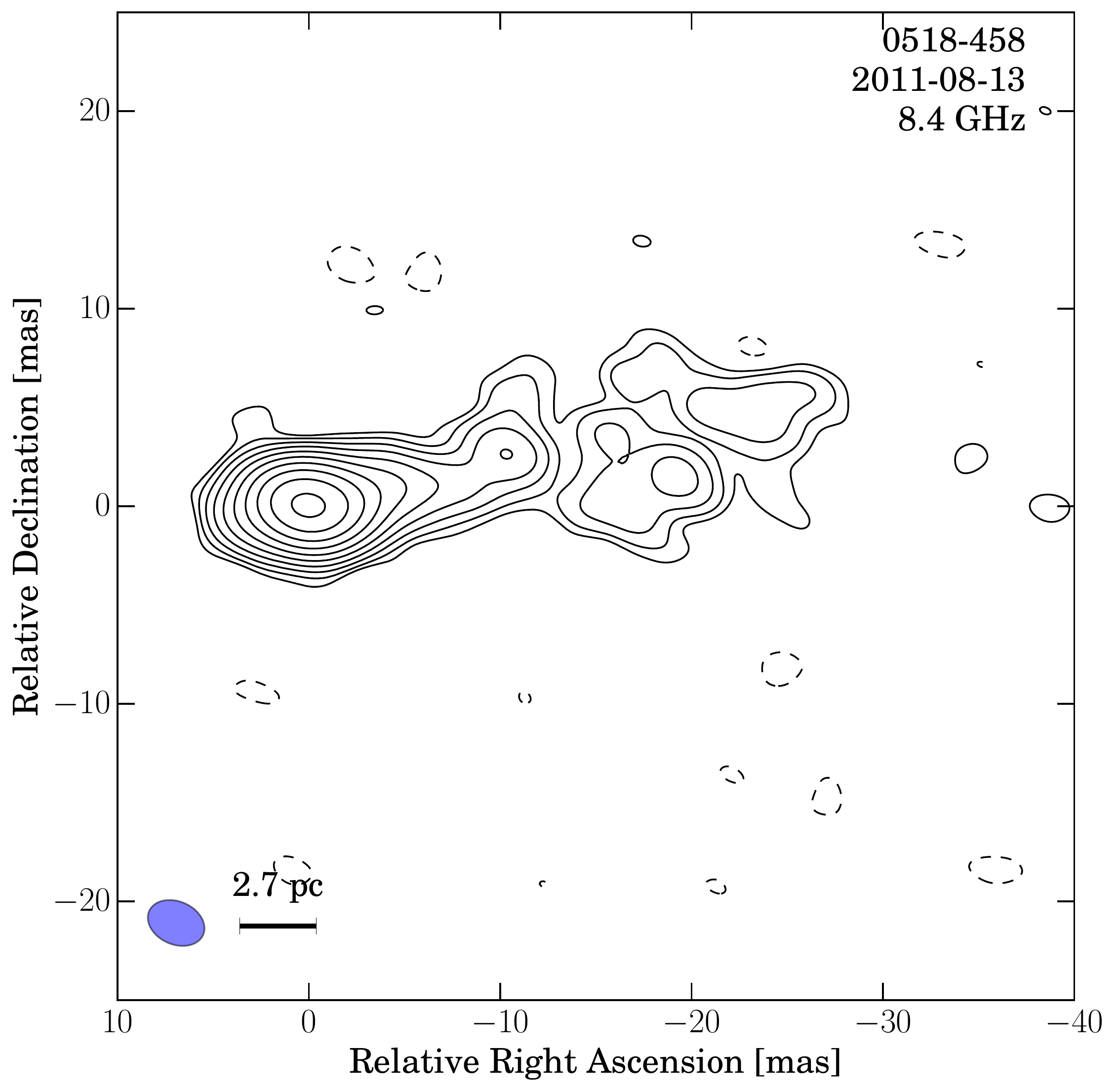}
\includegraphics[width=0.43\linewidth]{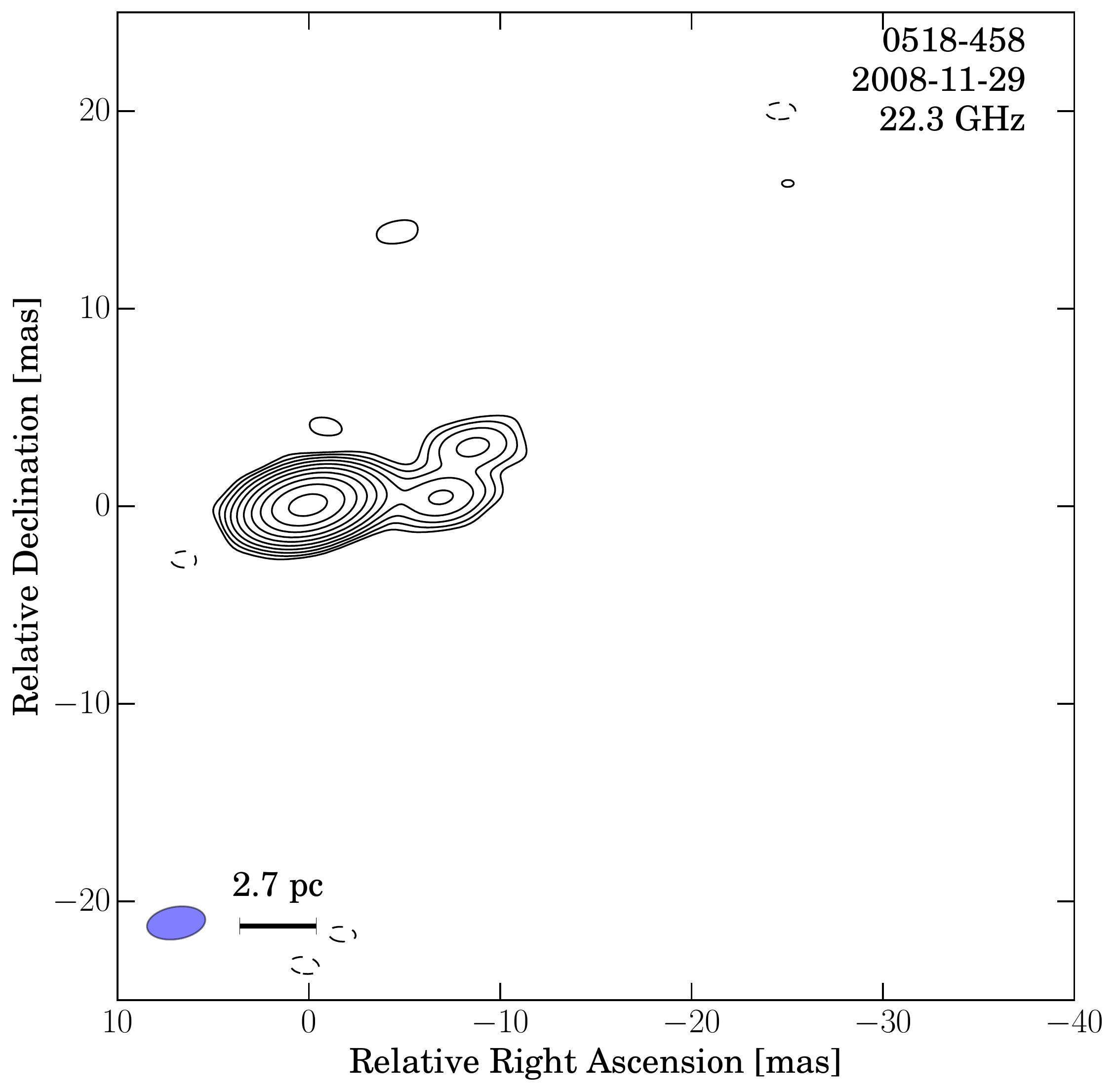}
\end{center}
\caption{Full-resolution images of Pictor~A. The map parameters for
  each epoch can be found in Table~\ref{pica_tab}. The blue ellipse
  represents the beam size, while the black line indicates the linear
  scale at the source's redshift. Contours increase in steps of two starting from 1.8, 3.0, 3.5, 5.5, 4.0, 2.0 times the noise level in each map, from top left to bottom right, respectively.}
\label{pica_full}
\end{figure*}

\begin{figure*}[!htbp]
\begin{center}
\includegraphics[width=0.43\linewidth]{0521-365_2007-11-10.pdf}
\includegraphics[width=0.43\linewidth]{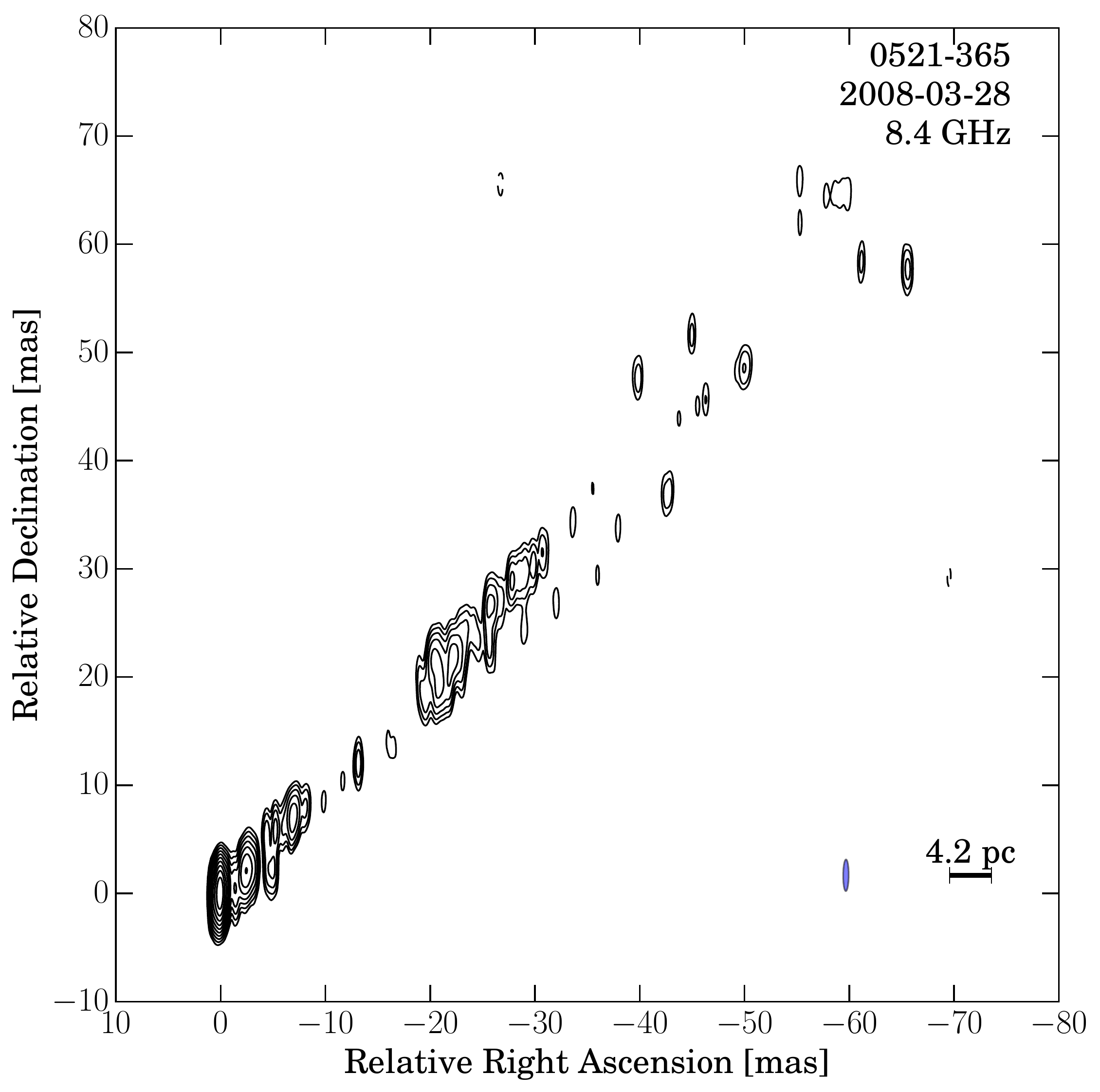}
\includegraphics[width=0.43\linewidth]{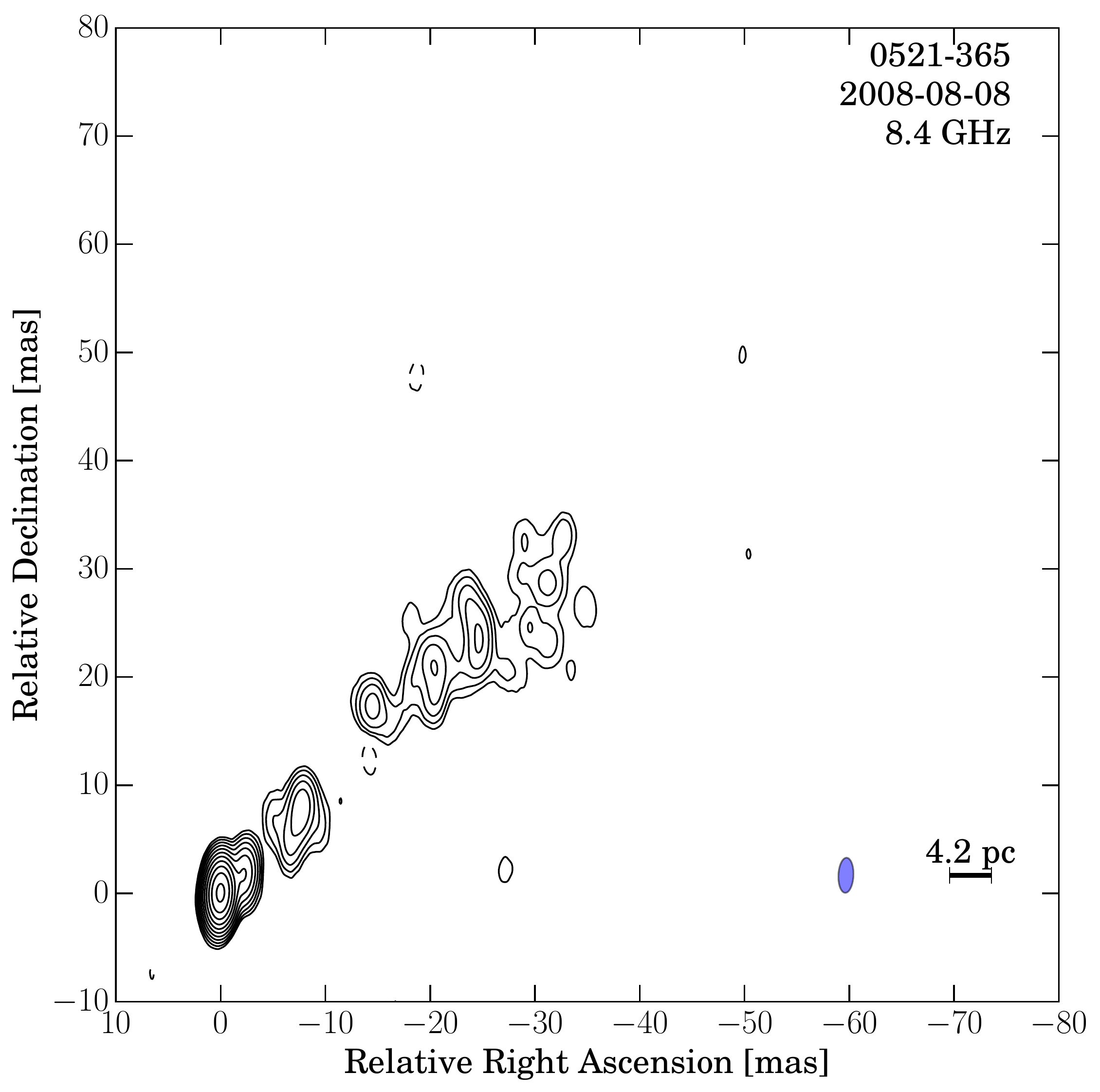}
\includegraphics[width=0.43\linewidth]{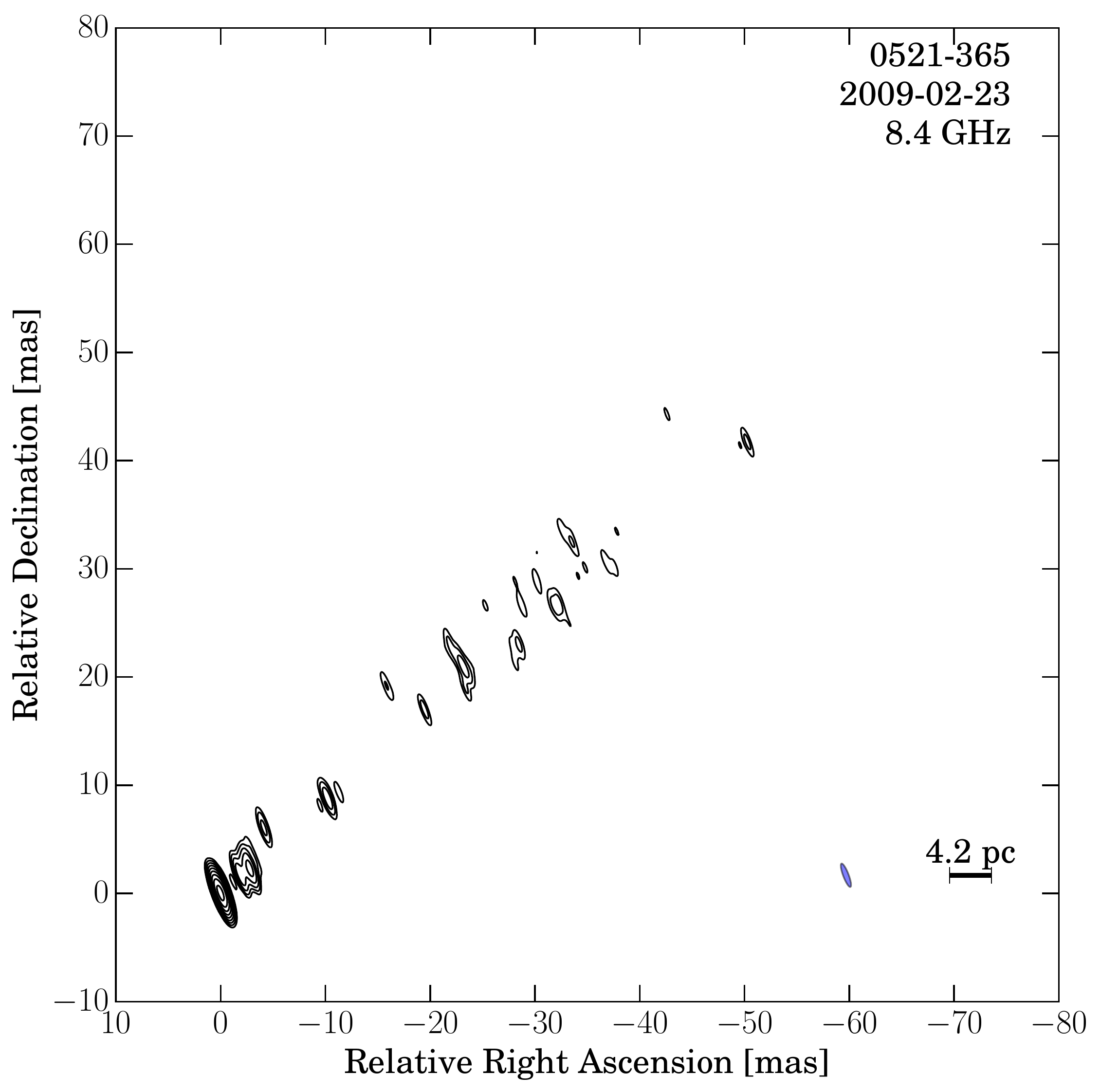}
\includegraphics[width=0.43\linewidth]{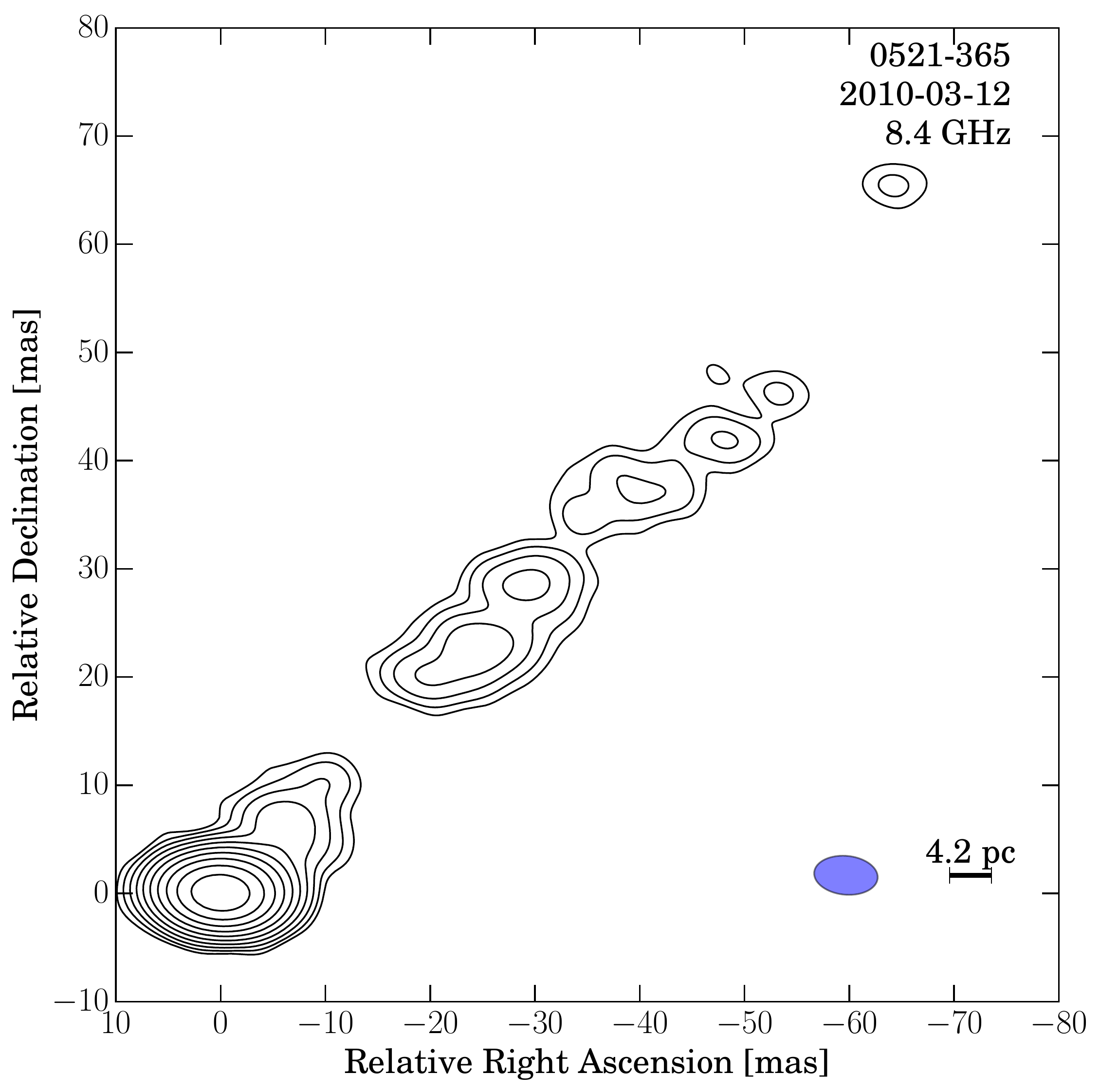}
\includegraphics[width=0.43\linewidth]{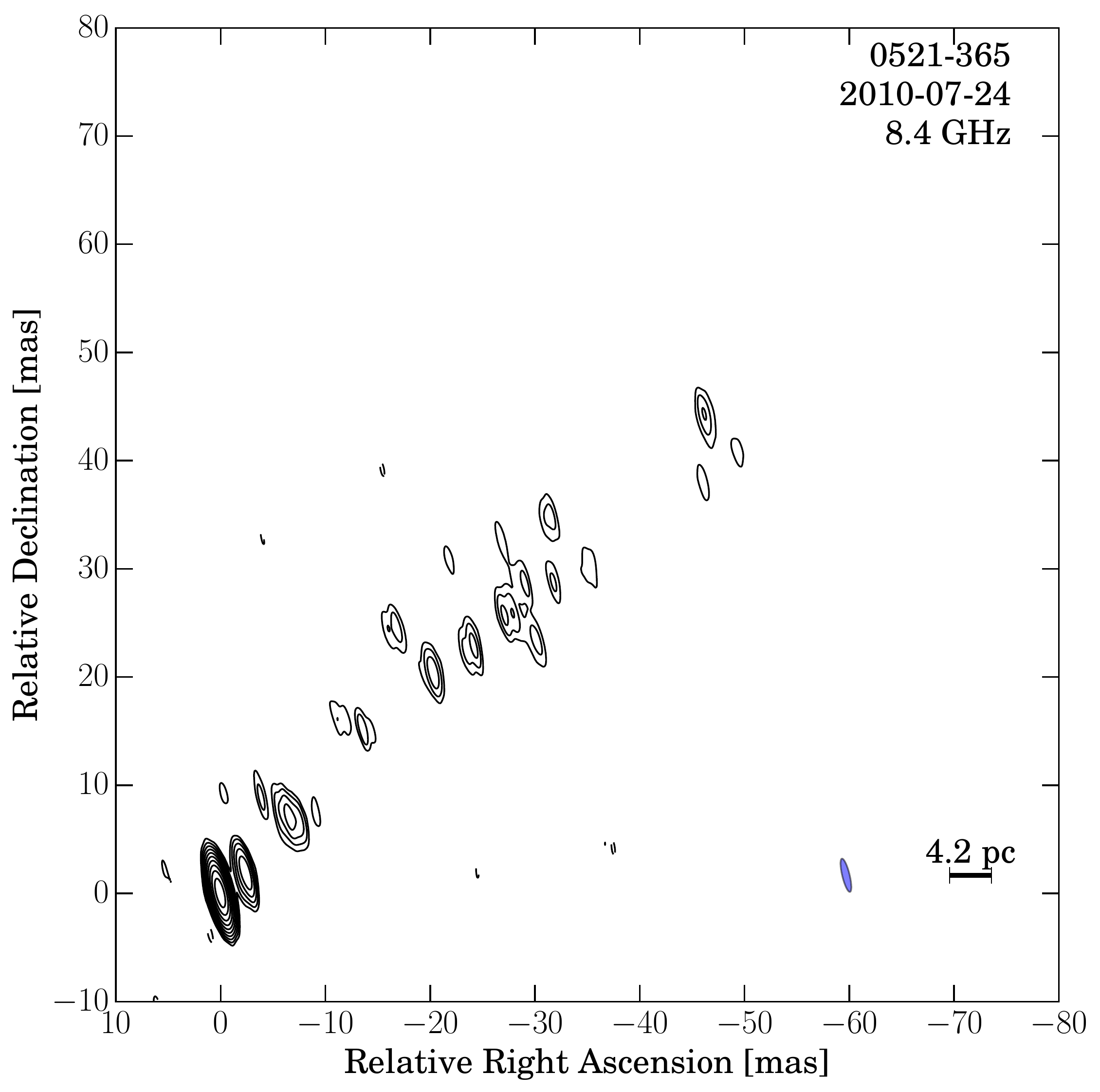}
\end{center}
\caption{Full-resolution images of PKS~0521$-$36. The map parameters for
  each epoch can be found in Table~\ref{0521_tab}. The blue ellipse
  represents the beam size, while the black line indicates the linear
  scale at the source's redshift. Contours increase in steps of two starting from 4.0, 3.0, 3.0, 3.0, 3.0, 5.0 times the noise level in each map, from top left to bottom right, respectively.}
\label{0521_full_a}
\end{figure*}
\renewcommand{\thefigure}{A.\arabic{figure} (Continued)}
\addtocounter{figure}{-1}
\begin{figure*}[!htbp]
\begin{center}
\includegraphics[width=0.43\linewidth]{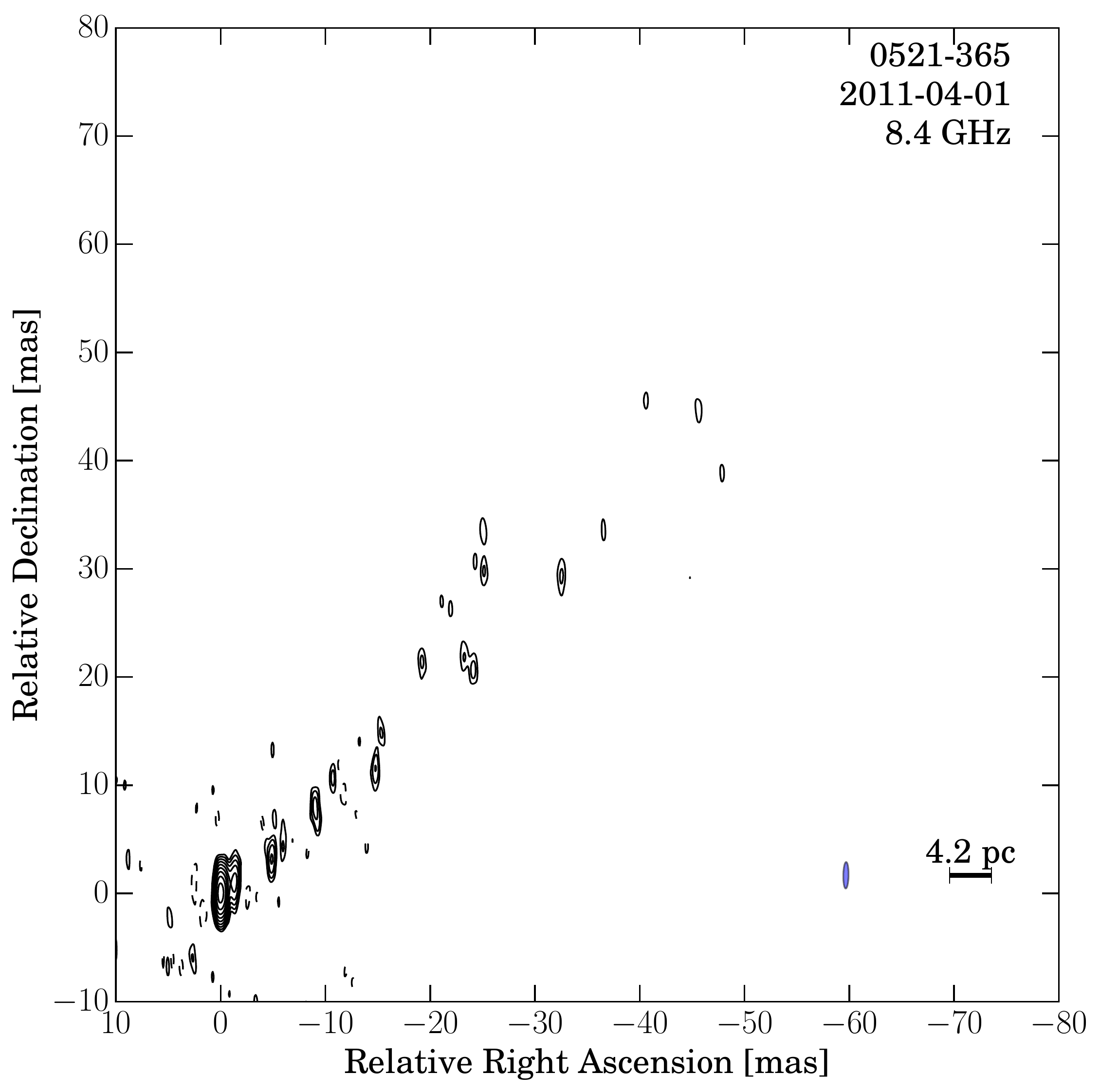}
\includegraphics[width=0.43\linewidth]{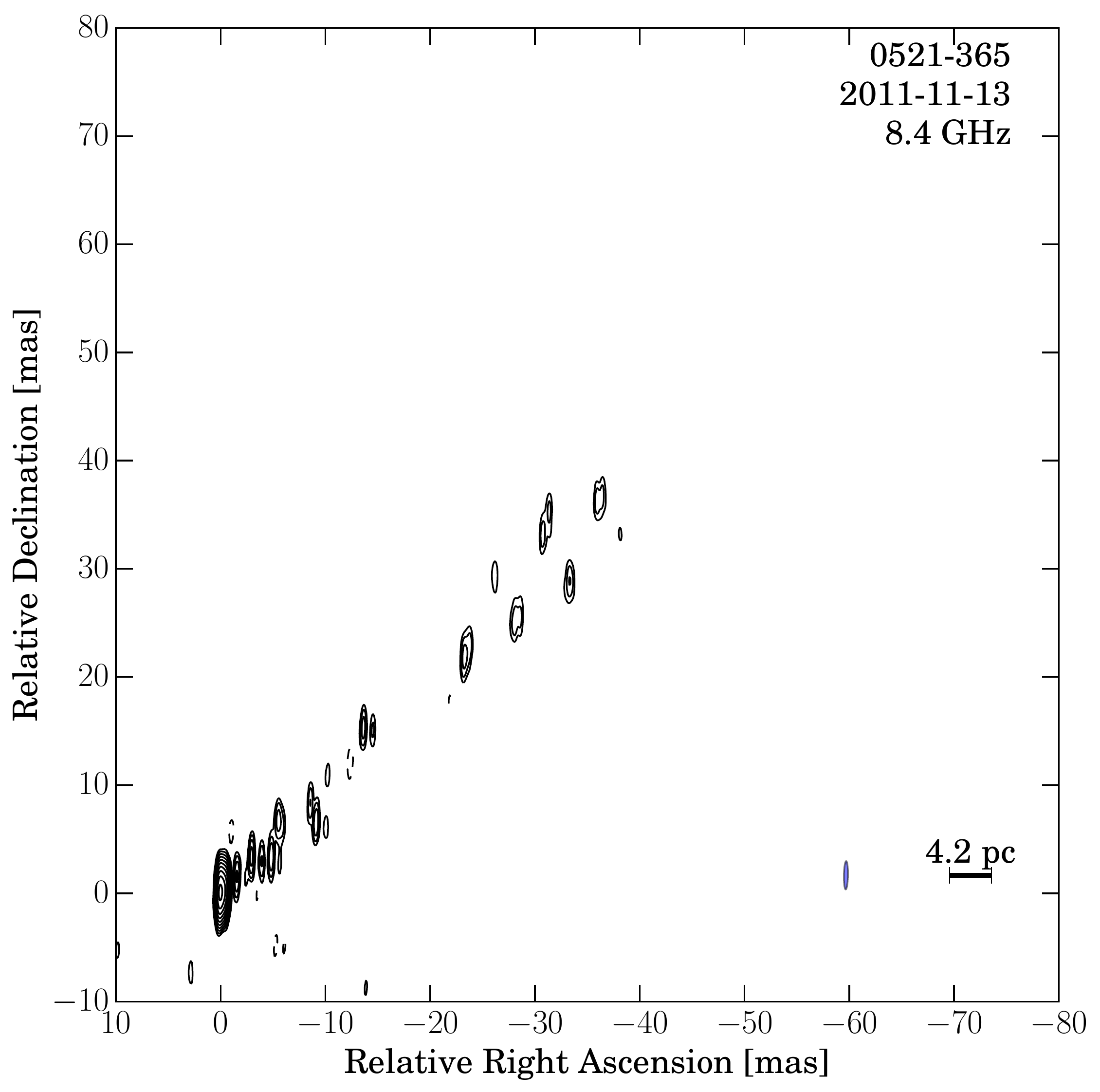}
\includegraphics[width=0.43\linewidth]{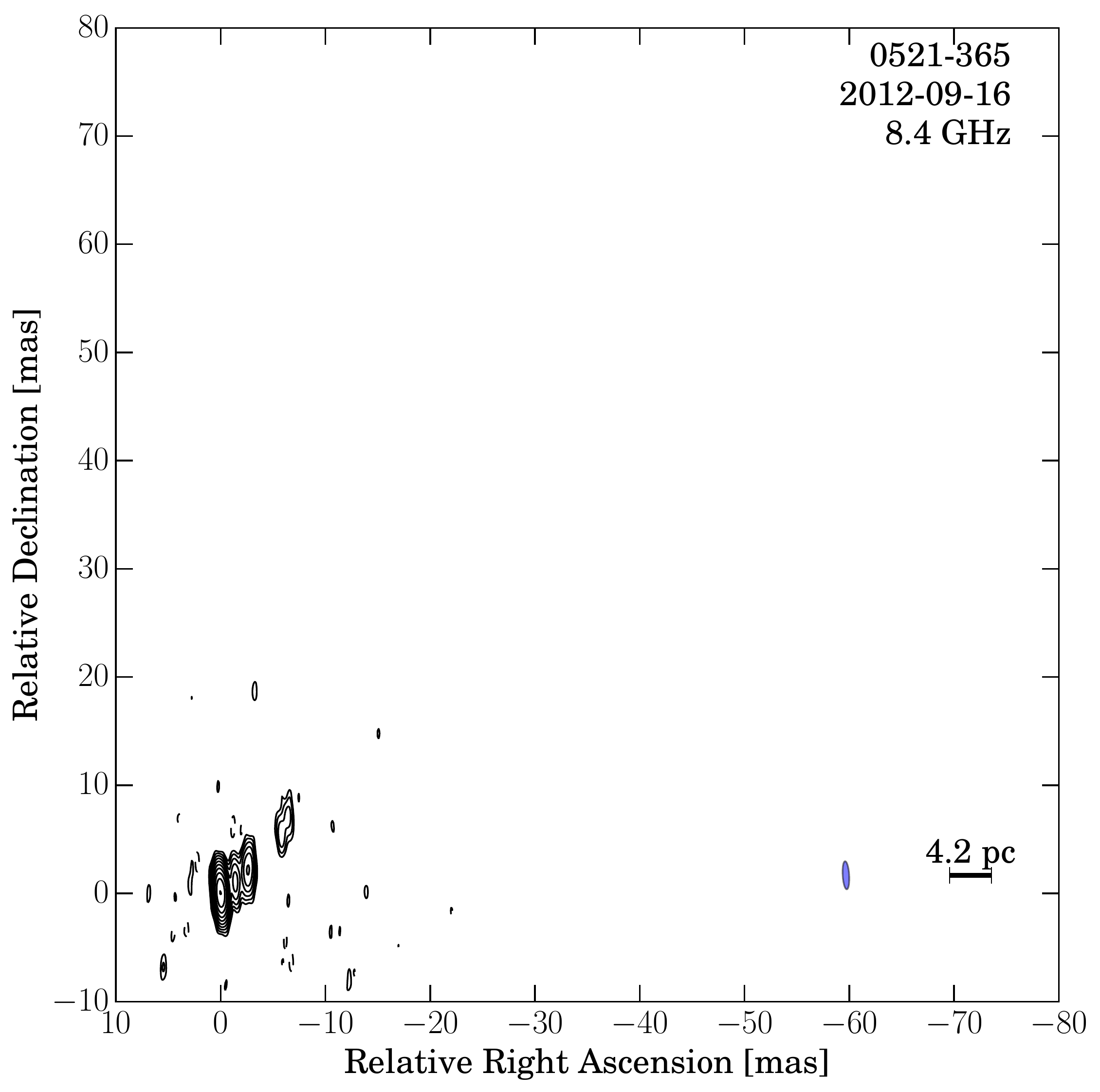}
\includegraphics[width=0.43\linewidth]{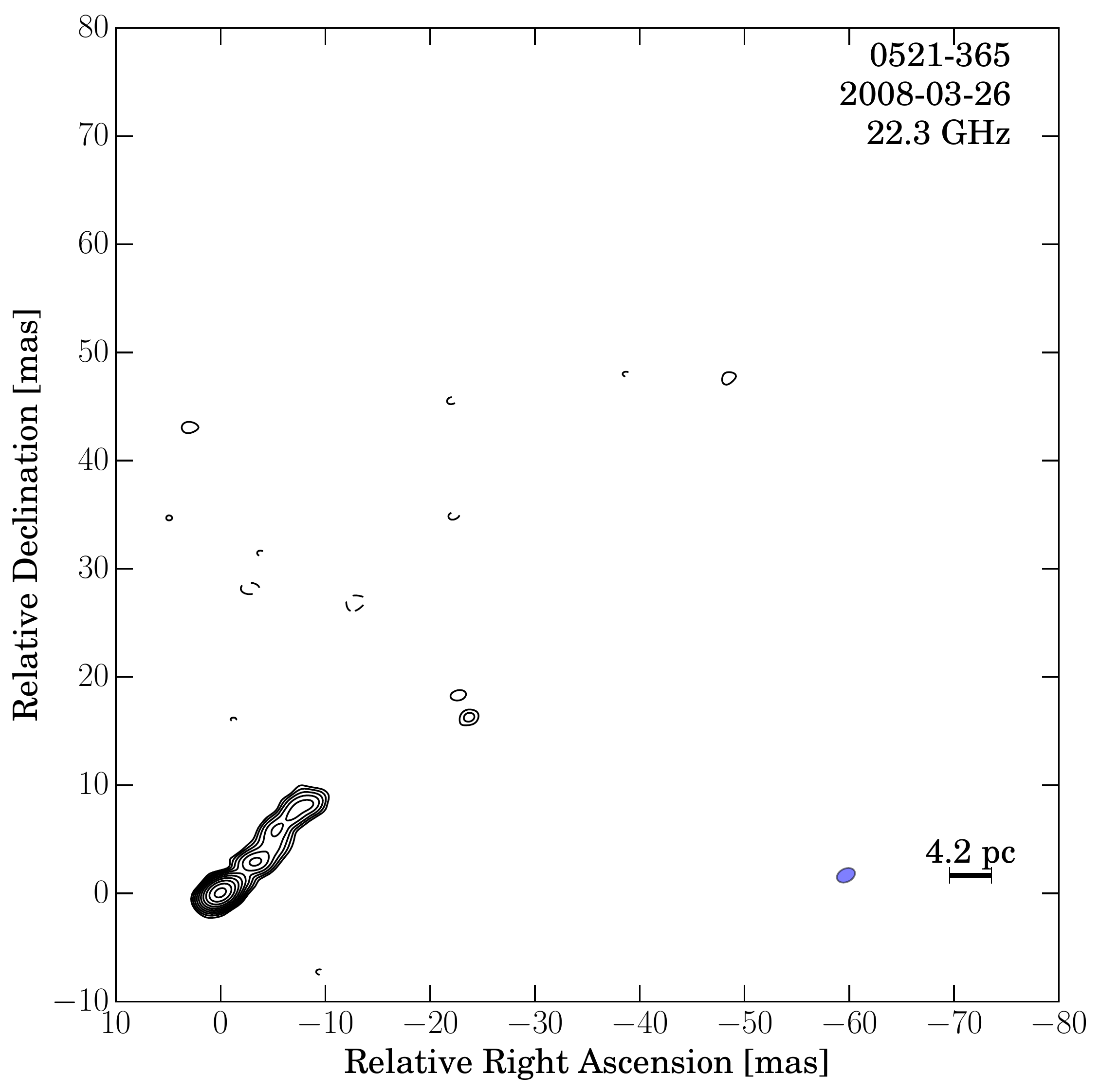}
\end{center}
\caption{Full-resolution images of PKS~0521$-$36 (continued). The map parameters for  each epoch can be found in Table~\ref{0521_tab}. The blue ellipse  represents the beam size, while the black line indicates the linear
  scale at the source's redshift. Contours increase in steps of two starting from 10.0, 4.0, 10.0, 1.0 times the noise level in each map, from top left to bottom right, respectively.}
\label{0521_full_b}
\end{figure*}
\renewcommand{\thefigure}{A.\arabic{figure}}
\clearpage
\begin{figure*}[!htbp]
\begin{center}
\includegraphics[width=0.43\linewidth]{0625-35_2007-11-10.pdf}
\includegraphics[width=0.43\linewidth]{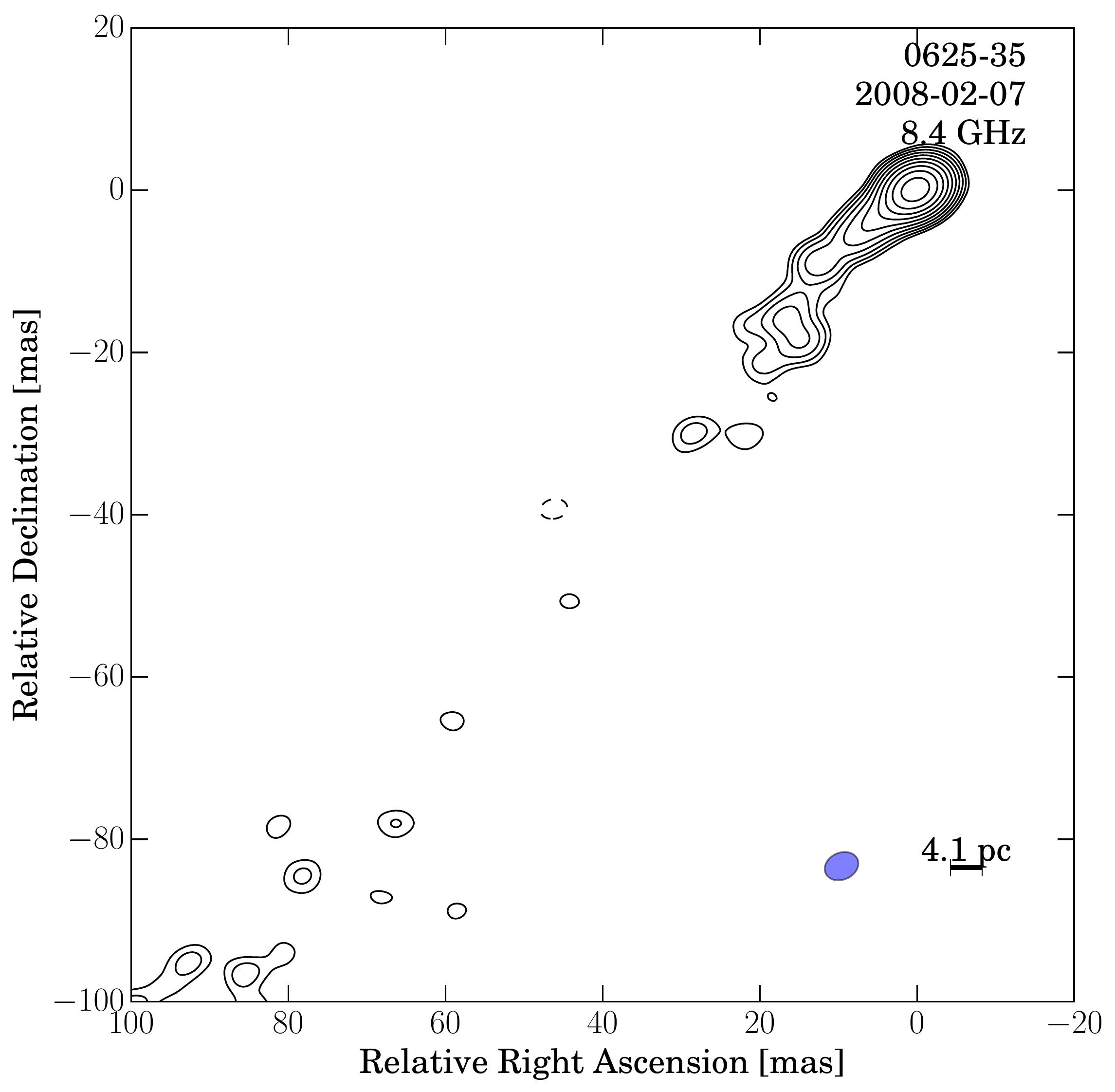}
\includegraphics[width=0.43\linewidth]{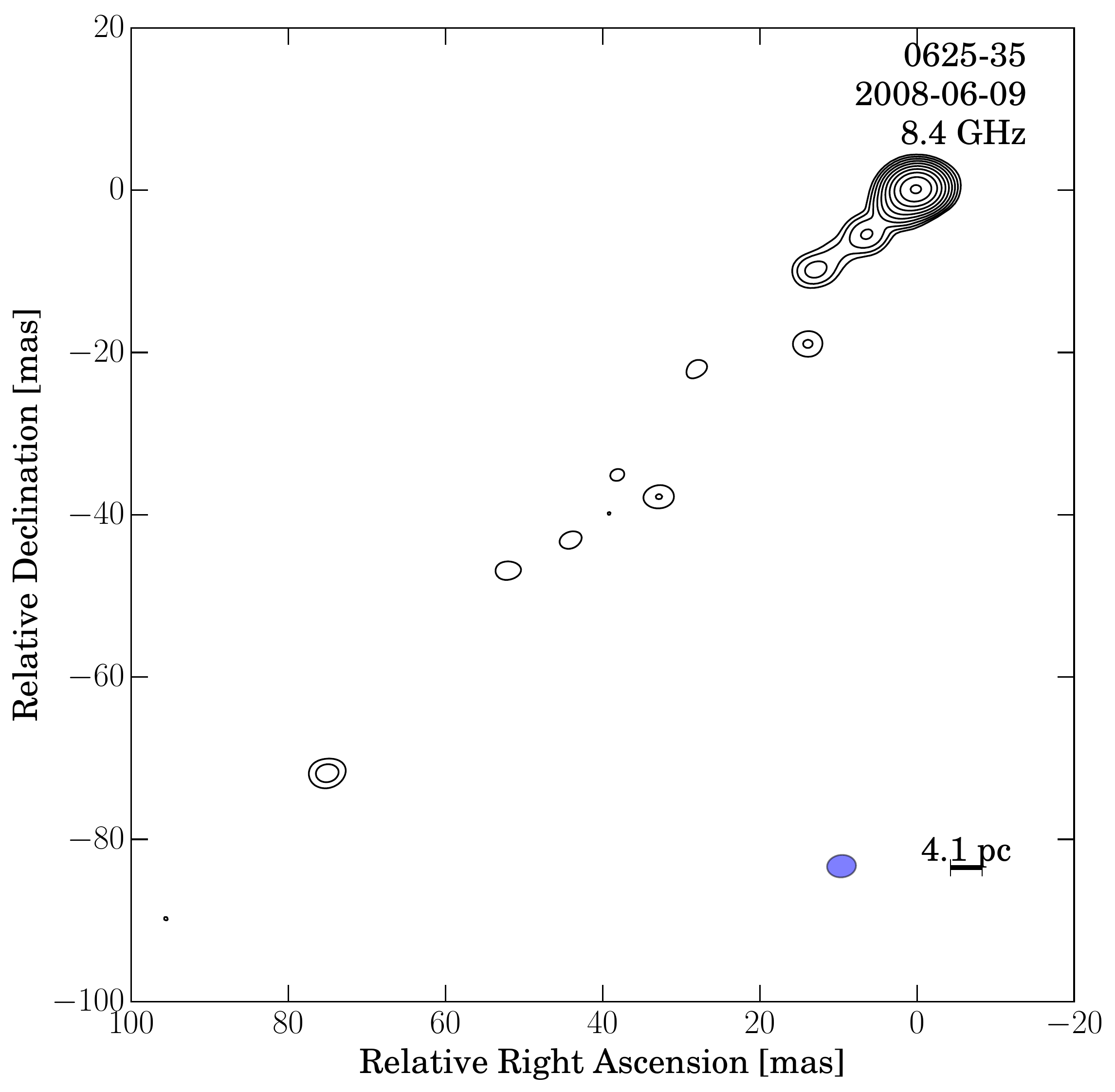}
\includegraphics[width=0.43\linewidth]{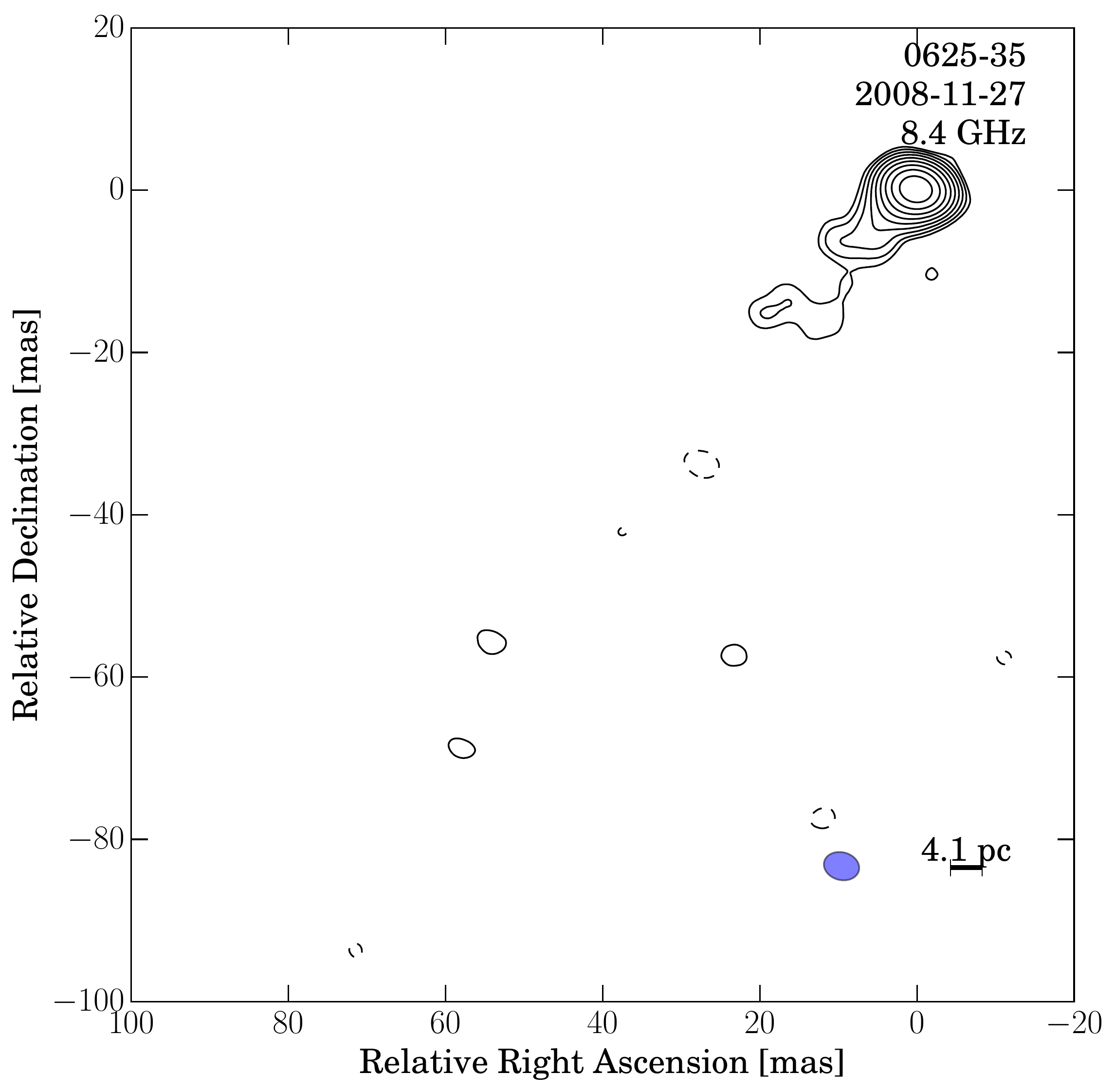}
\includegraphics[width=0.43\linewidth]{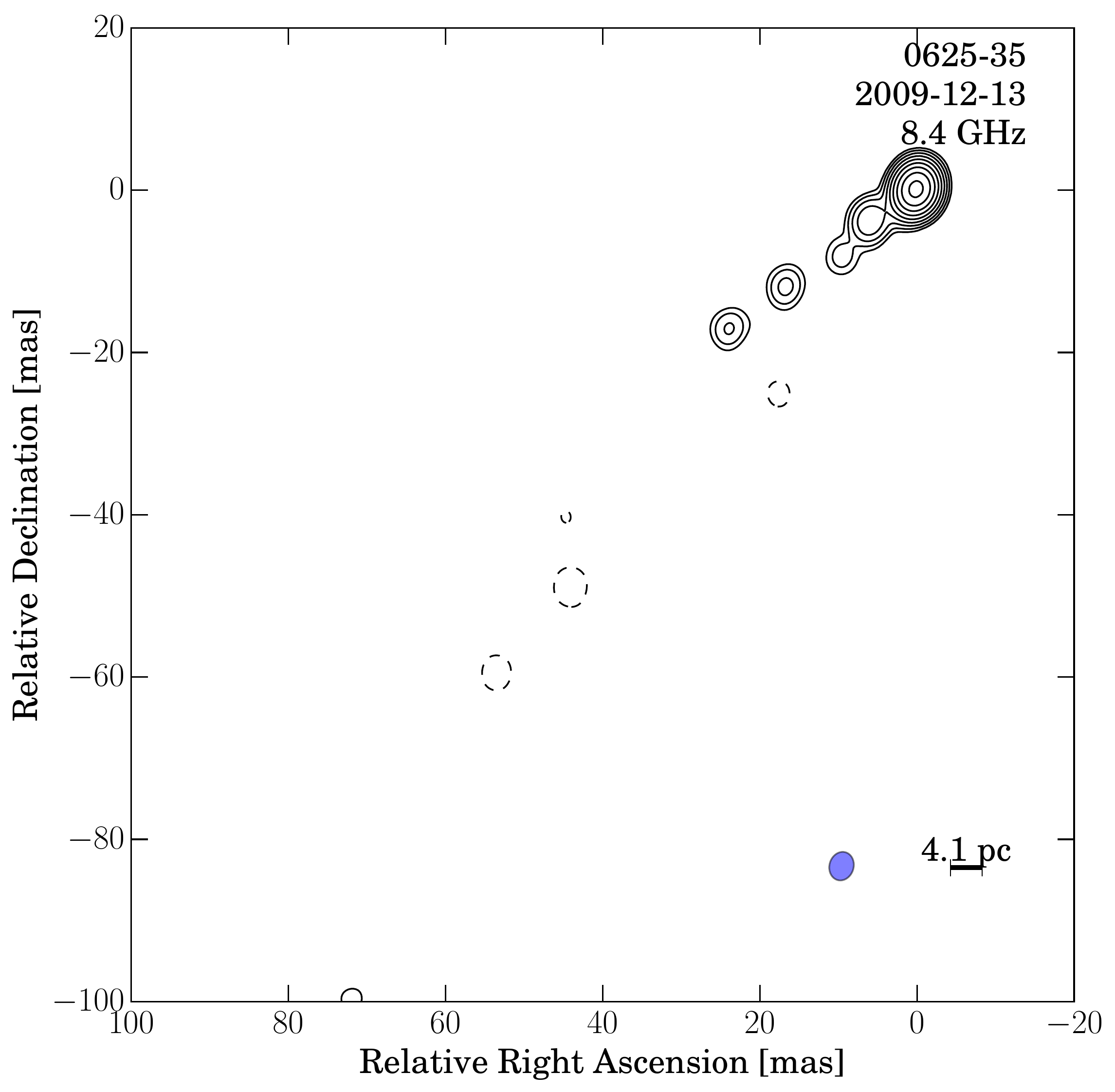}
\includegraphics[width=0.43\linewidth]{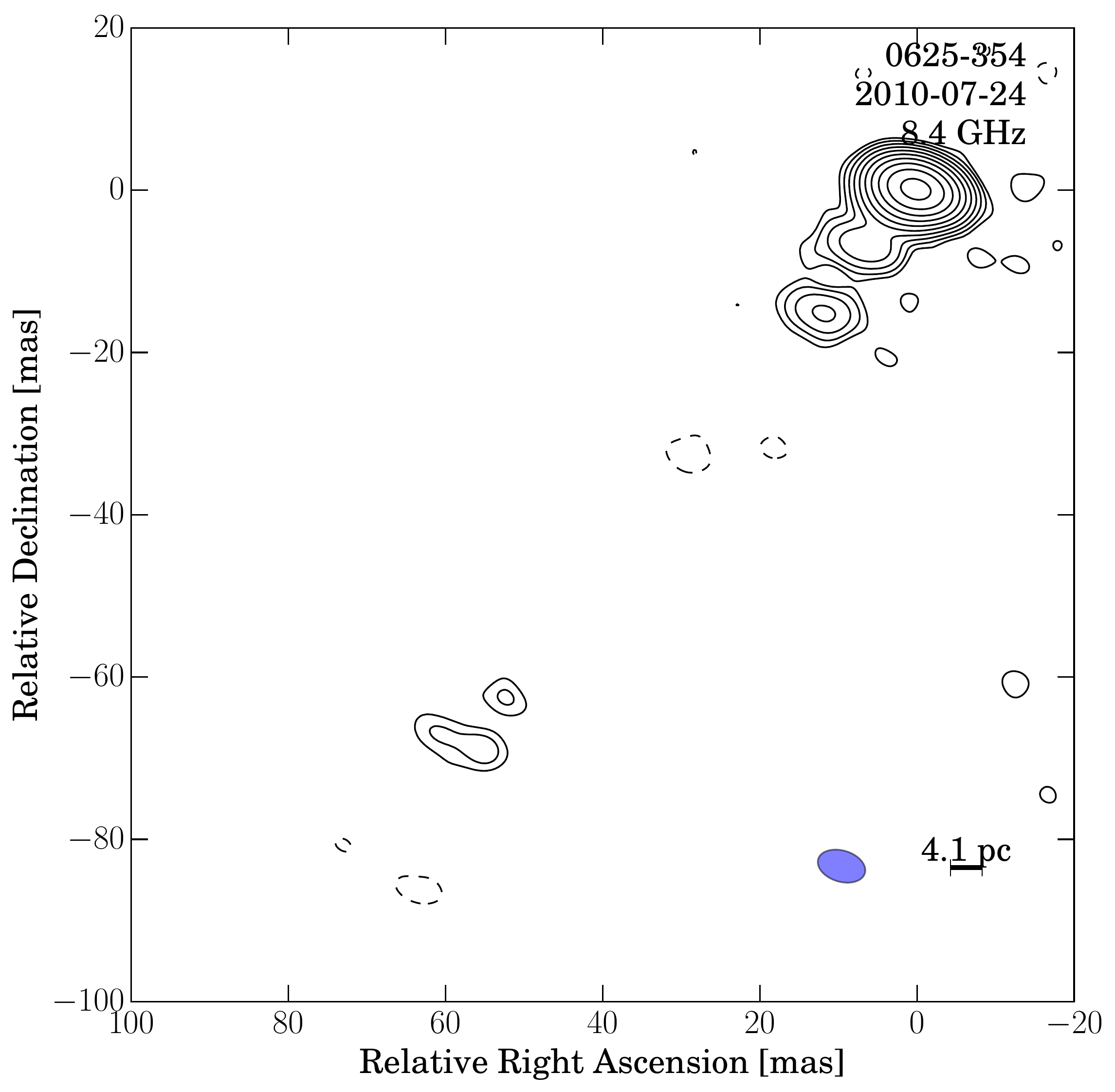}
\end{center}
\caption{Full-resolution images of PKS~0625$-$35. The map parameters for
  each epoch can be found in Table~\ref{0625_tab}. The blue ellipse
  represents the beam size, while the black line indicates the linear
  scale at the source's redshift. Contours increase in steps of two starting from 2.0, 2.0, 2.0, 4.0, 3.0, 4.0 times the noise level in each map, from top left to bottom right, respectively.}
\label{0625_full_a}
\end{figure*}
\renewcommand{\thefigure}{A.\arabic{figure} (Continued)}
\addtocounter{figure}{-1}
\begin{figure*}[!htbp]
\begin{center}
\includegraphics[width=0.43\linewidth]{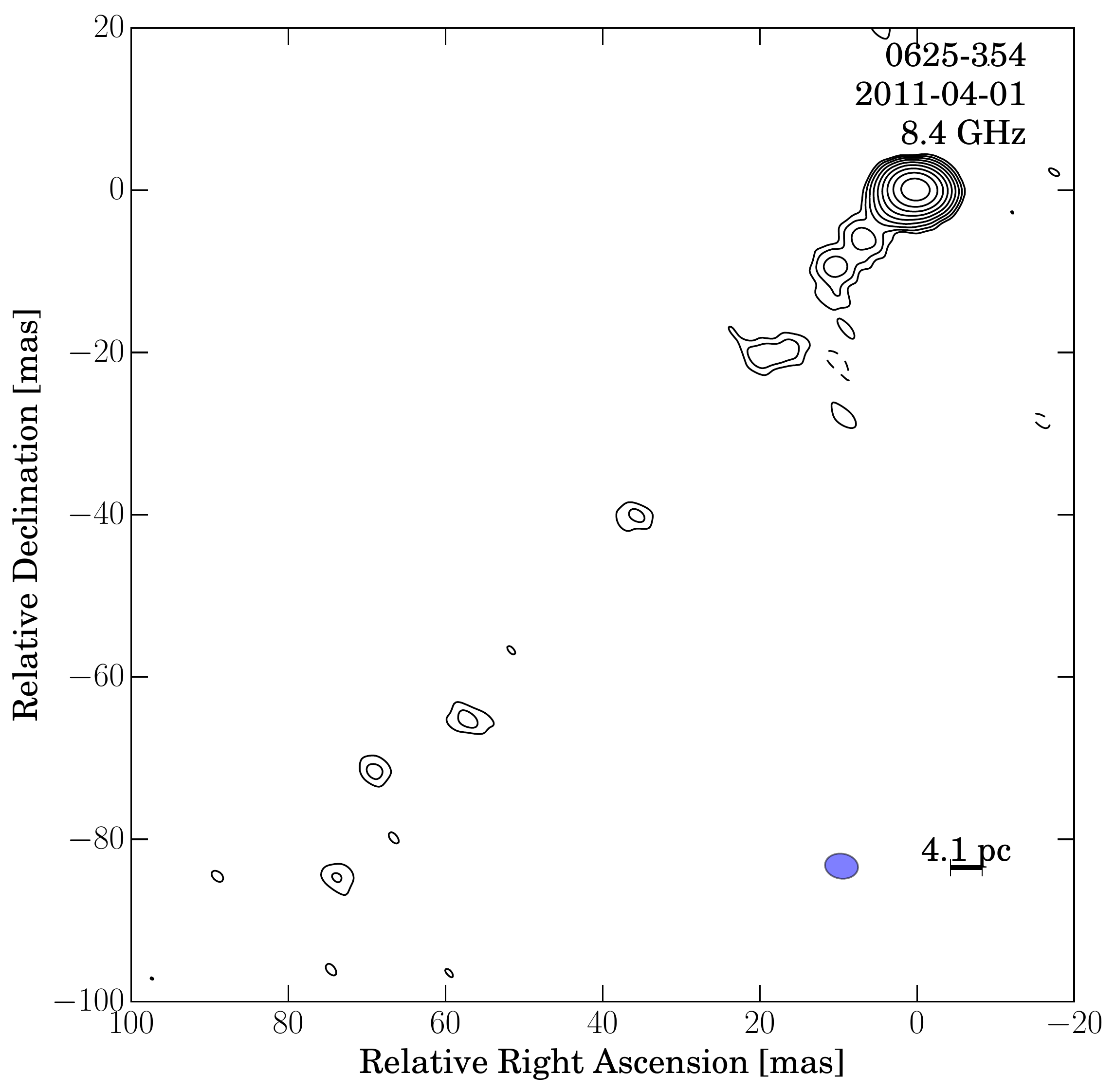}
\includegraphics[width=0.43\linewidth]{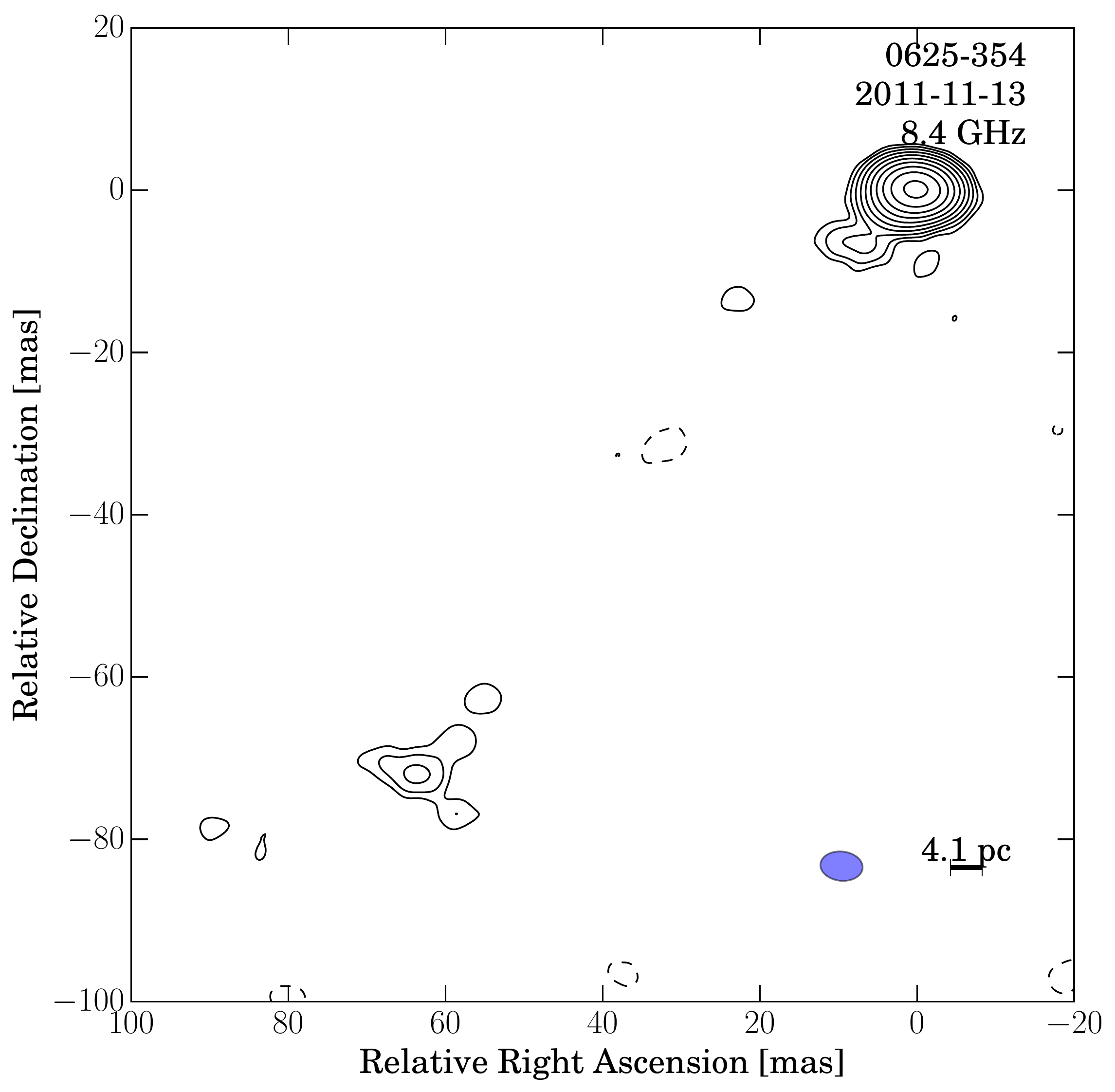}
\includegraphics[width=0.43\linewidth]{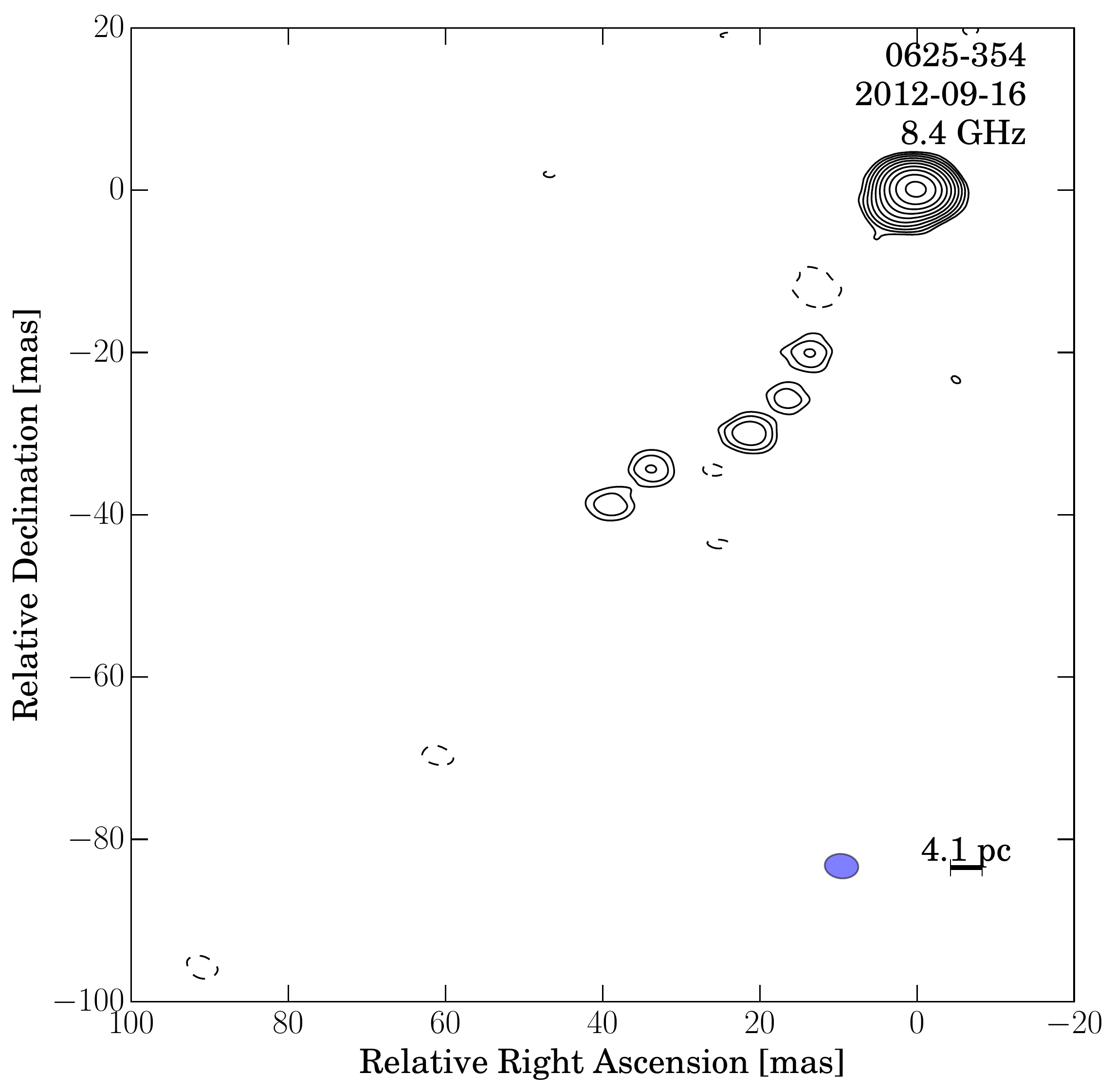}
\includegraphics[width=0.43\linewidth]{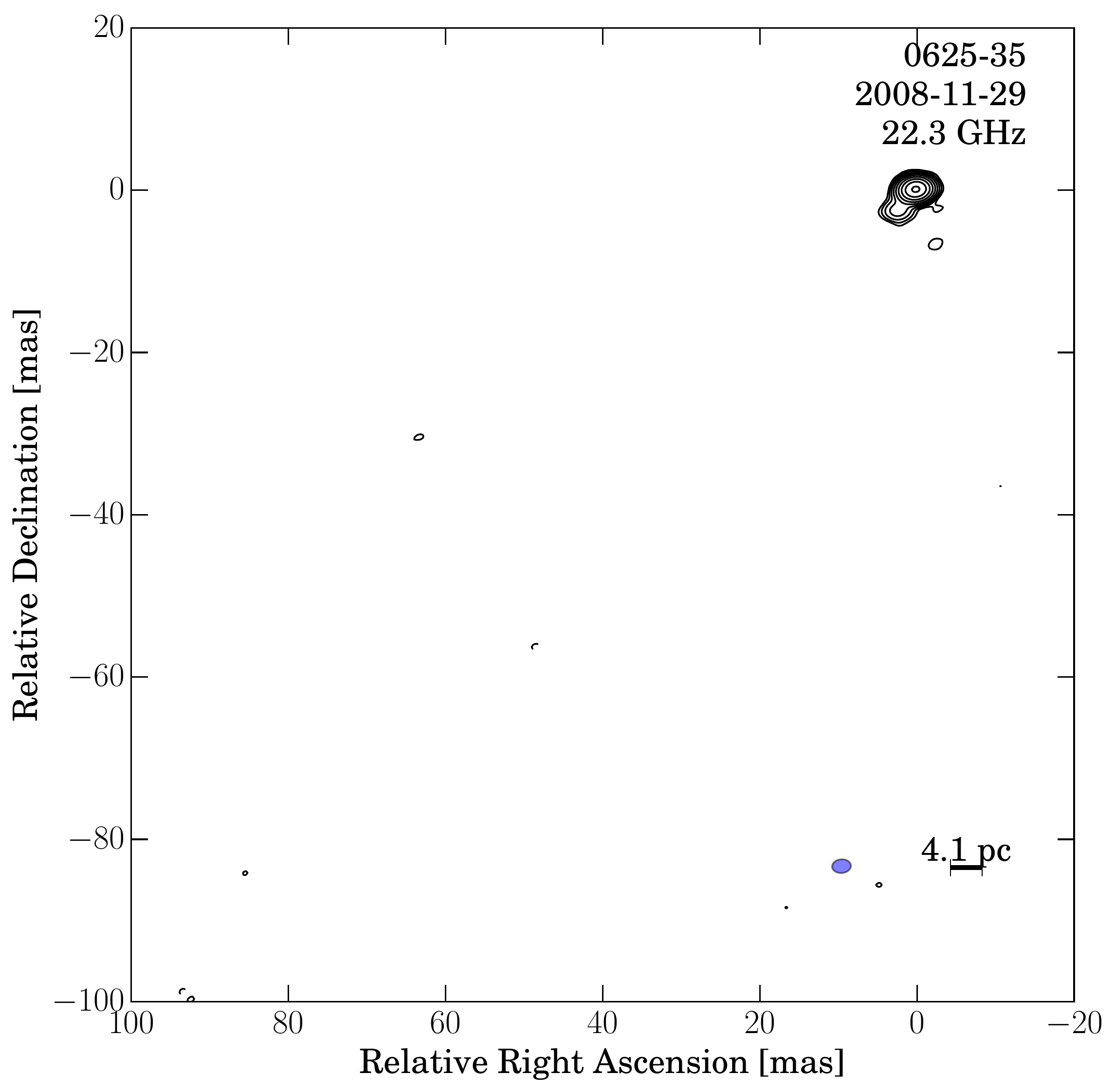}
\end{center}
\caption{Full-resolution images of PKS~0625$-$35 (continued). The map parameters for
  each epoch can be found in Table~\ref{0625_tab}. The blue ellipse
  represents the beam size, while the black line indicates the linear
  scale at the source's redshift. Contours increase in steps of two starting from 4.0, 2.0, 4.0, 2.0 times the noise level in each map, from top left to bottom right, respectively.}
\label{0625_full_b}
\end{figure*}
\renewcommand{\thefigure}{A.\arabic{figure}}

\begin{figure*}[!htbp]
\begin{center}
\includegraphics[width=0.43\linewidth]{1343-601_2011-07-21.pdf}
\includegraphics[width=0.43\linewidth]{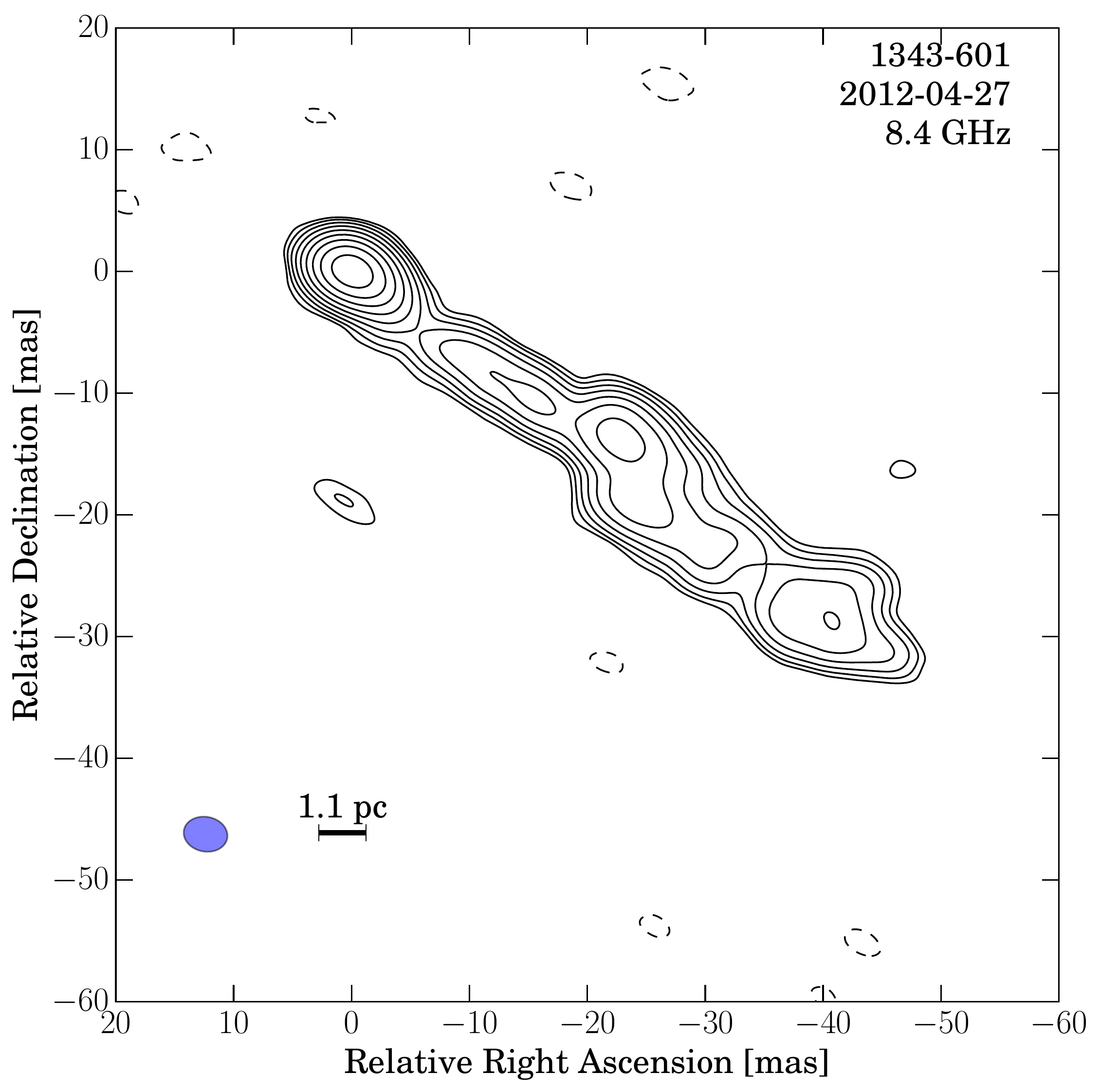}
\end{center}
\caption{Full-resolution images of Centaurus~B. The map parameters for
  each epoch can be found in Table~\ref{cenb_tab}. The blue ellipse
  represents the beam size, while the black line indicates the linear
  scale at the source's redshift. Contours increase in steps of two starting from 9.0 times the noise level in each map.}
\label{cenb_full}
\end{figure*}

\begin{figure*}[!htbp]
\begin{center}
\includegraphics[width=0.43\linewidth]{1718-649_2008-02-07.pdf}
\includegraphics[width=0.43\linewidth]{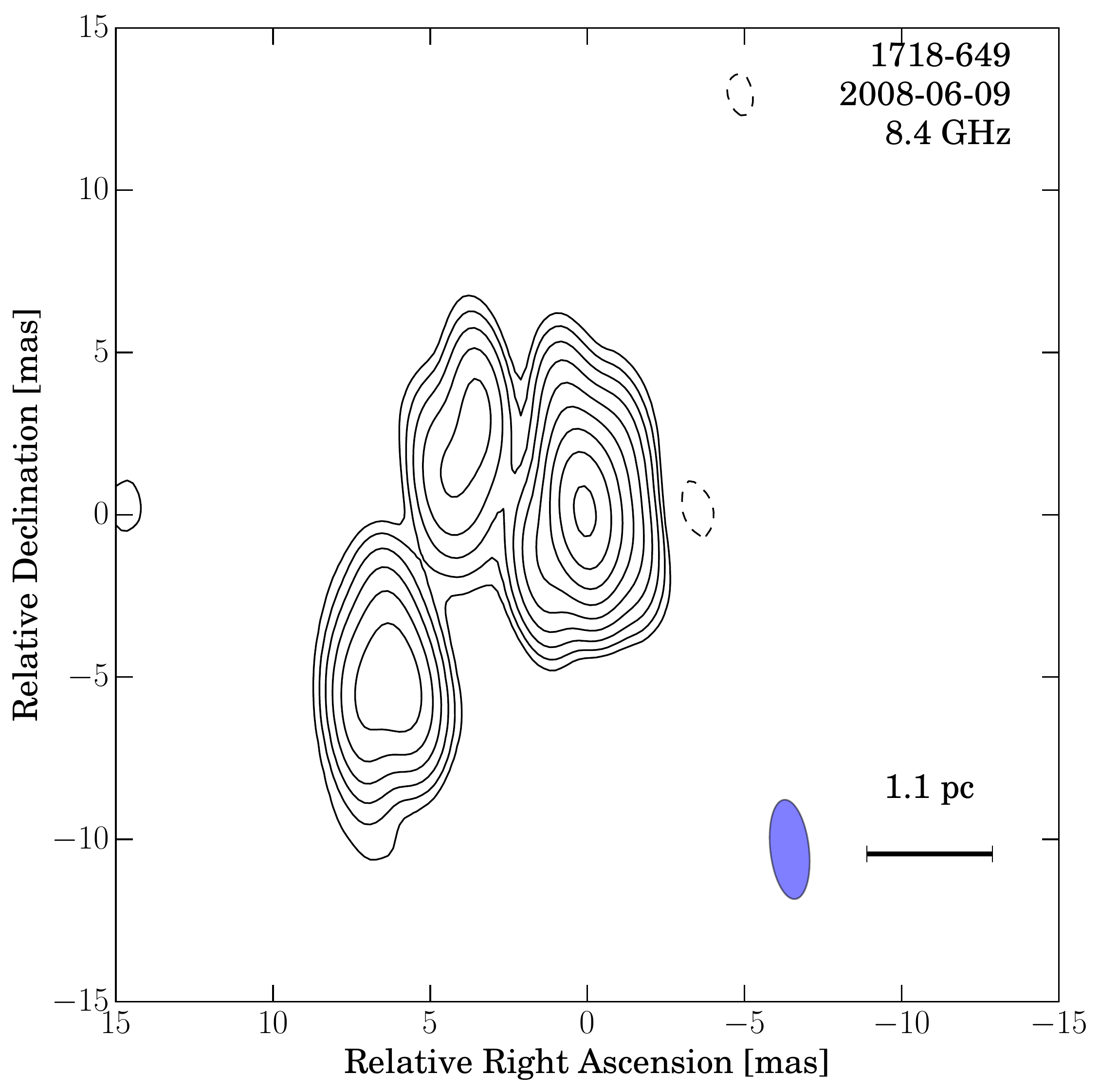}
\includegraphics[width=0.43\linewidth]{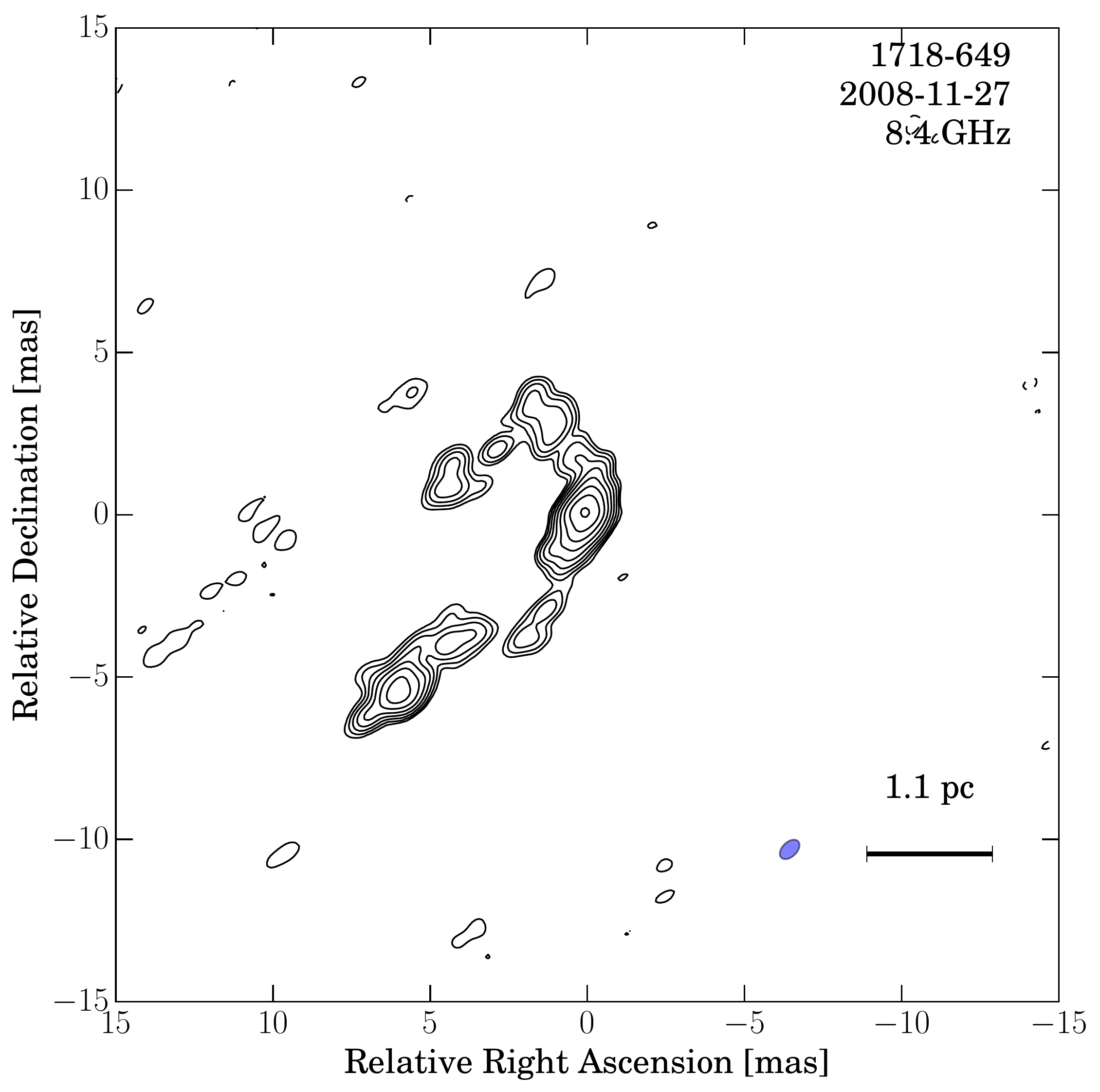}
\includegraphics[width=0.43\linewidth]{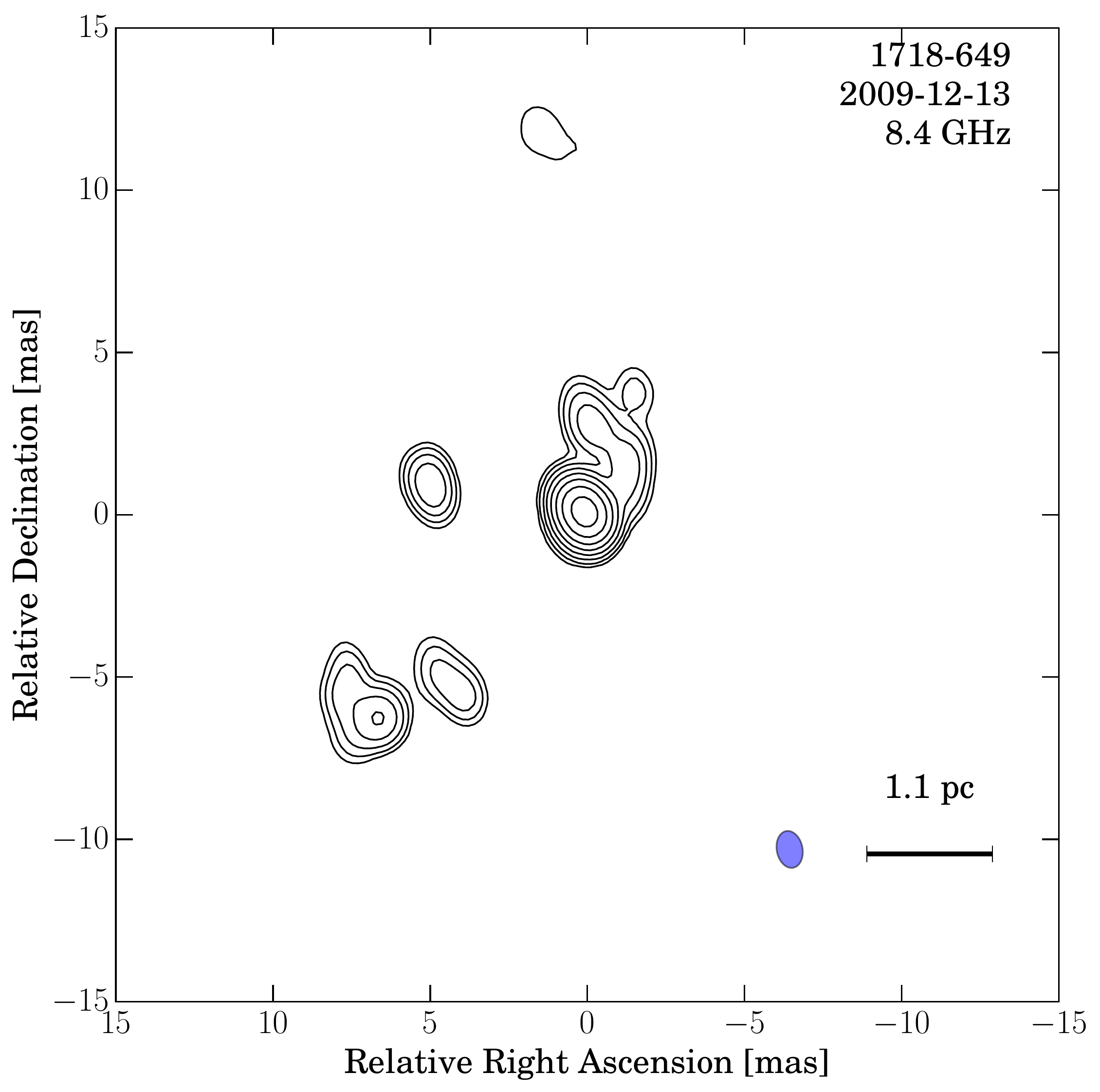}
\includegraphics[width=0.43\linewidth]{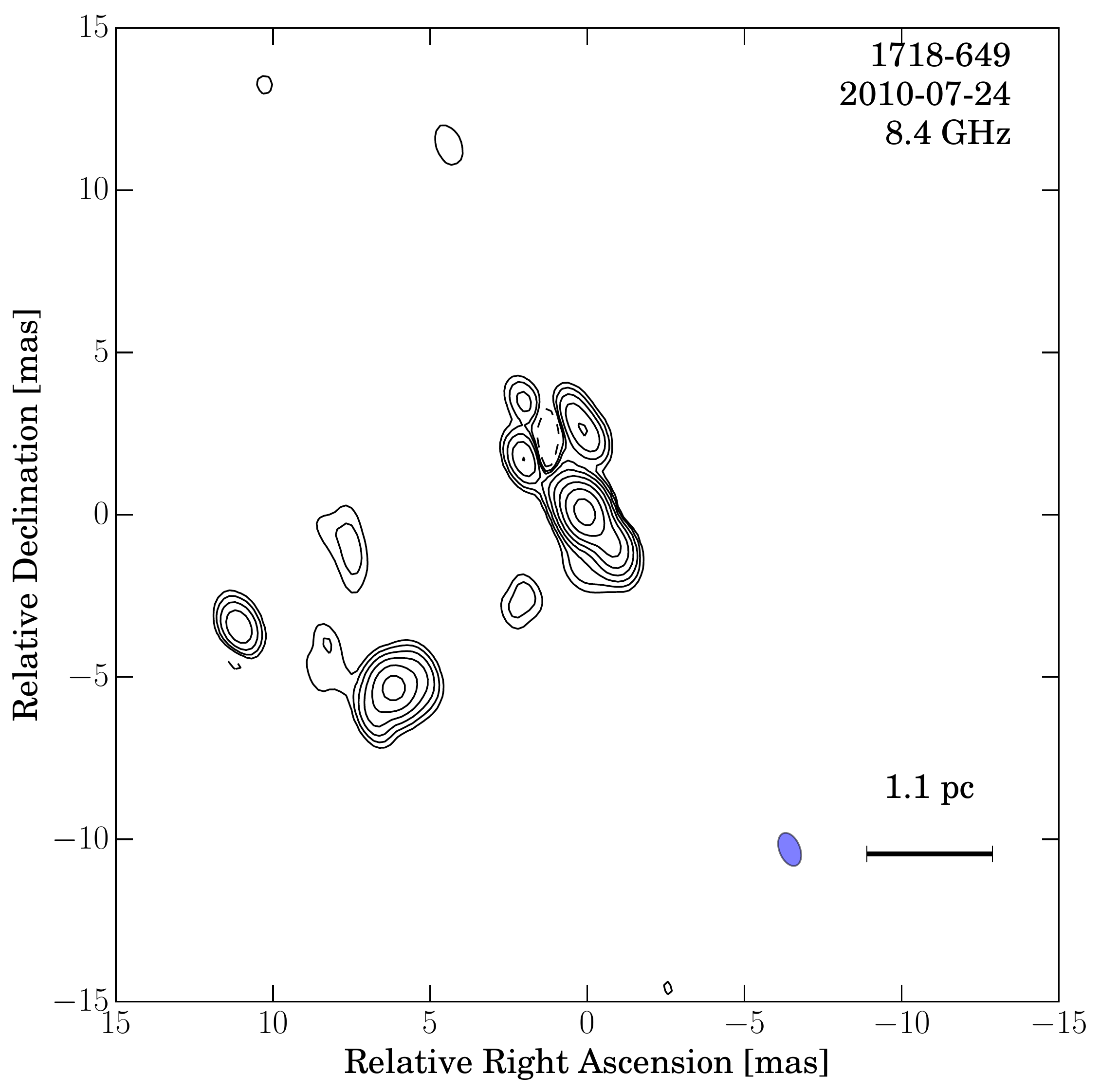}
\includegraphics[width=0.43\linewidth]{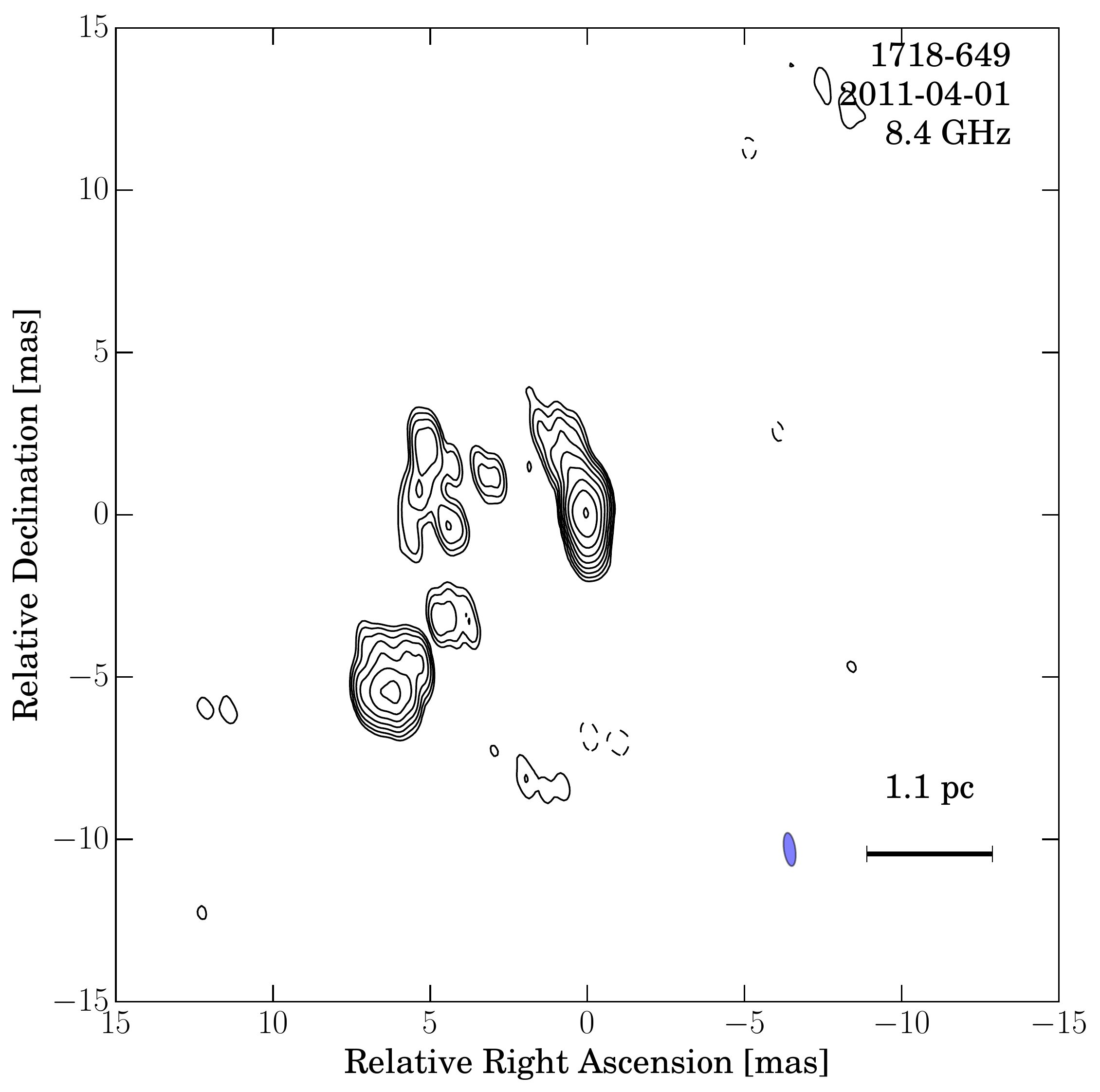}
\end{center}
\caption{Full-resolution images of PKS~1718$-$649. The map parameters for
  each epoch can be found in Table~\ref{1718_tab}. The blue ellipse
  represents the beam size, while the black line indicates the linear
  scale at the source's redshift. Contours increase in steps of two starting from 5.0, 5.0, 3.0, 10.0, 10.0, 5.0 times the noise level in each map, from top left to bottom right, respectively.}
\label{1718_full_a}
\end{figure*}
\renewcommand{\thefigure}{A.\arabic{figure} (Continued)}
\addtocounter{figure}{-1}
\begin{figure*}[!htbp]
\begin{center}
\includegraphics[width=0.43\linewidth]{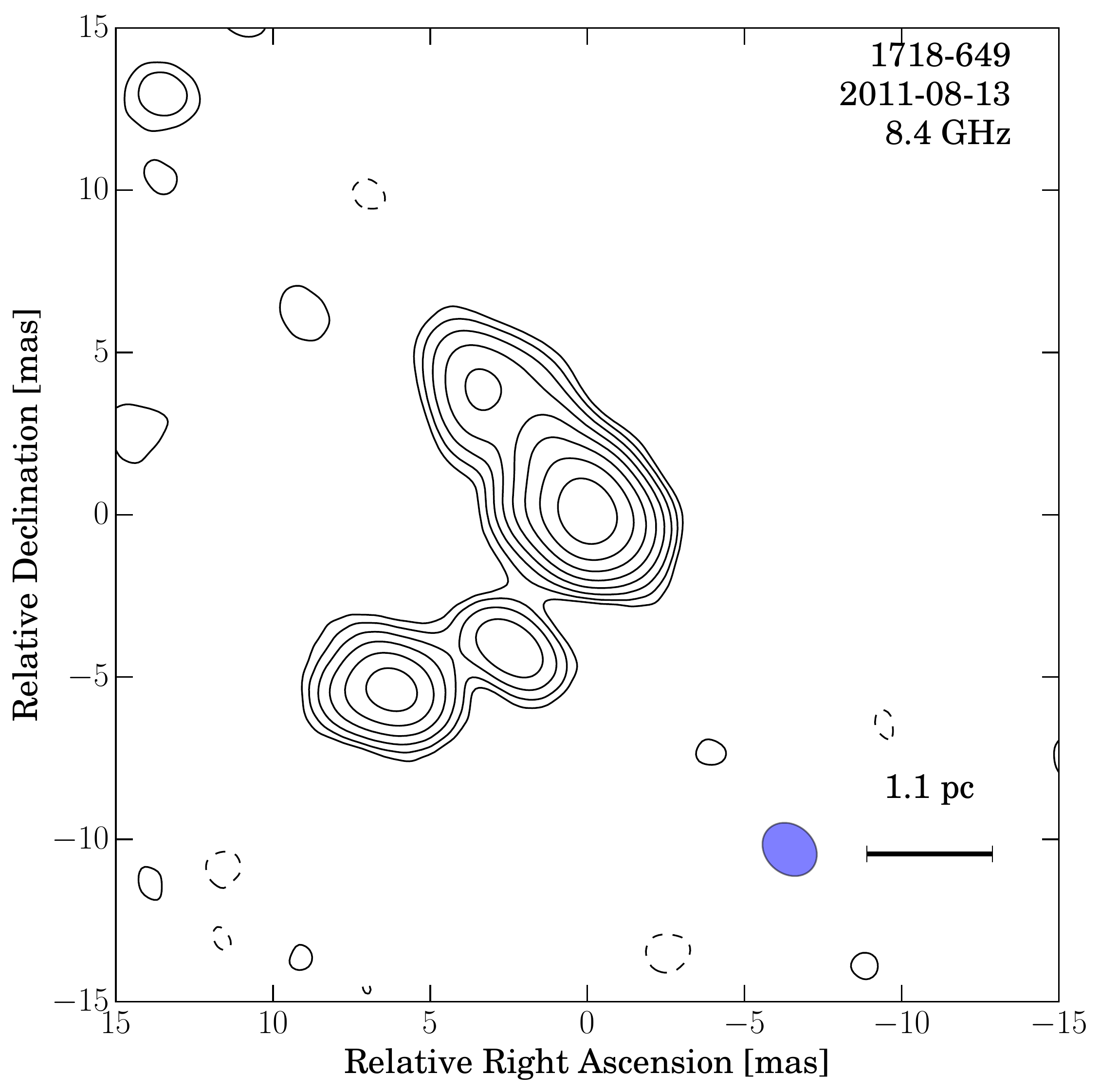}
\includegraphics[width=0.43\linewidth]{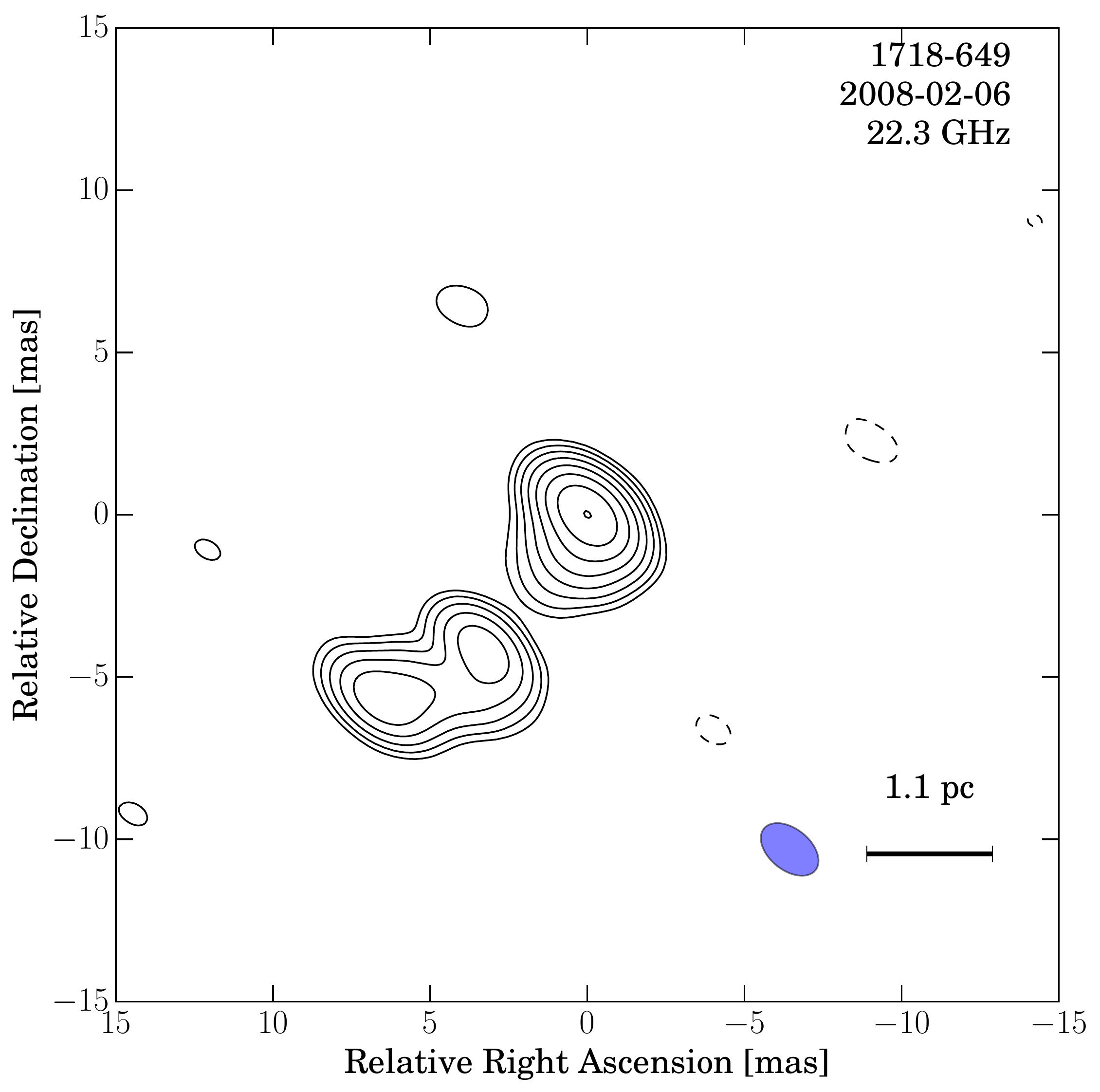}
\end{center}
\caption{Full-resolution images of PKS~1718$-$649 (continued). The map parameters for
  each epoch can be found in Table~\ref{1718_tab}. The blue ellipse
  represents the beam size, while the black line indicates the linear
  scale at the source's redshift. Contours increase in steps of two starting from 20.0 and 3.0 times the noise level.}
\label{1718_full_b}
\end{figure*}
\renewcommand{\thefigure}{A.\arabic{figure}}

\clearpage
\begin{table*}
\caption{Details of the 8.4\,GHz TANAMI observations of Pictor~A}             
\label{pica_tab}  
\begin{center}    
\begin{tabular}{c c c c c c c c }    
\hline\hline 
Obs.~date & Array configuration$^a$ & $S_\mathrm{total}$$^b$ & $S_\mathrm{peak}$$^b$ & RMS$^b$ & $b_\mathrm{maj}$$^c$ & $b_\mathrm{min}$$^c$ & P.A.$^c$ \\
(yyyy-mm-dd)  &                  & (Jy) & (Jy beam$^{-1}$) & (mJy beam$^{-1}$) & (mas) & (mas) & ($^\circ$) \\
\hline
2007-11-10 & AT-MP-HO-HH-CD-PKS        & 0.77 & 0.37 & 0.11 & 1.90 & 0.57 & 4.5\\
2008-06-09 & AT-MP-HO-HH-CD-PKS        & 0.80 & 0.48 & 0.24 & 2.62 & 0.71 & 5.4\\
2008-11-27 & TC-OH-AT-MP-HO-CD-PKS-DSS43 & 0.75 & 0.36 & 0.29 & 1.12 & 0.86 & 23.8\\
2010-07-24 & TC-AT-MP-HO-CD-PKS          & 0.81 & 0.49 & 0.30 & 3.71 & 1.28 & 25.2\\
2011-08-14 & YG-TC-AT-MP-HO-HH-CD-PKS-DSS43 & 0.86 & 0.58 & 0.31 & 3.13 & 2.25 & 68.1\\
\hline  
\end{tabular}
\end{center}
$^a$ See Table~\ref{array} for the antenna codes.\\
$^b$ Total flux density, peak brightness and RMS noise level in the
\texttt{CLEAN}-image. An error of 15\% is assumed (see Section~\ref{vlbi})\\
$^c$ Major and minor axes and position angle of restoring beam.\\
\end{table*}
\begin{table*}
\caption{Details of the 8.4\,GHz TANAMI observations of PKS~0521$-$36}             
\label{0521_tab}  
\begin{center}    
\begin{tabular}{c c c c c c c c }    
\hline\hline 
Obs.~date & Array configuration$^a$ & $S_\mathrm{total}$$^b$ & $S_\mathrm{peak}$$^b$ & RMS$^b$ & $b_\mathrm{maj}$$^c$ & $b_\mathrm{min}$$^c$ & P.A.$^c$ \\
(yyyy-mm-dd)  &                  & (Jy) & (Jy beam$^{-1}$) & (mJy beam$^{-1}$) & (mas) & (mas) & ($^\circ$) \\
\hline
2007-11-10 & AT-MP-HO-HH-CD-PKS  & 1.65 & 0.94 & 0.37 & 2.05 & 0.49 & 1.55\\
2008-03-28 & AT-MP-HO-HH-CD-PKS-DSS43  & 1.66 & 0.83 & 0.18 & 2.92 & 0.53 & $-$0.369\\
2008-08-08 & AT-MP-HO-HH-CD-PKS-DSS45  & 1.88 & 1.49& 0.30 & 3.24 & 1.43 & $-$2.59\\
2009-02-23 & AT-MP-HO-CD-PKS-TC-OH  & 1.59 & 0.89 & 0.36 & 2.31 & 0.52 & 21.6\\
2010-03-12 & AT-MP-HO-CD-PKS-DSS43  & 1.53 & 1.32 & 0.28 & 6.07 & 3.58 & 84.9\\
2010-07-24 & AT-MP-HO-CD-PKS-TC-DSS43   & 1.98 & 1.34 & 0.34 & 3.16 & 0.70 & 14.2\\
2011-04-01 & AT-MP-HO-HH-CD-PKS-DSS43-WW  & 3.07 & 1.83 & 1.14 & 2.44 & 0.50 & 0.97\\
2011-11-13 & AT-MP-HO-HH-CD-PKS-WW-DSS43-DSS45  & 2.77 & 1.55 & 0.51 & 2.61 & 0.39 & $-$1.03\\
2012-09-16 & AT-HO-HH-CD-PKS-DSS34-DSS45-KE-AK  & 2.16& 1.55 & 0.88 & 2.59 & 0.62 & 3.63\\
\hline  
\end{tabular}
\end{center}
$^a$ See Table~\ref{array} for the antenna codes.\\
$^b$ Total flux density, peak brightness and RMS noise level in the
\texttt{CLEAN}-image. An error of 15\% is assumed (see Section~\ref{vlbi}).\\
$^c$ Major and minor axes and position angle of restoring beam.\\
\end{table*}
\begin{table*}
\caption{Details of the 8.4\,GHz TANAMI observations of PKS~0625$-$35}             
\label{0625_tab}  
\begin{center}    
\begin{tabular}{c c c c c c c c }    
\hline\hline 
Obs.~date & Array configuration$^a$ & $S_\mathrm{total}$$^b$ & $S_\mathrm{peak}$$^b$ & RMS$^b$ & $b_\mathrm{maj}$$^c$ & $b_\mathrm{min}$$^c$ & P.A.$^c$ \\
(yyyy-mm-dd)  &                  & (Jy) & (Jy beam$^{-1}$) & (mJy beam$^{-1}$) & (mas) & (mas) & ($^\circ$) \\
\hline
2007-11-10 & AT-MP-HO-HH-CD-PKS & 0.35 & 0.30& 0.08 & 3.64 &
3.08 & 62.6\\
2008-02-07 & AT-MP-HO-CD-PKS & 0.46 & 0.37 & 0.07 & 4.41 &
3.32 & -68.4\\
2008-06-09 & AT-MP-HO-HH-CD-PKS & 0.35 & 0.31 & 0.07 & 3.68 &
2.77 & -85.0\\
2008-11-27 & TC-OH-AT-MP-HO-CD-PKS-DSS43 & 0.37 & 0.34 & 0.15 & 4.55 &
3.43 & 77.0\\
2009-12-14 & AT-MP-HO-CD-TC & 0.34 & 0.31 & 0.10 & 3.56 &
3.06 & -19.8\\
2010-07-24 & TC-AT-MP-HO-CD-PKS & 0.34 & 0.31 & 0.11 & 6.2 &
3.86 & 74.2\\
2011-04-01 & AT-MP-HO-HH-CD-PKS-DSS43-WW & 0.40 & 0.34 & 0.16 & 4.22 &
3.07 & 82.5\\
2011-11-13 & AT-MP-HO-HH-CD-PKS-WW-DSS43-DSS45 & 0.37 & 0.34 & 0.12 & 5.39 &
3.61 & 84.6\\
2012-09-16 & AT-HO-HH-CD-PKS-DSS34-DSS45-KE & 0.38 & 0.34 & 0.10 & 4.29 & 3.02 & 84.6\\
\hline  
\end{tabular}
\end{center}
$^a$ See Table~\ref{array} for the antenna codes.\\
$^b$ Total flux density, peak brightness and RMS noise level in the
\texttt{CLEAN}-image. An error of 15\% is assumed (see Section~\ref{vlbi}).\\
$^c$ Major and minor axes and position angle of restoring beam.\\
\end{table*}
\begin{table*}
\caption{Details of the 8.4\,GHz TANAMI observations of Centaurus~B.}             
\label{cenb_tab}  
\begin{center}    
\begin{tabular}{c c c c c c c c }    
\hline\hline 
Obs.~date & Array configuration$^a$ & $S_\mathrm{total}$$^b$ & $S_\mathrm{peak}$$^b$ & RMS$^b$ & $b_\mathrm{maj}$$^c$ & $b_\mathrm{min}$$^c$ & P.A.$^c$ \\
(yyyy-mm-dd)  &                  & (Jy) & (Jy beam$^{-1}$) & (mJy beam$^{-1}$) & (mas) & (mas) & ($^\circ$) \\
\hline
2011-07-22 & AT-MP-HO-HH-CD-PKS-DSS43-DSS34 & 2.35& 0.96 & 0.73 & 3.85 & 2.85 & 72.9\\
2012-04-27 & AT-MP-HO-CD-PKS & 2.09 & 0.77 & 0.46 & 3.73 & 2.85 & 80.3\\
\hline  
\end{tabular}
\end{center}
$^a$ See Table~\ref{array} for the antenna codes.\\
$^b$ Total flux density, peak brightness and RMS noise level in the
\texttt{CLEAN}-image. An error of 15\% is assumed (see Section~\ref{vlbi}).\\
$^c$ Major and minor axes and position angle of restoring beam.\\
\end{table*}
\begin{table*}
\caption{Details of the 8.4\,GHz TANAMI observations of PKS~1718$-$649}             
\label{1718_tab}  
\begin{center}    
\begin{tabular}{c c c c c c c c }    
\hline\hline 
Obs.~date & Array configuration$^a$ & $S_\mathrm{total}$$^b$ & $S_\mathrm{peak}$$^b$ & RMS$^b$ & $b_\mathrm{maj}$$^c$ & $b_\mathrm{min}$$^c$ & P.A.$^c$ \\
(yyyy-mm-dd)  &                  & (Jy) & (Jy beam$^{-1}$) & (mJy beam$^{-1}$) & (mas) & (mas) & ($^\circ$) \\
\hline
2008-02-07 & AT-MP-HO-CD-PKS & 3.9 & 1.13 & 1.12 & 1.94 & 0.45 & -10.2\\
2008-06-09 & AT-MP-HO-HH-CD-PKS & 3.26 & 1.64 & 1.80 & 3.08 & 1.23 & 6.77\\
2008-11-27 & TC-OH-AT-MP-HO-CD-PKS-DSS43 & 2.67 & 0.63 & 0.79 & 0.74 & 0.46 & -47.9\\
2009-12-13 & AT-MP-HO-CD-TC & 3.08 & 1.44 & 1.78 & 1.15 & 0.82 & 12.4\\
2010-07-24 & TC-AT-MP-HO-CD-PKS & 2.33 & 1.03 & 1.11 & 1.07 & 0.65 & 23.4\\
2011-04-01 & AT-MP-HO-HH-CD-PKS-DSS43-WW & 3.00 & 0.61 & 0.57 & 1.03 & 0.36 & 8.48\\
2011-08-14 & AT-MP-HO-CD-PKS-TC-DSS43-YG & 3.36 & 1.31 & 1.70 & 1.85 & 1.51 & 52.6\\
\hline  
\end{tabular}
\end{center}
$^a$ See Table~\ref{array} for the antenna codes.\\
$^b$ Total flux density, peak brightness and RMS noise level in the
\texttt{CLEAN}-image. An error of 15\% is assumed (see Section~\ref{vlbi}).\\
$^c$ Major and minor axes and position angle of restoring beam.\\
\end{table*}

\clearpage
\section{Extended kinematic analysis information}
\label{app:kin}
Here we include additional information illustrating the results of our kinematic analysis of the multi-epoch TANAMI data. Figures~\ref{kina_pica} through~\ref{kina_1718} show the multi-epoch images of TANAMI radio galaxies, with the corresponding component identification and tracking. Please note that, for ease of representation, the distance between the images at different epochs is not to scale in these figures. Moreover, the colored lines are not fits to the displayed component positions, but simple interpolations meant to clarify the identification and tracking of the different components.

Tables~\ref{pica_mods} through~\ref{1718_mods} list the flux density, radial distance, position angle and size for each circular Gaussian component identified in each source during the kinematic analysis. Note that all components have been shifted so that the core position is always at $(0,0)$ coordinates in all epochs. The position angle is given in the range $(\pi,-\pi)$, with the zero in the N-S direction (in image coordinates) and positive values in the counter-clockwise direction. Components that were not identified (blue crossed circles in Figures~\ref{kina_pica} through~\ref{kina_1718}) are not listed. Note that an apparent speed was fitted only for components detected in at least five epochs. 

\renewcommand{\thefigure}{B.\arabic{figure}}
\begin{figure*}[!!htbp]
\begin{center}
\includegraphics[width=0.6\linewidth]{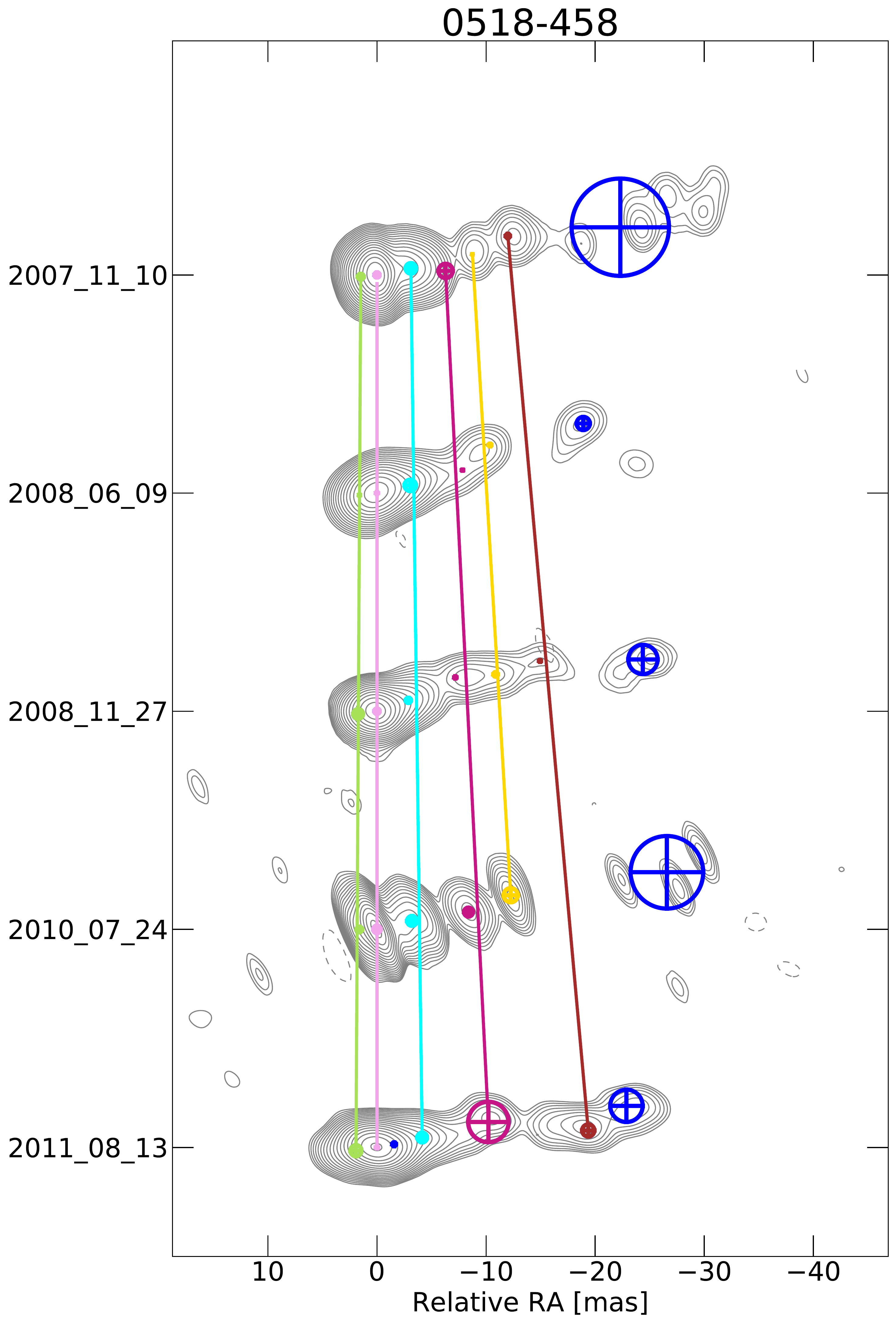}
\end{center}
\caption{Multi-epoch tapered images of Pictor A. The colored crossed circles represent the circular Gaussian components that have been fitted to the clean maps. The distance between the images at different epochs is not to scale. The colored lines are not fits to the displayed component positions, but simple interpolations meant to guide the eye.}
\label{kina_pica}
\end{figure*}
\begin{figure*}[!!htbp]
\begin{center}
\includegraphics[width=0.6\linewidth]{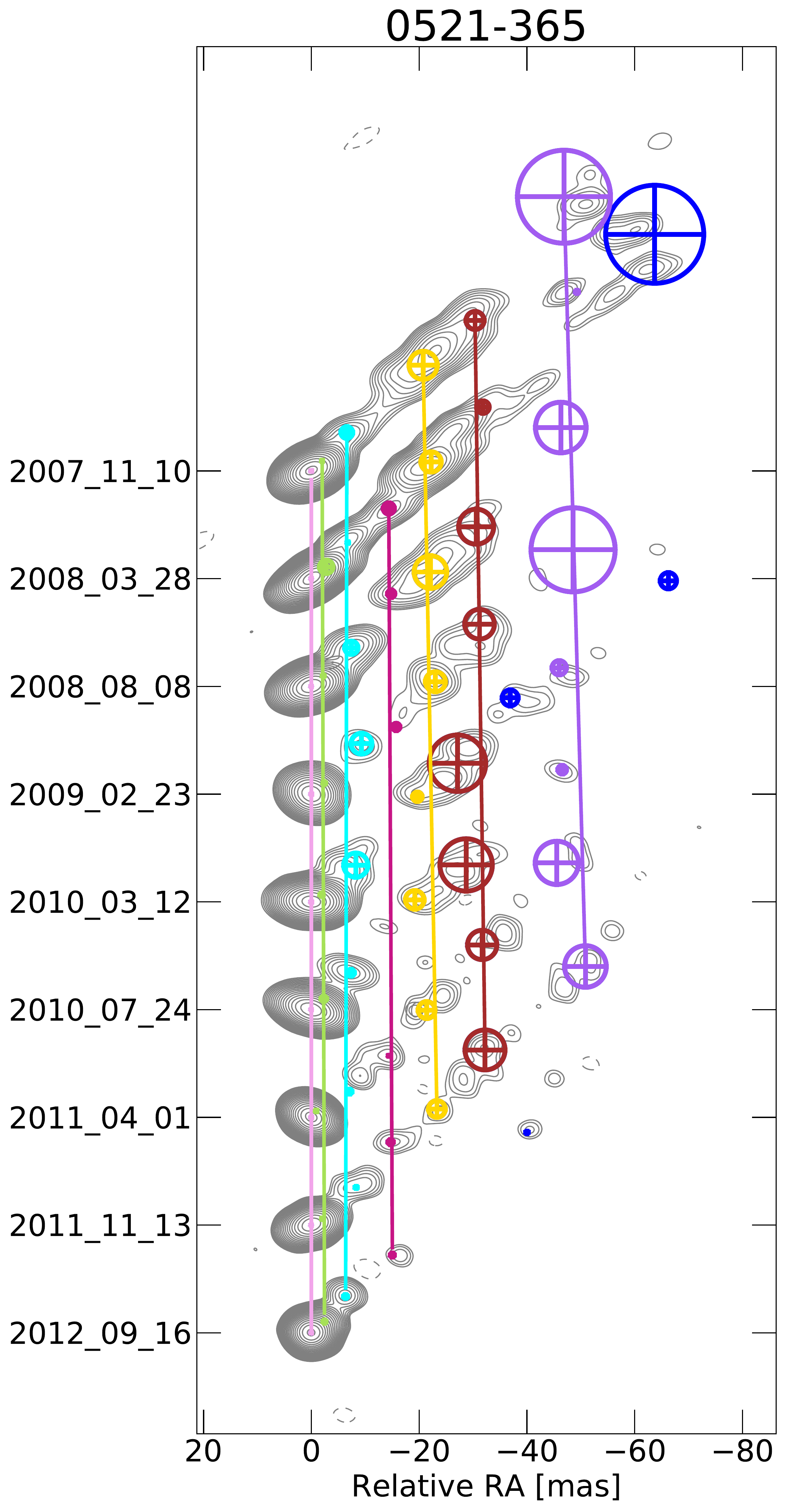}
\end{center}
\caption{Multi-epoch tapered images of PKS~0521$-$36. The colored crossed circles represent the circular Gaussian components that have been fitted to the clean maps. The distance between the images at different epochs is not to scale. The colored lines are not fits to the displayed component positions, but simple interpolations meant to guide the eye.}
\label{kina_0521}
\end{figure*}
\begin{figure*}[!!htbp]
\begin{center}
\includegraphics[width=0.4\linewidth]{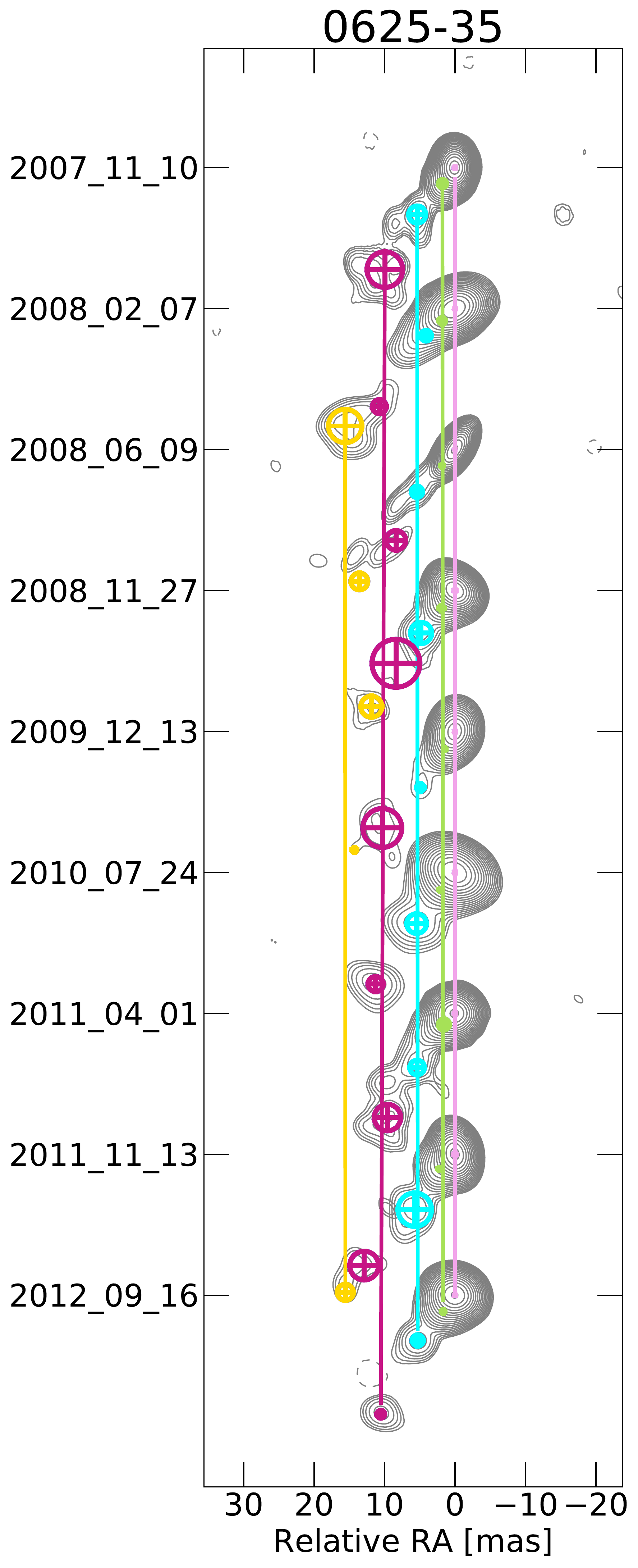}
\end{center}
\caption{Multi-epoch images of PKS~0625$-$35. The colored crossed circles represent the circular Gaussian components that have been fitted to the clean maps. The distance between the images at different epochs is not to scale. The colored lines are not fits to the displayed component positions, but simple interpolations meant to guide the eye.}
\label{kina_0625}
\end{figure*}
\begin{figure*}[!!htbp]
\begin{center}
\includegraphics[width=0.4\textwidth]{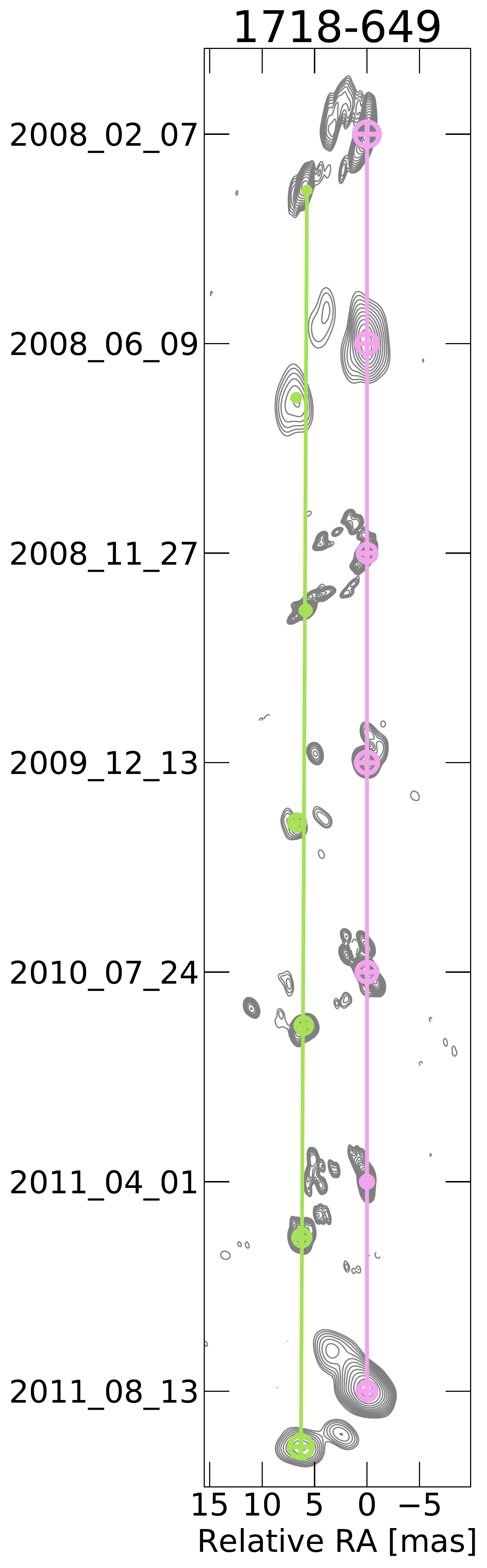}
\end{center}
\caption{Multi-epoch images of PKS~1718$-$649. The colored crossed circles represent the circular Gaussian components that have been fitted to the clean maps. The distance between the images at different epochs is not to scale. The colored lines are not fits to the displayed component positions, but simple interpolations meant to guide the eye.}
\label{kina_1718}
\end{figure*}

\begin{figure*}[!htbp]
\begin{center}
\includegraphics[width=0.495\linewidth]{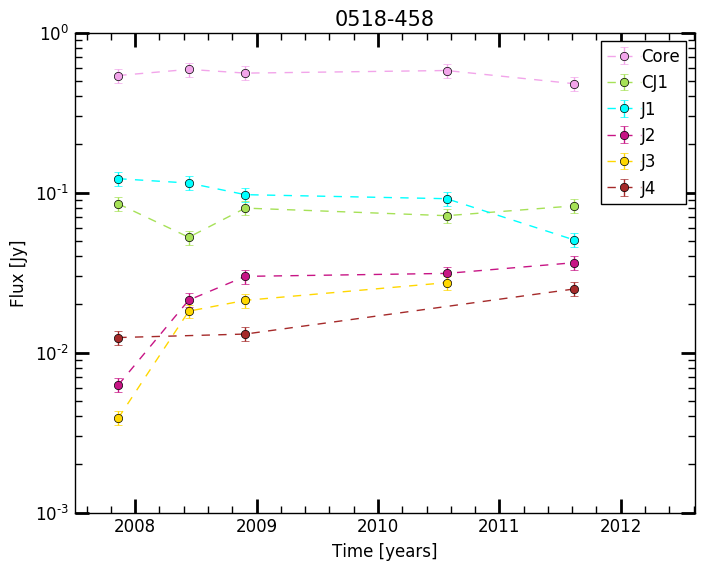}
\includegraphics[width=0.495\linewidth]{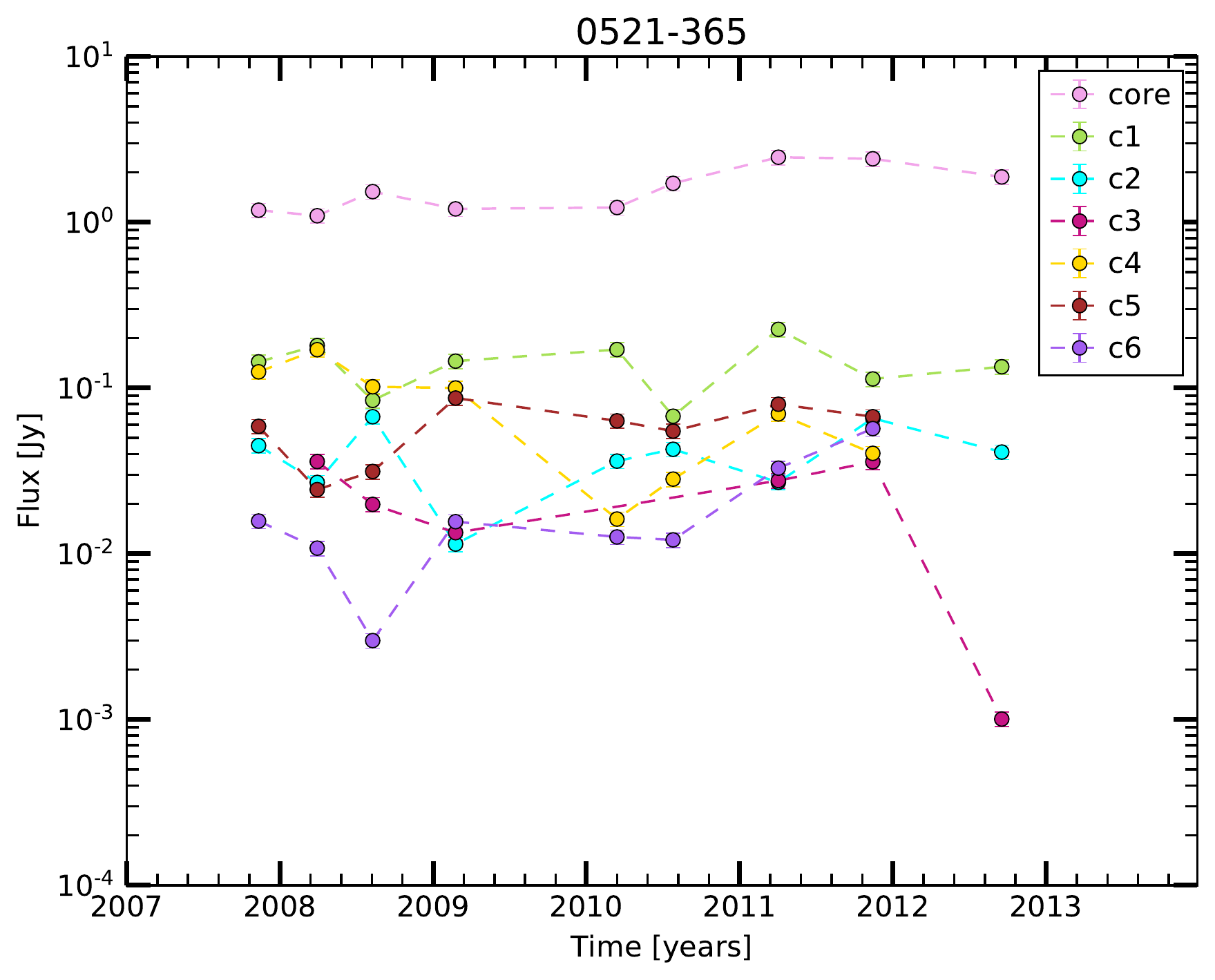}
\includegraphics[width=0.495\linewidth]{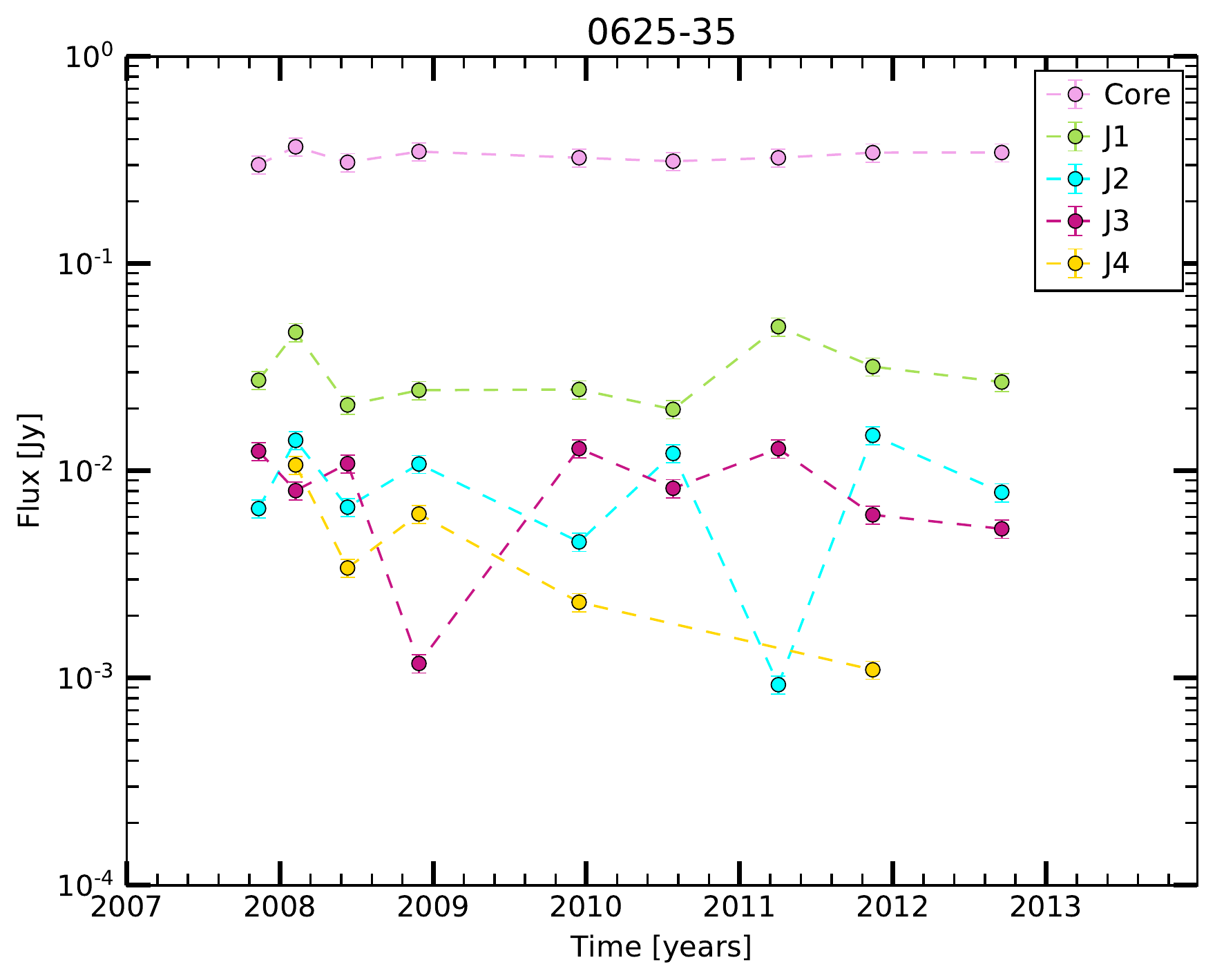}
\includegraphics[width=0.495\linewidth]{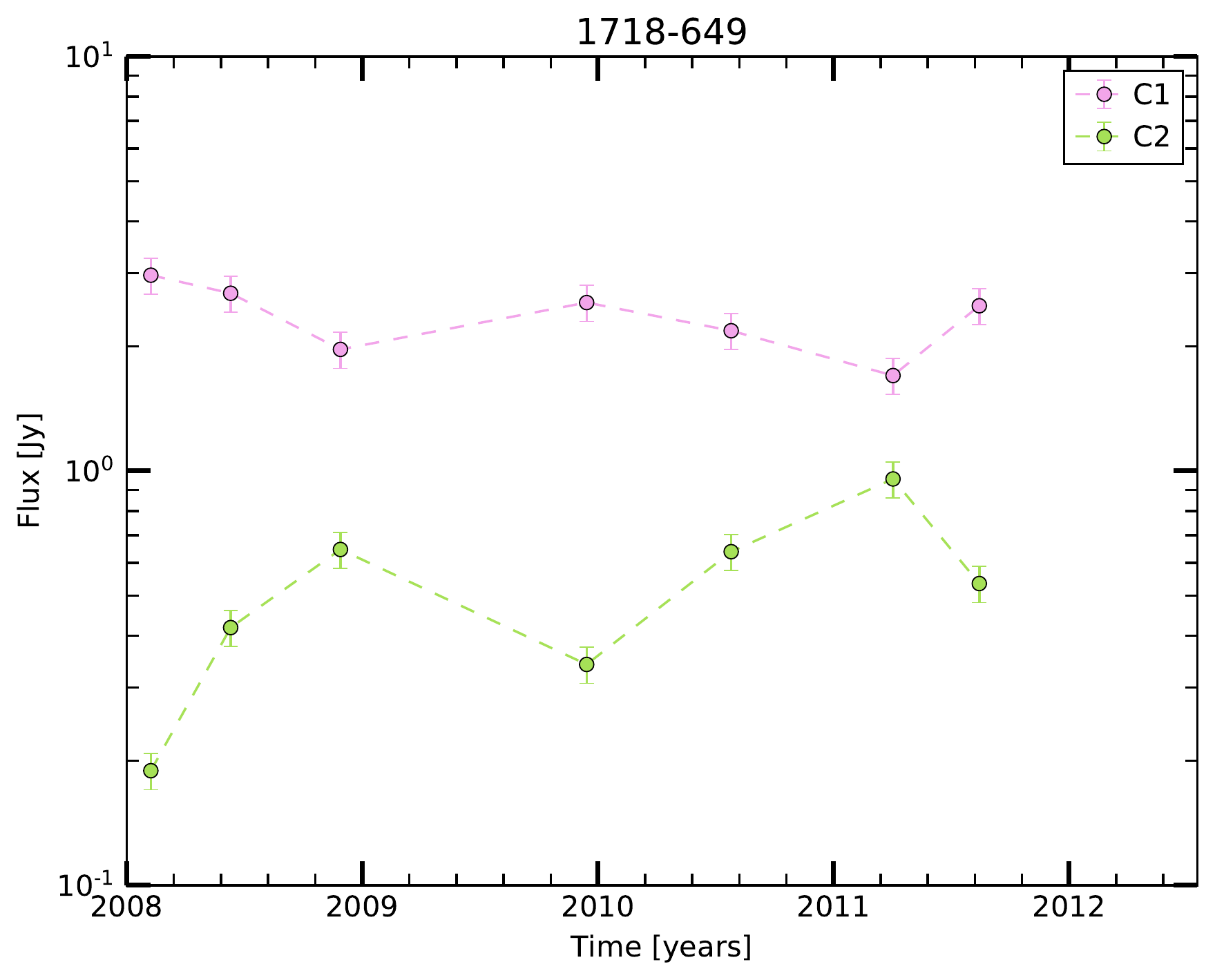}
\end{center}
\caption{Flux density evolution of the modelled jet features with time.
Top left to bottom right: Pictor~A, PKS~0521$-$36, PKS~0625$-$35, PKS~1718$-$649.}
\label{kin_flux}
\end{figure*}
\clearpage
\begin{table}
\caption{Difmap \texttt{Modelfit} parameters for the Gaussian components model of the TANAMI 8.4~GHz images of Pictor~A.}    
\label{pica_mods}  
\small
\begin{center}  
\begin{tabular}{llcccc}    
\hline\hline 
Epoch & ID & $S^a$ (Jy) & $d^b$ (mas) & $\phi^c$ (deg) & Size (mas)\\
\hline
2007-11-11 & Core & 0.538 & 0.00 & -56.87 & 0.24\\
 & CJ1 & 0.085 & 1.50 & 96.06 & 0.27\\
 & J1 & 0.122 & 3.17 & -78.93 & 0.48\\
 & J2 & 0.006 & 6.30 & -86.63 & 0.66\\
 & J3 & 0.004 & 8.95 & -77.84 & 0.03\\
 & J4 & 0.012 & 12.52 & -73.40 & 0.20\\
2008-06-10 & Core & 0.588 & 0.00 & 145.94 & 0.09\\
 & CJ1 & 0.052 & 1.61 & 96.84 & 0.05\\
 & J1 & 0.115 & 3.13 & -76.97 & 0.52\\
 & J2 & 0.021 & 8.12 & -74.94 & 0.06\\
 & J3 & 0.018 & 11.26 & -66.96 & 0.13\\
2008-11-28 & Core & 0.557 & 0.00 & 51.34 & 0.24\\
 & CJ1 & 0.080 & 1.75 & 98.34 & 0.45\\
 & J1 & 0.097 & 3.04 & -70.88 & 0.24\\
 & J2 & 0.030 & 7.83 & -66.70 & 0.11\\
 & J3 & 0.021 & 11.39 & -72.74 & 0.21\\
 & J4 & 0.013 & 15.64 & -72.87 & 0.09\\
2010-07-26 & Core & 0.579 & 0.00 & -64.70 & 0.34\\
 & CJ1 & 0.072 & 1.62 & 90.37 & 0.26\\
 & J1 & 0.091 & 3.30 & -76.47 & 0.45\\
 & J2 & 0.031 & 8.54 & -79.30 & 0.41\\
 & J3 & 0.027 & 12.65 & -75.67 & 0.65\\
2011-08-14 & Core & 0.478 & 0.00 & -16.23 & 0.10\\
 & CJ1 & 0.083 & 1.95 & 98.73 & 0.51\\
 & J1 & 0.051 & 4.24 & -77.55 & 0.45\\
 & J2 & 0.036 & 10.48 & -77.11 & 1.85\\
 & J4 & 0.025 & 19.43 & -85.40 & 0.56\\
\hline  
\end{tabular}
\end{center}
$^a$ Flux density.\\
$^b$ Radial distance from the core.\\
$^c$ Position angle.\\
\end{table}

\begin{table}
\caption{Difmap \texttt{Modelfit} parameters for the Gaussian components model of the TANAMI 8.4~GHz images of PKS~0521$-$36.}    
\label{0521_mods}  
\small

\begin{center}  
\begin{tabular}{llcccc}    
\hline\hline 
Epoch & ID & $S^a$ (Jy) & $d^b$ (mas) & $\phi^c$ (deg) & Size (mas)\\
\hline
2007-11-11 & Core & 1.184 & 0.00 & 158.21 & 0.08\\
 & J1 & 0.144 & 2.73 & -47.11 & 0.11\\
 & J2 & 0.045 & 9.70 & -42.68 & 1.09\\
 & J4 & 0.125 & 28.55 & -46.62 & 2.62\\
 & J5 & 0.059 & 41.27 & -47.40 & 1.68\\
 & J6 & 0.016 & 69.20 & -42.65 & 8.64\\
2008-03-30 & Core & 1.096 & 0.00 & -53.33 & 0.06\\
 & J1 & 0.181 & 3.52 & -51.41 & 1.27\\
 & J2 & 0.027 & 9.53 & -45.37 & 0.15\\
 & J3 & 0.036 & 19.40 & -47.81 & 1.04\\
 & J4 & 0.171 & 31.03 & -45.71 & 1.89\\
 & J5 & 0.024 & 45.04 & -45.01 & 1.16\\
 & J6 & 0.011 & 72.54 & -42.80 & 0.30\\
2008-08-09 & Core & 1.532 & 0.00 & 44.63 & 0.06\\
 & J1 & 0.084 & 2.93 & -48.61 & 0.19\\
 & J2 & 0.067 & 10.32 & -45.93 & 1.24\\
 & J3 & 0.020 & 22.71 & -40.73 & 0.69\\
 & J4 & 0.102 & 30.64 & -46.14 & 3.04\\
 & J5 & 0.031 & 42.60 & -45.89 & 3.25\\
 & J6 & 0.003 & 66.73 & -43.93 & 4.70\\
2009-02-23 & Core & 1.206 & 0.00 & -169.52 & 0.06\\
 & J1 & 0.145 & 3.15 & -49.35 & 0.26\\
 & J2 & 0.011 & 13.21 & -44.76 & 1.94\\
 & J3 & 0.013 & 20.09 & -51.59 & 0.69\\
 & J4 & 0.100 & 31.07 & -47.81 & 1.86\\
 & J5 & 0.087 & 44.38 & -44.70 & 2.75\\
 & J6 & 0.016 & 66.46 & -46.96 & 7.81\\
2010-03-14 & Core & 1.228 & 0.00 & -109.34 & 0.08\\
 & J1 & 0.171 & 2.31 & -55.09 & 0.31\\
 & J2 & 0.036 & 10.70 & -50.38 & 2.25\\
 & J4 & 0.016 & 27.71 & -45.24 & 0.89\\
 & J5 & 0.064 & 37.39 & -46.50 & 5.25\\
 & J6 & 0.013 & 63.28 & -46.62 & 1.32\\
2010-07-26 & Core & 1.718 & 0.00 & 35.86 & 0.04\\
 & J1 & 0.068 & 3.06 & -48.92 & 0.54\\
 & J2 & 0.043 & 10.03 & -47.08 & 0.63\\
 & J4 & 0.028 & 28.00 & -43.29 & 1.73\\
 & J5 & 0.055 & 39.33 & -46.91 & 4.86\\
 & J6 & 0.012 & 64.43 & -46.27 & 0.80\\
2011-04-03 & Core & 2.470 & 0.00 & -60.52 & 0.06\\
 & J1 & 0.226 & 1.51 & -37.52 & 0.15\\
 & J2 & 0.027 & 8.56 & -55.91 & 0.45\\
 & J3 & 0.028 & 18.39 & -51.54 & 0.12\\
 & J4 & 0.070 & 29.23 & -47.04 & 1.59\\
 & J5 & 0.080 & 45.05 & -44.73 & 2.72\\
 & J6 & 0.033 & 65.63 & -43.96 & 4.02\\
2011-11-14 & Core & 2.418 & 0.00 & -83.00 & 0.06\\
 & J1 & 0.114 & 2.38 & -60.16 & 0.16\\
 & J2 & 0.066 & 10.88 & -49.97 & 0.21\\
 & J3 & 0.036 & 21.31 & -43.67 & 0.48\\
 & J4 & 0.040 & 31.74 & -47.21 & 1.57\\
 & J5 & 0.067 & 45.78 & -44.74 & 3.72\\
 & J6 & 0.057 & 69.94 & -46.66 & 3.86\\
2012-09-17 & Core & 1.879 & 0.00 & -102.43 & 0.05\\
 & J1 & 0.135 & 3.21 & -49.55 & 0.31\\
 & J2 & 0.041 & 9.27 & -43.29 & 0.38\\
 & J3 & 0.001 & 20.85 & -46.15 & 0.38\\
\hline  
\end{tabular}
\end{center}
$^a$ Flux density.\\
$^b$ Radial distance from the core.\\
$^c$ Position angle.\\
\end{table}

\begin{table}
\caption{Difmap \texttt{Modelfit} parameters for the Gaussian components model of the TANAMI 8.4~GHz images of PKS~0625$-$35.}    
\label{0625_mods}
\small

\begin{center}  
\begin{tabular}{llcccc}    
\hline\hline 
Epoch & ID & $S^a$ (Jy) & $d^b$ (mas) & $\phi^c$ (deg) & Size (mas)\\
\hline
2007-11-11 & Core & 0.301 & 0.00 & 124.15 & 0.12\\
 & J1 & 0.027 & 2.87 & 141.47 & 0.61\\
 & J2 & 0.007 & 8.54 & 140.98 & 1.22\\
 & J3 & 0.013 & 17.55 & 145.39 & 2.53\\
2008-02-07 & Core & 0.367 & 0.00 & -97.56 & 0.04\\
 & J1 & 0.047 & 2.50 & 133.55 & 0.50\\
 & J2 & 0.014 & 5.62 & 132.94 & 0.77\\
 & J3 & 0.008 & 17.56 & 142.27 & 1.00\\
 & J4 & 0.011 & 22.81 & 136.83 & 2.38\\
2008-06-10 & Core & 0.309 & 0.00 & -65.20 & 0.02\\
 & J1 & 0.021 & 2.93 & 140.90 & 0.23\\
 & J2 & 0.007 & 8.06 & 137.90 & 0.81\\
 & J3 & 0.011 & 15.36 & 146.91 & 1.39\\
 & J4 & 0.003 & 23.11 & 144.02 & 1.20\\
2008-11-28 & Core & 0.348 & 0.00 & 112.20 & 0.15\\
 & J1 & 0.025 & 3.14 & 142.40 & 0.39\\
 & J2 & 0.011 & 7.71 & 141.29 & 1.51\\
 & J3 & 0.001 & 13.29 & 140.87 & 3.40\\
 & J4 & 0.006 & 20.31 & 144.18 & 1.51\\
2009-12-14 & Core & 0.325 & 0.00 & -104.25 & 0.07\\
 & J1 & 0.025 & 2.87 & 149.51 & 0.19\\
 & J2 & 0.004 & 9.34 & 148.14 & 0.54\\
 & J3 & 0.013 & 17.14 & 142.98 & 2.76\\
 & J4 & 0.002 & 22.05 & 139.62 & 0.34\\
2010-07-26 & Core & 0.312 & 0.00 & 15.51 & 0.09\\
 & J1 & 0.020 & 3.22 & 139.97 & 0.22\\
 & J2 & 0.012 & 9.07 & 143.04 & 1.40\\
 & J3 & 0.008 & 19.44 & 144.56 & 1.09\\
2011-04-03 & Core & 0.325 & 0.00 & 126.16 & 0.09\\
 & J1 & 0.050 & 2.23 & 134.81 & 0.80\\
 & J2 & 0.001 & 9.37 & 144.95 & 1.04\\
 & J3 & 0.013 & 17.62 & 147.02 & 1.88\\
2011-11-14 & Core & 0.344 & 0.00 & -58.08 & 0.22\\
 & J1 & 0.032 & 3.02 & 132.83 & 0.23\\
 & J2 & 0.015 & 9.71 & 143.93 & 2.38\\
 & J3 & 0.006 & 20.39 & 140.74 & 2.05\\
 & J4 & 0.001 & 25.02 & 141.50 & 1.15\\
2012-09-17 & Core & 0.344 & 0.00 & -34.11 & 0.06\\
 & J1 & 0.027 & 2.85 & 143.47 & 0.27\\
 & J2 & 0.008 & 8.31 & 140.43 & 0.80\\
 & J3 & 0.005 & 19.90 & 148.02 & 0.56\\
\hline  
\end{tabular}
\end{center}
$^a$ Flux density.\\
$^b$ Radial distance from the core.\\
$^c$ Position angle.\\
\end{table}
\begin{table}
\caption{Difmap \texttt{Modelfit} parameters for the Gaussian components model of the TANAMI 8.4~GHz images of PKS~1718$-$649.}    
\label{1718_mods}
\small

\begin{center}  
\begin{tabular}{llcccc}    
\hline\hline 
Epoch & ID & $S^a$ (Jy) & $d^b$ (mas) & $\phi^c$ (deg) & Size (mas)\\
\hline
2008-02-07 & C1 & 2.967 & 0.00 & -25.16 & 1.17\\
 & C2 & 0.189 & 7.86 & 133.24 & 0.29\\
2008-06-10 & C1 & 2.684 & 0.00 & -36.00 & 0.99\\
 & C2 & 0.419 & 8.50 & 127.48 & 0.31\\
2008-11-28 & C1 & 1.965 & 0.00 & -77.07 & 0.85\\
 & C2 & 0.646 & 8.00 & 133.06 & 0.44\\
2009-12-14 & C1 & 2.549 & 0.00 & -150.40 & 1.01\\
 & C2 & 0.341 & 8.79 & 130.54 & 0.73\\
2010-07-26 & C1 & 2.180 & 0.00 & 172.80 & 0.95\\
 & C2 & 0.638 & 7.89 & 130.27 & 0.75\\
2011-04-03 & C1 & 1.699 & 0.00 & -26.43 & 0.54\\
 & C2 & 0.957 & 8.20 & 130.70 & 0.77\\
2011-08-14 & C1 & 2.504 & 0.00 & 35.25 & 0.85\\
 & C2 & 0.535 & 8.23 & 130.07 & 1.08\\
\hline  
\end{tabular}
\end{center}
$^a$ Flux density.\\
$^b$ Radial distance from the core.\\
$^c$ Position angle.\\
\end{table}

\end{appendix}

\end{document}